\documentclass[10pt]{article}

\usepackage[top=1.18in,left=1in,right=1in,bottom=1in]{geometry}

\usepackage[utf8]{inputenc}
\usepackage{csquotes}
\usepackage[english]{babel}
\usepackage[backend=biber,defernumbers=true,sortcites=true,sorting=none,maxbibnames=5,maxcitenames=1,doi=true,style=nature]{biblatex}
\usepackage{xcolor}
\usepackage{url}
\usepackage[toc,title]{appendix}
\usepackage[hypertexnames=true]{hyperref}
\hypersetup{
    colorlinks=true,
    citecolor=blue,
    linkcolor=blue,
    filecolor=magenta,      
    urlcolor=blue,
}
\usepackage{nameref}
\usepackage{graphicx}
\usepackage{amsmath}
\usepackage{amssymb}
\usepackage[T1]{fontenc}
\usepackage{multicol}
\usepackage{multirow}
\usepackage{dcolumn}
\usepackage[font=footnotesize,labelfont=bf]{caption}
\usepackage[capitalise,nameinlink]{cleveref}
\usepackage{booktabs}
\usepackage{setspace}
\usepackage{rotating}
\usepackage{enumitem}
\setitemize{noitemsep,topsep=2pt,parsep=2pt,partopsep=0pt}
\usepackage[left,modulo]{lineno}
\usepackage[section]{placeins}
\usepackage{listings}
\usepackage{fancyvrb}
\fvset{fontsize=\scriptsize,frame=single,framesep=2mm,boxwidth=3in}

\usepackage{fancyhdr}
\fancypagestyle{firstpage}{%
    \fancyhf{}
    \fancyfoot[L]{Brattig Correia, R., de Araújo Kohler, L., Mattos, M.M. \textit{et al}. City-wide electronic health records reveal gender and age biases in administration of known drug–drug interactions. \textit{npj Digit. Med.} \textbf{2}, 74 (2019) doi:10.1038/s41746-019-0141-x}

}
\usepackage{etoc}
\etocsetnexttocdepth{subsection}

\definecolor{darkred}{rgb}{178,0,0}

\usepackage{titlesec}

\titleformat{\section}
  {\Large\bfseries\raggedright} % format
  {\thesection} % label
  {14pt} % sep
  {\Large}{} %beforesep %aftersep
\titlespacing*{\section}
    {0pt} %left
    {4ex} % beforesep
    {4ex} % aftersep
\bibliography{ddi-blumenau}

\begin{document}

\title{
    City-wide Electronic Health Records Reveal Gender and Age Biases in Administration of Known Drug-Drug Interactions
}
\author{
    \small
    \textbf{Rion Brattig Correia$^{1,2,3,*}$, Luciana P. de Araújo Kohler$^{4}$, Mauro M. Mattos$^{4}$, Luis M. Rocha$^{1,3,*}$} \\
    \small
    $^1$School of Informatics, Computing \& Engineering, Indiana University, Bloomington, IN 47408 USA \\
    \small
    $^2$CAPES Foundation, Ministry of Education of Brazil, Brasília, DF 70040-020, Brazil \\
    \small
    $^3$Instituto Gulbenkian de Ciência, Oeiras 2780-156, Portugal \\
    \small
    $^4$Universidade Regional de Blumenau (FURB), Blumenau, SC 89030-903, Brazil \\
    \small
    $^*$ correspondence to rocha@indiana.edu and rionbr@gmail.com
}
\date{}
\maketitle
\thispagestyle{firstpage}
\doublespacing

%
% Short Abstract (Max 250 words)
%
\begin{abstract}
\it
The occurrence of drug-drug-interactions (DDI) from multiple drug dispensations is a serious problem, both for individuals and health-care systems, since patients with complications due to DDI are likely to reenter the system at a costlier level. 
We present a large-scale longitudinal study (18 months) of the DDI phenomenon at the primary- and secondary-care level using electronic health records (EHR) from the city of Blumenau in Southern Brazil  (pop. $\approx 340,000$).
We found that 181 distinct drug pairs known to interact were dispensed concomitantly to 12\% of the patients in the city’s public health-care system. Further, 4\% of the patients were dispensed drug pairs that are likely to result in major adverse drug reactions (ADR)---with costs estimated to be much larger than previously reported in smaller studies. 
The large-scale analysis reveals that women have a 60\% increased risk of DDI as compared to men; the increase becomes 90\% when considering only DDI known to lead to major ADR. 
Furthermore, DDI risk increases substantially with age; patients aged 70-79 years have a 34\% risk of DDI when they are dispensed two or more drugs concomitantly. 
Interestingly, a statistical null model demonstrates that age- and female-specific risks from increased polypharmacy fail by far to explain the observed DDI risks in those populations, suggesting unknown social or biological causes. 
We also provide a network visualization of drugs and demographic factors that characterize the DDI phenomenon and demonstrate that accurate DDI prediction can be included in healthcare and public-health management, to reduce DDI-related ADR and costs.
\end{abstract}

\noindent \textbf{keywords}: drug-drug interactions, public-health, electronic health records, digital medicine, computational intelligence, network science, machine learning.

%
% Introduction
%
%\newpage
\pagestyle{plain}
\section{Introduction}
\label{ch:introduction}

Adverse drug reactions (ADR) from drug-drug interactions (DDI) is a well-known public health problem worldwide \cite{Davies:2004, Hajjar:2007, Becker:2007}. Most efforts to measure the scale of ADR from DDI focus on hospitalizations and emergency visits \cite{Pfaffenbach:2002, Camargo:2006, Rozenfeld:2007, Moura:2009, Silva:2010, Wu:2012, Okuno:2013} or literature meta-analysis \cite{Becker:2007, Cano:2009, Hakkarainen:2012}. Very few studies so far have been able to characterize this problem in primary and secondary care settings. Lack of access to longitudinal data from Electronic Health Records (EHR) of large populations continues to be the main barrier to measuring the prevalence of DDI and characterizing the phenomenon in medical care \cite{Grimson:2010, Percha:2013, Jensen:2012}.
For instance, Molden \textit{et al} \cite{Molden:2005} searched 43,500 patients in pharmacy databases in southeastern Norway, studying only DDI from CYP inhibitor-substrate drugs.
Pinto \textit{et al} \cite{Pinto:2014} studied DDI prevalence in a small cohort of forty elderly hypertensive patients in a primary health care unit in Brazil.
Iyer \textit{et al} \cite{Iyer:2014} mined 50 million clinical notes from the private EHR database STRIDE \cite{STRIDE}, to identify signals of unknown potential DDI from clinical text. While STRIDE contains EHR from multiple care levels, this analysis did not address the concomitant dispensation of pairs of drugs with \textit{known} DDI in primary- and secondary-care.
Lastly, Guthrie \textit{et al} \cite{Guthrie:2015} performed a repeated cross-sectional comparison of 84 days in 1995 and 2010, to study the increase in polypharmacy and DDI at the primary- and secondary-care level in the Tayside region of Scotland (pop. 405,721); DDI was defined according to the \textit{British National Formulary}, a private publication.
This study estimated that 13\% of adults ($\geq20$ y.o.) were prescribed a ``potentially serious'' known DDI in 2010, and that the number of drugs prescribed was the characteristic most predictive of DDI. Patients prescribed 15 or more drugs had an almost 27 fold DDI risk increase over those prescribed two to four drugs.
However, by using only 84-day windows, this analysis misses potential co-administrations from separate prescriptions made outside of the relatively short windows; it also analyzed prescription, rather than dispensation data.

Here we pursue a large-scale longitudinal study of the DDI phenomenon at the primary- and secondary-care levels in an entire city, using considerably larger time-windows and relying on public DDI and ADR standards.
We obtained 18 months of EHR data for the city of Blumenau in Southern Brazil (pop. 338,876), a city with a very high Human Development Index ($\text{HDI}{=}0.806$ \cite{AtlasBrasil})---at the level of the top quartile of countries according to this United Nations Development Programme index \cite{UN:HumanDev}.
Brazil has a universal public health-care system, and Blumenau possesses a city-wide Health Information System (HIS) with prescription and dispensation information for its entire population.
The analysis of Blumenau's EHR data is thus an opportunity to understand the DDI phenomenon in a highly developed city in a country where DDI is known to occur similarly to other nations \cite{Cano:2009, Okuno:2013}.
The study provides an understanding of both prevalence and bias in the dispensation of known DDI outside of hospital settings.
Dispensation data is only a surrogate for administration of DDI, as we are not certain that patients actually take the medications that are dispensed concomitantly. However, dispensation data can only be a better surrogate of administration than prescription data that was used in previous studies (e.g.\cite{Guthrie:2015}), as a prescription may ultimately not be dispensed.

From a public-health perspective, the concomitant administration of drugs with adverse interactions is of great concern \cite{Cano:2009, Camargo:2006, Okuno:2013}.
Since over 30\% of all ADR are thought to be caused by DDI \cite{Iyer:2014}, better identification and prediction of administration of known DDI in primary- and secondary-care could reduce the number of patients seeking urgent care in hospitals, resulting in substantial savings for health systems worldwide \cite{Becker:2007, Moura:2009, Percha:2013}. 
A systematic review from 2009 showed that the proportion of hospital inpatients with ADR (in general, not DDI only) ranged from 1.6 to 41.4\% \cite{Cano:2009}. Furthermore, an estimated 52\% (45\%) of ADR in outpatients (inpatients) were preventable \cite{Hakkarainen:2012}. 
In the elderly population alone ($>65$ y.o.), the yearly financial burden of ADR was estimated to reach \$11.9 million for the province of Ontario (pop. 12M) \cite{Wu:2012}, or about \$1 per capita, per year.
As we report below, the yearly cost of major DDI estimated from the Blumenau EHR dispensation data for the same age group is higher, at least \$2 per capita, per year, after adjusting for inflation and exchange rates---though for less stringent assumptions it can be as high as \$7 per capita, per year.
This suggests that the financial burden of DDI is more severe than previously thought.
Moreover, the rate of major DDI found to be dispensed in Blumenau is smaller than what was reported to be prescribed in Scotland \cite{Guthrie:2015}. Therefore the financial burden of DDI is likely higher in other health-care systems, especially those with older populations.

To characterize the significant factors in DDI, we study demographic variables such as gender and age, as well as the drugs involved in DDI in greater detail, and reveal previously unknown factors in this phenomenon.
We show that women in Blumenau are at a greater risk of being dispensed known DDI than men, with a 1.6 risk multiplier. 
This increased risk for females is not confounded by the larger number of women present in the data nor their age. The analysis also identifies the drug pairs that most lead to DDI in women which, surprisingly, are not attributable to female-specific medicines (e.g. hormone therapy).
We also demonstrate that there is a significant increase of DDI risk with age, reaching more than 30\% for adults over 65 years of age.
Importantly, using a statistical null model, we show that the age risk growth is not explained simply by the increase in polypharmacy in older age. This suggests that the specific drugs dispensed to older populations are more prone to DDI and/or that insufficient attention is paid to this phenomenon in primary care for this population.

While the number of drugs dispensed and the number of concomitant drug dispensations are the best predictors of DDI (previously only observed for number of drugs prescribed \cite{Guthrie:2015}), we show that these quantities by themselves are poor predictors of DDI. We look at demographic variables such as education and neighborhood affluence and show they do not play a significant role in the risk for DDI in our data. Other factors, however, play very significant roles, chiefly age, gender, and the specific drugs dispensed.
Indeed, we demonstrate that the automatic prediction of which patients are dispensed known DDI is quite accurate when those factors are included. This makes decision-support systems for predicting DDI risk in HIS not only feasible, but necessary to lower the rates of known DDI being dispensed.

To better understand which drugs are most involved in the DDI phenomenon, we integrate all DDI information of the Blumenau population into easy-to-visualize DDI networks. Looking at gender differences, for example, analysis of these networks identifies key drugs and interactions in the DDI phenomenon, and demonstrates that the higher DDI risk women face is not associated with any type of hormone therapy. 
Indeed, drugs that most contribute to the gender-disparity in DDI risk are not female-specific. This suggests there may be social or biological processes at play in primary- and secondary-care that lead to increased DDI risk for women. 
A full listing of the drugs that most contribute to the DDI observed in our study are presented in our DDI network analysis and accompanying tables.

%
% Results
%
\section*{Results}
\label{ch:results}

\subsection*{DDI Demographics, Severity, and Cost}
\label{ch:demographics}

Our analysis tallied $\Psi=1,025,754$ distinct drug pair co-administrations.
Almost $3\%$ of these, or $\Phi=26,524$, are known DDI and involve $75$ distinct drugs that participate in $|\Delta|=181$ observed distinct interaction drug pairs.
There is very strong linear relationship between volume of drug dispensation ($\alpha^{N})$ and DDI ($\Phi^{N}$) across neighborhoods ($N$) which fits a regression line almost perfectly ($R^2=.92$, $p<10^{-6}$); 
%see Figure \ref{fig:SI:hood-disp}-right in Supporting Information (SI).
see Supplementary Figure \ref{fig:SI:hood-disp}-right.
The distribution of these DDI pairs per severity class is detailed in Table \ref{table:ddi-severity}. 
A majority ($69\%$) are labeled \textit{Moderate}, although, worryingly, $22.5\%$ are classified as \textit{Major} DDI.
The observed DDI pairs were dispensed to $|U^{\Phi}|=15,527$ unique patients, which represent $12\%$ of the \textit{Pronto} patient population (and almost $5\%$ of the entire Blumenau population). Looking only at the adult \textit{Pronto} population, this number is raised to 15\% (15,336).
%
% Results: Severity Analysis
%
Almost $4\%$  of all \textit{Pronto} patients (5.01\% of adults) were administered a major DDI, and $9.58\%$ (12.15\% of adults) were administered a moderate DDI; these numbers represent $1.54\%$ and $3.75\%$ of the entire Blumenau population, respectively.
See Methods for precise definitions of symbols and formulae used in this section.

%
% Table 1 "major/moderate/minor" goes here
%
\begin{table}[ht]
    \centering
    \small
    \begin{tabular}{c|r r r r r }
    \toprule
severity $s$ &  $\Phi$ &  $|U^{\Phi}|$ & $|U^{\Phi}|/|U|$ & $|U^{\Phi}|/\Omega$ & $|U^{\Phi,[y>20]}|/|U^{[y>20]}|$ \\
\midrule
% UPDATED: 2018-11-06
% FILE: display_stats.py
\textit{Major}    &  5,968 (22.50\%) &  5,224 &  3.94\% &  1.54\% &  5.01\% \\
\textit{Moderate} & 18,335 (69.13\%) & 12,711 &  9.58\% &  3.75\% & 12.15\% \\
\textit{Minor}    &    542 (2.04\%)  &    528 &  0.4\%  &  0.16\% &  0.51\% \\
\textit{n/a}      &  1,679 (6.33\%)  &  1,493 &  1.12\% &  0.44\% &  1.43\% \\
\midrule
\textit{Major} $\lor$ \textit{Moderate} & 24,303 (91.63\%) & 15,030 & 11.32\% & 4.44\% & 14.35\% \\
\textit{Moderate} $\lor$ \textit{Minor} & 18,877 (71.17\%) & 12,791 &  9.64\% & 3.77\% & 12.22\% \\
\bottomrule
    \end{tabular}
    \caption{
        \textbf{Number and proportions of DDI observations and affected patients per DDI severity class}.
        Drugs or interactions identified in \emph{DrugBank} but not present in \textit{Drugs.com} are tallied as \textit{n/a}, see SM for details. First column: $\Phi$, number and proportion of observed DDI co-administrations.
        Second column: $|U^{\Phi}|$, number of patients affected by at least one DDI.
        Third and Fourth columns: proportion of patients from the \textit{Pronto} system and entire Blumenau populations, respectively.
        Fifth column: proportion of adult patients ($y\geq20$ y.o) from the pronto system.
        $\lor$ denotes the logical disjunction.
        Notice that the same patient may have been administered DDI of more than one severity class.
    }
    \label{table:ddi-severity}
\end{table}

% Results: Cost Analysis
We estimate the financial burden of DDI to Blumenau by evaluating how many of the 24,592 hospital admissions billed to this public health system in the same period \cite{SIH/SUS} were due to ADR from DDI.
This estimation relies on conjecturing what proportion ($p_h$) of patients who where dispensed a \textit{major} DDI are likely to have an ADR that requires hospitalization (details in Supplementary Information \S\ref{ch:SI:ddi-cost}).
We focus on the most conservative value from available literature \cite{Becker:2007} which yields $p_h = 2.68\%$, as well as on a less conservative estimate also previously reported \cite{Wu:2012} of $p_h = 8.35\%$.
The most conservative estimate leads to a cost of DDI-related hospitalization in Blumenau of over \$1M in the 18-month period, or a \textit{per capita} cost of \$2.03. The extrapolated costs to the state and the country are \$21M and \$565M, respectively (see Supplementary Tables \ref{table:SI:ddi-cost-state} and \ref{table:SI:ddi-cost-country}).
The less conservative estimate reaches a \textit{per capita} cost of \$6.33, or  \$3.2M, \$61M, and \$1.5B, for the city, state and country levels respectively.
However all of these conjectures are likely to err on the side of under-reporting emergency room admissions due to DDI or ADR, since this is a well-known problem in studies of this phenomenon \cite{Patrignani:2018, Gonzalez:2011, Alvarez:2013, Tatonetti:2012}.
Therefore, in Supplementary Information we also report cost estimates for various values of $p_h$, so that readers can judge what is an appropriate value to consider.

%
% Results: Drugs
%
\subsection*{Drugs Involved in Interactions}
\label{ch:results:common-ddi}

Table \ref{table:top-20-ddi} lists the top 20 DDI pairs, ordered by the rank product of their strength of DDI association, $\tau^{\Phi}_{i,j}$, with the number of patients they were administered to, $|U^{\Phi}_{i,j}|$. The complete list of DDI pairs, including the severity class and other measures, is provided in Supplementary Table \ref{table:SI:ListDDI-1-50} ordered by the number of affected patients (see also Supplementary Note \ref{ch:SI:interactions}).
$\tau_{i,j}$ is largest (smallest) for DDI pairs $(i,j)$ that are more (less) likely to be co-administered when either one of drugs $i$ or $j$ is administered. 
Computing the rank product between $\tau^{\Phi}_{i,j}$ and $|U^{\Phi}_{i,j}|$ identifies DDI pairs that are very prevalent in the population but which also tend to be co-administered.

Only 2\% of the observed DDI administrations are considered of \textit{minor} risk, affecting 542 patients. 
The highest ranked one (9\textsuperscript{th}) in Table \ref{table:top-20-ddi} is (Digoxin, Spironolactone) and it was administered to $|U^{\Phi}_{i,j}|=272$ patients (for $\langle \lambda^{u}_{i,j} \rangle = 140$ days on average); it leads to increased levels of Digoxin while decreasing the effect of Spironolactone.
The vast majority (almost 70\% per Table \ref{table:ddi-severity}) of observed DDI administrations fall in the \textit{moderate} risk class.
For instance, (Digoxin, Furosemide) can cause ``possible electrolyte variations and arrhythmia'' (4\textsuperscript{th},  $|U^{\Phi}_{i,j}|=385$, $\langle \lambda^{u}_{i,j} \rangle$ = 155).
Others, like the pair (Haloperidol, Biperiden; 2\textsuperscript{nd}, $|U^{\Phi}_{i,j}|=524$, $\langle \lambda^{u}_{i,j} \rangle$ = 243) give rise to various ADR, such as central nervous system depression and tardive dyskinesia; despite the known ADR this pair has been used clinically \cite{Drugs.com}, which explains the large value of $\tau^{\Phi}_{i,j} = 0.7$, meaning that these drugs are more likely to be co-administered.
In hot weather this DDI increases the risk of hyperthermia and heat stroke, and Blumenau has a humid subtropical climate with temperatures reaching 30$^{\circ}$C with 100\% humidity during summer.

(Omeprazole, Clonazepam) is the most frequent DDI pair observed, by a large margin to the second (5\textsuperscript{th}, $|U^{\Phi}_{i,j}|=5,078$, $\langle \lambda^{u}_{i,j} \rangle$ = 102). Omeprazole is used to treat acid reflux and other gastroesophageal problems, while Clonazepam is a benzodiazepine anti-epileptic. This prevalent dispensation requires particular attention to dosage since ``Omeprazole may increase the pharmacological effect and serum levels of certain benzodiazepines via hepatic enzyme inhibition'' \cite{Drugs.com, Caraco:1995}.
Similarly, (Acetylsalicylic Acid (ASA), Glyburide) is the top ranked pair in Table \ref{table:top-20-ddi} and very frequently dispensed (1\textsuperscript{st}, $|U^{\Phi}_{i,j}|=1,249$, $\langle \lambda^{u}_{i,j} \rangle$ = 141). This pair is especially problematic for diabetic patients since ``the salicylate increases the effect of sulfonylurea;'' It causes hypoglycemia by enhancing insulin sensitivity, particularly in patients with advanced age and/or renal impairment \cite{Drugs.com, Kubacka:1996}.

%
% Table 2 - "Top 20 DDI pairs" goes here
%
\begin{table}[ht]
    \centering
    \scriptsize
    \begin{tabular}{c|r r|c|c c|r|l}
        \toprule
        rankp(${\tau,U}$) & $\tau^{\Phi}_{i,j}$ & $|U^{\Phi}_{i,j}|$ & $\langle \lambda^{u}_{i,j} \rangle$ & $i$ & $j$ & $RRI^{F}_{i,j}$ & class \\
        \midrule
% UPDATED: 2018-04-26
% FILE: display_stats_ddi.py

 1 (2,4)   & 0.60 & 1249 & 141 $\pm$ 124 &           ASA &      Glyburide & 0.89 & Moderate \\
 2 (1,12)  & 0.70 &  524 & 243 $\pm$ 188 &   Haloperidol &      Biperiden & 0.62 & Moderate \\
 3 (4,11)  & 0.58 &  535 & 152 $\pm$ 132 &      Atenolol &      Glyburide & 1.22 & Moderate \\
 4 (3,17)  & 0.60 &  385 & 155 $\pm$ 125 &       Digoxin &     Furosemide & 0.61 & Moderate \\
 5 (62,1)  & 0.26 & 5078 & 102 $\pm$  95 &    Omeprazole &     Clonazepam & 2.28 & Moderate \\
 6 (8,16)  & 0.55 &  470 & 160 $\pm$ 133 &     Diltiazem &    Simvastatin & 1.27 &    Major \\
 7 (26,5)  & 0.45 & 1190 & 127 $\pm$ 127 & Amitriptyline &     Fluoxetine & 3.55 &    Major \\
 8 (82,2)  & 0.23 & 2117 &  53 $\pm$  74 &           ASA &      Ibuprofen & 1.42 &    Major \\
 9 (10,22) & 0.55 &  272 & 140 $\pm$ 114 &       Digoxin & Spironolactone & 0.58 &    Minor \\
10 (5,46)  & 0.57 &   95 & 140 $\pm$ 126 &   Propranolol &      Glyburide & 1.61 & Moderate \\
11 (15,18) & 0.50 &  377 & 143 $\pm$ 138 &    Fluoxetine &  Carbamazepine & 0.98 & Moderate \\
12 (91,3)  & 0.21 & 1460 &  54 $\pm$  77 &      Atenolol &      Ibuprofen & 1.88 & Moderate \\
13 (61,6)  & 0.27 &  999 &  87 $\pm$  86 &    Omeprazole &       Diazepam & 1.21 & Moderate \\
14 (16,26) & 0.49 &  226 & 151 $\pm$ 145 & Amitriptyline &  Carbamazepine & 0.99 & Moderate \\
15 (6,84)  & 0.56 &   25 & 157 $\pm$ 136 &     Diltiazem &     Amiodarone & 1.26 &    Major \\
16 (12,47) & 0.52 &   91 & 154 $\pm$ 142 &      Atenolol &      Diltiazem & 1.19 &    Major \\
17 (21,27) & 0.47 &  222 & 148 $\pm$ 139 &    Fluoxetine &        Lithium & 1.79 &    Major \\
18 (40,15) & 0.36 &  496 & 103 $\pm$  87 &           ASA &     Gliclazide & 0.78 &     None \\
19 (96,7)  & 0.20 &  892 &  56 $\pm$  61 &   Fluconazole &    Simvastatin & 2.63 &    Major \\
19 (14,48) & 0.50 &   90 & 161 $\pm$ 157 &    Imipramine &  Carbamazepine & 1.35 & Moderate \\
        \bottomrule
    \end{tabular}
    \caption{
        \textbf{Top 20 known DDI pairs}.
        Top 20 known DDI pairs $(i,j)$ by rank product (1\textsuperscript{st} column; individual rank in parenthesis) of the ranks of $\tau^{\Phi}_{i,j}$, the strength of DDI association from eq. \ref{eq:tau_ij}, and $|U^{\Phi}_{i,j}|$, the number of patients affected by the DDI (2\textsuperscript{nd} and 3\textsuperscript{rd} columns, respectively).
        Mean ($\pm$ s.d.) co-administration length, $\langle \lambda^{u}_{i,j} \rangle$, is shown in column 4 (in days) for each DDI pair $(i,j)$ whose English drug names are shown in columns 5 and 6.
        Relative gender risk of DDI pair co-administration, $RRI^{F}_{i,j}$ is shown in column 7.
        DDI severity classification, according to \textit{Drugs.com}, shown in column 8, with DDIs not found in \textit{Drugs.com} labeled as \textit{None}.
    }
    \label{table:top-20-ddi}
\end{table}

\textit{Major} DDI pairs represent 22.5\% of all observed DDI administrations per Table \ref{table:ddi-severity}.
The top 20 major DDI pairs are listed in Supplementary Table \ref{table:SI:top-20-major-ddi} and include:
\begin{itemize}
    
    \item (Diltiazem, Simvastatin), 6\textsuperscript{th}, $|U^{\Phi}_{i,j}|=470$, $\langle \lambda^{u}_{i,j} \rangle$ = 160, where ``Diltiazem increases the effect and toxicity of simvastatin'' possibly causing liver damage as a side effect \cite{You:2010};
      
    \item (Fluoxetine, Amitriptyline), 7\textsuperscript{th}, $|U^{\Phi}_{i,j}|=1,190$, $\langle \lambda^{u}_{i,j} \rangle$ = 127, where ``Fluoxetine increases the effect and toxicity of tricyclics'' \cite{Preskorn:1990}. The same ADR affects  (Fluoxetine, Imipramine), 23\textsuperscript{rd}, $|U^{\Phi}_{i,j}|=257$, and  (Fluoxetine, Nortriptyline), 33\textsuperscript{rd}, $|U^{\Phi}_{i,j}|=154$. 
    
    \item (ASA, Ibuprofen), 8\textsuperscript{th}, $|U^{\Phi}_{i,j}|=2,117$, $\langle \lambda^{u}_{i,j} \rangle$ = 53, where ``Ibuprofen reduces ASA cardioprotective effects''. In 2015 the European Medicines Agency issued an updated advice that occasional use of Ibuprofen should not affect the benefits of low-dose ASA \cite{EMA:2015}. Our analysis shows that patients were dispensed this pair concomitantly on average for 53 days ($\pm74$ s.d.), conflicting with occasional use. However, since these are common medications we cannot rule out the possibility they were dispensed to be taken as needed.
    
    \item (Fluoxetine, Lithium), 17\textsuperscript{th}, $|U^{\Phi}_{i,j}|=222$, $\langle \lambda^{u}_{i,j} \rangle$ = 148), where ``the SSRI increases serum levels of lithium'' potentiating the risk of serotonin syndrome, which is rare but serious and potentially fatal \cite{Drugs.com, Hadley:1989};
    
    \item (Fluconazole, Simvastatin), 19\textsuperscript{th}, $|U^{\Phi}_{i,j}|=892$, $\langle \lambda^{u}_{i,j} \rangle$ = 56), which leads to ``increased risk of myopathy/rhabdomyolysis''. Also from the azole class, Ketoconazole and Itraconazole are considered potent inhibitors generally causing less clinically significant interactions with Simvastatin than Fluconazole \cite{Drugs.com}. Both substitutes are available free of charge in the public health care system \cite{RENAME:2015}.

\end{itemize}

In addition, the top 20 DDI pairs ranked by a normalized drug ``footprint'' in the population are listed in Supplementary Table \ref{table:SI:top-20-ranked-gamma}.

%
% Results: Gender Risk
%
\subsection*{Gender Risk and DDI Networks}
\label{ch:results:gender-risk}

The set of patients who were co-administered known DDI was comprised of  $|U^{\Phi,\text{M}}|=4,793$ (30.54\%) males and $|U^{\Phi,\text{F}}|=10,734$ (69.46\%) females (see Supplementary Figure \ref{fig:SI:age}).
To understand whether this difference in the proportion of DDI per gender was due to \textit{Pronto} having more female patients ($59\%$), or because women tend to be prescribed more drugs in general \cite{Bjerrum:1998}, we computed two measures of relative risk of for women.
The \textit{relative risk of co-administration} for women is $RRC^{\text{F}} = 1.0653$ while their \textit{relative risk of interaction} is $RRI^{\text{F}}=1.5864$.
If the risks were equivalent for both genders, we would observe $RRC^{\text{M}} \approx RRC^{\text{F}} \approx 1$ and $RRI^{\text{M}} \approx RRI^{\text{F}} \approx 1$.
While the relative risk of drug co-administration is only slightly larger ($\approx 7\%$) for females, the relative risk of drug interaction is much larger ($\approx 59\%$).
This risk becomes even higher when we look only at the most dangerous severity class:
$RRI^{\text{F}}_{major}=1.8739$, while $RRI^{\text{F}}_{minor}=.8059$ (see Supplementary Table \ref{table:SI:rri-severity-gender}).
Removing female anti-contraceptive drugs only slighly lowers $RRI^{F}$ from $1.59$ to $1.55$.

%
% Figure "DDI network" goes here.
%
% Figure 1
\begin{figure}
    \centering
    \includegraphics[width=15cm]{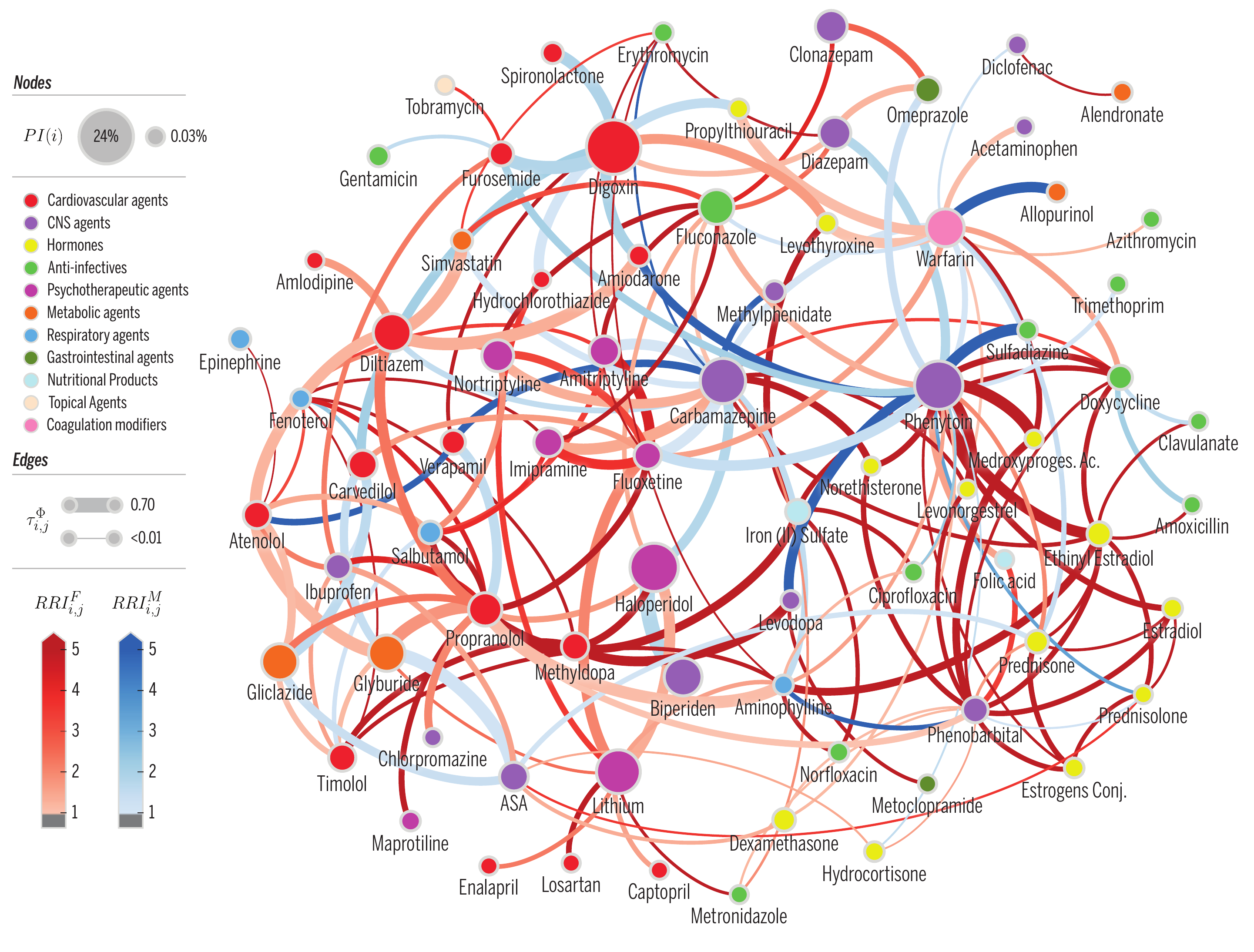}
    \caption{
        \textbf{DDI network}.
        A weighted version of network $\Delta$ where weights are defined by $\tau^{\Phi}_{i,j}$.
        \textbf{Nodes} denote drugs $i$ involved in at least one co-administration known to be a DDI.
        Node color represents the highest level of primary action class, as retrieved from Drugs.com (see legend).
        Node size represents the probability of interaction $PI(i)$, as defined in text.
        \textbf{Edge weights} are the values of $\tau^{\Phi}_{i,j}$ obtained from eq. \ref{eq:tau_ij}.
        \textbf{Edge colors} denote $RRI^{g}_{i,j}$, where $g \in \{ \text{M},\text{F} \}$, to identify DDI edges that are higher risk for females (blue) or males (red). Color intensity for $RRI^{g}_{i,j}$ varies in $[1,5]$; that is, values are clipped at 5.
    }
    \label{fig:ddi-network}
\end{figure}

To understand the DDI phenomenon at large as well as which drugs are most responsible for the higher risk of DDI women face over men, we also computed \emph{DDI networks} that characterize drug pairs according to measures of patient volume ($|U^\Phi_{i,j}|$) and DDI association strength ($\tau^{\Phi}_{i,j}$).
One of these networks is shown in Figure \ref{fig:ddi-network} (others shown in Supplementary Note \ref{ch:SI:ddi-network}).
The 75 drug nodes involved in DDI are colored by their primary action class.
Node size represents the \textit{probability of interaction} of a drug, $PI(i)$, with larger nodes identifying drugs most contributing to potential ADR from DDI.
To better grasp gender differences in the DDI phenomenon, edges are colored according to the \textit{relative risk of drug pair interaction for each gender}, $RRI^{g}_{i,j}$ with $g \in \{\text{F},\text{M}\}$, such that red (blue) edges denote increased DDI risk for women (men).

%
% Table 3 - "Top 10 known major DDI" goes here
%
\begin{table}[ht!]
    \centering
    \scriptsize
    \begin{tabular}{r c c r|r c c r}
        \toprule
        $|U^{\Phi,F}_{i,j}|$ & $i$ & $j$ & $RRI^{F}_{i,j}$ & 
        $|U^{\Phi,M}_{i,j}|$ & $i$ & $j$ & $RRI^{M}_{i,j}$ \\
        \midrule
% UPDATED: 2018-11-09
% FILE: display_stats_ddi.py
   13 &  Carbamazepine &  Ethinyl Estradiol & $\infty$ &  29 &       Digoxin &    Amiodarone &      1.78 \\
   13 & Levonorgestrel &      Carbamazepine & $\infty$ &  11 &    Diclofenac &      Warfarin &      1.19 \\
1,411 &            ASA &          Ibuprofen &     1.42 &  \multicolumn{4}{c}{-} \\
  992 &  Amitriptyline &         Fluoxetine &     3.55 &  \multicolumn{4}{c}{-} \\
  703 &    Fluconazole &        Simvastatin &     2.63 &  \multicolumn{4}{c}{-} \\
  209 &     Imipramine &         Fluoxetine &     3.08 &  \multicolumn{4}{c}{-} \\
  302 &      Diltiazem &        Simvastatin &     1.27 &  \multicolumn{4}{c}{-} \\
  159 &     Fluoxetine &            Lithium &     1.79 &  \multicolumn{4}{c}{-} \\
  122 &     Fluoxetine &      Nortriptyline &     2.70 &  \multicolumn{4}{c}{-} \\
   28 &    Propranolol &         Salbutamol &     6.61 &  \multicolumn{4}{c}{-} \\
        \bottomrule
    \end{tabular}
    \caption{
        \textbf{Top 10 known \emph{major} DDI pairs}.
        Top 10 known \emph{major} DDI pairs $(i,j)$ with increased risk of co-administration per gender, $g \in \{\text{M}, \text{F}\}$, which affected at least 10 patients of each gender.
        Rows ordered by the rank product of the ranks of $RRI^{g}_{i,j}$, the relative gender risk of co-administration, and $|U^{\Phi,g}_{i,j}|$, the number of patients of given gender affected by the DDI.
    }
    \label{table:top-10-major-ddi-gender}
\end{table}

Of the $|\Delta|=181$ DDI edges, $133$ are associated with an increased risk for women, whereas only $48$ denote an increased risk for men---a ratio of $2.8$.
Removing hormone therapy drugs from the network changes the number of edges associated with increased risk for women from $133/181=73.48\%$ to $116/158=73.42\%$; for men the ratio changes from $48/181=26.52\%$ to $42/158=26.58\%$.
In other words, there is virtually no change when hormone therapy drugs are removed from the network.
Looking at the subgraph comprised only of very gender-imbalanced pairs, $RRI^{g}_{i,j}>3$, we find 49 drugs in interactions that affected 3,327 women (4.28\% of female Pronto population), but only 13 drugs in interactions that affected 64 men (0.01\% of male Pronto population). The 65 (9) such interactions for women (men) contain 16 (3) that are considered \textit{major} (see also Supplementary Figure \ref{fig:SI:u-ddi-network}).
Table \ref{table:top-10-major-ddi-gender} shows the top \emph{major} DDI pairs per gender which affected at least 10 patients; interestingly, only two DDI pairs that affect at least 10 patients were observed with a higher relative risk of interaction for males---see Supplementary Tables \ref{table:SI:rrr-gender-females} and \ref{table:SI:rrr-gender-males} for full listings.

%
% Results: Age Risk
%
\subsection*{Age Risk}
\label{ch:results:age-risk}

\begin{figure}
    \centering
    % UPDATED: 2018-11-29
    % FILE: plot_rc_age.py, plot_ri_age_h0.py, plot_u_ddi_age_h0.py, etc
    %
    \includegraphics[width=.32\textwidth]{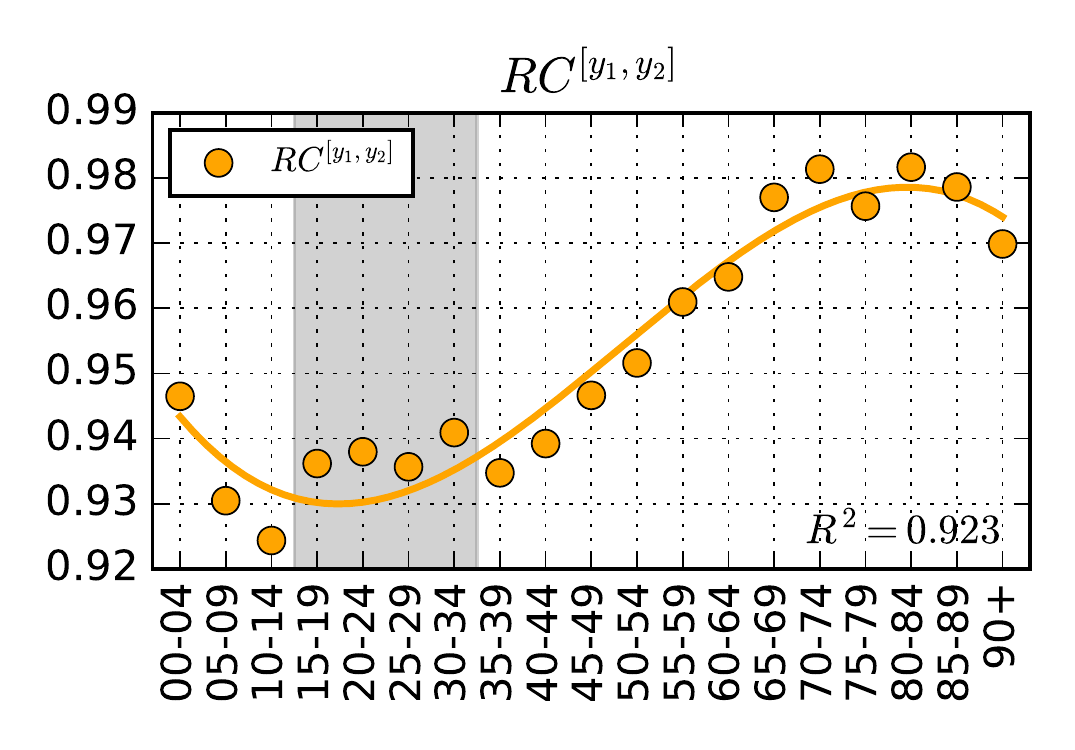}
    \includegraphics[width=.32\textwidth]{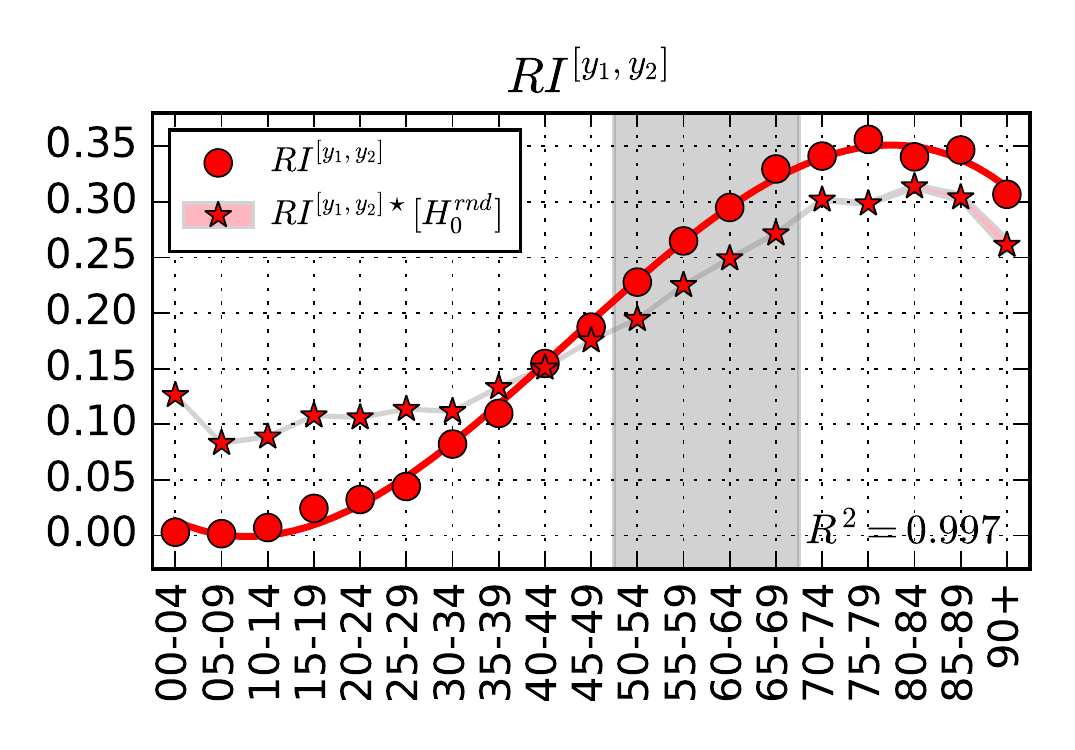} 
    \includegraphics[width=.32\textwidth]{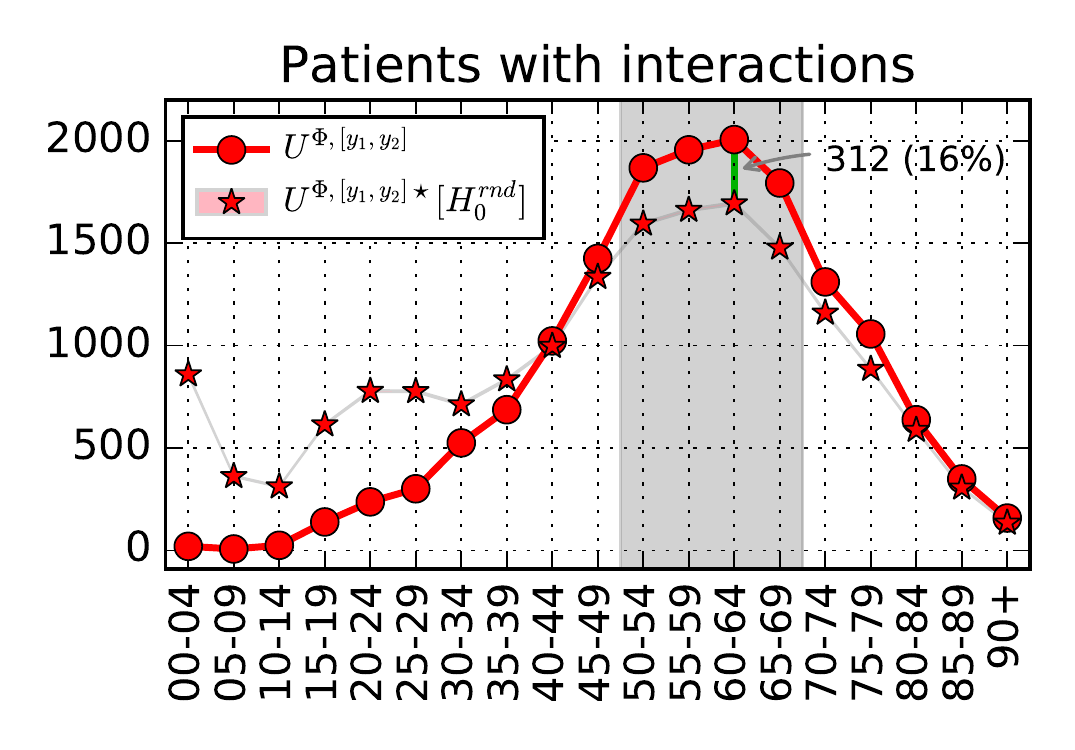} 
    \caption{
        \textbf{Risk of co-administration and interaction per age range}.
        \textbf{A \& B}.
        Co-administration ($RC^{[y_1, y_2]}$) and interaction risk ($RI^{[y_1, y_2]}$) per age group, computed via eq. \ref{eq:RIy}.
        Solid orange line is the cubic regression for $RC^{[y_1, y_2]}$ while solid red line is the cubic regression for $RI^{[y_1, y_2]}$ (linear and quadratic regressions in Supplementary Information).    
        \textbf{C}.
        Absolute number of patients with at least one co-administration known to be a DDI.
        For all plots, age groups $[\text{90,94}]$, $[\text{95,99})$ were aggregated into $[\text{90+}]$.
        Stars ($\star$) depict values computed from the null model, $H^{rnd}_0$, with background filling denoting the 95\% confidence interval based on 100 runs.
    }
    \label{fig:rc-ri-age}
\end{figure}

To investigate the role of age in DDI co-administration we calculated two additional measures, the \emph{risk of co-administration for age group}, $RC^{[y_1,y_2]}$, and the \emph{risk of interaction for age group}, $RI^{[y_1,y_2]}$.
If the number of DDI observed were proportional to the number of co-administrations, the latter quantity would be essentially flat across age groups (see eq. \ref{eq:RIy} in Methods). As shown in Figure \ref{fig:rc-ri-age}, center, $RI$ increases substantially for older age groups (see also Supplementary Table \ref{table:SI:prob-interactions-age}), varying from near zero for younger age groups to $0.35$ for groups over 70.
While there is some variation, $RC$ varies a lot less than $RI$---no more than $6\%$ across all age groups as seen Figure \ref{fig:rc-ri-age}-left (note the difference in scale).
This shows that risk of co-administration is largely proportional to the number of dispensed drugs, while risk of interaction seems to grow more than the increase in co-administrations (polypharmacy) observed with age.

%
% Figure "rc-ri-u" goes here.
%
% Figure 2

The risk of co-administration is overall quite high for all age groups ($RC^{[y_1,y_2]} \in [.92,.98]$), with increasing values as patients age. 
Patients dispensed at least two drugs are almost always being dispensed drugs concomitantly.
Conversely, the risk of interaction starts from almost nonexistent at age $[\text{0-14}]$ and reaches more than 25\% after the age of 55.

%
% Figure "img-coadmin-ddi-dist" goes here
%
% Figure 3
\begin{figure}
    \centering
    % UPDATED: 2017-01-29
    % FILE: coadmin-ddi-dist.py
    %
    \includegraphics[width=.9\linewidth]{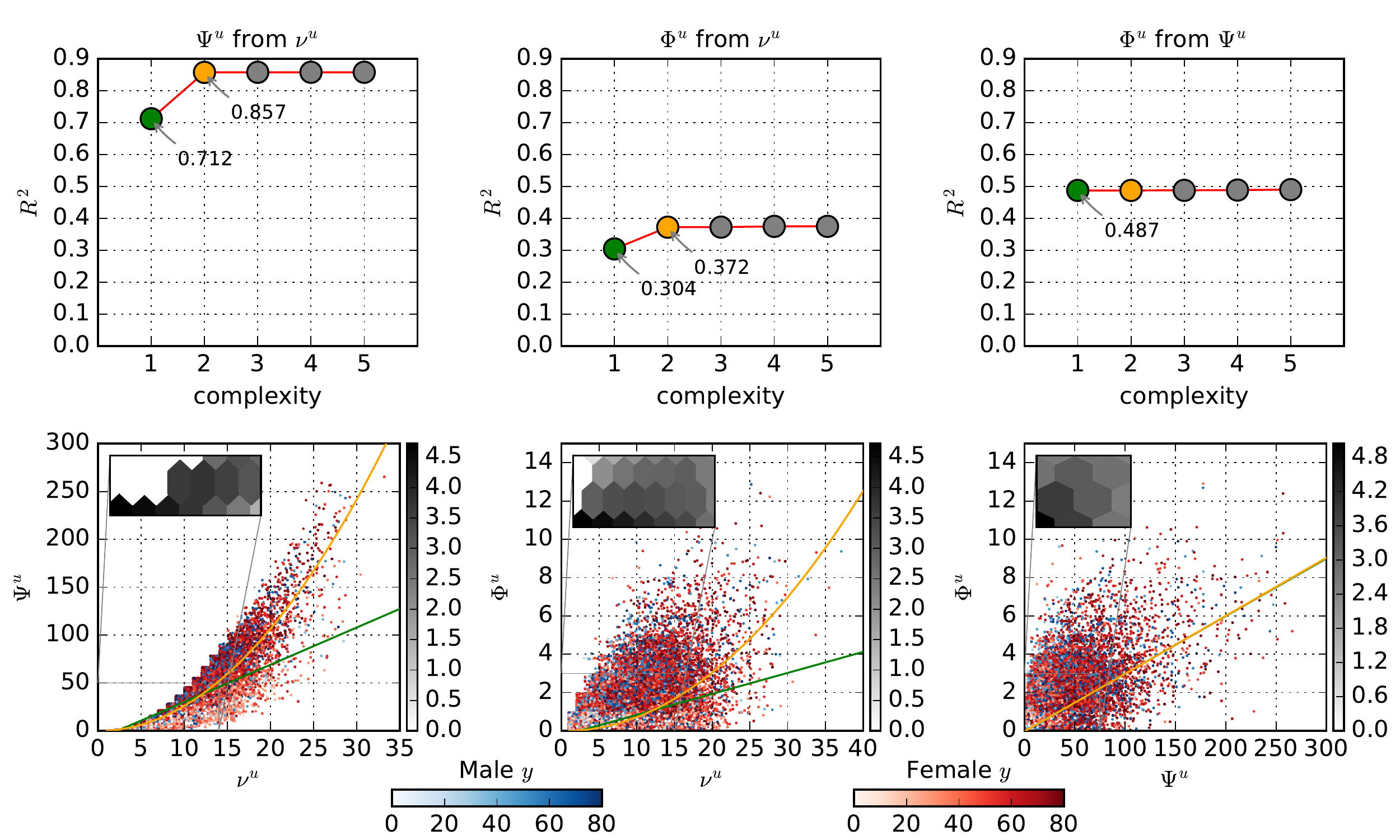}
    \caption{
        \textbf{Patients with their number of drugs dispensed $\nu^{u}$, co-administrations $\Psi^{u}$ and interactions $\Phi^{u}$}.
        \textbf{D, E \& F}.
        Each circle depicts a patient, with red (blue) circles denoting females (males). Color intensity denotes their age, with stronger red (blue) representing older women (men).
        To reduce circle overlay and enhance visualization, a uniform noise $\in [0,1]$ was added to both coordinates.
        Green and orange lines denotes linear and quadratic regressions, respectively.
        Inserts with Hexagonal log-bins are included to better depict the density of patients close to the origin.
        \textbf{A, B \& C}. Pareto fronts comparing regression results ($R^2$) at increasing regression model complexity. For example, complexity 1 and 2 denote a linear and quadratic regression, respectively.
    }
    \label{fig:coadmin-ddi-dist}
\end{figure}

The relationship among the number of drugs dispensed ($\nu^{u}$), co-administrations ($\Psi^{u}$), and interactions ($\Phi^{u}$) for all users is shown in Figure \ref{fig:coadmin-ddi-dist}.
While there is a strong nonlinear (quadratic) relationship between $\nu^{u}$ and $\Psi^{u}$ (Fig. \ref{fig:coadmin-ddi-dist}-D), there is no evidence of a nonlinear relationship between $\Psi^{u}$ and $\Phi^{u}$ (Fig. \ref{fig:coadmin-ddi-dist}-F), which could explain the observed growth of RI with age---which implies that interactions grow faster than co-administrations with age.
In contrast to previous reports \cite{Guthrie:2015}, co-administrations ($\Psi^{u}$) predict interactions ($\Phi^{u}$) better than number of drugs prescribed ($\nu^{u}$), though neither do so particularly well.

\begin{figure}
    \centering
    % UPDATED: 2018-11-29
    % FILE: plot_rrc_rri.py
    %
    \includegraphics[width=.37\textwidth]{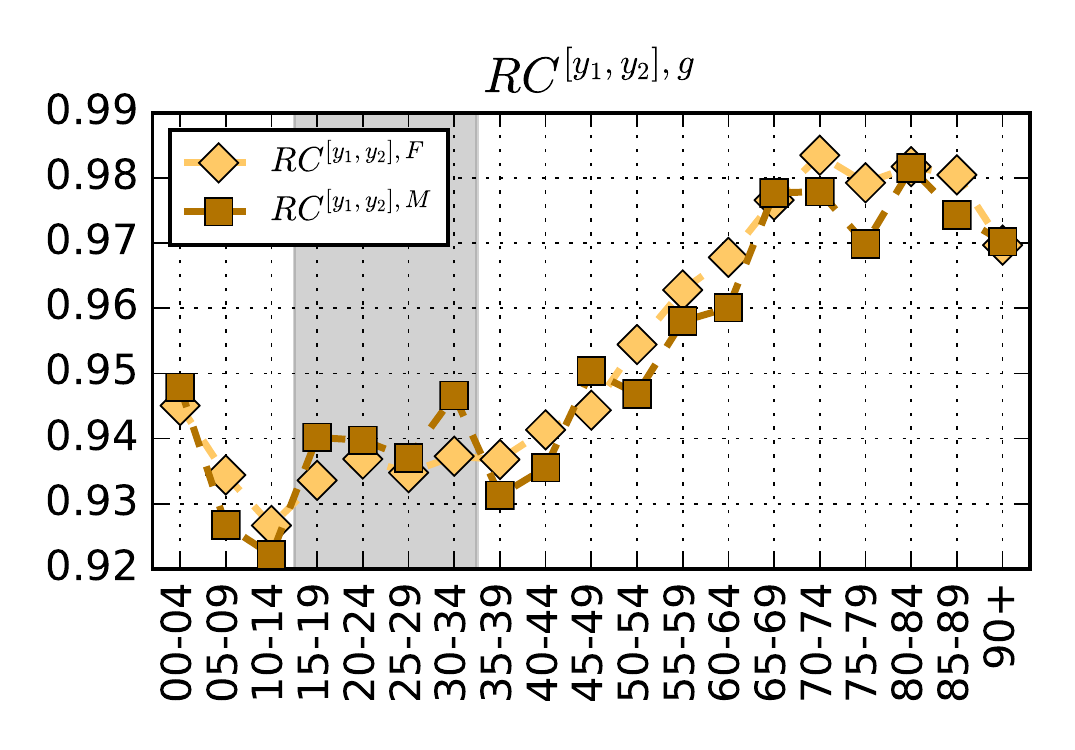}
    \includegraphics[width=.37\textwidth]{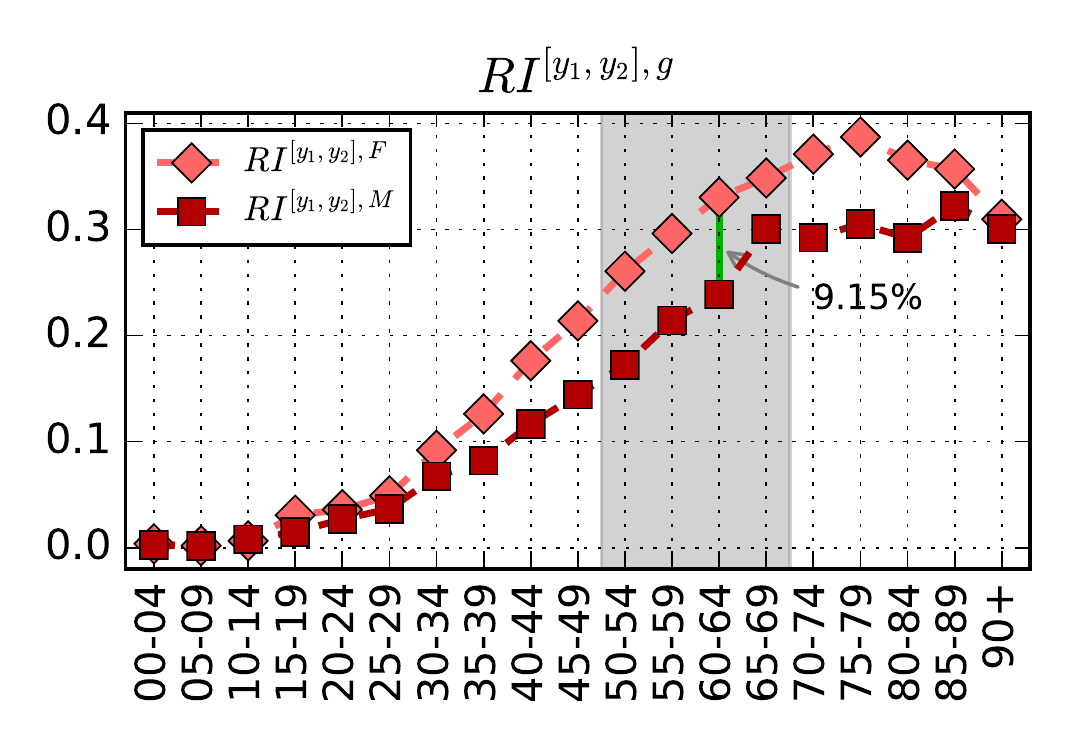}\\
    \includegraphics[width=.44\textwidth]{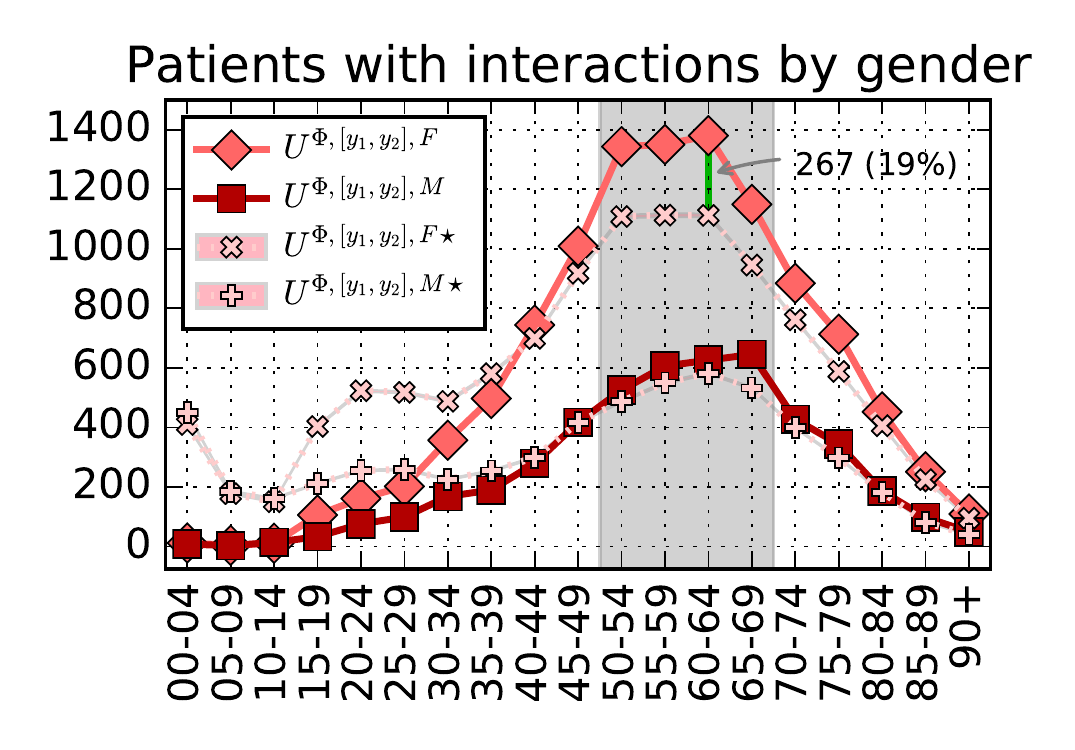} \\
    \includegraphics[width=.37\textwidth]{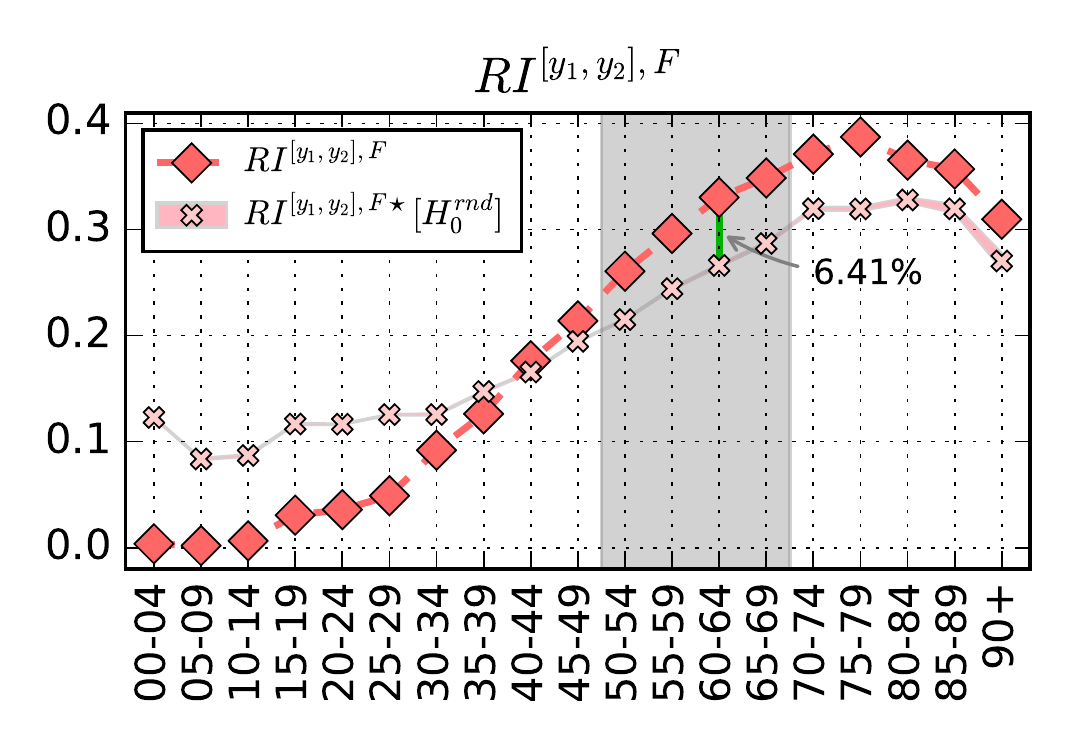}  
    \includegraphics[width=.37\textwidth]{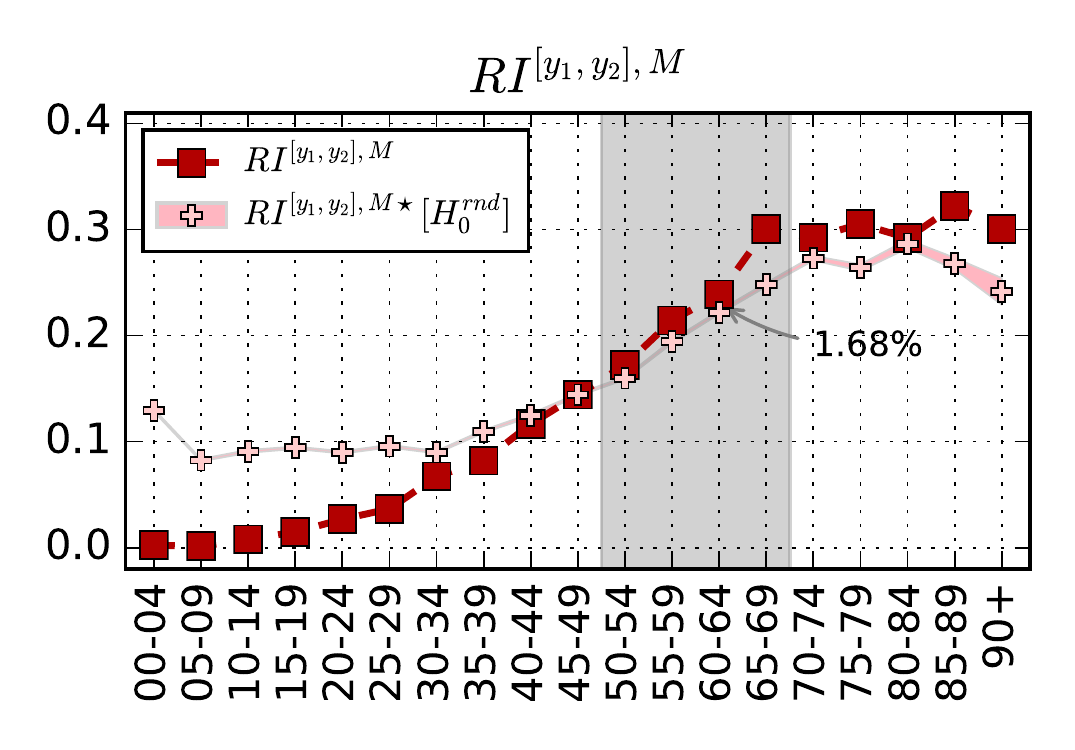}
    \caption{
        \textbf{Risk of co-administration and interaction per age range and gender}.
        \textbf{A}.
        Risk of co-administration per age group and gender, $RC^{[y_1,y_2],g}$.
        \textbf{B}
        Risk of interaction per age group and gender, $RI^{[y_1,y_2],g}$.
        \textbf{C}.
        Absolute number of patients with at least one known DDI co-administration, per age and gender $U^{\Phi,[y_1,y_2],g}$.
        \textbf{D \& E}.
        Female and male risk of interaction per age group and gender, $RI^{[y_1,y_2],F}$ (D) and $RI^{[y_1,y_2],M}$ (E).
        For all plots, age groups $[\text{90,94}], [\text{95,99]}, [\text{90},\text{90}+]\}$ were aggregated into $[\text{90+}]$.
        Stars ($\star$) depict values computed from the null model, $H^{rnd}_0$, with background filling denoting the 95\% confidence interval based on 100 runs.
        Shaded areas identify specific age groups mentioned in the main manuscript.
    }
    \label{fig:rc-ri-age-gender}
\end{figure}

%
%Null Model
%
To further investigate whether factors other than increase in co-administration cause the increase of DDI risk with age, we developed a statistical null model; values reported for the null model are identified with a star ($\star$) and associated $95\%$ confidence intervals (for 100 runs) in Figure \ref{fig:rc-ri-age}.
The idea is to explore if the growth of $RI^y$ is an expected phenomenon of increased polypharmacy with age, which necessarily results in a combinatorial increase of possible drug pairs that can interact. 
The null model was not able to reproduce the observed behavior of $RI^y$ ($X^2=2840.6$, $p<.01$), especially for older and younger ages (see Figures \ref{fig:rc-ri-age} and \ref{fig:rc-ri-age-gender} and Supplementary Information \S \ref{ch:SI:null-models} for additional details).

We observe that for younger ages, $RI^{[0,29]}$ is much lower than the model's predicted $RI^{[0,29]\star}$ (Fig. \ref{fig:rc-ri-age}-b); the same is true for the number of patients affected (Fig. \ref{fig:rc-ri-age}-c).
The largest discrepancies between model and real data occur at this age range, especially $[\text{0,4}]$ and $[\text{20,24}]$.
However, this expected behavior is inverted for ages $[\text{50+}]$, with the transition occurring around age $[\text{40,44}]$ (Fig. \ref{fig:rc-ri-age}-b).
For older ages, the largest discrepancies between model and reality occur for age groups in $[\text{50,70}]$, where the predicted number of patients with DDI ($|U^{\Phi\star}|$) for age group $[\text{60,64}]$ is 16\% lower than what is observed (see Fig. \ref{fig:rc-ri-age}-c).

%
% Figure "rc-ri-u-ri-ri" goes here
%
% Figure 4

We additionally parse age risk by gender by computing $RC^{[y_1,y_2],g}$ and $RI^{[y_1,y_2],g}$, shown in Figure \ref{fig:rc-ri-age-gender} (see also Supplementary Tables \ref{table:SI:prob-interactions-age-male} and \ref{table:SI:prob-interactions-age-female}).
Both genders have overall similar risk of co-administration in all age groups.
Even during childbearing age, the co-administration risk is similar for the numbers of drugs dispensed, even if slightly larger for females (see filling in Fig. \ref{fig:rc-ri-age-gender}-top-left).
Interestingly, for $RI^{[y_1,y_2],g}$ a clear difference between genders occurs \textit{after} childbearing age, maximized between 50 and 69 years-old (see filling in Fig. \ref{fig:rc-ri-age-gender}-top-right and absolute number of patients in Fig. \ref{fig:rc-ri-age-gender}-middle). 
The gender difference in $RI$ appears after the age of $35$, reaching more than a 9\% difference for age group $[\text{60,64}]$.

Figure \ref{fig:rc-ri-age-gender} \textbf{d}-\textbf{e} show the null model's gender risk of interaction $RI^{[y_1,y_2],g\star}$, in comparison to observed values, $RI^{[y_1,y_2],g}$, for men (\textbf{d}) and women (\textbf{e}), respectively.
For both genders, we still observe that the real $RI$ for children and young adults ($[\text{0,34}]$) is well below the null model.
However, the transition observed for older age is much more pronounced for women.
In fact, after age 40, observed male $RI$ is largely consistent with the null model, while female risk is higher.

\subsection*{Prediction of Patients with DDI}
\label{ch:classification}

We computed several multiple regression (MR) models. These show that the inclusion of additional variables does not improve much at all the prediction of the variance of $\Phi^{u}$. 
For instance, a MR with both $\nu^{u}$ and $\Psi^{u}$ leads only to very marginal increase in the explained variance of $\Phi^{u}$: adjusted $R^2=0.492$.
Adding higher order, nonlinear models also does not improve upon the original regression between $\Psi^{u}$ and $\Phi^{u}$.
Even the inclusion of demographic variables in MR models does not lead to improvement of $R^2$ for $\Phi^{u}$---we analyzed many neighborhood-level variables such as average income, robbery, theft, suicide, transit crime, trafficking and rape rates.
Restricting the analysis to the subset of patients who reported education, and using it as an independent categorical variable also yields no improvement (see Supplementary Information \S \ref{ch:SI:multiple-regression} for MR and ANOVA details).

Interestingly, even the inclusion of gender as a categorical variable, does not improve $R^2$ for $\Phi^{u}$.
At first glance, this seems a somewhat counter intuitive result, given the observed high risk of DDI for females in comparison to males.
However, the MR analysis revealed that even though women certainly face a much greater risk of DDI, the number of DDI pairs they are administered ($\Phi^{u}$) is on average similar to that of men, and both have large variance of $\Phi^{u}$ (see Supplementary Figure \ref{fig:SI:age-means}). Thus, while gender clearly is a very strong factor in the risk of \emph{at least one} DDI, it is not a good predictor of the \emph{specific number} of interactions per patient.

Therefore, we sought to answer the question of how well we can automatically predict patients with at least one DDI (not the number of interactions per patient)?
Using binary classifiers we are able to achieve very good performance on this task. Classifiers perform well above null models, with MCC $\approx 0.7$ and excellent AUC scores: AUC ROC $\approx 0.97$ and AUC P/R $\approx 0.83$.

\section*{Discussion}
\label{ch:discussion}

Our 18-month longitudinal analysis of EHR data of the entire city of Blumenau allowed us to study the DDI problem in primary and secondary care in greater detail and for a longer period of time than what has been hitherto possible.
In summary, the DDI phenomenon is stable across the city, and proportional to population size---demonstrating no major inequalities due to income, education, crime, or other neighborhood social factors, which suggests a balanced and fair access to medical care in Blumenau. Our analysis revealed that $\approx 12\%$ of all patients of the \textit{Pronto} HIS where administered known DDI, which represents $5\%$ of the entire Blumenau population. If we consider only the adult population, $\approx 15\%$ were dispensed a known DDI (more than $6\%$ of the Blumenau adult population).
Looking at the type of DDI, we observe that 4\% of all patients (5\% of adults) were dispensed a \textit{major} DDI likely to result in a very serious ADR---almost 2\% of the city's population.

% Comparing to Tayside
Given the lack of similar studies, we cannot directly compare the rate of DDI severity observed in Blumenau to other public health systems. The Tayside study (with a smaller, 84 day observation window) reported a rate of $13\%$ ``potentially serious'' DDI for adult patients \cite{Guthrie:2015}. This severity class was derived from the \textit{British National Formulary}, a private publication we do not have access to. If this severity is similar to the \textit{Drugs.com} major DDI class, then Blumenau has a considerably lower rate of this type of DDI than Tayside---5\% to 13\%. If, on the other hand, ``potentially serious'' encompasses both the major and moderate \textit{Drugs.com} DDI classes, then the rates observed in Blumenau are similar to those observed in Tayside---14.35\% to 13\%.

We uncovered $181$ DDI pairs that most likely could have been prevented \cite{Hakkarainen:2012}.
These drugs known to interact were nonetheless dispensed for co-administration to 15,527 people, including more than five thousand who were administered a \textit{major} DDI, likely to require medical attention.
%
% Cost Analysis
In addition to the human suffering caused, patient hospitalization due to \textit{major} DDI may lead to a large financial burden to health-care systems.
All our estimates lead to very substantial costs for the various levels of government, suggesting that the financial burden of DDI is at least double what was previously reported---\$1 per capita in Ontario \cite{Wu:2012}---even when considering the most conservative estimate of the proportion of hospitalizations that derive from co-administration of known major DDIs.
Thus, our large-scale longitudinal analysis suggests that previous estimates based on smaller studies likely underestimate the cost of the DDI phenomenon.

We provide comprehensive lists of the DDI pairs uncovered in the data, allowing others to look at specific drugs of interest. The data can be seen from different angles, such as the volume of people affected or the likelihood that certain drugs are co-administered.
These include common medications such as proton-pump inhibitors (Omeprazole), anti-depressants (Fluoxetine), or common analgesics (Ibuprofen), as well as not so common drugs (e.g. Erythromycin).
It is noteworthy that the DDI co-administration of CYP(3A4 and 2D6) inhibitors with their respective enzymes substrates was often found in our results.
From our dataset CYP[3A4] inhibitors include Omeprazole, Fluconazole and Erythromycin and their respective substrates include Clonazepam, Simvastatin and Carbamazepine.
Recently, the FDA included a comparison list \cite{FDA:093664} of \textit{in vitro} and clinical inhibitors, inducers and substrates for CYP-mediated metabolisms.
In agreement with previous work \cite{Molden:2005}, our analysis revealed several such DDI, including the most common DDI pair in our data (Omeprazole, Clonazepam).
Many other major interactions, while not ranked at the top, are nonetheless of concern due to severe ADR. For instance, in 2011 the FDA issued a warning \cite{FDA:256581} contraindicating the concomitant use of Simvastatin with Erythromycin, due to increased risk of myopathy by ``possibly increasing the statin toxicity''. Still, our analysis identified 10 patients concomitantly administrating this major DDI (117\textsuperscript{th}, $|U^{\Phi}_{i,j}|=10$), also known for its increased risk of liver damage and a rare but serious condition of rhabdomyolysis that involves the breakdown of skeletal muscle tissue \cite{Drugs.com, Itakura:2003}.

% Network
Our network representation also allows us to integrate, summarize and visualize the DDI phenomenon.
The analysis of the network itself also reveals nodes with largest degree, that is, drugs that participate in more known DDI. The top ones, participating in over 10 distinct DDI are: \textit{Phenytoin}, \textit{Carbamazepine}, Phenobarbital, Propranolol, \textit{Warfarin}, Aminophylline, Fluoxetine, Fluconazole (see Supplementary Table \ref{table:SI:ddi-network-louvain} for others). Drugs in italic have both high degree and high $PI$, meaning they interact with many other drugs and are also more likely to interact with some other drug when dispensed.
The network also allows us to investigate the roles of individual drugs and DDI pairs, in relation to others.
For instance, Phenytoin, an anti-seizure medication, is the drug with largest degree and node size: it interacts with 24 other drugs, granting it the highest total degree strength, $\sum_{j} \tau^{\Phi}_{ij}=6.51$; 1 in 5 times that Phenytoin is co-administered with another drug it leads to an interaction, $PI(Phenytoin)=0.2$; and it also has the largest betweeness centrality ($0.30$) \cite{Freeman:1977}, thus acting as bridge between other drugs with known DDI.

% Gender Risk
Our characterization of the significant demographic factors in the DDI phenomenon, shows that women in Blumenau are at a strikingly greater risk of being dispensed known DDI than men, with a 1.6 risk multiplier.
In other words, women in the Blumenau's \emph{Pronto} system have an almost 60\% increased risk over men of being dispensed a DDI, but only a 6.5\% increased risk of being dispensed drugs concomitantly.
When only \textit{major} DDI are considered the risk multiplier is even higher: 1.9. That is, women have almost double the risk of men of being dispensed a \textit{major} known DDI.
It is noteworthy that we pursued a relative risk analysis for all age groups, showing that females face a greater or similar risk of DDI than males in all age groups, with substantially higher risk observed after 50 years of age.
For instance, in age group $[\text{60-64}]$, 1 in 3 women who are dispensed two or more drugs concomitantly face a known DDI, whereas that ratio is less than 1 in 4 for men for the same age group (see Figure \ref{fig:rc-ri-age-gender}).
Therefore increased risk for females is not confounded by the larger number of women present in the data nor their age.

% Age Risk
It is known that age is also a factor in predicting the number of prescribed drugs \cite{Bjerrum:1998}, especially because of increased co-morbidity in older patients.
Our analysis shows that one in every four patients over 55 is likely to be face a known DDI when co-dispensed two or more drugs.
The risk of interaction for older age groups of both genders is also severe, reaching more than 30\% for adults over 70 years of age in comparison to younger age groups. 
While a greater risk for older age groups is expected due to increased polypharmacy with age, a comparison of the observed risk with a null model accounting for random polypharmacy (and preserving same number of co-administrations per age) shows that it does not explain the high levels of interactions older age groups face.
This can be contrasted with the almost nonexistent number and risk of interactions in children, which are considerably lower than what the null model predicts for polypharmacy at that age.
It is very surprising, indeed shocking, that there are more cases (and increased risk) of DDI in older age than random (age-conditioned) dispensation of drugs would yield.
We would expect all age groups to have fewer cases than a random null model, but this is only observed for younger age groups.

The null model also revealed an additional gender bias, as older women clearly have a \textit{worse-than-random}, while older men have a more \textit{similar-to-random} risk of DDI in most age groups. In fact, deviation from the null model in older age is mostly explained by increased risk for females. In contrast, younger age groups of both genders have much \textit{better-than-random} risk of DDI.

These observed gender and age risks suggest two possible hypothesis: specific drugs dispensed to women or older populations are more dangerous; and/or that not as much attention to DDI in primary care is reserved for these populations.
The fact that the specific drugs dispensed greatly improve the automatic prediction of patients with DDI favors the first hypothesis, but given the age and gender risks observed, it is also clear that the same DDI-prone drugs are administered differently between genders and across age groups.
This second hypothesis is strengthened by the fact that removing female-specific hormone therapy from the the DDI network of Figure \ref{fig:ddi-network} barely reduces the DDI gender risk (from 59 to 55\%).
Indeed, the DDI pairs with increased risk for women traverse all drug classes and are not gender-specific, ranging from cardiovascular to central nervous systems agents.

While it was already known that drugs withdrawn from the market for ADR presented greater risks for women \cite{GAO:01:286R}, our study demonstrates that women (and older populations) in Blumenau also face a higher risk of being dispensed known DDI.
It could be that in older age groups (especially for women) there are fewer alternative drugs (with fewer adverse reactions) in the Blumenau public system, either because they are more expensive or simply because they are not available anywhere, thus forcing the prescription of known DDI.
These and other possibilities warrant further study outside the scope of the present article.
For instance, would the introduction of newer and costlier drugs into the public system, overcome the financial and human burden of current DDI levels?
Nonetheless, since medical care should in principle provide a \textit{better-than-random} risk of DDI for all age groups and genders, our results suggest that factors of a social, biological, or medical-care nature are at play at the primary- and secondary-care levels and should be further studied everywhere.

The performance achieved by our classifiers demonstrates that a useful computational intelligence pipeline can be devised to flag patients for further assessment by a primary care physician, pharmacist, public official, or even to request a home visit from a community health agent. 
Existing prescription alert systems already warn against known DDI, still, these are evidently being prescribed in worrying numbers.  
This could be because there are good medical reasons to prescribe certain drug combinations despite known DDI risk, or because drugs may be prescribed by distinct physicians, who may not be aware of or check previous prescriptions, or simply dismiss HIS alerts \cite{Slight:2013}---perhaps due to physician alert fatigue \cite{Tucker:2012}.

To be useful, personalized alert systems based on the type of predictions produced by our classifiers
do not necessarily need to be added to prescription systems.
Indeed, their utility lies not in identifying known DDI pairs---as those are already by definition available via formularies, web resources like \textit{Drugbank}, or prescription HIS---but rather in identifying \textit{patients} at greater risk of being prescribed DDI in the future, or \textit{subpopulation}s and \textit{comorbidities} that for social or biomedical reasons face greater risk of DDI.
Thus, they should be more useful for those involved in integrating and managing the care of individual patients or the entire public-health system. 
Those are decisions that each public-health system will have to weight.
Still, our work demonstrates that a personalized alert system for DDI is accurate and can be used to reduce the DDI phenomenon not only in future versions of the \emph{Pronto} HIS, but in other cities that have observed high levels of DDI---e.g. the Tayside region, in Scotland \cite{Guthrie:2015}.
In future work we intend to add such a pipeline to \textit{Pronto} as well as utilize new data sources such as social media, since \emph{Pronto} already includes such patient handles. Indeed, such data may allow early-warning signal detection of adverse events and DDI \cite{Correia:2016, MacLeod:2016}.

Large-scale analyses of EHR to establish the prevalence of known DDI are rare. Most studies are obtained from small populations in hospital settings, so they vary by a large margin \cite{Cano:2009, Hakkarainen:2012, Rozenfeld:2007, Codagnone:2010}. 
Our study of the entire city of Blumenau at the primary- and secondary-care level offers an important new large-scale measurement of the DDI phenomenon in a public health-care system---a baseline that can be compared to other worldwide locations beyond Brazil, as EHR data becomes available.
For instance, are the gender and age risk levels we observed similar in other primary- and secondary-care settings? Are there cultural or public/private differences? Will the health systems of other cities also prove to be unaffected by neighborhood and income levels, etc?

Our large-scale epidemiological analysis demonstrates that an integrated data- and network-science approach to public health can uncover biases in the DDI phenomenon as well as yield tools capable of issuing accurate DDI prediction per patient. Both outcomes contribute to preventing ADR from DDI and thus may lead to a significant positive impact on the quality of life of patients and finances of public-health systems.
Moreover, the gender and age risks of DDI we discovered, should inform physicians and other health professionals anywhere that such factors are important in the drug management of their patients. We expect the results to increase awareness of those risks we uncovered.

%
% Methods
%
\section*{Methods}
\label{ch:data-methods}

%
% Data
% 
\subsection*{Data}
\label{ch:methods:data}

Eighteen months of drug dispensing data (Jan 2014-Jun 2015) were gathered from the \emph{Pronto} HIS \cite{Mattos:2015, LDTT:Pronto} (see Supplemental Note \ref{ch:SI:Pronto} for a system description).
Drugs reported in this system are available via medical prescription only, free of charge, and dispensed to citizens of Blumenau (population $\Omega=338,876$ \cite{IBGE}) during the observation period.
Doctors prescribe medications by selecting drug and dosage via the HIS.
Low-cost drugs can generally be directly dispensed at the primary-care facilities, whereas specialized and higher-cost medication is distributed in three central facilities across the city. All drugs are dispensed by pharmacists who must select in \emph{Pronto} the drug and quantity to be dispensed, allowing the length of administration to be estimated. 
It must be noted that patients are not required to retrieve drugs from the public system. They can buy prescribed medications from private pharmacies at their own expense, without such transactions being recorded in \emph{Pronto}.
However, there is no incentive to pay more at private pharmacies for the same medication. Indeed, our analysis indicates that use of \emph{Pronto} is similar across all neighborhoods of Blumenau, irrespective of their average income (see Supplementary Figure \ref{fig:SI:age-bairro-heat}).

EHR were anonymized at the source and only drug dispensation and demographic variables, including gender, age, neighborhood, marital status and educational level, were kept.
Methods were performed in accordance with guidelines and regulations.
All patient consent was handled at the source prior to the anonymization and outside of the responsibility of this team.
Nonetheless, this study was approved by Indiana University's Institutional Review Board (IRB).
Drug names originally in Portuguese were converted to English, disambiguated and matched to their DrugBank ID (e.g., \textit{Cefalexina 500mg Comprimido} and \textit{Cefalexina 250MG/5ml Suspensão Oral} were matched to Chlorphenamine, DBID \texttt{DB01114}).
Medications with multiple drug compounds (e.g., Amoxicillin 500mg \& Clavulanate 125mg) were split into their constituent individual drugs.
Other dispensed substances (e.g., infant formula milk or vitamin complexes) unmatched to DrugBank were discarded. In total, 122 unique drugs were keep for analysis.
Because we have no means to know whether patients actually took the dispensed drugs, our analysis assumes that drugs dispensed were administered.

Throughout the year of 2014 and the first six months of 2015, Blumenau's \emph{Pronto} HIS registered $1,573,678$ distinct drug interval administrations, dispensed to $|U| = 132,722$ distinct patients---39.17\% of the city population.
The male/female proportions are 41.5/58.5\%, respectively.
Of the 46\% who declared their education level, a large proportion (46.77\%) reported having incomplete elementary school and 20.49\% had finished high school or above (see Supplementary Figure \ref{fig:patient-attrib} for details).
$|U^{\nu\geq2}|=104,811$ patients, corresponding to 78.97\% of the \textit{Pronto} patient population, were dispensed two or more distinct drugs in the period; only this set could have been dispensed known DDI.

%
% Figure "img-patient-attrib" goes here
%
% Figure 5
\begin{figure}
    \centering
    \includegraphics[width=11cm]{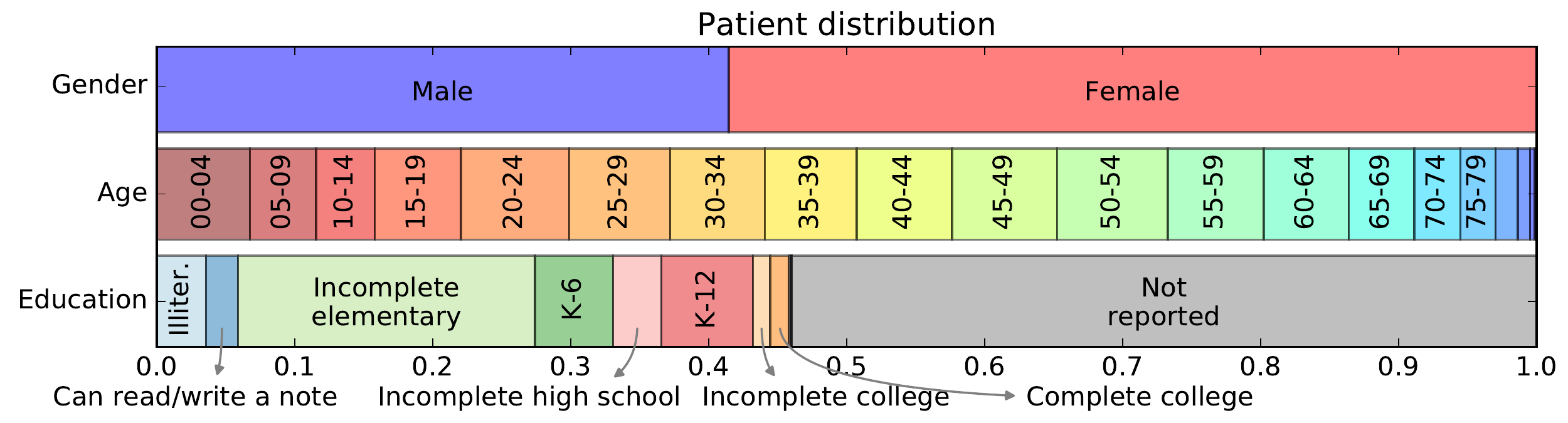}
    \caption{
        \textbf{Distribution of patients given gender, age and education level}.
        In total $|U^{\text{M}}|=55,032$ (41.46\%) were males and $|U^{\text{F}}|=77,690$ (58.54\%) were females.
        On education, a majority $|U^{e=\varnothing}|=71,662$ (53.99\%) did not report their education level. $|U^{e-}|=48,547$ (36,58\%) declared having at most some high school education whereas $|U^{e+}|=12,513$ (9,43\%) had completed high school education or above.
        On age, patients $|U^{y=[\text{20,24}]}|=10,382$ (7,82\%) and $|U^{y=[\text{50,54}]}|=10,650$ (8,02\%) accounted for the two largest age groups.
        Labels K-6 and K-12 are \textit{Completed elementary} and \textit{Completed high school} education, respectively.
        Labels for age $y\geq80$ and education level above \textit{Completed college} not shown.
    }
    \label{fig:patient-attrib}
\end{figure}

%
% Methods
%
\subsection*{Methods}
\label{ch:methods:methods}

\begin{figure}
    \centering
    % UPDATED: 2017-01-19
    % FILE: img-graph-timeline.ai
    %
    \includegraphics[width=12cm]{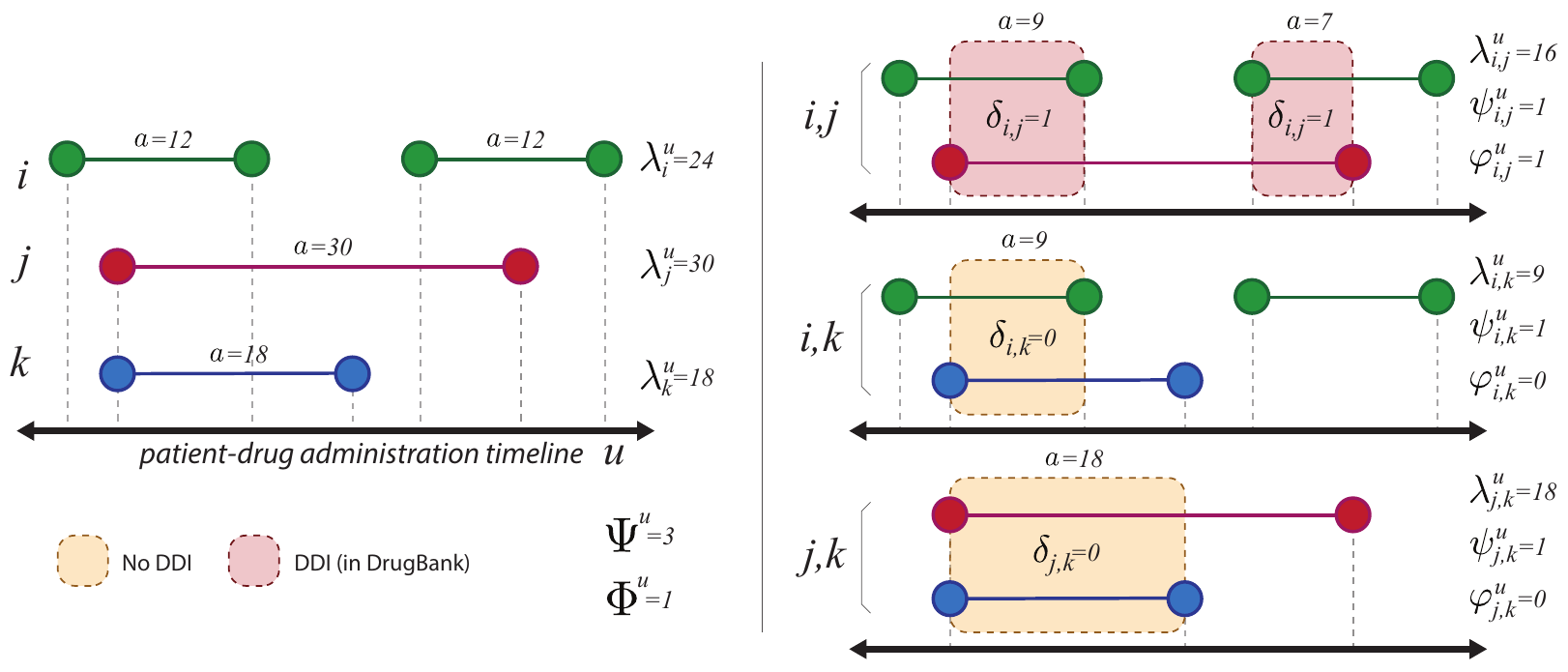}
    \caption{
        \textbf{Diagram of co-administration and interaction computation}.
        \textbf{A}. A hypothetical patient-drug dispensing timeline with three drugs ($i$,$j$ \& $k$). Drug administration length ($a$, in days, $n$) are shown for each dispensation.
        \textbf{B}. The three possible pairwise comparisons $(i,j)$, $(i,k)$ and $(j,k)$ between the dispensed drugs are shown with their co-administration overlap marked with either an orange (no known DDI) or red (known DDI) background.
    }
    \label{fig:timeline}
\end{figure}

A drug interaction between a pair of drugs is measured if both drugs were concomitantly administered \emph{and} the pair is identified as a known DDI in the 2011 version of \textit{DrugBank}, an open-source drug database containing DDI information \cite{DrugBank}.
Figure \ref{fig:timeline} displays a co-administration timeline example.
More formally (see also Table \ref{table:notation} and Supplemental Note \ref{ch:SI:notation}), let us denote patients by $u \in U$ and drugs by  $i,j \in D$ ($|D| = 122$); $U_i \in U$ is the subset of users who were dispensed drug $i$, $D^{u} \subseteq D$ is the subset of drugs administered to patient $u$, and $\nu^u \equiv |D^{u}|$ is the number of distinct drugs dispensed to patient $u$. Patients can be administered a drug $i$ multiple times in the observation period, therefore
$A^u_i \equiv \{ a_n^{i,u} \}$ denotes the set of distinct administration intervals $a$ of drug $i$ to patient $u$, where $a \in \mathbb{N}$ is measured in days ($n$). $\alpha^u_i = |A^u_i|$ and $\lambda^{u}_{i} = \sum_n a_n^{i,u}$ denote the number of times and total number of days drug $i$ was administered to patient $u$, respectively.

%
% Figure "img-graph-timeline" goes here
%
% Figure 6

Similarly, $\alpha^u_{i,j}$ and $\lambda^{u}_{i,j}$ denote the number of times and total number of days (\textit{co-administration length}) drugs $i$ and $j$ were co-administered to patient $u$, respectively (see Supplementary Information \S\ref{ch:SI:computation} for more details of co-administration measurement).
To identify the \textit{co-administration} of drug pair $(i,j)$ to patient $u$  we define a Boolean variable $\psi^{u}_{i,j} \in \{0,1\}$ as:

\begin{equation}
    \psi^{u}_{i,j} = \big( \lambda^{u}_{i,j} > 0 \big) 
\end{equation}

\noindent a logical variable measuring whether patient $u$ co-administered drug pair $(i,j)$ for at least one day.
Next, we define a symmetrical binary map $\Delta : D \times D \to \{0,1\}$ to indicate whether drug pair $(i,j) \in D \times D$ is ($\delta_{i,j}=1$) a known DDI in \textit{DrugBank}, or not ($\delta_{i,j}=0$).
Thus, to flag the co-administration of a \textit{known drug interaction}
$(i,j)$ to patient $u$ we similarly define a Boolean variable $\varphi^{u}_{i,j} \in \{0,1\}$ as:

\begin{equation}
    \varphi^{u}_{i,j} = \big( \psi^{u}_{i,j} = 1  \land \delta_{i,j}=1 \big).
\end{equation}

For each DDI pair observed, literature references and a severity score $s \in \{major, moderate, minor, n/a\}$ were retrieved from \emph{Drugs.com} \cite{Drugs.com}.
From these values, other quantities and sets are computed per patient $u$, drug $i$ or drug pair $(i,j)$ as listed in Table \ref{table:notation}.

%
% Table "Co-admin / interaction quantities" goes here
%
\begin{table}[h!]
    \centering
    \small
    {\renewcommand\arraystretch{1.25}
    \begin{tabular}{l|l}
        \toprule
quantity notation & number of \\
\midrule
$\nu^{u} \equiv |D^{u}|$ & distinct drugs dispensed to patient $u$.\\
$\Psi^{u} = \sum\limits_{i,j \in D^{u}} \psi^{u}_{i,j}$ & co-administrations to patient $u$. \\
$\Psi_{i,j} = \sum\limits_{u \in U} \psi^{u}_{i,j}$ & co-administrations of drug pair $(i,j)$ to all patients. \\
$\Phi^{u} = \sum\limits_{i,j \in D^{u}} \varphi^{u}_{i,j}$ & co-administrations of known DDI pairs to patient $u$. \\
$\Phi_{i,j} = \sum\limits_{u \in U} \varphi^{u}_{i,j}$ & \multicolumn{1}{p{8cm}}{co-administrations of known DDI pair $(i,j)$ to all patients.} \\

\midrule
subset notation  & subset of patients \\
\midrule

$U^{\nu>x} = \{ u \in U : \nu^{u} > x \}$ & \multicolumn{1}{p{7cm}}{who had at least $x \in \mathbb{N}$ drug administrations.} \\
$U^{\Psi} = \{ u \in U : \Psi^{u} > 0 \}$ & \multicolumn{1}{p{7cm}}{who had at least $1$ co-administration.} \\
$U^{\Psi}_{i,j} = \{ u \in U : \psi^{u}_{i,j} = 1 \}$ & \multicolumn{1}{p{7cm}}{who co-administered drug pair $(i,j)$.} \\
$U^{\Phi} = \{ u \in U : \Phi^{u} > 0 \}$ & \multicolumn{1}{p{7cm}}{who had at least $1$ known DDI.} \\
$U^{\Phi}_{i,j} = \{ u \in U : \varphi^{u}_{i,j} = 1 \}$ & \multicolumn{1}{p{7cm}}{who co-administered known DDI pair $(i,j)$.} \\
$U^{g} = \{ u \in U : gender(u)=g \} , \; g \in \{ \text{M},\text{F} \} $ & \multicolumn{1}{p{7cm}}{per gender.} \\
$U^{[y_1,y_2]} = \{ u \in U : age(u) \in [y_1,y_2] \}, y_1,y_2 \in \mathbb{N}$ & \multicolumn{1}{p{7cm}}{per age bracket.} \\ 
$U^{N}  =  \{ u \in U : neighborhood(u) \in N \}, N \in \mathbb{N}$ & \multicolumn{1}{p{7cm}}{per neighborhood.} \\
$U^{E}  =  \{ u \in U : education(u) \geq E \} , \; E \in \mathbb{N}$ & \multicolumn{1}{p{7cm}}{per education level. $U^{E=\varnothing}$ is the subset of patients who did not report their education level.} \\
        \bottomrule
        \multicolumn{2}{l}{From these subsets we also denote their possible intersections by combining the appropriate sub and superscripts.}
    \end{tabular}}
    \caption{
        Co-administration and interaction quantities and subsets used throughout the analysis.
    }
    \label{table:notation}
\end{table}

The drug pairs $(i,j)$ with the largest ``footprint'' in the population, are the pairs that maximize $|U_{i,j}^{\Psi}|$.
Out of these most co-administered pairs, we are naturally most interested in those that are known DDI and thus maximize $|U_{i,j}^{\Phi}|$.
A normalized version of this measure is computed as

\begin{equation}
    \gamma_{i,j}^{\Phi} = \frac{|U_{i,j}^{\Phi}|}{ |U_{i}| },
    \label{eq:gamma_ij}
\end{equation}

\noindent which conditions the number of users co-administered known DDI pair $(i,j)$ on the number of users that are administered drug $i$. This measure is not symmetrical:  $\gamma_{i,j}^{\Phi} \neq  \gamma_{j,i}^{\Phi}$. Maximizing it yields DDI pairs $(i,j)$ that tend to be co-administered to patients who are administered either $i$ or $j$ independently; see Supplementary Table \ref{table:SI:top-20-ranked-gamma} for top 20 such DDI pairs.

Another facet of the DDI phenomenon we can observe is related to the co-administration length of drug pairs ($\lambda^{u}_{i,j}$). 
A normalized version is computed as: $\tau^{u}_{i,j} = \lambda^{u}_{i,j} / (\lambda^{u}_{i} + \lambda^{u}_{j} - \lambda^{u}_{i,j})$, where $\tau \in [0,1]$.
This symmetric proximity measure \cite{Simas:2015} allows us to distinguish drug pairs that tend to be co-administered to patient $u$ only simultaneously ($\tau^{u}_{i,j} \to 1$), or with small temporal overlap ($\tau^{u}_{i,j} \to 0$). 
A normalized measure for the entire patient population is then computed as:

\begin{equation}
    \tau^{\Psi}_{i,j} = \frac{ \displaystyle\sum_{u \in U_{i,j}^{\Psi}} \tau^{u}_{i,j} }{ |U_{i,j}^{\Psi}| }
    \label{eq:tau_ij}
\end{equation}

\noindent This proximity measure defines a weighted graph $T^{\Psi}$ \cite{Simas:2015} on set $D$; the graph's edges, $\tau^{\Psi}_{i,j} \in [0,1]$, link drugs that were co-administered in the patient population.
$\tau^{\Psi}_{i,j}$ is larger when drug pairs $(i,j)$ tend to be co-administered when either $i$ or $j$ is administered (correlated), and smaller otherwise (independent).
Therefore, $\tau^{\Psi}_{i,j}$ is a measure of \textit{the strength of drug association} in the data for drug pairs $(i,j)$;
high values can pick drug pairs dispensed together for known comorbidities, which physicians should be aware of, as well as for unknown cormobidities (especially involving distinct specialists prescribing drugs independently). Since we do not know the underlying comorbidities, we cannot separate the two cases with this dataset.
However, to focus on the DDI phenomenon (for known and unknown comorbidity), we obtain a subgraph $T^{\Phi}$, restricted to known DDI pairs by computing $\tau^{\Phi}_{i,j} = \tau_{i,j}^{\Psi} . \delta_{i,j}$; thus,  $T^{\Phi}$ is a weighted version of $\Delta$.

%
% Methods: Gender Risk
%
\subsection*{Gender Risk}
\label{ch:methods:gender-risk}

The \emph{relative risk of co-administration} ($RRC$) for women is computed as the ratio of the conditional probabilities of patients being dispensed at least one pair of drugs concomitantly, given gender: 

\begin{equation}
    RRC^{\text{F}} = \frac{ P(\Psi^u>0 \, | \, u \in U^{\text{F}}) }{P(\Psi^u>0 \, | \, u \in U^{\text{M}}) }
    = \frac{ |U^{\Psi,\text{F}} \, | \, / |U^{\text{F}}| }{ |U^{\Psi,\text{M}} \, | \, / |U^{\text{M}}| }
\label{eq:RRCg}
\end{equation}

\noindent Naturally, the same risk for males is computed as $RRC^{\text{M}} = 1/RRC^{\text{F}}$. 
Similarly, we also computed the \textit{relative risk of interaction} ($RRI$) for women as:

\begin{equation}
    RRI^{\text{F}} = \frac{ P(\Phi^u >0 \, | \, u \in U^{\text{F}}) }{ P(\Phi^u >0 \, | \, u \in U^{\text{M}}) } 
    = \frac{ |U^{\Phi,\text{F}} \, | \, / |U^{\text{F}}| }{ |U^{\Phi,\text{M}} \, | \, / |U^{\text{M}}| }
\label{eq:RRIg}
\end{equation}

\noindent with $RRI^{\text{M}} = 1/RRI^{\text{F}}$.

%
% Methods: DDI Network
%
\subsection*{DDI Network}

The DDI Network is a weighted version of graph $\Delta$ where edge weights between drugs $i,j$ (nodes in graph) are the values $\tau^{\Phi}_{i,j}$ obtained from eq. \ref{eq:tau_ij}---yielding a proximity between drug pairs according to their co-occurrence in DDI co-administrations when either drug is administered (a symmetrical measure of strength of association/correlation \cite{Simas:2015}).
Node size represents the \textit{probability of interaction} for drug $i$:

\begin{equation}
    PI(i) =  \frac{\sum_{j} \Phi_{i,j}}{\sum_{j} \Psi_{i,j}}
    \label{eq:PI} 
\end{equation}

\noindent which denotes the propensity of drug $i$ to be involved in a DDI with all drugs it is co-administered with in the data (see Supplementary Table \ref{table:SI:ddi-network-louvain} for values); larger nodes thus identify more dangerous drugs in the sense that they most contribute to potential ADR from DDI in our data.

To better grasp gender differences in the DDI phenomenon, edges are colored according to the \textit{relative risk of drug pair interaction for each gender}:
$RRI^{g}_{i,j}$ where $g \in \{\text{M},\text{F}\}$.
These quantities are computed for each DDI pair $(i,j)$ via eq. \ref{eq:RRIg}, but using $\Phi_{i,j}^u$ (number of co-administrations of known DDI pair $(i,j)$ to patient $u$) instead of $\Phi^u$. 
Naturally, $RRI^{F}_{i,j} = 1/RRI^{M}_{i,j}$.
If $RRI^{F}_{i,j} > 1$, the edge is colored in red with intensity proportional to $RRI^{F}_{i,j}$, otherwise the edge is colored in blue with intensity proportional to $RRI^{M}_{i,j}$ (see legend).
Therefore, increased DDI risk for women (men) is identified by darker red (blue) edges. Supplementary Tables \ref{table:SI:rrr-gender-females} and \ref{table:SI:rrr-gender-males} show the $RRI^{g}_{i,j}$ values for the top most gender imbalanced DDI pairs per gender.

For some results we remove the following contraceptive drugs: Ethinyl Estradiol, Estradiol, Norethisterone, Levonorgestrel and Estrogens Conjugated.

%
% Methods: Age Risk
%
\subsection*{Age Risk}
\label{ch:methods:age-risk}

To investigate the role of age in known DDI co-administration, we aggregated patients into age groups and computed the risk of specific age groups to be dispensed a known DDI for the amount of co-administrations observed for that age group.
Thus, a \emph{risk of interaction for age group} $[y_1,y_2]$ is calculated as

\begin{equation}
    RI^{[y_1,y_2]} = \frac{ P(\Phi^u >0 \, | \, u \in U^{[y_1,y_2]})}{P(\Psi^u >0 \, | \, u \in U^{[y_1,y_2]}) } 
    \quad ,
\label{eq:RIy} 
\end{equation}

\noindent which can be interpreted as the probability of being dispensed a known DDI given the expected number of co-administrations for a patient in a specific age range $[y_1,y_2]$.
A \emph{Risk of Co-administration for age group} $[y_1,y_2]$, $RC^{[y_1,y_2]}$, is similarly computed, but using $\nu^{u} \geq 2$---the number of patients with at least $2$ drug administrations---instead of $\Psi^u$.
This is interpreted as the probability of being concomitantly dispensed two or more drugs (co-administration), when a patient of a given age group is dispensed two or more drugs in the full observation period.
Additionally, we also parse age risk by gender by computing $RI^{[y_1,y_2],g}$ for each gender $g \in \{M,F\}$ using eq. \ref{eq:RIy}, but for users $u \in U^{[y_1,y_2],g}$. Similarly, $RC^{[y_1,y_2],g}$ is computed for the risk of co-administration per age and gender.

%
% Methods: Null Model
%
\subsection*{Null Model}
\label{ch:methods:null-model}

The null model, $H^{rnd}_{0}$, aims to capture the expected increase in $RI^y$ with age, given the observed polypharmacy and gender for each specific age group.
Thus, the model's assumption is that all drugs that were in reality dispensed in a given age group are dispensed at random with the same overall frequency of co-administration for that age group.
Specifically, for each co-administration observed in the data for an age group $[y_1,y_2]$, the null model draws random drug pairs $(i,j)$ from the set of all drugs observed for that age group, $D^{[y_1,y_2]}$.
The random drug pairs are subsequently checked for DDI status in \textit{DrugBank}, just like the original analysis.
This way, the null model has exactly the same number of co-administration occurrences for each age group and gender, but randomly shuffled drug pairs---and only the drugs dispensed for a certain age are randomly shuffled for that age group (additional details in Supplementary Note \ref{ch:SI:null-models}).

%
% Methods: Machine Learning Classifier
%
\subsection*{Machine Learning Classifiers}
\label{ch:methods:machine-learning}

We trained linear kernel Support Vector Machine (SVM) \cite{Boser:1992} and Logistic Regression (LR) \cite{Cox:1966} classifiers using stratified 4-fold cross-validation to ensure generalization performance (additional details in Supplementary Note \ref{ch:SI:classification}).
Age, gender, number of drugs ($\nu^u$) and co-administrations ($\Phi^u$) were used as demographic variables features.
In addition, all $|D|=122$ drugs in the data are used as binary features, whereby if patient $u$ was administered drug $i$ that feature is set to $1$ and to $0$ otherwise; this allows classifiers to be trained on which drugs, and drug combinations, are most likely to be involved in DDI.

The trained classifiers are compared to two ``coin-toss'' null models, one unbiased where each class has equal probability, and a biased one based on estimated class frequency. A third, more elaborate null model classifier, finds the best age cutoff for each gender, from which all patients above the cutoff age are considered as having a DDI. This last ``age-gender'' null model represents a baseline comparison of the best we could do if only gender and age were given for each patient.
To assess the performance of all classifiers, in Supplementary Note \ref{ch:SI:classification} we report several measures. Here, we focus on the Matthew's Correlation Coefficient (MCC) \cite{Matthews:1975}, which is regarded as an ideal measure of the quality of binary classification in unbalanced scenarios such as this \cite{Baldi:2000}. We also report two other measures widely used in machine learning classifier performance, the area under the receiver operating charactistic curve (AUC ROC), and the area under the precision and recall curve (AUC P/R).

Other classifiers, feature selection and cross-validation techniques can be used to increase performance, but such gains when studying the DDI phenomenon do not typically lead to substantial performance increases \cite{Kolchinsky:2015}, so such optimization is beyond the scope of this article.

%
% Code Availability
%
\section*{Code Availability Statement}

Custom python and R scripts are available in \url{github.com/rionbr/DDIBlumenau}.

%
% Data Availability
%
\section*{Data Availability Statement}

The anonymized data that support the findings of this study are available from the city of Blumenau, Brazil. Restrictions apply to the availability of these data, as they may contain information that could compromise research participant privacy through de-anonymization. The data were used under a license agreement, and so are not publicly available, but are however available from the authors upon reasonable request and with permission of the city of Blumenau. All data tables and aggregates are available in appropriate electronic form.

%
% Acknowledgments
%
\section*{Acknowledgments}

The authors would like to thank:
The LDTT team,
Ana Paula Zanette,
David Wild, 
João Carlos,
Deborah Rocha,
Jaqueline Elias,
Johan Bollen, and
Marijn ten Thij.
We are also thankful for insightful reviewer comments on earlier versions of this manuscript.
RBC was supported by CAPES Foundation, grant 18668127, Instituto Gulbenkian de Ciência (IGC), Indiana University Precision Health to Population Health (P2P) Study, and the National Institutes of Health, National Library of Medicine Program, grant 1R01LM012832-01.
LMR was partially funded by the National Institutes of Health, National Library of Medicine Program, grants 01LM011945-01 and 1R01LM012832-01, by a Fulbright Commission fellowship, and by NSF-NRT grant 1735095 ``Interdisciplinary Training in Complex Networks and Systems.''
The funders had no role in study design, data collection and analysis, decision to publish, or preparation of the manuscript.

%
% Competing interests
%
\section*{Competing Interests}

The Authors declare no Competing Financial or Non-Financial Interests.

%
% Author Contributions
%
\section*{Author Contributions}

RBC and LMR conceived the research strategy. RBC conducted the analysis. RBC, LPA and MMM acquired the data. All authors discussed the results. RBC and LMR wrote the final manuscript. All authors approved the final manuscript.

% Paper References
\printbibliography

%
% Include SI in the same file
%

\singlespacing

%
% Supplemental Information
%
\clearpage
\addcontentsline{toc}{part}{Supplemental Information} % needed to that \locatableofcontents works
\newrefsection
\begin{center}
    \LARGE
    Supplemental Information for\\
    ``City-wide Electronic Health Records Reveal Gender and Age Biases in Administration of Known Drug-Drug Interactions''\\
    \vspace{2ex}
    \small
    \textbf{Rion Brattig Correia$^{1,2,3,*}$, Luciana P. de Araújo Kohler$^{4}$, Mauro M. Mattos$^{4}$, Luis M. Rocha$^{1,3,*}$} \\
    \vspace{1ex}
    \small
    $^1$School of Informatics, Computing \& Engineering, Indiana University, Bloomington, IN 47408 USA \\
    \small
    $^2$CAPES Foundation, Ministry of Education of Brazil, Brasília, DF 70040-020, Brazil \\
    \small
    $^3$Instituto Gulbenkian de Ciência, Oeiras 2780-156, Portugal \\
    \small
    $^4$Universidade Regional de Blumenau (FURB), Blumenau, SC 89030-903, Brazil \\
    \small
    $^*$ correspondence to rocha@indiana.edu and rionbr@gmail.com
\end{center}

%
% Make all Table and Figure numbers start with 'Supplementary [Table/Figure/]'
%
%\setcounter{chapter}{1}
\setcounter{section}{0}
\setcounter{equation}{0}
\setcounter{table}{0}
\setcounter{figure}{0}
\setcounter{page}{1}
%\makeatletter
%
\renewcommand{\tablename}{Supplementary Table}
\renewcommand{\figurename}{Supplementary Figure}
%
%\renewcommand{\thesection}{S\arabic{section}}
%\renewcommand{\theequation}{S\arabic{equation}}
%\renewcommand{\thetable}{S\arabic{table}}
%\renewcommand{\thefigure}{S\arabic{figure}}
%\renewcommand{\thelstlisting}{S\arabic{lstlisting}}

%% Custom \chapter and \section
% \chapter
\titleformat{\part}
  {\LARGE\bfseries} % format
  {} % label
  {0pt} %sep
  {\LARGE}{} %beforesep %aftersep
\titlespacing*{\part}
    {0pt} %left
    {6ex} % beforesep
    {6ex} % aftersep
    %[<right>] % after

% \section
\titleformat{\section}
  {\Large\bfseries\raggedright} % format
  {Supplementary Note \thesection} % label
  {14pt} % sep
  {\Large}{} %beforesep %aftersep
\titlespacing*{\section}
    {0pt} %left
    {4ex} % beforesep
    {4ex} % aftersep
    %[<right>]

%
% Table of Contents
%
\begingroup
    \vspace{6ex}
    \etocruledstyle[2]{}
    \localtableofcontents
    \label{toc:si}
\endgroup

%
% Imports
%
\pagebreak

%\part{Supplementary Notes}

%
% Pronto
%

\section{Pronto: academia, government and patients}
\label{ch:SI:Pronto}

This section explores some important details about the health information system that made the data presented in this paper possible and tries to enlighten for the broader impact of such projects.

\subsection{The need for a city-wide HIS}

Apart from systems nationally developed for specific health attention policies---vital statistics, mortality, epidemiology, diabetes, etc---Brazil has no universal electronic health record (EHR) country-wide \cite{Paim:2011,MS:2012}. Only secondary care (specialists) or high-cost procedures are fed into federal HIS that contains user identification with their national health card (\textit{Cartão Nacional de Saúde}; CNS), even though the majority of services are performed at primary care \cite{Paim:2011}. The CNS, initiated in 1999 \cite{Cunha:2002}, was the initial step towards a unified EHR, but several difficulties were met along these now 20 years of the program \cite{Hexsel:2002, Franco:2003, Lorenzetti:2014}. Thus, it is currently not possible to follow patients across systems---specially those that only access primary care---or request to the system their medical record. At the city level, most big cities have enough funds to buy specialized, private HIS, to develop and implement EHR along with an intra-city public health development plan for its citizens. On the other hand, the vast majority of small and mid-cities (there are 5,336 cities with less than 100.000 inhabitants \cite{IBGE}) hardly have financial access to the same costly solutions, and most of the information still transits on paper. Still, the necessity to manage several primary care installations, hundreds of health professional agendas and input city-level information into federal HIS takes place.

To address this need in the city of Blumenau, southern Brazil, the municipal government and the regional university (\textit{Universidade Regional de Blumenau}; FURB) joined forces to develop their own open-source HIS to collect and store EHR for its citizens.
The system, named \emph{Pronto}, was built by the Laboratory of Technology Development and Transfer (\textit{Laboratório de Desenvolvimento e Transferência de Tecnologia}; LDTT \cite{LDTT:Pronto}), a small transdisciplinary \cite{Choi:2006} team of professors and students from diverse fields of research---such as compute science, nursing, medicine, dentistry, psychology, communication, and others---at FURB. This enterprise, bridging academia, government and private sector \cite{Etzkowitz:2010}, spun off several scientific quests \cite{Araujo:2014, Araujo:2015, Mattos:2015, LDTT:Pronto} in order to enhance the quality of life of patients in Blumenau, broadening FURB's societal impact, and enabling patients to experience outcomes of scientific research first hand.

After development and deployment, the technology was transferred to the private sector under public bidding regulations, and continues to this day to serve as Blumenau's public health care system under municipal oversight. Pronto is currently used in all health institutions throughout the city. From primary to specialized care---hospitals have their own specialized system---and drug dispensing units. The system currently maintains health professional agendas, integrated medical and dental records, and drug prescription and dispensing across more than 30 health care units.

Doctors prescribe medications by selecting drug and dosage in the electronic system. Low-cost drugs can generally be directly dispensed at the primary-care facilities, whereas specialized and higher-cost medication are distributed in three central facilities across the city. All drugs are dispensed by pharmacists who must select in \emph{Pronto} the drug and quantity to be dispensed, allowing the length of administration to be estimated. There is no pill manipulation as all drugs are dispensed in their original sealed packaging. The database also stores inventory information---in case of drug recall, for instance, patients can be contacted in regards to a specific drug lot.

Pronto runs on a custom built, decentralized database model, where each individual health unit has its own database and a master-to-master replication takes place asynchronously, a design feature due to the unreliable network infrastructure in rural areas. Included in the technical challenges faced by the developing team were also the heterogeneous data feed from multiple health professionals; the transformation from such data data into insightful knowledge to diverse stakeholders, and the constant adaptation to match city as well as state and federal regulations.

A city-wide unified EHR enables a variety of scientific research questions, but most importantly, it permits a holistic approach to public health care. Since patients may enter the system at different clinics throughout the city, the system enables their EHR to be present whenever they go. Moreover, since different health professionals have access to the EHR, health is not only seen at the individual level but also from the family social structure perspective, a government defined strategy for primary care \cite{PNAB:2012}.

Another important system design concept, in line with recent international recommendations for HIS \cite{TheLancet:2018}, is that all data models are either patient- or family-centered. This means that all health professionals feed data into a model that enables information exchange to better support decisions focused either on the individual or their family. For instance, leaving professional restriction on data privacy aside, let's say patient John D., a young teenager living with his parents initiates a fluoxetine treatment prescribed by a local physician under complains of anxiety. From their conversations on a follow up examination, the doctor decides to ask the patient to check with a psychiatrist. The psychiatrist then detects traces of acute depression. Within the system the specialist has access to the community health agent notes (a type of family health specialist who monthly surveys households around a primary care health center) who, couple weeks back, checked on the family and reported that the household provider lost his/her job. With better social and medical characterization of the problem, all three professionals can now provide a more accurate, personalized, and systematic treatment to the teenager. Furthermore, caring for the family well-being as a whole, and the long-term health of other family members, the physician requests that a social worker be included in the case.
This example, albeit simplistic, demonstrates the potential of a holistic approach to public health, focused on prevention, and enabled by a city-wide EHR.

\subsection{Drug dispensation in Blumenau}
\label{ch:SI:Pronto:dispensation}

\begin{figure}[!h]
    \begin{center}
    % UPDATED: 2017-01-10
    % FILE: plot_map_brasil.py
    \includegraphics[width=.39\textwidth]{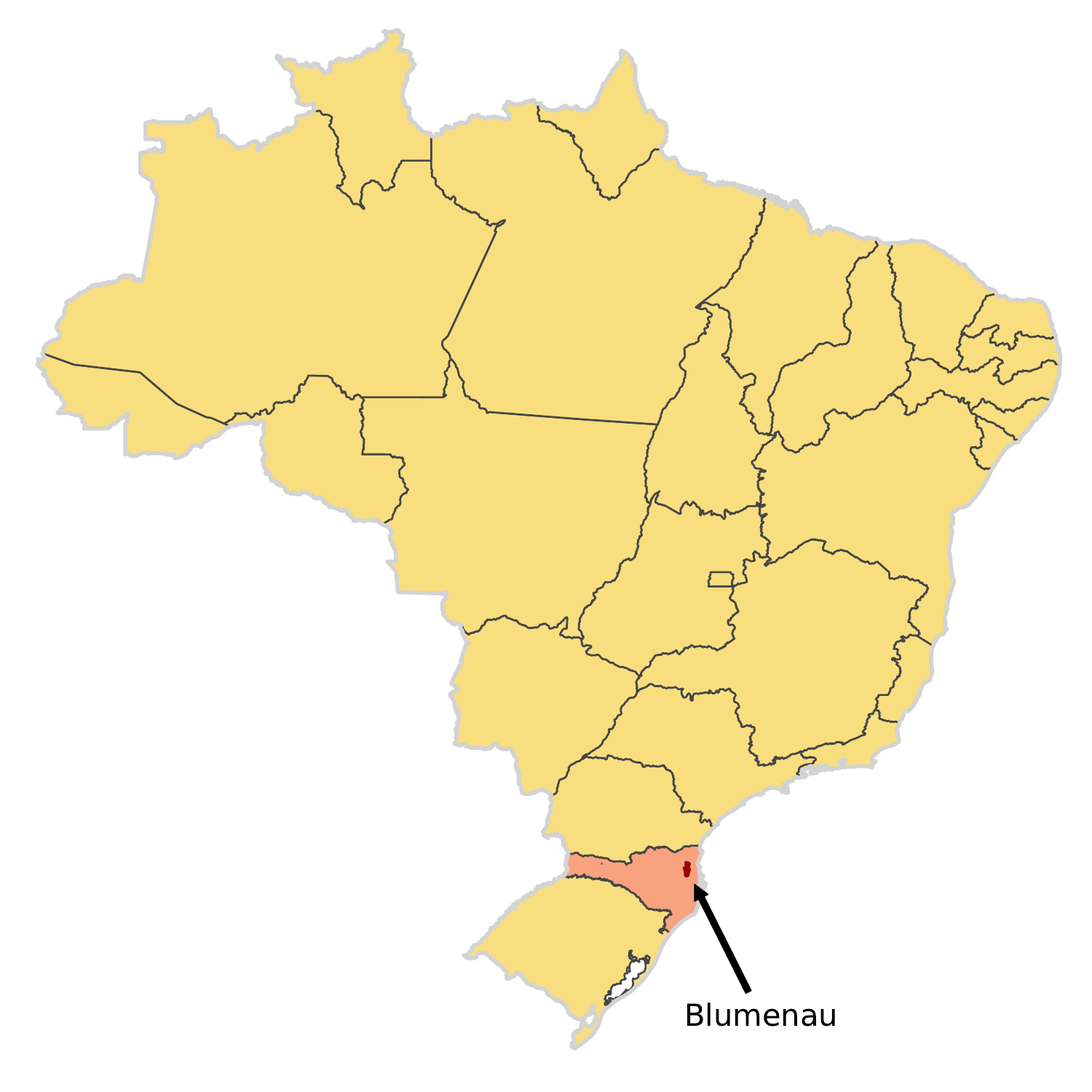}
    % UPDATED: 2017-01-10
    % FILE: plot_map_blumenau.py
    \includegraphics[width=.59\textwidth]{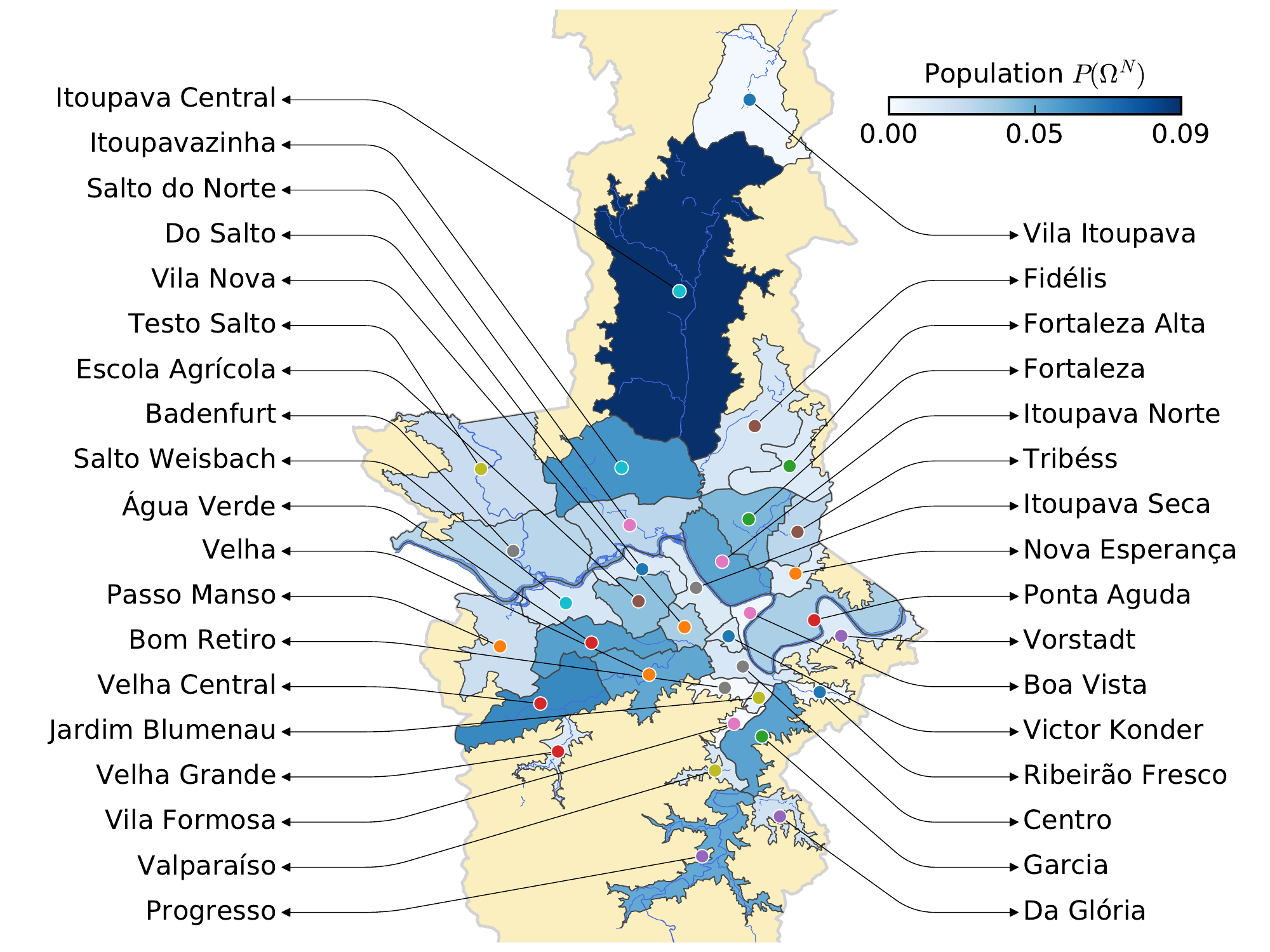}
    \caption{
        (left) Political map of Brazil with state borders. Arrow denotes city of \textit{Blumenau} in the state of \textit{Santa Catarina}.
        (right) Political map of Blumenau with neighborhoods ($N$) annotated and mapped to city population, $P(\Omega^{N})$.
        Cartographic shapes from IBGE \cite{IBGE}.
    }
    \label{fig:SI:Blumenau}
    \end{center}
\end{figure}

In order to provide readers some geographical context, Supplementary Figure \ref{fig:SI:Blumenau} shows the location of Blumenau in Brazil as well as the city neighborhoods with their individual population density.

The monthly drug dispensation in the city of Blumenau can be seen in Supplemental Figure \ref{fig:SI:drug-dispensation}.
We conjecture that the smaller number of dispensed medication during summer months (Dec-Feb) are due to a difference portion of the city population taking mandatory 30-days vacations yearly, which are usually split in two 10-days vacations during the summer months, and another 10-days during the winter months. The Atlantic Ocean coast, only a 40 minutes drive east, is a common destination for Blumenau citizens on weekends and holidays. Carnival (\textit{Carnaval}), which is usually held at the end of February, also draws many citizens for a 1 week vacation on the coast.

\begin{figure}
    \centering
    % UPDATED: 2017-01-16
    % FILE: plot_drugs_month_temp.py
    \includegraphics[width=8cm]{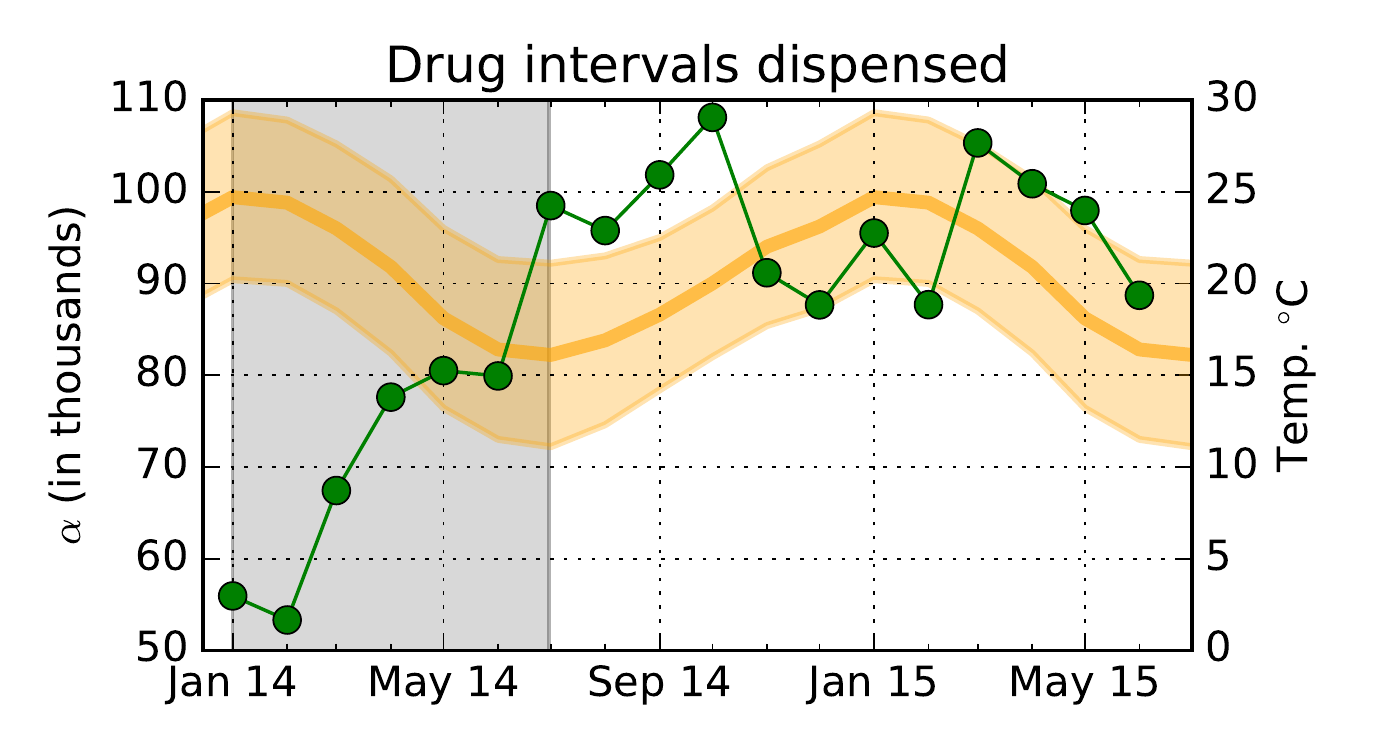}
    \caption{
        Total number (in thousands) of drug intervals dispensed ($\alpha$) monthly in the city of Blumenau.
        Orange fill shows average temperature range in Blumenau (in $^{\circ}$C).
        There is no correlation (0.06) between drug dispensation (in non shaded area) and average temperature in the same period.
        Grey area shows months in which the \emph{Pronto} HIS was under field deployment.}
    \label{fig:SI:drug-dispensation}
\end{figure}

Since Brazil also has private health care and pharmaceutical systems, patients of the public system are often thought to be from lower economical classes, a hypothesis we investigated.

Indeed, the proportion of \textit{Pronto} patients for most age brackets in the four richest neighborhoods---namely \textit{Jardim Blumenau}, \textit{Bom Retiro}, \textit{Victor Konder} and \textit{Vila Formosa}---are significantly smaller than in other neighborhoods(\textit{t}-test, $p<3^{-20}$). This strongly suggests that patients from the richest neighborhoods use the public drug dispensation system much less than equivalent groups from other areas (see Supplemental Figure \ref{fig:SI:age-bairro-heat}).

The only exception to this pattern was found for females age 45-74 from \textit{Bom Retiro} and \textit{Victor Konder} (2\textsuperscript{nd} and 3\textsuperscript{rd} richest neighborhoods, respectively), who, while using the system less than the same group in other neighborhoods, do use it significantly more than those from the richest neighborhood, \textit{Jardim Blumenau} (\textit{t}-test, $p<2^{-6}$). This suggests that these two higher-income neighborhoods have a population of older women who uses the public health care system. This may be an interesting phenomenon warranting further sociological studies.

\begin{figure}
    \centering
    % UPDATED: 2017-01-16
    % FILE: plot_age_hood_income.py
    \includegraphics[width=10cm]{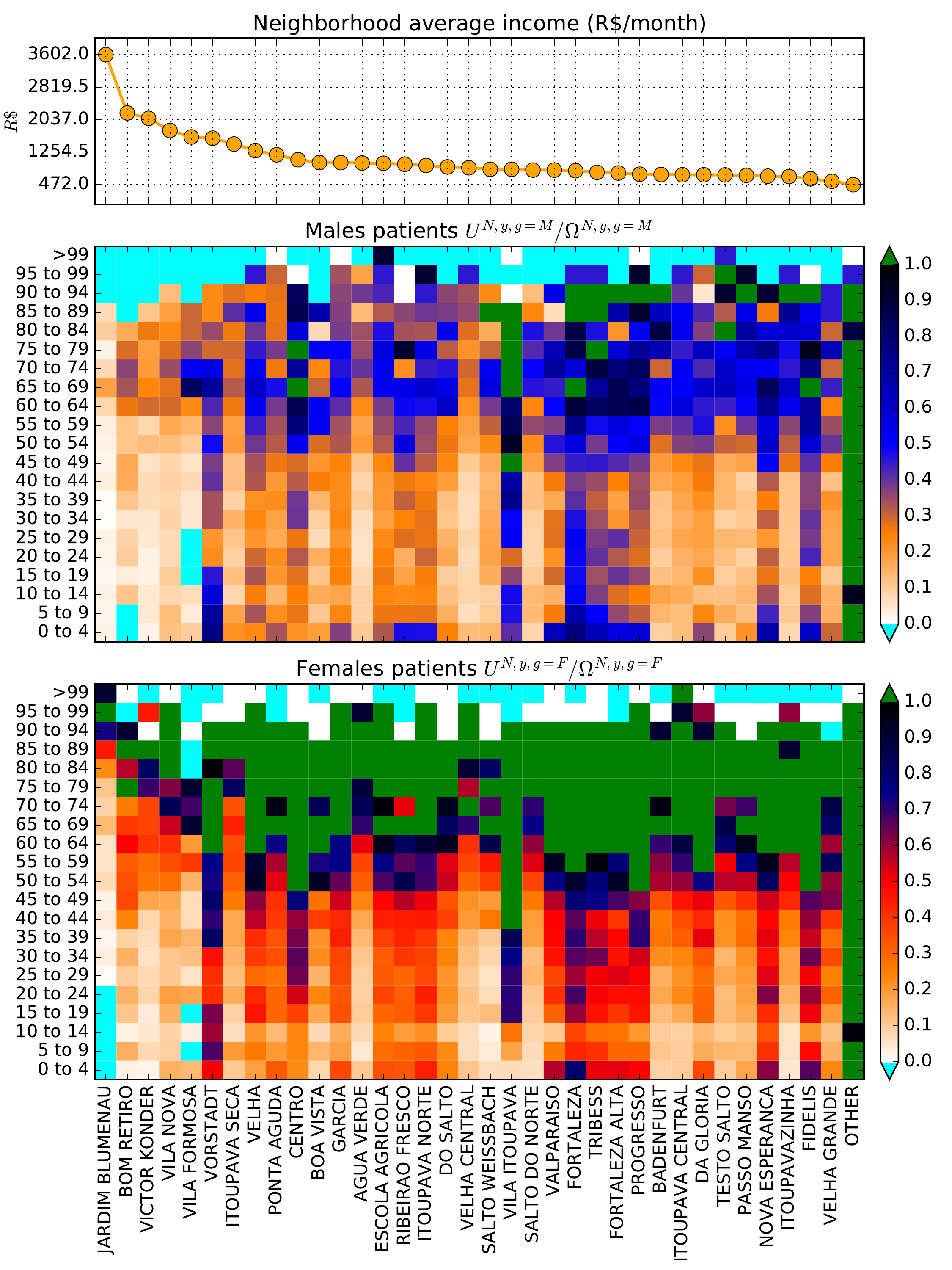}
    \caption{
        \textbf{Top.} Neighborhood average income in Brazilian Reais (R\$) \cite{IBGE}.
        \textbf{Middle \& bottom.} Age-neighborhood bins of male (middle; $U^{N,y,g=M} / \Omega^{N,y,g=M}$) and female (bottom; $U^{N,y,g=F} / \Omega^{N,y,g=F}$) patients registered in \textit{Pronto} with at least one drug dispensed and matched to DrugBank.
        Each bin is a probability-like value of patients normalized by official census population data collected and defined by IBGE \cite{IBGE}.
        Green bins represent values above 1, meaning our data has more patients than IBGE\cite{IBGE} census data.
        Conversely, cyan bins represent values where our data contains no patient.
    }
    \label{fig:SI:age-bairro-heat}
\end{figure}

\subsection{Patient education}
\label{ch:SI:Pronto:patient-education}

To place the education numbers acquired via \emph{Pronto} system in perspective, we gathered data from the Atlas Brasil Blumenau\footnote{\url{http://atlasbrasil.org.br/2013/pt/perfil_m/blumenau_sc}}, a United Nations Program for the Development of Brazil (PNUD).

In 2010 the city of Blumenau reported that the proportion of children age 5-6 in school was 88.41\%.
For the same year the proportion of children age 11-13 attending the last years of elementary school was 90.41\%.
The proportion of teenagers age 15-17 having completed elementary school was 72,34\%.
And the proportion of young adults age 18-20 who completed high school was 51.38\%.
Nationally these number were 91.12\%, 84,86\%, 57.24\% and 41.01\%, respectively.
The average length of study for children in school age was 10.81 years for Blumenau and 9.97 for the state of Santa Catarina.
The number of adults, age 18 or older, who completed elementary school was 65.88\% for Blumenau and 54.92\% for the state.
Considering adults age 25 or older: 2,13\% were illiterate, 61,55\% completed elementary school, 41,22\% completed high school and 15,49\%, completed college.
Nationally, these proportions are 11,82\%, 50,75\%, 35,83\% and 11,27\%, respectively.

Below is the self-reported education distribution for unique patients of \textit{Pronto}.
Education level is requested upon registration or profile update and no documents are required. However, staff in health centers are trained to retrieve the best response from patients without their embarrassment---by displaying a card with enumerated answers asking them to respond the according letter.

\begin{table}
    \centering
    \scriptsize
    \begin{tabular}{l|r|r r|r r}
        \toprule
$E$                     &    $U^{E}$ &     \% &  Ac.  \% &     \% &   Ac. \% \\
\midrule
% UPDATED: 2017-01-16
% FILE: display_stats.py
Cant read/write        &       4,720 & 0.0356 &   0.0356 & 0.0773 &   0.0773 \\
Can read/write a note  &       3,104 & 0.0234 &   0.0590 & 0.0508 &   0.1281 \\
Incomplete elementary  &      28,557 & 0.2152 &   0.2741 & 0.4677 &   0.5958 \\
Complete elementary    &       7,516 & 0.0566 &   0.3307 & 0.1231 &   0.7189 \\
Incomplete high school &       4,650 & 0.0350 &   0.3658 & 0.0762 &   0.7951 \\
Complete high school   &       8,797 & 0.0663 &   0.4321 & 0.1441 &   0.9391 \\
Incomplete college     &       1,654 & 0.0125 &   0.4445 & 0.0271 &   0.9662 \\
Complete college       &       1,823 & 0.0137 &   0.4583 & 0.0299 &   0.9961 \\
Espec./Residency       &         192 & 0.0014 &   0.4597 & 0.0031 &   0.9992 \\
Masters                &          26 & 0.0002 &   0.4599 & 0.0004 &   0.9997 \\
Doctoral               &          21 & 0.0002 &   0.4601 & 0.0003 &   1.0000 \\
\hline
Not reported           &      71,662 & 0.5399 &   1.0000 &        &          \\
\midrule
Total                  &     132,722 & 1.0000 &          &        &          \\
        \bottomrule
    \end{tabular}
    \label{table:SI:education-all}
    \caption{Education level of \emph{Pronto} patients}
\end{table}

\begin{table}
    \centering
    \scriptsize
    \begin{tabular}{l|r|r r|r r}
        \toprule
$E$                     &    $U^{E}$ &     \% &  Ac.  \% &     \% &   Ac. \% \\
\midrule
% UPDATED: 2017-01-16
% FILE: display_stats.py
Cant read/write        &       1,245 & 0.0134 &   0.0134 & 0.0257 &   0.0257 \\
Can read/write a note  &       2,552 & 0.0274 &   0.0408 & 0.0528 &   0.0785 \\
Incomplete elementary  &      23,983 & 0.2577 &   0.2985 & 0.4957 &   0.5742 \\
Complete elementary    &       6,733 & 0.0723 &   0.3708 & 0.1392 &   0.7134 \\
Incomplete high school &       3,126 & 0.0336 &   0.4044 & 0.0646 &   0.7780 \\
Complete high school   &       7,544 & 0.0811 &   0.4855 & 0.1559 &   0.9340 \\
Incomplete college     &       1,233 & 0.0132 &   0.4987 & 0.0255 &   0.9594 \\
Complete college       &       1,732 & 0.0186 &   0.5173 & 0.0358 &   0.9952 \\
Espec./Residency       &         187 & 0.0020 &   0.5194 & 0.0039 &   0.9991 \\
Masters                &          25 & 0.0003 &   0.5196 & 0.0005 &   0.9996 \\
Doctoral               &          18 & 0.0002 &   0.5198 & 0.0004 &   1.0000 \\
\hline
Not reported           &     44,690  & 0.4802 &   1.0000 &        &          \\
\midrule
Total                  &      93,068 & 1.0000 &          &        &          \\
        \bottomrule
    \end{tabular}
    \label{table:SI:education>age24}
    \caption{Education level of \emph{Pronto} patients age 25 or older}
\end{table}

%
% Computation
%
\section{Computation details}
\label{ch:SI:computation}

This section describes variables used in the main manuscript. It also details computations in order to facilitate replication. A quick symbol reference can be seen in Supplementary Note \ref{ch:SI:notation}. All computations were done in python using custom built scripts.

In our analysis, patients are denoted by $u \in U$ and drugs by $i,j \in D$; $U_i \in U$ is the subset of users who were dispensed drug $i$, $D^{u} \subseteq D$ is the subset of drugs dispensed to patient $u$, and $\nu^u \equiv |D^{u}|$ is the number of distinct drugs dispensed to patient $u$.
Drugs are dispensed to patients in administration intervals $a = \left(i, t_s, t_f\right)$, where $t_{s}$ and $t_{f}$ are the start and end times (in days; $t \in \mathbb{N}$) of drug administration, and $\left(t_{f}-t_{s}\right)$ represents the total length of administration, respectively.
The total number of drug $i$ intervals dispensed to patient $u$ is $\alpha^{u}_{i} = |A^{u}_{i}|$, where $A^{u}_{i} \equiv \{ a^{i,u}_{n} \}$ is the set of administration intervals for patient $u$ of drug $i$ in the data, with $n=1,\dots,\alpha^{u}_{i}$.

\textbf{Administration length.}
The total number of days patient $u$ administered drug $i$ (possibly over $n$ distinct dispensations) is then computed as

\begin{equation}
    \lambda^{u}_{i} = \sum_{1}^{n} a^{i,u}_{n}
    \quad .
    \label{eq:SI:len_i}
\end{equation}

\textbf{Co-administration length.}
For each drug pair $(i,j)$ administered to patient $u, \forall i,j \in D^u$, we identify the possible length of administration overlap between administrations of both drugs, $A^{u}_{i} \equiv \{ a^{i,u}_{n} \}$ and $A^{u}_{j} \equiv \{ a^{j,u}_{m} \}$, assuming without loss of generality that $t_{s,n} \leq t_{s,m}$, as

%. This is done by computing the \emph{number of co-administrations} to patient $u$, that is, the  number of times the drug pair $(i,j)$ was dispensed concomitantly to patient $u$ as

\begin{equation}
    \lambda^{u}_{i,j} = \sum_{\substack{a_n \in A^u_i \\ a_m \in A^u_j}}
    \begin{cases}
        \left(t_{f,n} - t_{s,m}\right) \; , & \text{iff } \left(t_{f,n} < f_{f,m}\right) \\
        \left(t_{f,m} - t_{s,m}\right) \, , & \text{otherwise}
    \end{cases}
    \quad .
    \label{eq:SI:len_ij}
\end{equation}

\textbf{Co-administrations.} To be able to discriminate patients with a specific co-administration, and to compute how many were prescribed such co-administration, we define 

\begin{equation}
    \psi^{u}_{i,j} = \big( \lambda^{u}_{i,j} > 0 \big)
    \quad ,
\end{equation}

\noindent a logical variable measuring whether patient $u$ had at least one day of co-administration between drug pair $(i,j)$; $\psi^{u}_{i,j} \in \{0,1\}$.
Then, the total number of co-administrations, per patient $u$ or drug pair $(i,j)$ is calculated as

\begin{equation}
    \Psi^{u} = \sum_{i,j \in D^{u}} \psi^{u}_{i,j} \equiv |U^{\Psi}|
    \quad , \quad
    \Psi_{i,j} = \sum_{u \in U} \psi^{u}_{i,j} \equiv |U^{\Psi}_{i,j}|
    \quad .
\end{equation}

\textbf{Interactions.} 
Next, we define a symmetrical binary map, also known as a symmetrical graph, $\Delta : D \times D \to \{0,1\}$ on set $D$ indicating if a drug pair $(i,j) \in D \times D$ has ($\delta_{i,j}=1$) a known DDI in DrugBank, or not ($\delta_{i,j}=0$).
%Therefore, a co-administration of drug pair $(i,j)$ known to be a DDI is observed in the data for patient $u$, is computed as:
Then, to discriminate patients with a known DDI we define

\begin{equation}
    \varphi^{u}_{i,j} = \big( \psi^{u}_{i,j} = 1  \land \delta_{i,j}=1 \big)
    \quad ,
\end{equation}

\noindent a logical variable measuring whether patient $u$ had at least one day of co-administration between drug pair $(i,j)$ \emph{and} this drug pair is listed in DrugBank as a known DDI; $\varphi^{u}_{i,j} \in \{0,1\}$.
The total number of co-administrations of known DDI, per patient $u$ or drug pair $(i,j)$ is calculated as

\begin{equation}
    \Phi^{u} = \sum_{i,j \in D^{u}} \varphi^{u}_{i,j} \equiv |U^{\Phi}|
    \quad , \quad
    \Phi_{i,j} = \sum_{u \in U} \varphi^{u}_{i,j} \equiv |U^{\Phi}_{i,j}|
    \quad .
\end{equation}

\textbf{Normalized interactions.} To identify the drug pairs ($i,j$) with the largest ``footprint'' in the population, we compute the pairs that are most co-administered in the population: those pairs that maximize $|U^{\Psi}_{i,j}|$. Out of these, we are naturally most interested in the drug pairs that are known DDI and are most co-administered: those that maximize $|U^{\Phi}_{i,j}|$. Two asymmetrical normalized versions of this measure are computed as

\begin{equation}
    \gamma^{\Phi}_{i,j} = \frac{ |U^{\Phi}_{i,j}| }{ |U_{i}| }
    \quad , \quad
    \gamma^{\Phi}_{j,i} = \frac{ |U^{\Phi}_{i,j}| }{ |U_{j}| }
    \quad ,
\end{equation}

\noindent which conditions the number of users co-administered drug pair ($i,j$) on the number of users that are administering either drug, $i$ or $j$.

\textbf{Normalized lengths.} To obtain a normalized value of co-administration length, we also define

\begin{equation}
    \tau^{u}_{i,j} = \frac{ \lambda^{u}_{i,j} }{ \lambda^{u}_{i} + \lambda^{u}_{j} - \lambda^{u}_{i,j} }
    \quad ,
    \label{eq:SI:tau^u_ij}
\end{equation}

\noindent where $\tau^{u}_{i,j} \in [0,1]$, and can be thought of a probability---or a Jaccard measure where values indicate a proximity \cite{Rocha:MyLibraryLANL:2005, Simas:2015, Correia:2016}---of having drug pair $(i,j)$ co-administered in relation to each drug's individual length of administration, for patient $u$.

Intuitively, if patient $u$ always administers drugs $i$ and $j$ simultaneously, $\tau^{u}_{i,j} \to 1$. Conversely, drug pairs with small co-administration overlap have $\tau^{u}_{i,j} \to 0$.
A normalized measure for the entire population is computed as 

\begin{equation}
    \tau^{\Psi}_{i,j} = \frac{ \displaystyle\sum_{u \in U^{\Psi}_{i,j}} \tau^{u}_{i,j} }{ |U^{\Psi}_{i,j}| }
    \label{eq:SI:tau_ij}
    \quad ,
\end{equation}

%$T^{u}$ can be though of a per patient measure of the degree to which her/his co-administrations are known DDI.
%In contract, 
%\noindent where $\tau^{\Psi}_{i,j}$ measures the degree to which a specific drug pair $(i,j)$, known to be a DDI or not, is usually co-prescribed for all patients in the public health care system.

\noindent where this proximity measure defines a weighted graph $T^{\Phi}$ \cite{Simas:2015} on set $D$ where edges are $\tau^{\Psi}_{i,j} \in [0,1]$ and link drugs that were co-administered in the population. $\tau^{\Psi}_{i,j}$ is larger when drug pairs ($i,j$) tend to be co-administered when either $i$ or $j$ is administered (correlated), and smaller otherwise (independent).
To obtain a subgraph $T^{\Phi}$, restricted to known DDI pairs, we compute $\tau^{\Psi}_{i,j} \times \delta_{i,j}$; thus $T^{\Phi}$ is a weighted version of $\Delta$.
In practice, due to computational complexity, we only compute $\tau^{\Psi}_{i,j}$ for drug pairs known to be a DDI ($\varphi^{u}_{i,j}>0$).

\textbf{Drug classes.} For drug pairs co-administered and known to be DDI, we gathered their respective drug class hierarchy from Drugs.com \cite{Drugs.com}. In the main manuscript we used the top level of this hierarchy to distinguish different types of drugs (i.e., cardiovascular agents, hormones, etc). For example, Fluoxetine\footnote{https://www.drugs.com/fluoxetine.html} has the following hierarchy: ``Psychotherapeutic agents'', ``Antidepressants'', and ``Selective serotonin reuptake inhibitors'', where the base class used was ``Psychotherapeutic agents''.

%
% Notations
%
\clearpage
\section{Notation and symbol reference}
\label{ch:SI:notation}

For quick reference, this chapter lists symbols used in the main manuscript and supplemental information.

\begin{table}[!hbt]
    \small
    \centering
    {\renewcommand\arraystretch{1.25}
    \begin{tabular}{c|l l l}
    symbol & description \\
    \toprule
    
    $U,u$ & \multicolumn{2}{p{10cm}}{
        Set of patients $u \in U \subset \mathbb{N}$ to whom at least one drug matched to DrugBank was dispensed.} \\
    
    $D,i$ & \multicolumn{2}{p{10cm}}{
        Set of drugs $i \in D \subset \mathbb{N}$ available for dispensation in the public health care system of Blumenau; $D^{u} \subseteq D$ is the subset of drugs administered to patient $u$. } \\
    
    $A^u_i \equiv \big\{a^{i,u}_{n} \big\}, \; a_n= \left(i, t_s, t_f\right)_n $ & \multicolumn{2}{p{10cm}}{
        Set of administration intervals $a_n$. Each interval is defined as a n-tuple comprised of drug $i$ and its administration start $t_s$ (the dispensation), end time $t_f$, and administration length $t_{t} = (t_{f}-t_{s})$, where $t \in \mathbb{N}$ (in days, $n$). } \\
    
    $\alpha^{u}_{i}$ & \multicolumn{2}{p{10cm}}{
        Number of drug intervals $a_n$ (dispensations) to patient $u$. $\alpha^{u}_{i} = |A^{u}_{i}|$ } \\

    $\nu^{u}$ & \multicolumn{2}{p{10cm}}{
        Number of distinct drugs dispensed to patient $u$. $\nu^{u} \equiv |D^{u}|$ } \\

    $\lambda^{u}_{i}$ & \multicolumn{2}{p{10cm}}{
        Administration length (in days) of drug $i$ (across possibly multiple dispensations) for patient $u$. } \\

    $\lambda^{u}_{i,j}$ & \multicolumn{2}{p{10cm}}{
        Co-administration length (in days) of drug pair $(i,j)$ (across possibly multiple dispensations) for patient $u$. See Supplementary Equation \ref{eq:SI:len_ij} for overlap computation. } \\
    
    $\Delta, \delta_{i,j}$ & \multicolumn{2}{p{10cm}}{
        $\Delta : D \times D \to {0,1}$ is a symmetrical binary relation (symmetrical graph) on set $D$, denoting the drug pairs having ($\delta_{i,j}=1$) a known DDI in DrugBank \cite{DrugBank}, or not ($\delta_{i,j}=0$); $\delta_{i,j} = \delta_{j,i}$ , $\delta_{i,i}=0$ (no self relation).  } \\

    $\tau^{u}_{i,j}$ & \multicolumn{2}{p{10cm}}{
        Normalized length of co-administration between drugs $i$ and $j$ for patient $u$. $\tau^{u}_{i,j} \in [0,1]$ is a Jaccard measure between the number of days drug pair $(i,j)$ was co-administered (intersection) divided by the number each drug, $i$ and $j$ was administered individually (union). $\tau^{u}_{i,j}=\lambda^{u}_{i,j}/(\lambda^{u}_{i}+\lambda^{u}_{j}-\lambda^{u}_{i,j})$ } \\
    
    $\Omega$ & \multicolumn{2}{p{10cm}}{
        The total city population. Also $\Omega^{N}$ and $\Omega^{y,g}$ are population numbers for a specific neighborhood $N$ or for a certain age group $y$ and gender $g$, respectively.} \\
    
    \bottomrule
    \end{tabular}}
    \caption{Basic symbols used in the paper}
    \label{table:symbols-basic}
\end{table}

\begin{table}[!hbt]
    \small
    \centering
    {\renewcommand\arraystretch{1.25}
    \begin{tabular}{c|l l l}
    symbol & description \\
    \toprule
    
    $g$ & \multicolumn{2}{p{13cm}}{
        Gender where $g \in \{\text{M},\text{F}\}$.
        } \\
    
    $y$ & \multicolumn{2}{p{13cm}}{
        Age where $y \in \mathbb{N}$. $y$ can also be grouped into age intervals (e.g., $y^{[\text{0-4}]}$, $y^{[\text{5-9}]}$, ..., $y^{[\text{95-99}]}$, $y^{[\text{>99}]}$) following \textcite{IBGE} convention.
        } \\

    $N$ & \multicolumn{2}{p{13cm}}{
        Neighborhood $N \in \mathbb{N}$ in the city of Blumenau.
        } \\

    $E$ & \multicolumn{2}{p{13cm}}{
        Education levels $N \in \mathbb{N}$ following \textcite{IBGE} convention.
        }\\
    
    $s$ & \multicolumn{2}{p{13cm}}{
        DDI severity based on Drugs.com \cite{Drugs.com} classification. $s \in \{\text{major}, \text{moderate}, \text{minor}, \text{*}, \text{none}\}$
        } \\

    \bottomrule
\end{tabular}}
\caption{Symbols used in indexing.}
\label{table:symbols-patient-quantifier}
\end{table}

\begin{table}[!hbt]
    \small
    \centering
    {\renewcommand\arraystretch{1.25}
    \begin{tabular}{c|l l l}
    symbol & description \\
    \toprule

    $\alpha$ & \multicolumn{2}{p{13cm}}{
        Total number of administration intervals dispensed. $\alpha=\sum_{u \in U, i \in D^{u}} \alpha^{u}_{i}$
        } \\

    \midrule

    $\psi^{u}_{i,j}$ & \multicolumn{2}{p{13cm}}{
        Logical variable denoting whether drug pair $(i,j)$ was co-administered by patient $u$. $\psi^{u}_{i,j}=(\lambda^{u}_{i,j}>0)$
        } \\

    $\Psi^{u}$ & \multicolumn{2}{p{13cm}}{
        Number of distinct co-administrations for patient $u$. $\Psi^{u} = \sum_{i,j \in D^{u}} \psi^{u}_{i,j}$
        } \\

    $\Psi_{i,j}$ & \multicolumn{2}{p{13cm}}{
        Number of distinct co-administrations between drug pair $(i,j)$ for all patients. $\Psi_{i,j} = \sum_{u \in D^{u}} \psi^{u}_{i,j} \equiv |U_{i,j}|$
        } \\

    $\Psi$ & \multicolumn{2}{p{13cm}}{
        Total number of drug co-administrations. $\Psi = \sum_{u} \Psi^{u} = \sum_{i,j} \Psi_{i,j}$
        } \\

    \midrule

    $\varphi^{u}_{i,j}$ & \multicolumn{2}{p{13cm}}{
        Logical variable denoting whether drug pair $(i,j)$ was co-administered to patient $u$ \emph{and} the pair is listed in DrugBank as a known DDI. $\varphi^{u}_{i,j}=(\psi^{u}_{i,j}>0 \land \delta_{i,j}=1)$
        } \\

    $\Phi^{u}$ & \multicolumn{2}{p{13cm}}{
        Number of distinct co-administration for patient $u$ known to be a DDI. $\Phi^{u} = \sum_{i,j \in D^{u}} \varphi^{u}_{i,j}$
        } \\

    $\Phi_{i,j}$ & \multicolumn{2}{p{13cm}}{
        Number of distinct co-administrations known to be a DDI between drug pair $(i,j)$, for all patients. $\Phi_{i,j} = \sum_{u \in U} \varphi^{u}_{i,j}$
        } \\

    $\Phi$ & \multicolumn{2}{p{13cm}}{
        Total number of distinct drug interaction pairs. $\Phi = \sum_{u} \Phi^{u} = \sum_{i,j} \Phi_{i,j}$
        } \\

    \midrule

    $U^{\nu>x}$ & \multicolumn{2}{p{13cm}}{
        Subset of patients who had at least $x \in \mathbb{N}$ distinct drugs administrations. $U^{\nu>x} = \{ u \in U : \nu^{u}>x \}$.
        } \\

    $U^{\Psi}$ & \multicolumn{2}{p{13cm}}{
        Subset of patients who had at least $1$ drug co-administration. $U^{\Psi}=\{ u \in U : \Psi^{u}>0 \}$.
        } \\
    
    $U^{\Psi}_{i,j}$ & \multicolumn{2}{p{13cm}}{
        Subset of patients who were co-administered drug pair ($i,j$). $U^{\Psi}_{i,j} \equiv \{u \in U : \psi^{u}_{i,j}=1 \}$.
        } \\

    $U^{\Phi}$ & \multicolumn{2}{p{13cm}}{
        Subset of patients who had at least $1$ known DDI. $U^{\Phi}=\{ u \in U : \Phi^{u}>0 \}$.
        } \\

    $U^{\Phi}_{i,j}$ & \multicolumn{2}{p{13cm}}{
        Subset of patients who were co-administered known DDI pair ($i,j$). $U^{\Phi}_{i,j} \equiv \{u \in U : \varphi^{u}_{i,j}=1 \}$.
        } \\

    %$U^{\Phi}_{s}$ & \multicolumn{2}{p{13cm}}{
    %    Subset of patients who were co-administered known DDI pair of severity ($s$). $U^{\Phi}_{i,j} \equiv \{u \in U : \varphi^{u}_{i,j}=1 \}$.
    %   } \\

    $U^{g}$ & \multicolumn{2}{p{13cm}}{
        Subset of patients per gender $g$. $U^{g} \equiv \{u \in U : gender(u) = g \}$.
        } \\

    $U^{[y_1,y_2]}$ & \multicolumn{2}{p{13cm}}{
        Subset of patients per age bracket $[y_1,y_2]$. $U^{[y_1,y_2]} \equiv \{u \in U : age(u) \in [y_1,y_2] \}$.
        } \\

    $U^{N}$ & \multicolumn{2}{p{13cm}}{
        Subset of patients per neighborhood $N$. $U^{N} \equiv \{u \in U : neighborhood(u) \in \mathbb{N} \}$.
        }\\

    $U^{E}$ & \multicolumn{2}{p{13cm}}{
        Subset of patients per education level $E$. $U^{E=\varnothing}$ is the subset of patients who did not report their education level.
        %$U^{E-}$ and $U^{E+}$ are the subset of patients who declared having at most some high school education and those that have completed high school or above, respectively.
        }\\

    \midrule
    
    $\gamma^{\Psi}_{i,j} , \gamma^{\Psi}_{j,i}$ & \multicolumn{2}{p{13cm}}{
        Normalized number of patients that were co-administering drug pair $(i,j)$. $\gamma^{\Psi}_{i,j} = { |U^{\Psi}_{i,j}| }/{ |U_{i}| }$ and $\gamma^{\Psi}_{j,i} = { |U^{\Psi}_{i,j}| }/{ |U_{j}|}$. Note $\gamma^{\Psi}_{i,j} \neq \gamma^{\Psi}_{j,i}$. } \\

    $\gamma^{\Phi}_{i,j} , \gamma^{\Phi}_{j,i}$ & \multicolumn{2}{p{13cm}}{
        Normalized number of patients that were co-administering drug pair $(i,j)$, known to be a DDI.  $\gamma^{\Phi}_{i,j} = { |U^{\Phi}_{i,j}| }/{ |U_{i}| }$ and $\gamma^{\Phi}_{j,i} = { |U^{\Phi}_{i,j}| }/{ |U_{j}|}$. Note $\gamma^{\Phi}_{i,j} \neq \gamma^{\Phi}_{j,i}$. } \\

    $\tau^{\Psi}_{i,j}$ & \multicolumn{2}{p{13cm}}{
        Normalized length of co-administration of drug pair $(i,j)$, for all patients. $\tau^{\Psi}_{i,j} = { \sum_{u \in U^{\Psi}_{i,j}} \tau^{u}_{i,j} }/{ |U_{i,j}^{\Psi}| }$ } \\
    
    $\tau^{\Phi}_{i,j}$ & \multicolumn{2}{p{13cm}}{
        Normalized length of co-administration of drug pair $(i,j)$, known to be a DDI, for all patients. $\tau^{\Phi}_{i,j} = \tau^{\Psi}_{i,j} \times \delta_{i,j}$ } \\

    \bottomrule
    \end{tabular}}
    \caption{Administration, co-administration and interaction symbols}
    \label{table:symbols-inter}
\end{table}

\clearpage

\begin{table}[!t]
    \small
    \centering
    {\renewcommand\arraystretch{1.25}
    \begin{tabular}{c|l l l}
    symbol & description \\
    \toprule
    
    $RRC^{F}$ , $RRC^{M}$ & \multicolumn{2}{p{11cm}}{
        Relative risk of co-administration for women and men, respectively. For computation details see Supplementary Note \ref{ch:SI:risk-measures}. }\\
    
    $RRI^{F}$ , $RRI^{M}$ & \multicolumn{2}{p{11cm}}{
        Relative risk of interaction for women and men, respectively. For computation details see Supplementary Note \ref{ch:SI:risk-measures}. }\\
    
    $PI(i)$ & \multicolumn{2}{p{11cm}}{
        Probability of interaction, or the propensity of a drug $i$ to be involved in a DDI with all drugs it is co-administered with in the data. $PI(i) = \sum_{j} \Phi_{i,j}/ \sum_{j} \Psi_{i,j}$ } \\

    \bottomrule
    \end{tabular}}
    \caption{Relative Risks and Probabilities}
    \label{table:symbols-rrr}
\end{table}

%
% Interactions
%
\FloatBarrier
\pagebreak
\section{Drug Interactions}
\label{ch:SI:interactions}

This section lists DDI found in the analysis. Data source for these interactions were retrieved from \url{http://wifo5-04.informatik.uni-mannheim.de/drugbank/}. This dataset was last updated in 2011 and it contains the DrugBank ID for each pair of drugs and a textual description of the interaction. The latest (version 5.0) version of the DrugBank database includes a much larger number of interaction although much of the interaction at the top of the list could not be validated from a second source, namely Drugs.com \cite{Drugs.com}. Thus we opted for a more conservative approach with fewer number of overall unique interaction that we could attribute a severity score from a second data source.

From Drugs.com \cite{Drugs.com}, the description of each severity score is as follow:
\begin{itemize}
    \item \emph{Major}: Highly clinically significant. Avoid combinations; the risk of the interaction outweighs the benefit.
    \item \emph{Moderate}: Moderately clinically significant. Usually avoid combinations; use it only under special circumstances.
    \item \emph{Minor}: Minimally clinically significant. Minimize risk; assess risk and consider an alternative drug, take steps to circumvent the interaction risk and/or institute a monitoring plan.
\end{itemize}

Note that some interactions present in DrugBrank were not found in Drugs.com. These are marked as \textit{None}.

\begin{sidewaystable}
    \centering
    \tiny
    \begin{tabular}{c r|c r|c|c c r|l ll}
        \toprule
        rank$_{\Phi}$ & $|U_{i,j}^{\Phi}|$ & $\gamma_{i,j}^{\Phi}$ & $\tau_{i,j}^{\Phi}$ & $\langle \lambda^{u}_{i,j} \rangle$ & $i$ & $j$ & $RRI^{F}_{i,j}$ & severity & interaction \\
        \midrule
% UPDATED: 2018-11-07
% FILE: display_stats_ddi.py
  1 & 5078 & 0.19 & 0.26 & 102 $\pm$  95 &          Omeprazole &          Clonazepam &   2.28 & Moderate &                      Omeprazole increases the effect of benzodiazepine \\
  2 & 2117 & 0.18 & 0.23 &  53 $\pm$  74 &                 ASA &           Ibuprofen &   1.42 &    Major &                         Ibuprofen reduces ASA cardioprotective effects \\
  3 & 1460 & 0.20 & 0.21 &  54 $\pm$  77 &            Atenolol &           Ibuprofen &   1.88 & Moderate &                             Risk of inhibition of renal prostaglandins \\
  4 & 1249 & 0.10 & 0.60 & 141 $\pm$ 124 &                 ASA &           Glyburide &   0.89 & Moderate &                    The salicylate increases the effect of sulfonylurea \\
  5 & 1190 & 0.19 & 0.45 & 127 $\pm$ 127 &       Amitriptyline &          Fluoxetine &   3.55 &    Major &             Fluoxetine increases the effect and toxicity of tricyclics \\
  6 &  999 & 0.04 & 0.27 &  87 $\pm$  86 &          Omeprazole &            Diazepam &   1.21 & Moderate &                      Omeprazole increases the effect of benzodiazepine \\
  7 &  892 & 0.14 & 0.20 &  56 $\pm$  61 &         Fluconazole &         Simvastatin &   2.63 &    Major &                              Increased risk of myopathy/rhabdomyolysis \\
  8 &  752 & 0.06 & 0.12 &  30 $\pm$  50 &                 ASA &       Dexamethasone &   1.30 & Moderate &                 The corticosteroid decreases the effect of salicylates \\
  9 &  627 & 0.10 & 0.16 &  46 $\pm$  54 &         Fluconazole &          Clonazepam &   3.40 &     None &                             Increases the effect of the benzodiazepine \\
 10 &  609 & 0.07 & 0.19 &  48 $\pm$  93 &          Prednisone &                 ASA &   0.94 & Moderate &                 The corticosteroid decreases the effect of salicylates \\
 11 &  535 & 0.07 & 0.58 & 152 $\pm$ 132 &            Atenolol &           Glyburide &   1.22 & Moderate &                The beta-blocker decreases the symptoms of hypoglycemia \\
 12 &  524 & 0.50 & 0.70 & 243 $\pm$ 188 &         Haloperidol &           Biperiden &   0.62 & Moderate &          Anticholinergic inc. risk of psychosis and tardive dyskinesia \\
 13 &  501 & 0.21 & 0.25 &  44 $\pm$  62 &         Propranolol &           Ibuprofen &   3.42 & Moderate &                             Risk of inhibition of renal prostaglandins \\
 14 &  500 & 0.15 & 0.20 &  52 $\pm$  75 &          Furosemide &           Ibuprofen &   1.93 & Moderate & NSAID decreases diuretic and antihypertensive effects of loop diuretic \\
 15 &  496 & 0.04 & 0.36 & 103 $\pm$  87 &                 ASA &          Gliclazide &   0.78 &     None &                    The salicylate increases the effect of sulfonylurea \\
 16 &  470 & 0.63 & 0.55 & 160 $\pm$ 133 &           Diltiazem &         Simvastatin &   1.27 &    Major &                       Increases the effect and toxicity of simvastatin \\
 17 &  385 & 0.59 & 0.60 & 155 $\pm$ 125 &             Digoxin &          Furosemide &   0.61 & Moderate &                        Possible electrolyte variations and arrhythmias \\
 18 &  377 & 0.03 & 0.50 & 143 $\pm$ 138 &          Fluoxetine &       Carbamazepine &   0.98 & Moderate &                                  Increases the effect of carbamazepine \\
 19 &  364 & 0.17 & 0.28 & 110 $\pm$ 106 &       Carbamazepine &         Simvastatin &   0.94 & Moderate &                                     Decreases the effect of the statin \\
 20 &  355 & 0.03 & 0.26 &  86 $\pm$  84 &          Fluoxetine &         Propranolol &   4.76 & Moderate &                      The SSRI increases the effect of the beta-blocker \\
 21 &  284 & 0.04 & 0.27 &  66 $\pm$  57 &       Levothyroxine &   Iron (II) Sulfate &   4.59 & Moderate &                             Iron decreases absorption of levothyroxine \\
 22 &  272 & 0.42 & 0.55 & 140 $\pm$ 114 &             Digoxin &      Spironolactone &   0.58 &    Minor &      Increased digoxin levels and decreased effect with spironolactone \\
 23 &  257 & 0.16 & 0.42 & 123 $\pm$ 130 &          Imipramine &          Fluoxetine &   3.08 &    Major &             Fluoxetine increases the effect and toxicity of tricyclics \\
 24 &  245 & 0.04 & 0.19 &  42 $\pm$  40 &         Fluconazole &       Amitriptyline &   4.25 & Moderate &       The imidazole increases the effect and toxicity of the tricyclic \\
 25 &  244 & 0.01 & 0.22 &  57 $\pm$  77 &       Acetaminophen &            Warfarin &   1.07 &    Minor &                       Acetaminophen increases the anticoagulant effect \\
 26 &  226 & 0.04 & 0.49 & 151 $\pm$ 145 &       Amitriptyline &       Carbamazepine &   0.99 & Moderate &                   The tricyclics increases the effect of carbamazepine \\
 27 &  222 & 0.02 & 0.47 & 148 $\pm$ 139 &          Fluoxetine &             Lithium &   1.79 &    Major &                             The SSRI increases serum levels of lithium \\
 28 &  201 & 0.03 & 0.34 & 107 $\pm$  95 &            Atenolol &          Gliclazide &   1.09 &     None &                The beta-blocker decreases the symptoms of hypoglycemia \\
 29 &  186 & 0.18 & 0.43 & 142 $\pm$ 156 &         Haloperidol &       Carbamazepine &   0.62 & Moderate &                      Carbamazepine decreases the effect of haloperidol \\
 30 &  179 & 0.08 & 0.09 &  10 $\pm$   6 &   Ethinyl Estradiol &         Amoxicillin & 126.09 & Moderate &      Anti-infectious agent could decrease effect of oral contraceptive \\
 31 &  173 & 0.27 & 0.41 & 109 $\pm$  96 &             Digoxin &          Carvedilol &   0.53 & Moderate &                          Carvedilol increases levels/effect of digoxin \\
 32 &  155 & 0.02 & 0.22 &  68 $\pm$  80 &       Amitriptyline &          Salbutamol &   2.83 & Moderate &                     The tricyclic increases the sympathomimetic effect \\
 33 &  154 & 0.02 & 0.45 & 144 $\pm$ 122 &       Levothyroxine &            Warfarin &   1.05 & Moderate &                     Thyroid hormones increase the anticoagulant effect \\
 33 &  154 & 0.01 & 0.33 &  94 $\pm$  92 &          Fluoxetine &       Nortriptyline &   2.70 &    Major &             Fluoxetine increases the effect and toxicity of tricyclics \\
 35 &  149 & 0.28 & 0.32 & 115 $\pm$ 109 &           Phenytoin &          Omeprazole &   0.80 & Moderate &                           Omeprazole increases the effect of hydantoin \\
 36 &  148 & 0.14 & 0.49 & 168 $\pm$ 160 &         Haloperidol &             Lithium &   1.31 &    Major &                      Possible extrapyramidal effects and neurotoxicity \\
 37 &  147 & 0.02 & 0.23 &  60 $\pm$  76 &            Atenolol &          Salbutamol &   1.37 & Moderate &                                                             Antagonism \\
 38 &  130 & 0.00 & 0.08 &  27 $\pm$  45 &           Ibuprofen &             Lithium &   2.08 & Moderate &                            The NSAID increases serum levels of lithium \\
 39 &  123 & 0.00 & 0.16 &  31 $\pm$  43 &           Ibuprofen &          Carvedilol &   0.88 & Moderate &                             Risk of inhibition of renal prostaglandins \\
 40 &  117 & 0.18 & 0.46 & 126 $\pm$ 127 &             Digoxin & Hydrochlorothiazide &   0.95 & Moderate &                        Possible electrolyte variations and arrhythmias \\
 41 &  116 & 0.02 & 0.14 &   9 $\pm$   7 &         Norfloxacin &   Iron (II) Sulfate &   6.14 & Moderate &                                  Formation of non-absorbable complexes \\
 42 &  103 & 0.16 & 0.43 & 113 $\pm$ 110 &             Digoxin &       Levothyroxine &   1.50 & Moderate &                   The thyroid hormones decreases the effect of digoxin \\
 42 &  103 & 0.01 & 0.23 &  76 $\pm$  80 &          Fluoxetine &          Carvedilol &   1.50 & Moderate &                      The SSRI increases the effect of the beta-blocker \\
 44 &  102 & 0.01 & 0.04 &   4 $\pm$   4 &          Diclofenac &         Alendronate &   9.61 & Moderate &                                     Increased risk of gastric toxicity \\
 45 &  101 & 0.01 & 0.25 &  92 $\pm$  81 &          Fluoxetine &            Warfarin &   1.46 & Moderate &                         The SSRI increases the effect of anticoagulant \\
 46 &   95 & 0.04 & 0.57 & 140 $\pm$ 126 &         Propranolol &           Glyburide &   1.61 & Moderate &                The beta-blocker decreases the symptoms of hypoglycemia \\
 47 &   91 & 0.01 & 0.52 & 154 $\pm$ 142 &            Atenolol &           Diltiazem &   1.19 &    Major &                                          Increased risk of bradycardia \\
 48 &   90 & 0.06 & 0.50 & 161 $\pm$ 157 &          Imipramine &       Carbamazepine &   1.35 & Moderate &                    The tricyclic increases the effect of carbamazepine \\
 49 &   89 & 0.01 & 0.17 &  36 $\pm$  32 &         Fluconazole &            Diazepam &   2.16 & Moderate &                             Increases the effect of the benzodiazepine \\
 50 &   84 & 0.01 & 0.09 &  15 $\pm$  26 &          Prednisone &   Ethinyl Estradiol &  58.79 & Moderate &            The estrogenic agent increases the effect of corticosteroid \\ 
        \bottomrule
    \end{tabular}
    \caption{
        DDI list 1-50.
        Complete list of known DDI pairs $(i,j)$ by rank of $U_{i,j}^{\Phi}$, the number of patients affects by the DDI (1\textsuperscript{st} and 2\textsuperscript{nd} columns, respectively).
        The normalized drug pair footprint in the population ($\gamma_{i,j}^{\Phi}$) as well as the normalized co-administration length ($\tau_{i,j}^{\Phi}$), are shown in columns 3 and 4, respectively.
        Mean ($\pm$ s.d.) co-administration length, $\langle \lambda^{u}_{i,j} \rangle$, is shown in column 5 (in days) for each DDI pair $(i,j)$ whose English drug names are shown in columns 6 and 7.
        The relative gender risk of DDI pair co-administration, $RRI^{F}_{i,j}$, shown in column 8.
        DDI severity classification, according to \textit{Drugs.com}, shown in column 9; DDIs or drugs not found in \textit{Drugs.com} are labeled as \textit{None} or \textit{*}, respectively.
        Drug pair interaction, according to \textit{DrugBank}, shown in column 10.
        Continues on Supplementary Table \ref{table:SI:ListDDI-51-100}.
    }
    \label{table:SI:ListDDI-1-50}
\end{sidewaystable}

\begin{sidewaystable}
    \centering
    \tiny
    \begin{tabular}{c r|c r|c|c c r|l ll}
        \toprule
        rank$_{\Phi}$ & $|U_{i,j}^{\Phi}|$ & $\gamma_{i,j}^{\Phi}$ & $\tau_{i,j}^{\Phi}$ & $\langle \lambda^{u}_{i,j} \rangle$ & $i$ & $j$ & $RRI^{F}_{i,j}$ & severity & interaction \\
        \midrule
% UPDATED: 2018-11-07
% FILE: display_stats_ddi.py
 51 &   71 & 0.13 & 0.47 & 169 $\pm$ 151 &           Phenytoin &          Fluoxetine &   0.73 & Moderate &                           Fluoxetine increases the effect of phenytoin \\
 51 &   71 & 0.01 & 0.08 &  15 $\pm$  19 &            Atenolol &           Fenoterol &   2.64 &        * &                                                             Antagonism \\
 53 &   69 & 0.02 & 0.16 &  10 $\pm$   8 &       Ciprofloxacin &   Iron (II) Sulfate &   4.18 & Moderate &                                  Formation of non-absorbable complexes \\
 54 &   63 & 0.17 & 0.35 &  33 $\pm$  28 &          Methyldopa &   Iron (II) Sulfate &  21.60 & Moderate &                      Iron decreases the absorption of dopa derivatives \\
 54 &   63 & 0.08 & 0.34 & 110 $\pm$ 118 &           Diltiazem &          Amlodipine &   1.52 & Moderate &                        Increases the effect and toxicity of amlodipine \\
 56 &   60 & 0.01 & 0.19 &  49 $\pm$  95 &          Prednisone &            Warfarin &   0.76 & Moderate &                     The corticosteroid alters the anticoagulant effect \\
 56 &   60 & 0.01 & 0.12 &  28 $\pm$  43 &       Amitriptyline &           Fenoterol &   2.83 &        * &                     The tricyclic increases the sympathomimetic effect \\
 58 &   59 & 0.01 & 0.15 &  34 $\pm$  34 &         Fluconazole &       Carbamazepine &   1.03 & Moderate &                                  Increases the effect of carbamazepine \\
 59 &   58 & 0.05 & 0.03 &   5 $\pm$  10 &      Hydrocortisone &                 ASA &   1.35 & Moderate &                 The corticosteroid decreases the effect of salicylates \\
 60 &   57 & 0.01 & 0.15 &  50 $\pm$  48 &         Fluconazole &          Imipramine &   7.37 & Moderate &       The imidazole increases the effect and toxicity of the tricyclic \\
 60 &   57 & 0.02 & 0.35 &  96 $\pm$  96 &           Glyburide &          Carvedilol &   0.73 & Moderate &                The beta-blocker decreases the symptoms of hypoglycemia \\
 62 &   52 & 0.08 & 0.49 & 118 $\pm$ 114 &             Digoxin &          Amiodarone &   0.56 &    Major &                             Amiodarone increases the effect of digoxin \\
 63 &   51 & 0.00 & 0.23 &  93 $\pm$  90 & Hydrochlorothiazide &             Lithium &   2.90 &    Major &                The thiazide diuretic increases serum levels of lithium \\
 64 &   48 & 0.00 & 0.16 &  65 $\pm$  74 &           Enalapril &             Lithium &   1.91 & Moderate &                    The ACE inhibitor increases serum levels of lithium \\
 65 &   47 & 0.04 & 0.46 & 135 $\pm$ 109 &         Allopurinol &            Warfarin &   0.19 & Moderate &                         Allopurinol increases the anticoagulant effect \\
 66 &   44 & 0.03 & 0.16 &  51 $\pm$  61 &          Imipramine &          Salbutamol &   3.19 & Moderate &                     The tricyclic increases the sympathomimetic effect \\
 67 &   43 & 0.08 & 0.40 & 144 $\pm$ 153 &           Phenytoin &            Diazepam &   0.62 & Moderate & Possible increased levels of the hydantoin, decrease of benzodiazepine \\
 68 &   41 & 0.00 & 0.21 &  82 $\pm$  74 &            Losartan &             Lithium &   4.13 & Moderate &                             Losartan increases serum levels of lithium \\
 69 &   39 & 0.04 & 0.42 &  82 $\pm$  64 &          Gliclazide &          Carvedilol &   0.67 &     None &                The beta-blocker decreases the symptoms of hypoglycemia \\
 69 &   39 & 0.14 & 0.15 &  30 $\pm$  46 &             Timolol &           Ibuprofen &   1.42 &     None &                             Risk of inhibition of renal prostaglandins \\
 69 &   39 & 0.06 & 0.42 & 130 $\pm$ 122 &       Nortriptyline &       Carbamazepine &   1.26 & Moderate &                    The tricyclic increases the effect of carbamazepine \\
 72 &   36 & 0.05 & 0.05 &  15 $\pm$  14 &       Phenobarbital &       Dexamethasone &   1.11 & Moderate &             The barbiturate decreases the effect of the corticosteroid \\
 73 &   31 & 0.01 & 0.04 &   9 $\pm$   8 &        Prednisolone &                 ASA &   2.95 & Moderate &                 The corticosteroid decreases the effect of salicylates \\
 73 &   31 & 0.01 & 0.20 &  48 $\pm$  66 &         Propranolol &          Salbutamol &   6.61 &    Major &                                                             Antagonism \\
 75 &   30 & 0.00 & 0.09 &  10 $\pm$   3 &         Clavulanate &   Ethinyl Estradiol &    inf &     None &      Anti-infectious agent could decrease effect of oral contraceptive \\
 76 &   28 & 0.04 & 0.22 &  83 $\pm$  86 &             Digoxin &            Diazepam &   1.09 & Moderate &                     The benzodiazepine increases the effect of digoxin \\
 77 &   27 & 0.01 & 0.30 &  81 $\pm$  73 &         Propranolol &          Gliclazide &   2.02 &     None &                The beta-blocker decreases the symptoms of hypoglycemia \\
 77 &   27 & 0.03 & 0.19 &  11 $\pm$   6 &         Doxycycline &   Ethinyl Estradiol &    inf & Moderate &      Anti-infectious agent could decrease effect of oral contraceptive \\
 77 &   27 & 0.05 & 0.05 &  10 $\pm$   6 &           Phenytoin &       Ciprofloxacin &   0.49 & Moderate &                           Ciprofloxacin decreases the hydantoin effect \\
 77 &   27 & 0.01 & 0.05 &   7 $\pm$   3 &       Carbamazepine &       Metronidazole &   1.68 & Moderate &                    Metronidazole increases the effect of carbamazepine \\
 77 &   27 & 0.00 & 0.07 &   8 $\pm$   7 &          Prednisone &           Estradiol &    inf & Moderate &            The estrogenic agent increases the effect of corticosteroid \\
 82 &   26 & 0.00 & 0.20 &  44 $\pm$  70 &         Fluconazole &       Nortriptyline &   5.43 & Moderate &       The imidazole increases the effect and toxicity of the tricyclic \\
 82 &   26 & 0.05 & 0.07 &  14 $\pm$  12 &           Phenytoin &       Dexamethasone &   1.13 & Moderate &          The enzyme inducer decreases the effect of the corticosteroid \\
 84 &   25 & 0.03 & 0.56 & 157 $\pm$ 136 &           Diltiazem &          Amiodarone &   1.26 &    Major &                       Increased risk of cardiotoxicity and arrhythmias \\
 85 &   24 & 0.01 & 0.17 &  15 $\pm$  10 &         Propranolol &           Fenoterol &   3.54 &        * &                                                             Antagonism \\
 85 &   24 & 0.01 & 0.29 & 100 $\pm$  85 &       Carbamazepine &            Warfarin &   0.99 & Moderate &                                     Decreases the anticoagulant effect \\
 85 &   24 & 0.00 & 0.05 &   3 $\pm$   2 &          Diclofenac &            Warfarin &   0.84 &    Major &                           The NSAID increases the anticoagulant effect \\
 85 &   24 & 0.04 & 0.15 &  29 $\pm$  45 &           Phenytoin &          Prednisone &   1.72 & Moderate &          The enzyme inducer decreases the effect of the corticosteroid \\
 89 &   23 & 0.03 & 0.47 & 152 $\pm$ 143 &           Diltiazem &         Propranolol &   2.01 &    Major &                                          Increased risk of bradycardia \\
 89 &   23 & 0.00 & 0.16 &  36 $\pm$  44 &         Fluconazole &         Haloperidol &   1.33 &    Major &         The imidazole increases the effect and toxicity of haloperidol \\
 91 &   22 & 0.05 & 0.20 &  19 $\pm$  28 &     Estrogens Conj. &          Prednisone &    inf & Moderate &            The estrogenic agent increases the effect of corticosteroid \\
 91 &   22 & 0.00 & 0.07 &   9 $\pm$   4 &       Ciprofloxacin &            Warfarin &   1.02 &    Major &                       The quinolone increases the anticoagulant effect \\
 93 &   21 & 0.01 & 0.08 &  10 $\pm$   6 &          Tobramycin &          Furosemide &   3.01 &    Major &                                                  Increased ototoxicity \\
 93 &   21 & 0.02 & 0.33 &  94 $\pm$ 116 &      Chlorpromazine &         Propranolol &   1.77 & Moderate &                                         Increased effect of both drugs \\
 95 &   19 & 0.00 & 0.07 &  26 $\pm$  35 &          Prednisone &       Phenobarbital &   1.53 & Moderate &             The barbiturate decreases the effect of the corticosteroid \\
 95 &   19 & 0.00 & 0.07 &  10 $\pm$   7 &       Ciprofloxacin &       Aminophylline &   1.21 &    Major &                     The quinolone increases the effect of theophylline \\
 97 &   18 & 0.00 & 0.13 &  33 $\pm$  44 &         Fluconazole &            Warfarin &   0.89 &    Major &                                     Increases the anticoagulant effect \\
 98 &   17 & 0.03 & 0.24 &  50 $\pm$  48 &       Nortriptyline &          Salbutamol &   1.70 & Moderate &                     The tricyclic increases the sympathomimetic effect \\
 98 &   17 & 0.01 & 0.26 & 107 $\pm$  85 &         Propranolol &       Phenobarbital &   1.01 & Moderate &       The barbiturate decreases the effect of metabolized beta-blocker \\
100 &   16 & 0.02 & 0.24 &  46 $\pm$  29 &         Haloperidol &         Propranolol &   1.56 & Moderate &                                         Increased effect of both drugs \\
        \bottomrule
    \end{tabular}
    \caption{
        DDI list 51-100.
        See Supplementary Table \ref{table:SI:ListDDI-1-50} for column description.
        Continues on Supplementary Table \ref{table:SI:ListDDI-101-150}.
    }
    \label{table:SI:ListDDI-51-100}
\end{sidewaystable}

\begin{sidewaystable}
    \centering
    \tiny
    \begin{tabular}{c r|c r|c|c c r|l ll}
        \toprule
        rank$_{\Phi}$ & $|U_{i,j}^{\Phi}|$ & $\gamma_{i,j}^{\Phi}$ & $\tau_{i,j}^{\Phi}$ & $\langle \lambda^{u}_{i,j} \rangle$ & $i$ & $j$ & $RRI^{F}_{i,j}$ & severity & interaction \\
        \midrule
% UPDATED: 2018-11-07
% FILE: display_stats_ddi.py
100 &   16 & 0.00 & 0.07 &   6 $\pm$   3 &        Azithromycin &            Warfarin &   1.18 & Moderate &                                     Increases the anticoagulant effect \\
100 &   16 & 0.00 & 0.08 &  16 $\pm$  21 &       Metronidazole &             Lithium &   4.96 & Moderate &             Metronidazole increases the effect and toxicity of lithium \\
100 &   16 & 0.03 & 0.31 &  94 $\pm$  83 &           Phenytoin &          Furosemide &   0.55 &    Minor &                       The hydantoin decreases the effect of furosemide \\
104 &   15 & 0.01 & 0.13 &  20 $\pm$  17 &          Carvedilol &           Fenoterol &   0.47 &        * &                                                             Antagonism \\
105 &   14 & 0.02 & 0.21 &   5 $\pm$   3 &         Doxycycline &         Amoxicillin &   0.53 & Moderate &                                          Possible antagonism of action \\
105 &   14 & 0.00 & 0.12 &   9 $\pm$   7 &          Furosemide &          Gentamicin &   0.71 &    Major &                                                  Increased ototoxicity \\
105 &   14 & 0.00 & 0.23 &  63 $\pm$  59 &         Fluconazole &           Phenytoin &   1.28 & Moderate &                                      Increases the effect of hydantoin \\
105 &   14 & 0.03 & 0.23 &  88 $\pm$  71 &           Phenytoin &            Warfarin &   0.94 & Moderate &                        Increased hydantoin levels and risk of bleeding \\
109 &   13 & 0.01 & 0.14 &  35 $\pm$  26 &       Carbamazepine &   Ethinyl Estradiol &    inf &    Major &     This product might cause a slight decrease of contraceptive effect \\
109 &   13 & 0.01 & 0.14 &  35 $\pm$  26 &      Levonorgestrel &       Carbamazepine &    inf &    Major &                       Carbamazepine decreases the contraceptive effect \\
109 &   13 & 0.01 & 0.56 & 122 $\pm$ 113 &         Propranolol &          Methyldopa &   8.50 &    Major &                                           Possible hypertensive crisis \\
109 &   13 & 0.05 & 0.19 &  55 $\pm$  59 &             Timolol &           Glyburide &   1.13 & Moderate &                The beta-blocker decreases the symptoms of hypoglycemia \\
113 &   12 & 0.01 & 0.18 &  72 $\pm$  98 &           Captopril &             Lithium &   1.42 & Moderate &                    The ACE inhibitor increases serum levels of lithium \\
114 &   11 & 0.01 & 0.08 &  33 $\pm$  63 &          Imipramine &           Fenoterol &   7.08 &        * &                     The tricyclic increases the sympathomimetic effect \\
114 &   11 & 0.01 & 0.24 &  57 $\pm$  46 &     Methylphenidate &       Carbamazepine &   0.07 &     None &               Carbamazepine could reduce the effect of methylphendiate \\
114 &   11 & 0.00 & 0.15 &  14 $\pm$  21 &         Norfloxacin &       Aminophylline &   7.08 & Moderate &                     The quinolone increases the effect of theophylline \\
117 &   10 & 0.02 & 0.04 &  11 $\pm$   4 &        Erythromycin &         Simvastatin &   2.83 &    Major &                   The macrolide possibly increases the statin toxicity \\
118 &    9 & 0.00 & 0.03 &   7 $\pm$   1 &        Prednisolone &       Phenobarbital &   0.89 & Moderate &             The barbiturate decreases the effect of the corticosteroid \\
118 &    9 & 0.00 & 0.29 &  62 $\pm$  46 &          Folic acid &           Phenytoin &   1.42 & Moderate &                           Folic acid decreases the levels of hydantoin \\
118 &    9 & 0.00 & 0.11 &  16 $\pm$   9 &      Metoclopramide &            Levodopa &   5.67 & Moderate &                        Levodopa decreases the effect of metoclopramide \\
118 &    9 & 0.00 & 0.29 &  72 $\pm$ 128 &       Carbamazepine &      Norethisterone &    inf &    Major &       This product may cause a slight decrease of contraceptive effect \\
118 &    9 & 0.03 & 0.15 &  51 $\pm$  91 &             Timolol &          Salbutamol &   0.89 &    Major &                                                             Antagonism \\
123 &    8 & 0.01 & 0.22 &  67 $\pm$  36 &           Diltiazem &       Carbamazepine &   0.71 &    Major &                                  Increases the effect of carbamazepine \\
123 &    8 & 0.00 & 0.04 &   6 $\pm$   2 &       Metronidazole &       Phenobarbital &   1.18 & Moderate &                  The barbiturate decreases the effect of metronidazole \\
123 &    8 & 0.02 & 0.33 &  53 $\pm$  40 &          Methyldopa &          Salbutamol &   4.96 & Moderate &                                            Increased arterial pressure \\
123 &    8 & 0.00 & 0.25 &  82 $\pm$  54 &       Carbamazepine &       Aminophylline &   0.71 & Moderate &                      Increases or decreases the effect of theophylline \\
127 &    7 & 0.01 & 0.16 &  62 $\pm$  97 &           Phenytoin &        Trimethoprim &   0.94 & Moderate &                         Trimethoprim increases the effect of hydantoin \\
127 &    7 & 0.00 & 0.23 & 122 $\pm$ 123 &         Propranolol &         Maprotiline &   4.25 &    Minor &                    Propranolol increases the serum levels of cisapride \\
127 &    7 & 0.01 & 0.11 &  25 $\pm$  15 &       Nortriptyline &           Fenoterol &   1.77 &        * &                     The tricyclic increases the sympathomimetic effect \\
127 &    7 & 0.02 & 0.09 &  13 $\pm$   8 &          Methyldopa &           Fenoterol &   4.25 &     None &                                            Increased arterial pressure \\
127 &    7 & 0.01 & 0.20 &  11 $\pm$   4 &         Doxycycline &   Iron (II) Sulfate &    inf & Moderate &                                  Formation of non-absorbable complexes \\
132 &    5 & 0.00 & 0.55 &  82 $\pm$  86 &         Propranolol &       Aminophylline &   1.06 &    Major &              Antagonism of action and increased effect of theophylline \\
132 &    5 & 0.04 & 0.45 & 221 $\pm$ 207 &    Propylthiouracil &            Warfarin &   1.06 & Moderate &   The anti-thyroid agent causes variations in the anticoagulant effect \\
132 &    5 & 0.01 & 0.09 &   9 $\pm$   4 &         Doxycycline &       Carbamazepine &   2.83 & Moderate &                 The anticonvulsant decreases the effect of doxycycline \\
132 &    5 & 0.02 & 0.22 &  34 $\pm$  26 &             Timolol &          Gliclazide &   1.06 &     None &                The beta-blocker decreases the symptoms of hypoglycemia \\
132 &    5 & 0.00 & 0.17 &  10 $\pm$   6 &        Prednisolone &   Ethinyl Estradiol &    inf & Moderate &        The estrogenic agent increases the effect of the corticosteroid \\
132 &    5 & 0.00 & 0.31 & 162 $\pm$ 120 &  Medroxyproges. Ac. &       Phenobarbital &    inf & Moderate &                    The enzyme inducer decreases the effect of hormones \\
132 &    5 & 0.01 & 0.35 & 107 $\pm$ 123 &           Phenytoin &          Amiodarone &   0.18 & Moderate &                           Amiodarone increases the effect of hydantoin \\
132 &    5 & 0.00 & 0.21 & 104 $\pm$ 153 &          Folic acid &       Phenobarbital &   2.83 & Moderate &                      Folic acid decreases the effect of anticonvulsant \\
140 &    4 & 0.00 & 0.16 &  40 $\pm$  29 &      Levonorgestrel &       Phenobarbital &    inf &    Major &                   Phenobarbital decreases the effect of levonorgestrel \\
140 &    4 & 0.00 & 0.26 &  72 $\pm$  61 &            Atenolol &           Verapamil &   0.00 &    Major &                                         Increased effect of both drugs \\
140 &    4 & 0.01 & 0.14 &  10 $\pm$   2 &     Estrogens Conj. &        Prednisolone &    inf & Moderate &            The estrogenic agent increases the effect of corticosteroid \\
140 &    4 & 0.00 & 0.14 &   4 $\pm$   3 &         Doxycycline &         Clavulanate &   0.71 &     None &                                          Possible antagonism of action \\
140 &    4 & 0.00 & 0.17 &  62 $\pm$  44 &       Phenobarbital &       Aminophylline &   0.24 & Moderate &                   The barbiturate decreases the effect of theophylline \\
145 &    3 & 0.01 & 0.18 &  97 $\pm$  94 &     Estrogens Conj. &       Phenobarbital &    inf & Moderate &                    The enzyme inducer decreases the effect of hormones \\
145 &    3 & 0.00 & 0.32 & 136 $\pm$ 117 &   Ethinyl Estradiol &       Aminophylline &    inf & Moderate &    The contraceptive increases the effect and toxicity of theophylline \\
145 &    3 & 0.00 & 0.20 &  53 $\pm$  16 &   Ethinyl Estradiol &       Phenobarbital &    inf &    Major &       This product may cause a slight decrease of contraceptive effect \\
145 &    3 & 0.00 & 0.22 &  45 $\pm$  43 &  Medroxyproges. Ac. &            Warfarin &    inf &     None &                        The agent increases the effect of anticoagulant \\
145 &    3 & 0.00 & 0.01 &   2 $\pm$   0 &      Hydrocortisone &       Phenobarbital &   1.42 & Moderate &             The barbiturate decreases the effect of the corticosteroid \\
145 &    3 & 0.00 & 0.23 &  11 $\pm$   9 &         Doxycycline &            Warfarin &   1.42 & Moderate &                    The tetracycline increases the anticoagulant effect \\
        \bottomrule
    \end{tabular}
    \caption{
        DDI list 101-150.
        See Supplementary Table \ref{table:SI:ListDDI-1-50} for column description.
        Continues on Supplementary Table \ref{table:SI:ListDDI-151-181}.
    }
    \label{table:SI:ListDDI-101-150}
\end{sidewaystable}

\begin{sidewaystable}
    \centering
    \tiny
    \begin{tabular}{c r|c r|c|c c r|l ll}
        \toprule
        rank$_{\Phi}$ & $|U_{i,j}^{\Phi}|$ & $\gamma_{i,j}^{\Phi}$ & $\tau_{i,j}^{\Phi}$ & $\langle \lambda^{u}_{i,j} \rangle$ & $i$ & $j$ & $RRI^{F}_{i,j}$ & severity & interaction \\
        \midrule
% UPDATED: 2018-11-07
% FILE: display_stats_ddi.py
145 &    3 & 0.01 & 0.12 &  23 $\pm$  13 &           Phenytoin &       Aminophylline &   1.42 & Moderate &                                      Decreased effect of both products \\
145 &    3 & 0.01 & 0.43 &  40 $\pm$  57 &          Methyldopa &            Levodopa &    inf &    Minor &               Methyldopa increases the effect and toxicity of levodopa \\
145 &    3 & 0.00 & 0.12 &  40 $\pm$  19 &             Digoxin &           Verapamil &    inf & Moderate &                              Verapamil increases the effect of digoxin \\
145 &    3 & 0.00 & 0.15 &  58 $\pm$  56 &       Aminophylline &             Lithium &   1.42 & Moderate &                         Theophylline decreases serum levels of lithium \\
145 &    3 & 0.01 & 0.10 &   9 $\pm$   4 &           Phenytoin &        Prednisolone &   0.35 & Moderate &          The enzyme inducer decreases the effect of the corticosteroid \\
145 &    3 & 0.01 & 0.43 & 185 $\pm$  98 &           Phenytoin &            Levodopa &   0.00 & Moderate &                         The hydantoin decreases the effect of levodopa \\
157 &    2 & 0.00 & 0.06 &   6 $\pm$   1 &           Estradiol &        Prednisolone &    inf & Moderate &            The estrogenic agent increases the effect of corticosteroid \\
157 &    2 & 0.01 & 0.13 &  31 $\pm$   0 &             Timolol &       Aminophylline &    inf &    Major &              Antagonism of action and increased effect of theophylline \\
157 &    2 & 0.00 & 0.24 &  62 $\pm$  53 &           Phenytoin &           Estradiol &    inf & Moderate &                The enzyme inducer decreases the effect of the hormones \\
157 &    2 & 0.00 & 0.08 &  23 $\pm$  11 &         Doxycycline &       Phenobarbital &    inf & Moderate &                 The anticonvulsant decreases the effect of doxycycline \\
157 &    2 & 0.00 & 0.20 &  79 $\pm$  30 &      Norethisterone &       Phenobarbital &    inf &    Major &       This product may cause a slight decrease of contraceptive effect \\
157 &    2 & 0.00 & 0.20 &  79 $\pm$  30 &           Estradiol &       Phenobarbital &    inf & Moderate &                    The enzyme inducer decreases the effect of hormones \\
157 &    2 & 0.00 & 0.42 & 102 $\pm$ 110 &         Propranolol &           Verapamil &   0.71 &    Major &                                         Increased effect of both drugs \\
157 &    2 & 0.01 & 0.03 &   2 $\pm$   0 &             Timolol &           Fenoterol &    inf &        * &                                                             Antagonism \\
157 &    2 & 0.00 & 0.02 &   2 $\pm$   0 &           Phenytoin &      Hydrocortisone &   0.71 & Moderate &          The enzyme inducer decreases the effect of the corticosteroid \\
157 &    2 & 0.00 & 0.53 & 288 $\pm$ 213 &           Phenytoin &  Medroxyproges. Ac. &    inf & Moderate &                The enzyme inducer decreases the effect of the hormones \\
157 &    2 & 0.00 & 0.24 &  62 $\pm$  53 &           Phenytoin &      Norethisterone &    inf &    Major &       This product may cause a slight decrease of contraceptive effect \\
157 &    2 & 0.00 & 0.42 & 274 $\pm$ 218 &             Digoxin &    Propylthiouracil &   0.71 & Moderate &                  The antithyroid agent increases the effect of digoxin \\
169 &    1 & 0.00 & 0.00 &   2 $\pm$   0 &            Atenolol &         Epinephrine &    inf & Moderate &                                         Hypertension, then bradycardia \\
169 &    1 & 0.00 & 0.49 & 179 $\pm$   0 &           Phenytoin &   Ethinyl Estradiol &    inf &    Major &       This product may cause a slight decrease of contraceptive effect \\
169 &    1 & 0.00 & 0.30 & 117 $\pm$   0 &         Haloperidol &          Methyldopa &    inf & Moderate &           Methyldopa increases haloperidol effect or risk of psychosis \\
169 &    1 & 0.00 & 0.49 & 179 $\pm$   0 &           Phenytoin &      Levonorgestrel &    inf &    Major &                           Phenytoin decreases the contraceptive effect \\
169 &    1 & 0.00 & 0.51 &  31 $\pm$   0 &           Phenytoin &        Sulfadiazine &   0.00 & Moderate &                      The sulfonamide increases the effect of hydantoin \\
169 &    1 & 0.00 & 0.02 &   4 $\pm$   0 &        Erythromycin &       Aminophylline &    inf & Moderate &        The macrolide increases the effect and toxicity of theophylline \\
169 &    1 & 0.00 & 0.24 &  12 $\pm$   0 &             Timolol &          Methyldopa &    inf &    Major &                                           Possible hypertensive crisis \\
169 &    1 & 0.00 & 0.05 &   6 $\pm$   0 &        Erythromycin &       Carbamazepine &   0.00 &    Major &                    The macrolide increases the effect of carbamazepine \\
169 &    1 & 0.00 & 0.06 &   2 $\pm$   0 &        Erythromycin &            Diazepam &    inf & Moderate &               The macrolide increases the effect of the benzodiazepine \\
169 &    1 & 0.00 & 0.03 &   9 $\pm$   0 &        Erythromycin &          Fluoxetine &    inf & Moderate &                 Possible serotoninergic syndrome with this combination \\
169 &    1 & 0.00 & 0.25 &  15 $\pm$   0 &           Phenytoin &         Doxycycline &    inf & Moderate &                 The anticonvulsant decreases the effect of doxycycline \\
169 &    1 & 0.00 & 0.06 &  29 $\pm$   0 &           Phenytoin &     Estrogens Conj. &    inf & Moderate &                The enzyme inducer decreases the effect of the hormones \\
169 &    1 & 0.00 & 0.31 & 124 $\pm$   0 &       Carbamazepine &           Verapamil &   0.00 &    Major &                        Verapamil increases the effect of carbamazepine \\
        \bottomrule
    \end{tabular}
    \caption{
        DDI list 151-181.
        See Supplementary Table \ref{table:SI:ListDDI-1-50} for column description.
    }
    \label{table:SI:ListDDI-151-181}
\end{sidewaystable}

\pagebreak

\begin{table}
    \centering
    \scriptsize
        \begin{tabular}{c r|r r|c|c c|r l}
            \toprule
            rank$_{\Phi}$ & $|U_{i,j}^{\Phi}|$ & $\gamma_{i,j}^{\Phi}$ & $\tau_{i,j}^{\Phi}$ & $\langle \lambda^{u}_{i,j} \rangle$ & $i$ & $j$ & $RRI^{F}_{i,j}$ & severity  \\
            \midrule
% UPDATED: 2018-11-07
% FILE: display_stats_ddi.py
 2 & 2117 & 0.18 & 0.23 &  53 $\pm$  74 &                 ASA &     Ibuprofen &  1.42 & Major \\
 5 & 1190 & 0.19 & 0.45 & 127 $\pm$ 127 &       Amitriptyline &    Fluoxetine &  3.55 & Major \\
 7 &  892 & 0.14 & 0.20 &  56 $\pm$  61 &         Fluconazole &   Simvastatin &  2.63 & Major \\
16 &  470 & 0.63 & 0.55 & 160 $\pm$ 133 &           Diltiazem &   Simvastatin &  1.27 & Major \\
23 &  257 & 0.16 & 0.42 & 123 $\pm$ 130 &          Imipramine &    Fluoxetine &  3.08 & Major \\
27 &  222 & 0.02 & 0.47 & 148 $\pm$ 139 &          Fluoxetine &       Lithium &  1.79 & Major \\
33 &  154 & 0.01 & 0.33 &  94 $\pm$  92 &          Fluoxetine & Nortriptyline &  2.70 & Major \\
36 &  148 & 0.14 & 0.49 & 168 $\pm$ 160 &         Haloperidol &       Lithium &  1.31 & Major \\
47 &   91 & 0.01 & 0.52 & 154 $\pm$ 142 &            Atenolol &     Diltiazem &  1.19 & Major \\
62 &   52 & 0.08 & 0.49 & 118 $\pm$ 114 &             Digoxin &    Amiodarone &  0.56 & Major \\
63 &   51 & 0.00 & 0.23 &  93 $\pm$  90 & Hydrochlorothiazide &       Lithium &  2.90 & Major \\
73 &   31 & 0.01 & 0.20 &  48 $\pm$  66 &         Propranolol &    Salbutamol &  6.61 & Major \\
84 &   25 & 0.03 & 0.56 & 157 $\pm$ 136 &           Diltiazem &    Amiodarone &  1.26 & Major \\
85 &   24 & 0.00 & 0.05 &   3 $\pm$   2 &          Diclofenac &      Warfarin &  0.84 & Major \\
89 &   23 & 0.03 & 0.47 & 152 $\pm$ 143 &           Diltiazem &   Propranolol &  2.01 & Major \\
89 &   23 & 0.00 & 0.16 &  36 $\pm$  44 &         Fluconazole &   Haloperidol &  1.33 & Major \\
91 &   22 & 0.00 & 0.07 &   9 $\pm$   4 &       Ciprofloxacin &      Warfarin &  1.02 & Major \\
93 &   21 & 0.01 & 0.08 &  10 $\pm$   6 &          Tobramycin &    Furosemide &  3.01 & Major \\
95 &   19 & 0.00 & 0.07 &  10 $\pm$   7 &       Ciprofloxacin & Aminophylline &  1.21 & Major \\
97 &   18 & 0.00 & 0.13 &  33 $\pm$  44 &         Fluconazole &      Warfarin &  0.89 & Major \\
            \bottomrule
        \end{tabular}
    \caption{
        Top 20 \emph{major} DDI pairs $(i,j)$ by rank of $|U_{i,j}^{\Phi}|$, the number of patients affects by the DDI (1\textsuperscript{st} and 2\textsuperscript{nd} columns, respectively).
        The normalized drug pair footprint in the population ($\gamma_{i,j}^{\Phi}$) as well as the normalized co-administration length ($\tau_{i,j}^{\Phi}$), are shown in columns 3 and 4, respectively.
        Mean ($\pm$ s.d.) co-administration length, $\langle \lambda^{u}_{i,j} \rangle$, is shown in column 5 (in days) for each DDI pair $(i,j)$ whose English drug names are shown in columns 6 and 7.
        The relative gender risk of DDI pair co-administration, $RRI^{F}_{i,j}$, is shown in column 8.
        DDI severity classification, according to \textit{Drugs.com}, shown in column 9.
    }
    \label{table:SI:top-20-major-ddi}
\end{table}

\begin{table}
    \centering
    \scriptsize
        \begin{tabular}{c r r|r r|c c|r l}
            \toprule
            rankp($\gamma$) & $\gamma_{i,j}^{\Phi}$ & $\gamma_{j,i}^{\Phi}$ & $|U_{i,j}^{\Phi}|$ & $\langle \lambda^{u}_{i,j} \rangle$ & $i$ & $j$ & $RRI^{F}_{i,j}$ & severity  \\
            \midrule
% UPDATED: 2018-11-07
% FILE: display_stats_ddi.py
 1 & 0.50 & 0.61 &  524 & 243 $\pm$ 188 &   Haloperidol &      Biperiden &  0.62 &  Moderate \\
 2 & 0.59 & 0.12 &  385 & 155 $\pm$ 125 &       Digoxin &     Furosemide &  0.61 &  Moderate \\
 3 & 0.19 & 0.36 & 5078 & 102 $\pm$  95 &    Omeprazole &     Clonazepam &  2.28 &  Moderate \\
 4 & 0.10 & 0.50 & 1249 & 141 $\pm$ 124 &           ASA &      Glyburide &  0.89 &  Moderate \\
 5 & 0.42 & 0.14 &  272 & 140 $\pm$ 114 &       Digoxin & Spironolactone &  0.58 &     Minor \\
 6 & 0.63 & 0.02 &  470 & 160 $\pm$ 133 &     Diltiazem &    Simvastatin &  1.27 &     Major \\
 7 & 0.27 & 0.15 &  173 & 109 $\pm$  96 &       Digoxin &     Carvedilol &  0.53 &  Moderate \\
 8 & 0.04 & 0.44 &  496 & 103 $\pm$  87 &           ASA &     Gliclazide &  0.78 &      None \\
 9 & 0.04 & 0.31 &  999 &  87 $\pm$  86 &    Omeprazole &       Diazepam &  1.21 &  Moderate \\
 9 & 0.19 & 0.09 & 1190 & 127 $\pm$ 127 & Amitriptyline &     Fluoxetine &  3.55 &     Major \\
11 & 0.14 & 0.16 &  148 & 168 $\pm$ 160 &   Haloperidol &        Lithium &  1.31 &     Major \\
12 & 0.07 & 0.22 &  535 & 152 $\pm$ 132 &      Atenolol &      Glyburide &  1.22 &  Moderate \\
13 & 0.18 & 0.08 &  186 & 142 $\pm$ 156 &   Haloperidol &  Carbamazepine &  0.62 &  Moderate \\
14 & 0.18 & 0.06 & 2117 &  53 $\pm$  74 &           ASA &      Ibuprofen &  1.42 &     Major \\
14 & 0.20 & 0.04 & 1460 &  54 $\pm$  77 &      Atenolol &      Ibuprofen &  1.88 &  Moderate \\
16 & 0.02 & 0.24 &  222 & 148 $\pm$ 139 &    Fluoxetine &        Lithium &  1.79 &     Major \\
17 & 0.03 & 0.18 &  201 & 107 $\pm$  95 &      Atenolol &     Gliclazide &  1.09 &      None \\
18 & 0.01 & 0.26 &  154 &  94 $\pm$  92 &    Fluoxetine &  Nortriptyline &  2.70 &     Major \\
19 & 0.28 & 0.00 &  149 & 115 $\pm$ 109 &     Phenytoin &     Omeprazole &  0.80 &  Moderate \\
20 & 0.03 & 0.17 &  377 & 143 $\pm$ 138 &    Fluoxetine &  Carbamazepine &  0.98 &  Moderate \\
            \bottomrule
        \end{tabular}
    \caption{
        Top 20 known DDI pairs $(i,j)$ by rank product (1\textsuperscript{st} column) of the ranks of $\gamma_{i,j}^{\Phi}$ and $\gamma_{j,i}^{\Phi}$, the normalized drug pair footprint in the population (1\textsuperscript{st}, 2\textsuperscript{nd} and 3\textsuperscript{rd} columns, respectively).
        The number of patients affected by the drug pair, $|U_{i,j}^{\Phi}|$, is shown in column 4.
        Mean ($\pm$ s.d.) co-administration length, $\langle \lambda^{u}_{i,j} \rangle$, is shown in column 5 (in days) for each DDI pair $(i,j)$ whose English drug names are shown in columns 6 and 7.
        The relative gender risk of DDI pair co-administration, $RRI^{F}_{i,j}$, is shown in column 8.
        DDI severity classification, according to \textit{Drugs.com}, shown in column 9; DDIs or drugs not found in \textit{Drugs.com} are labeled as \textit{None} or \textit{*}, respectively. 
    }
    \label{table:SI:top-20-ranked-gamma}
\end{table}

\begin{table}[!htb]
    \centering
    \scriptsize
        \begin{tabular}{c r|r r|c c|r l}
            \toprule
            rank$_{\tau}$ & $\tau_{i,j}^{\Phi}$ & $|U_{i,j}^{\Phi}|$ & $\langle \lambda^{u}_{i,j} \rangle$ & $i$ & $j$ & $RRI^{F}_{i,j}$ & severity  \\
            \midrule
% UPDATED: 2018-11-07
% FILE: display_stats_ddi.py
 1 & 0.70 &  524 & 243 $\pm$ 188 &   Haloperidol &          Biperiden &  0.62 & Moderate \\
 2 & 0.60 & 1249 & 141 $\pm$ 124 &           ASA &          Glyburide &  0.89 & Moderate \\
 3 & 0.60 &  385 & 155 $\pm$ 125 &       Digoxin &         Furosemide &  0.61 & Moderate \\
 4 & 0.58 &  535 & 152 $\pm$ 132 &      Atenolol &          Glyburide &  1.22 & Moderate \\
 5 & 0.57 &   95 & 140 $\pm$ 126 &   Propranolol &          Glyburide &  1.61 & Moderate \\
 6 & 0.56 &   25 & 157 $\pm$ 136 &     Diltiazem &         Amiodarone &  1.26 &    Major \\
 7 & 0.56 &   13 & 122 $\pm$ 113 &   Propranolol &         Methyldopa &  8.50 &    Major \\
 8 & 0.55 &  470 & 160 $\pm$ 133 &     Diltiazem &        Simvastatin &  1.27 &    Major \\
 9 & 0.55 &    5 &  82 $\pm$  86 &   Propranolol &      Aminophylline &  1.06 &    Major \\
10 & 0.55 &  272 & 140 $\pm$ 114 &       Digoxin &     Spironolactone &  0.58 &    Minor \\
11 & 0.53 &    2 & 288 $\pm$ 213 &     Phenytoin & Medroxyproges. Ac. &   inf & Moderate \\
12 & 0.52 &   91 & 154 $\pm$ 142 &      Atenolol &          Diltiazem &  1.19 &    Major \\
13 & 0.51 &    1 &  31 $\pm$   0 &     Phenytoin &       Sulfadiazine &  0.00 & Moderate \\
14 & 0.50 &   90 & 161 $\pm$ 157 &    Imipramine &      Carbamazepine &  1.35 & Moderate \\
15 & 0.50 &  377 & 143 $\pm$ 138 &    Fluoxetine &      Carbamazepine &  0.98 & Moderate \\
16 & 0.49 &  226 & 151 $\pm$ 145 & Amitriptyline &      Carbamazepine &  0.99 & Moderate \\
17 & 0.49 &   52 & 118 $\pm$ 114 &       Digoxin &         Amiodarone &  0.56 &    Major \\
18 & 0.49 &    1 & 179 $\pm$   0 &     Phenytoin &     Levonorgestrel &   inf &    Major \\
18 & 0.49 &    1 & 179 $\pm$   0 &     Phenytoin &  Ethinyl Estradiol &   inf &    Major \\
20 & 0.49 &  148 & 168 $\pm$ 160 &   Haloperidol &            Lithium &  1.31 &    Major \\
            \bottomrule
        \end{tabular}
    \caption{
        Top 20 known DDI pairs $(i,j)$ by rank of $\tau_{i,j}^{\Phi}$, the normalized co-administration length (1\textsuperscript{st} and 2\textsuperscript{nd} columns, respectively).
        The number of patients affected by the drug pair, $|U_{i,j}^{\Phi}|$, is shown in column 3.
        Mean ($\pm$ s.d.) co-administration length, $\langle \lambda^{u}_{i,j} \rangle$, is shown in column 4 (in days) for each DDI pair $(i,j)$ whose English drug names are shown in columns 5 and 6.
        The relative gender risk of DDI pair co-administration, $RRI^{F}_{i,j}$, shown in column 7.
        DDI severity classification, according to \textit{Drugs.com}, shown in column 8; DDIs or drugs not found in \textit{Drugs.com} are labeled as \textit{None} or \textit{*}, respectively. 
    }
    \label{table:SI:top-20-ranked-tau}
\end{table}
% Severity
\FloatBarrier
\subsection{Interactions per severity}
\label{ch:SI:interactons-per-severity}

In this section, Supplementary Table \ref{table:SI:severity-ddi-users} shows the number of individual interactions and unique users, both per severity of interaction.

Note that some interactions present in DrugBrank were not found in Drugs.com. These are marked as \textit{None}. The drug \textit{Fenoterol} (brand name \textit{Berotec} in Brazil) was not found in Drugs.com. These interactions were summed separately and are shown with an asterisk (\textit{*}).

\begin{table}[!hbt]
    \centering
    \scriptsize
    \begin{tabular}{l|r r r r}
        \toprule
        severity $s$ & $\Phi_{s}$ & $|U^{\Phi}_{s}|$ & $|U^{\Phi}_{s}|/|U|$ & $|U^{\Phi}_{s}|/|\Omega|$ \\
        \midrule
% UPDATED: 2018-04-25
% FILE: display_stats.py
\textit{Major}    &  5,968 (22.50\%) &  5,224 & 3.94\% & 1.54\% \\
\textit{Moderate} & 18,335 (69.13\%) & 12,711 & 9.58\% & 3.75\% \\
\textit{Minor}    &    542 (02.04\%) &    528 & 0.40\% & 0.16\% \\
\textit{None}     &  1,489 (05.61\%) &  1,314 & 0.99\% & 0.39\% \\
\textit{*}        &    190 (00.72\%) &    179 & 0.13\% & 0.05\% \\
\midrule
Total             & 26,524 (100\%)   & 19,956 &     -\% &     -\% \\
        \bottomrule
    \end{tabular}
    \caption{
        The 2\textsuperscript{nd} column lists the numbers of interactions, $\Phi_{s}$, per DDI severity class (1\textsuperscript{st} column); percentages of interactions per class are shown in parenthesis.
        Drugs or interactions identified in \emph{DrugBank} but not present in Drugs.com are tallied as \textit{None}.
        Interactions for \textit{Berotec} tallied as \textit{*}.
        The 3\textsuperscript{rd} column lists the number of patients affected by at least one interaction $|U^{\Phi}_{s}|$, per DDI severity.
        Fourth and fifth columns lists the proportion of patients in each DDI severity class for the \textit{Pronto} system and the entire Blumenau populations, respectively.
        Notice that the same patient may have been administered DDI of more than one severity type.
    }
    \label{table:SI:severity-ddi-users}
\end{table}

% Gender
\FloatBarrier
\subsection{Interactions per gender}
\label{ch:SI:interactions-per-gender}

In this section, Supplementary Table \ref{table:SI:age-pop-ddi-gender} shows the number of individual interactions and unique users per gender.

\begin{table}[!hbt]
    \centering
    \scriptsize
    \begin{tabular}{c|r r r r}
        \toprule
        gender $g$ & $\Phi^{g}$ & $|U^{\Phi,g}|$ & $|U^{\Phi,g}|/|U|$ & $|U^{\Phi,g}|/|\Omega|$ \\
        \midrule
% UPDATED: 2018-04-25
% FILE: display_stats.py
Male   &   8,100 (30.54\%) &   4,793 &  3.61\% &  1.41\% \\
Female &  18,424 (69.46\%) &  10,734 &  8.09\% &  3.17\% \\
\midrule
Total  &  26,524 (100\%)   &  15,527 & 11.70\% &  4.58\% \\
        \bottomrule
    \end{tabular}
    \caption{
        The 2\textsuperscript{nd} column lists the numbers of interactions, $\Phi$, per gender (1\textsuperscript{st} column); percentages of interactions per gender shown in parenthesis.
        The 3\textsuperscript{rd} column lists the number of patients affected by at least one interaction per gender, $|U^{\Phi,g}|$.
        The 4\textsuperscript{th} and 5\textsuperscript{th} columns show the proportion of patients in each gender for the \textit{Pronto} system and entire Blumenau populations, respectively.
    }
    \label{table:SI:age-pop-ddi-gender}
\end{table}
% Age
\FloatBarrier
\clearpage
\subsection{Interactions per age}
\label{ch:SI:interactions-per-age}

Supplementary Table \ref{table:SI:age-pop-ddi-dispensed} shows the number of individual interaction and unique users per age group.

\begin{table}[!hbt]
    \centering
    \scriptsize
    \begin{tabular}{c|r r r r}
        \toprule
        age y & $\Phi^{y}$ & $|U^{\Phi,y}|$ & $|U^{\Phi,y}|/|U|$ & $|U^{\Phi,y}|/|\Omega|$ \\
        \midrule
% UPDATED: 2018-04-25
% FILE: display_stats.py
00-04 &    23 (0.09\%)  &    20 & 0.02\% & 0.01\% \\
05-09 &     7 (0.03\%)  &     7 & 0.01\% & 0.00\% \\
10-14 &    29 (0.11\%)  &    25 & 0.02\% & 0.01\% \\
15-19 &   172 (0.65\%)  &   139 & 0.10\% & 0.04\% \\
20-24 &   311 (1.17\%)  &   237 & 0.18\% & 0.07\% \\
25-29 &   433 (1.63\%)  &   301 & 0.23\% & 0.09\% \\
30-34 &   771 (2.91\%)  &   525 & 0.40\% & 0.15\% \\
35-39 & 1,097 (4.14\%)  &   687 & 0.52\% & 0.20\% \\
40-44 & 1,581 (5.96\%)  & 1,023 & 0.77\% & 0.30\% \\
45-49 & 2,332 (8.79\%)  & 1,426 & 1.07\% & 0.42\% \\
50-54 & 3,128 (11.79\%) & 1,868 & 1.41\% & 0.55\% \\
55-59 & 3,447 (13.00\%) & 1,956 & 1.47\% & 0.58\% \\
60-64 & 3,508 (13.23\%) & 2,006 & 1.51\% & 0.59\% \\
65-69 & 3,254 (12.27\%) & 1,794 & 1.35\% & 0.53\% \\
70-74 & 2,417 (9.11\%)  & 1,311 & 0.99\% & 0.39\% \\
75-79 & 1,978 (7.46\%)  & 1,057 & 0.80\% & 0.31\% \\
80-84 & 1,143 (4.31\%)  &   638 & 0.48\% & 0.19\% \\
85-89 &   620 (2.34\%)  &   349 & 0.26\% & 0.10\% \\
90-94 &   205 (0.77\%)  &   117 & 0.09\% & 0.03\% \\
95-99 &    49 (0.18\%)  &    27 & 0.02\% & 0.01\% \\
>99   &    19 (0.07\%)  &    14 & 0.01\% & 0.00\% \\
\midrule
Total & 26,524 (100\%)  & 15,527 & 11.70\% &  4.58\% \\
        \bottomrule
    \end{tabular}
    \caption{
        The 2\textsuperscript{nd} column lists the numbers of interactions, $\Phi^{y}$, per age range (1\textsuperscript{st} column); percentages of interactions per age range shown in parenthesis.
        The 3\textsuperscript{rd} column lists the number of patients affected by at least one interaction per age range, $|U^{\Phi,y}|$.
        The 4\textsuperscript{th} and 5\textsuperscript{th} columns show the proportion of patients in each age range from the \textit{Pronto} system and entire Blumenau populations, respectively.
    }
    \label{table:SI:age-pop-ddi-dispensed}
\end{table}
% Age & Gender
\FloatBarrier
\subsection{Interaction per age and gender}
\label{ch:SI:interactions-per-age-and-gender}

\begin{figure}[!hbt]
    \centering
    % UPDATED: 2017-01-26
    % FILE: plot_age_gender.py
    \includegraphics[width=9cm]{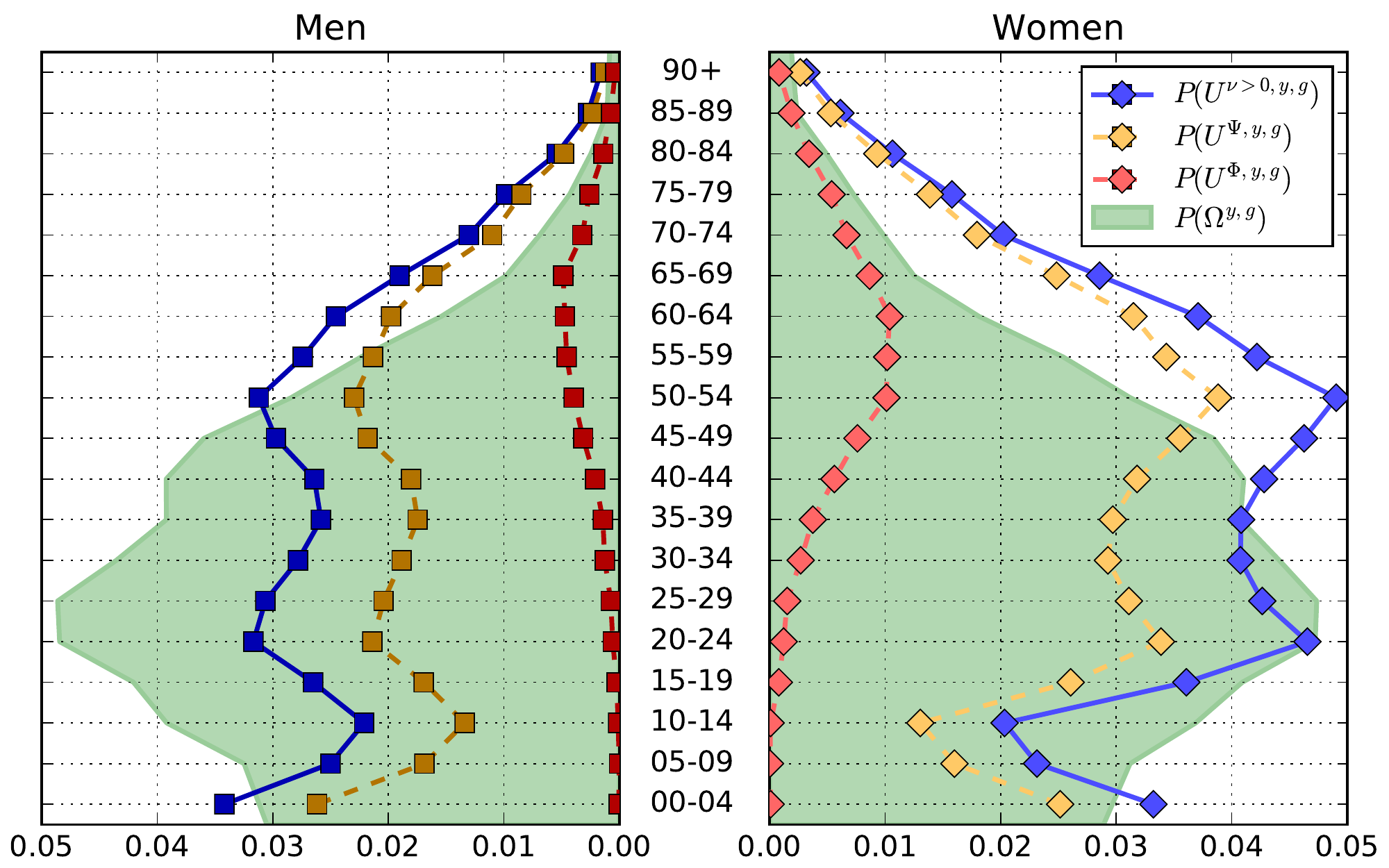}
    \caption{
        The joint probability a patient was dispensed at least one drug $P(U^{\nu>0,y,g})$, had co-administrations $P(U^{\Psi,y,g})$, or had a DDI $P(U^{\Phi,y,g})$, given age range ($[y_1,y_2]$) and gender ($g$), are shown in blue, orange and red lines, respectively.
        Values for age group $y\geq\text{90}$ were aggregated for plotting.
        Population distribution for Blumenau $P(\Omega^{y,g})$ is shown as a green fill.
        A Kolmogorov-Smirnov test cannot reject the hypothesis that both the female and male distribution of patients with at least one co-administration known to be DDI ($U^{\Phi,y,g}$) are drawn from the same underlying continuous distribution ($KS=.3810$, $p\text{-value}=.0706$).
    }
    \label{fig:SI:age}
\end{figure}

\begin{figure}[!hbt]
    \centering
    % UPDATED: 2017-01-16
    % FILE: plot_age_means.py
    \includegraphics[width=14cm]{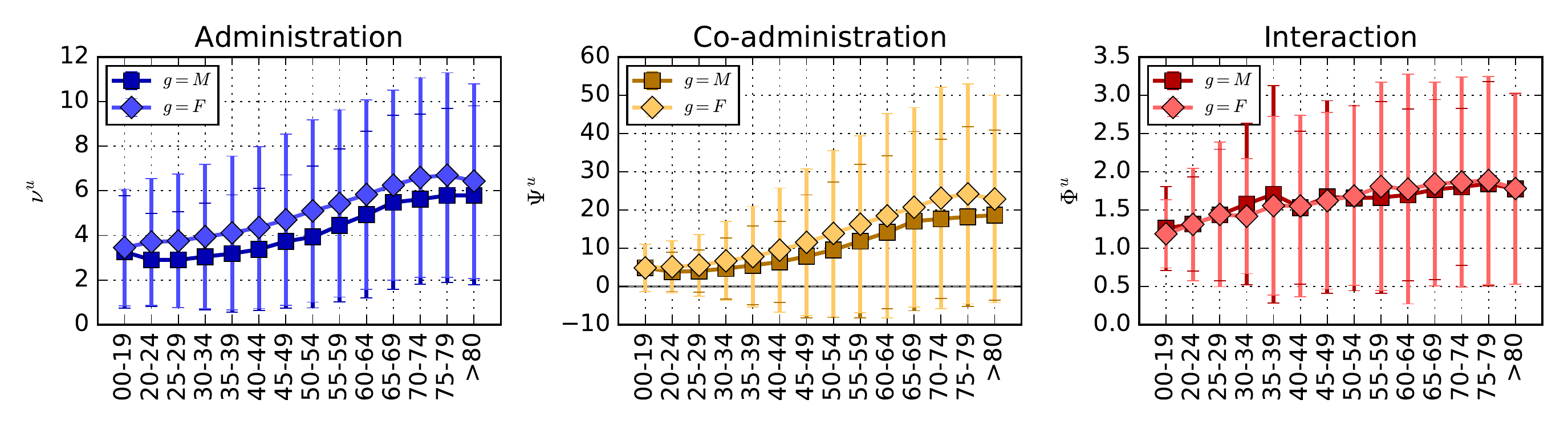}
    \caption{
        \textbf{Left.} Mean number of drugs dispensed ($\nu^u$) to patients in each age group.
        \textbf{Middle.} Mean number of drug pairs co-administered ($\Psi^u$) by patients in each age group.
        \textbf{Right.} Mean number of drug pairs known to be a DDI ($\Phi^u$) co-administered by patients in each age group.
        Numbers for male and female patients shown in lighter and darker colors, respectively.
        In all plots vertical bars denote the standard deviation.
    }
    \label{fig:SI:age-means}
\end{figure}

\begin{figure}[!hbt]
    \centering
    % UPDATED: 2018-11-06
    % FILE: plot_u_coadmin_age.py | plot_u_coadmin_age_gender.py
    \includegraphics[width=.42\textwidth]{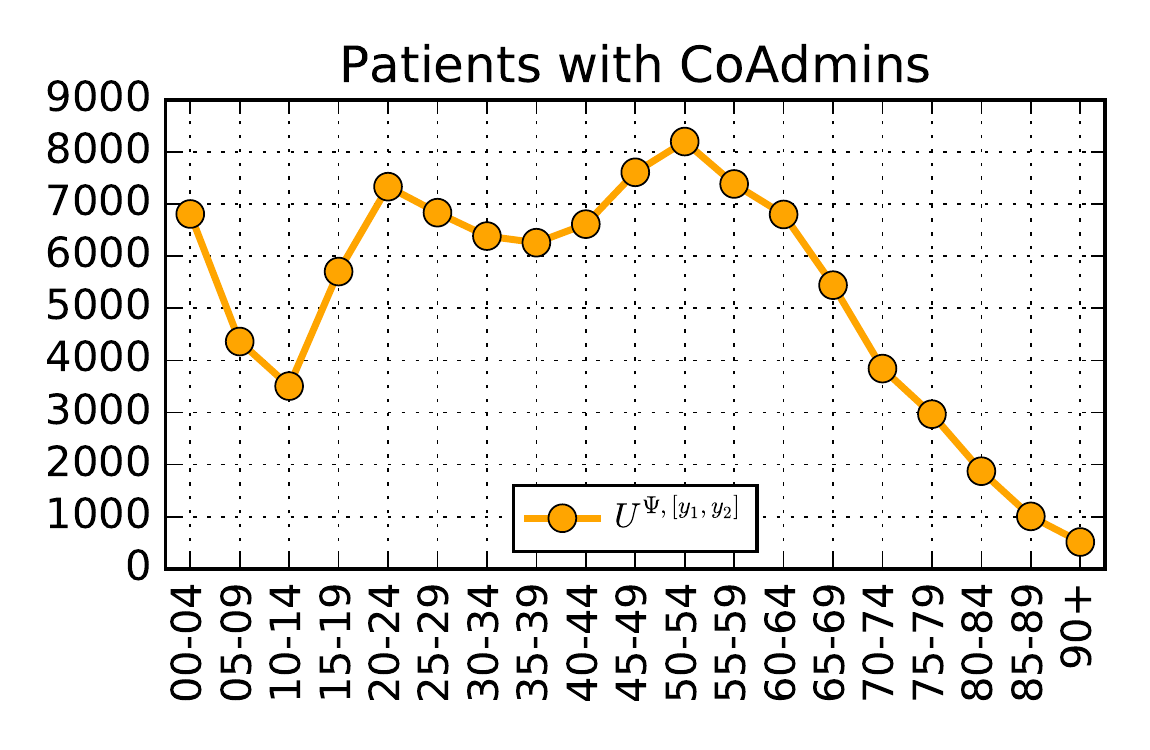}
    \includegraphics[width=.42\textwidth]{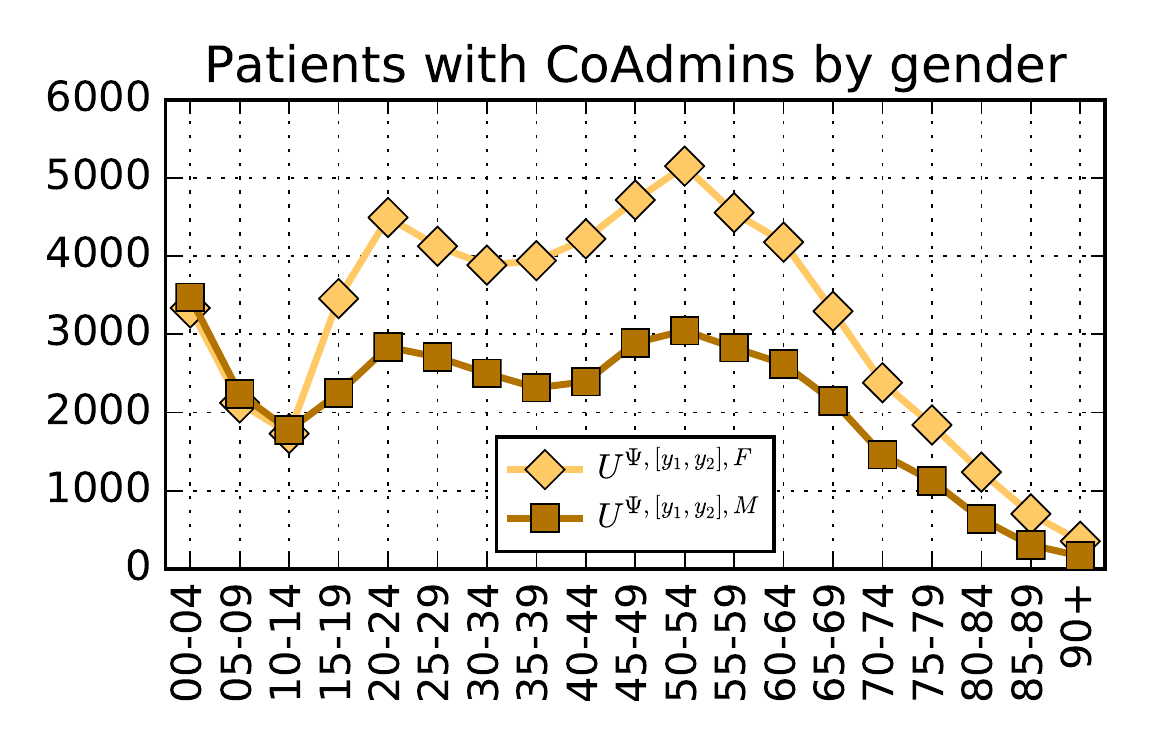}
    \caption{
        \textbf{Left.}
        Absolute number of patients with at least one co-administration per age group, $|U^{\Psi,[y_1,y_2]}|$.
        \textbf{Right.}
        Absolute number of patients with at least one co-administration per age group and gender, $|U^{\Psi,[y_1,y_2],g}|$.
    }
    \label{fig:SI:u-age-gender}
\end{figure}

%
% Relative Risk Ratios (RRC/RRI)
%
\FloatBarrier
\section{Risk and Relative Risk measures}
\label{ch:SI:risk-measures}

Risk and Relative Risk measures are computed based on the number of patients in specific groups.
Here we detail the computation of all measures used throughout the main manuscript.

The relative risk of co-administration and interaction for women, are computed as

\begin{equation}
    RRC^{\text{F}} = \frac{ P(\Psi^u>0 \, | \, u \in U^{\text{F}}) }{P(\Psi^u>0 \, | \, u \in U^{\text{M}}) }
    = \frac{ |U^{\Psi,\text{F}} \, | \, / |U^{\text{F}}| }{ |U^{\Psi,\text{M}} \, | \, / |U^{\text{M}}| }
    ; \;
    RRI^{\text{F}} = \frac{ P(\Phi^u >0 \, | \, u \in U^{\text{F}}) }{ P(\Phi^u >0 \, | \, u \in U^{\text{M}}) } 
    = \frac{ |U^{\Phi,\text{F}} \, | \, / |U^{\text{F}}| }{ |U^{\Phi,\text{M}} \, | \, / |U^{\text{M}}| }.
\end{equation}

\noindent Similarly, $RRC^{M} = 1/RRC^{F}$ and $RRI^{M} = 1/RRI^{F}$.
In the main manuscript we mentioned the computation of $RRI^{F}$ without contraceptive drugs. The drug removed in this computation were \textit{Ethinyl Estradiol}, \textit{Estradiol}, \textit{Norethisterone}, \textit{Levonorgestrel} and \textit{Estrogens Conjugated}.

The relative risk of an interaction at a certain DDI severity level, $s \in \{\text{major}, \text{moderate}, \text{minor}, \text{none}, \text{*}\}$, given gender, is computed as

\begin{equation}
    RRI^{F}_{s} = \frac{ P(\Phi^u_s>0 \, | \, u \in U^{\text{F}})}{ P(\Phi^u_s>0 \, | \, u \in U^{\text{M}}) }
    = \frac{ |U^{\Phi,\text{F}}_{s}| / |U^\text{F}| }{ |U^{\Phi,\text{M}}_{s}| \ |U^{\text{M}}| }
    \quad .
\end{equation}

The relative risk of an interaction between two drugs, given gender, is computed as

\begin{equation}
    RRI^{F}_{i,j} = \frac{ P(\Phi^{u}_{i,j}>0 \, | \, u \in U^{\text{F}}) }{ P(\Phi^{u}_{i,j}>0 \, | \, u \in U^{\text{M}}) }
    = \frac{ |U^{\Phi,\text{F}}_{i,j}| / |U^{\text{F}}| }{ |U^{\Phi,\text{M}}_{i,j}| / |U^{M}| }.
\end{equation}

The relative risk of co-administration and interaction, given number of dispensed drugs, are computed as

\begin{equation}
    RRC^{\nu=x} = \frac{ P(\Psi^{u} \, | \, u \in U^{\nu=x}) }{ P(\Psi^{y} \, | \, u \in U^{\nu=2}) }
    = \frac{ |U^{\Psi,\nu=x}| / |U^{\nu=x}| }{ |U^{\Psi,\nu=2}| / |U^{\nu=2}| }
    ; \;
    RRI^{\nu=x} = \frac{ P(\Phi^{u} \,| \, u \in U^{\nu=x}) }{ P(\Phi^{u} \, | \, u \in U^{\nu=2}) }
    = \frac{ |U^{\Phi,\nu=x}| / |U^{\nu=x}| }{ |U^{\Phi,\nu=2}| / |U^{\nu=2}| }.
\end{equation}

The risk of co-administration and interaction, given age group, are computed as

\begin{equation}
    RC^{[y_1,y_2]} = \frac{ P(\Psi^u >0 \, | \, u \in U^{[y_1,y_2]}) }{ P(\nu^u \geq 2 \, | \, u \in U^{[y_1,y_2]}) }
    = \frac{ |U^{\Psi,[y_1,y_2]}| }{ |U^{\nu \geq 2,[y_1,y_2]}| }
    ; \;
    RI^{[y_1,y_2]} = \frac{ P(\Phi^u >0 \, | \, u \in U^{[y_1,y_2]}) }{ P(\Psi^u >0 \, | \, u \in U^{[y_1,y_2]}) }
    = \frac{ |U^{\Psi,[y_1,y_2]}| }{ |U^{\Phi,[y_1,y_2]}| }.
\end{equation}

Note $RC^{[y_1,y_2]}$ and $RI^{[y_1,y_2]}$ can be also interpreted as probabilities.
Similarly, we also compute the risk of co-administration and interaction, per age group and gender as

\begin{equation}
    RC^{g,[y_1,y_2]} = \frac{ P(\Psi^u >0 \, | \, u \in U^{g,[y_1,y_2]}) }{ P(\nu^u \geq 2 \, | \, u \in U^{g,[y_1,y_2]}) }
    = \frac{ |U^{\Psi,g,[y_1,y_2]}| }{ |U^{\nu \geq 2,g,[y_1,y_2]}| }
    ; \;
    RI^{g,[y_1,y_2]} = \frac{ P(\Phi^u >0 \, | \, u \in U^{g,[y_1,y_2]}) }{ P(\Psi^u >0 \, | \, u \in U^{g,[y_1,y_2]}) }
    = \frac{ |U^{\Psi,g,[y_1,y_2]}| }{ |U^{\Phi,g,[y_1,y_2]}| }.
\end{equation}
% Gender
\FloatBarrier
\subsection{Relative Risk per gender}
\label{ch:SI:relative-risk-ratios-gener}

\begin{table}[!hbt]
    \centering
    \scriptsize
    \begin{tabular}{c|rrrr|rrr|rr}
        \toprule
        $g$ &      $|U^{g}|$ & $|U^{\nu \geq 2}|$ & $|U^{\Psi,g}|$ & $|U^{\Phi,g}|$ & $RRC^{F}$ & $RRI^{F}$ \\
        \midrule
        % 
% UPDATED: 2018-04-25
% FILE: display_stats.py
Male   &  55,032 & 41,922 &  39,723 &   4,793 &  1.0000 &  1.0000  \\
Female &  77,690 & 62,889 &  59,738 &  10,734 &  1.0653 &  1.5864  \\
        \bottomrule
    \end{tabular}
    \caption{
        Absolute number of patients and relative risk measures per gender ($g$, 1\textsuperscript{st} column).
        Columns 2 through 5 lists, per gender, absolute numbers of:
            patients ($|U^{g}|$),
            patients with at least $2$ administrations ($|U^{\nu \geq 2}|$),
            patients with at least one co-administration ($|U^{\Psi,g}|$), and
            patients with at least one known DDI co-administration ($|U^{\Phi,g}|$).
        Relative Risk for women for both co-administration ($RRC^{F}$) and known DDI co-administration ($RRI^{F}$) are listed in columns 6 and 7, respectively.
    }
    \label{table:SI:rrr-gender}
\end{table}

\begin{table}[!hbt]
    \centering
    \scriptsize
    %\rotatebox{90}{
        \begin{tabular}{c|rrc|cc|rl}
        \toprule
        rankp($RRI,U^{\text{F}})$ & $|U^{\Phi,M}|$ & $|U^{\Phi,F}|$ & $\langle \lambda^{u}_{i,j} \rangle$ & $i$ & $j$ & $RRI^{F}_{i,j}$ & severity \\
        \midrule
% UPDATED: 2018-11-07
% FILE: display_stats.py
 1 (1,49)   &    0 &   30 &  10 $\pm$   3 &        Clavulanate &  Ethinyl Estradiol &    inf &     None \\
 2 (1,51)   &    0 &   27 &   8 $\pm$   7 &         Prednisone &          Estradiol &    inf & Moderate \\
 2 (1,51)   &    0 &   27 &  11 $\pm$   6 &        Doxycycline &  Ethinyl Estradiol &    inf & Moderate \\
 4 (1,57)   &    0 &   22 &  19 $\pm$  28 &    Estrogens Conj. &         Prednisone &    inf & Moderate \\
 5 (1,71)   &    0 &   13 &  35 $\pm$  26 &      Carbamazepine &  Ethinyl Estradiol &    inf &    Major \\
 5 (1,71)   &    0 &   13 &  35 $\pm$  26 &     Levonorgestrel &      Carbamazepine &    inf &    Major \\
 7 (76,1)   & 1204 & 3874 & 102 $\pm$  95 &         Omeprazole &         Clonazepam &   2.28 & Moderate \\
 8 (1,83)   &    0 &    9 &  72 $\pm$ 128 &      Carbamazepine &     Norethisterone &    inf &    Major \\
 9 (1,89)   &    0 &    7 &  11 $\pm$   4 &        Doxycycline &  Iron (II) Sulfate &    inf & Moderate \\
10 (1,94)   &    0 &    5 & 162 $\pm$ 120 & Medroxyproges. Ac. &      Phenobarbital &    inf & Moderate \\
10 (1,94)   &    0 &    5 &  10 $\pm$   6 &       Prednisolone &  Ethinyl Estradiol &    inf & Moderate \\
12 (1,98)   &    0 &    4 &  40 $\pm$  29 &     Levonorgestrel &      Phenobarbital &    inf &    Major \\
12 (1,98)   &    0 &    4 &  10 $\pm$   2 &    Estrogens Conj. &       Prednisolone &    inf & Moderate \\
14 (1,102)  &    0 &    3 &  97 $\pm$  94 &    Estrogens Conj. &      Phenobarbital &    inf & Moderate \\
14 (1,102)  &    0 &    3 &  40 $\pm$  57 &         Methyldopa &           Levodopa &    inf &    Minor \\
14 (1,102)  &    0 &    3 &  45 $\pm$  43 & Medroxyproges. Ac. &           Warfarin &    inf &     None \\
14 (1,102)  &    0 &    3 &  40 $\pm$  19 &            Digoxin &          Verapamil &    inf & Moderate \\
14 (1,102)  &    0 &    3 &  53 $\pm$  16 &  Ethinyl Estradiol &      Phenobarbital &    inf &    Major \\
14 (1,102)  &    0 &    3 & 136 $\pm$ 117 &  Ethinyl Estradiol &      Aminophylline &    inf & Moderate \\
20 (1,111)  &    0 &    2 &  23 $\pm$  11 &        Doxycycline &      Phenobarbital &    inf & Moderate \\
20 (1,111)  &    0 &    2 &  79 $\pm$  30 &     Norethisterone &      Phenobarbital &    inf &    Major \\
20 (1,111)  &    0 &    2 & 288 $\pm$ 213 &          Phenytoin & Medroxyproges. Ac. &    inf & Moderate \\
20 (1,111)  &    0 &    2 &  62 $\pm$  53 &          Phenytoin &     Norethisterone &    inf &    Major \\
20 (1,111)  &    0 &    2 &  62 $\pm$  53 &          Phenytoin &          Estradiol &    inf & Moderate \\
20 (1,111)  &    0 &    2 &  79 $\pm$  30 &          Estradiol &      Phenobarbital &    inf & Moderate \\
20 (1,111)  &    0 &    2 &   2 $\pm$   0 &            Timolol &          Fenoterol &    inf &        * \\
20 (1,111)  &    0 &    2 &  31 $\pm$   0 &            Timolol &      Aminophylline &    inf &    Major \\
20 (1,111)  &    0 &    2 &   6 $\pm$   1 &          Estradiol &       Prednisolone &    inf & Moderate \\
29 (1,124)  &    0 &    1 & 117 $\pm$   0 &        Haloperidol &         Methyldopa &    inf & Moderate \\
29 (1,124)  &    0 &    1 &   2 $\pm$   0 &           Atenolol &        Epinephrine &    inf & Moderate \\
29 (1,124)  &    0 &    1 &  12 $\pm$   0 &            Timolol &         Methyldopa &    inf &    Major \\
29 (1,124)  &    0 &    1 &  29 $\pm$   0 &          Phenytoin &    Estrogens Conj. &    inf & Moderate \\
29 (1,124)  &    0 &    1 &   2 $\pm$   0 &       Erythromycin &           Diazepam &    inf & Moderate \\
29 (1,124)  &    0 &    1 &   4 $\pm$   0 &       Erythromycin &      Aminophylline &    inf & Moderate \\
29 (1,124)  &    0 &    1 & 179 $\pm$   0 &          Phenytoin &  Ethinyl Estradiol &    inf &    Major \\
29 (1,124)  &    0 &    1 & 179 $\pm$   0 &          Phenytoin &     Levonorgestrel &    inf &    Major \\
29 (1,124)  &    0 &    1 &  15 $\pm$   0 &          Phenytoin &        Doxycycline &    inf & Moderate \\
29 (1,124)  &    0 &    1 &   9 $\pm$   0 &       Erythromycin &         Fluoxetine &    inf & Moderate \\
39 (104,2)  &  706 & 1411 &  53 $\pm$  74 &                ASA &          Ibuprofen &   1.42 &    Major \\
40 (59,4)   &  198 &  992 & 127 $\pm$ 127 &      Amitriptyline &         Fluoxetine &   3.55 &    Major \\
41 (83,3)   &  400 & 1060 &  54 $\pm$  77 &           Atenolol &          Ibuprofen &   1.88 & Moderate \\
42 (75,5)   &  189 &  703 &  56 $\pm$  61 &        Fluconazole &        Simvastatin &   2.63 &    Major \\
43 (62,7)   &  108 &  519 &  46 $\pm$  54 &        Fluconazole &         Clonazepam &   3.40 &     None \\
44 (61,9)   &   86 &  415 &  44 $\pm$  62 &        Propranolol &          Ibuprofen &   3.42 & Moderate \\
45 (52,12)  &   46 &  309 &  86 $\pm$  84 &         Fluoxetine &        Propranolol &   4.76 & Moderate \\
46 (38,17)  &    1 &  178 &  10 $\pm$   6 &  Ethinyl Estradiol &        Amoxicillin & 126.09 & Moderate \\
47 (117,6)  &  369 &  630 &  87 $\pm$  86 &         Omeprazole &           Diazepam &   1.21 & Moderate \\
48 (53,14)  &   38 &  246 &  66 $\pm$  57 &      Levothyroxine &  Iron (II) Sulfate &   4.59 & Moderate \\
49 (81,10)  &  134 &  366 &  52 $\pm$  75 &         Furosemide &          Ibuprofen &   1.93 & Moderate \\
49 (54,15)  &   35 &  210 &  42 $\pm$  40 &        Fluconazole &      Amitriptyline &   4.25 & Moderate \\
51 (110,8)  &  265 &  487 &  30 $\pm$  50 &                ASA &      Dexamethasone &   1.30 & Moderate \\
52 (64,16)  &   48 &  209 & 123 $\pm$ 130 &         Imipramine &         Fluoxetine &   3.08 &    Major \\
53 (47,23)  &   12 &  104 &   9 $\pm$   7 &        Norfloxacin &  Iron (II) Sulfate &   6.14 & Moderate \\
54 (41,27)  &    7 &   95 &   4 $\pm$   4 &         Diclofenac &        Alendronate &   9.61 & Moderate \\
55 (39,29)  &    1 &   83 &  15 $\pm$  26 &         Prednisone &  Ethinyl Estradiol &  58.79 & Moderate \\
56 (115,11) &  197 &  338 & 152 $\pm$ 132 &           Atenolol &          Glyburide &   1.22 & Moderate \\
57 (40,35)  &    2 &   61 &  33 $\pm$  28 &         Methyldopa &  Iron (II) Sulfate &  21.60 & Moderate \\
58 (71,20)  &   31 &  124 &  68 $\pm$  80 &      Amitriptyline &         Salbutamol &   2.83 & Moderate \\
59 (112,13) &  168 &  302 & 160 $\pm$ 133 &          Diltiazem &        Simvastatin &   1.27 &    Major \\
60 (84,18)  &   63 &  159 & 148 $\pm$ 139 &         Fluoxetine &            Lithium &   1.79 &    Major \\
%61 (73,21)  &   32 &  122 &  94 $\pm$  92 &         Fluoxetine &      Nortriptyline &   2.70 &    Major \\
%62 (43,40)  &    5 &   52 &  50 $\pm$  48 &        Fluconazole &         Imipramine &   7.37 & Moderate \\
%63 (78,24)  &   33 &   97 &  27 $\pm$  45 &          Ibuprofen &            Lithium &   2.08 & Moderate \\
%64 (57,36)  &   10 &   59 &  10 $\pm$   8 &      Ciprofloxacin &  Iron (II) Sulfate &   4.18 & Moderate \\
%65 (46,50)  &    3 &   28 &  48 $\pm$  66 &        Propranolol &         Salbutamol &   6.61 &    Major \\
%66 (126,19) &   97 &  147 &  57 $\pm$  77 &      Acetaminophen &           Warfarin &   1.07 &    Minor \\
%67 (105,24) &   50 &   97 &  60 $\pm$  76 &           Atenolol &         Salbutamol &   1.37 & Moderate \\
%68 (77,33)  &   22 &   67 &  36 $\pm$  32 &        Fluconazole &           Diazepam &   2.16 & Moderate \\
%69 (125,21) &   79 &  122 & 107 $\pm$  95 &           Atenolol &         Gliclazide &   1.09 &     None \\
%70 (58,46)  &    6 &   35 &  82 $\pm$  74 &           Losartan &            Lithium &   4.13 & Moderate \\
        \bottomrule
        \end{tabular}
    %}
    \caption{    
        Top 60 known DDI pairs $(i,j)$ most imbalanced for \emph{females}, sorted by rank product (1\textsuperscript{st} column; individual rank in parenthesis) of $RRI^{F}_{i,j}$, the relative gender risk of DDI pair co-administration, and $|U^{\Phi,F}|$, the number of women affected by the DDI (7\textsuperscript{nd} and 3\textsuperscript{rd} columns, respectively).
        The number of men ($|U^{\Phi,M}|$) affected is shown in column 2.
        Mean ($\pm$ s.d.) co-administration length, $\langle \lambda^{u}_{i,j} \rangle$, is shown in column 4 (in days) for each DDI pair $(i,j)$ whose English drug names are shown in columns 5 and 6.
        DDI severity classification, according to \textit{Drugs.com}, shown in column 8; DDIs or drugs not found in \textit{Drugs.com} are labeled as \textit{None} or \textit{*}, respectively. 
    }
    \label{table:SI:rrr-gender-females}
\end{table}

\begin{table}
    \centering
    \scriptsize
    %\rotatebox{90}{
        \begin{tabular}{c|rrc|cc|rl}
        \toprule
        rankp$(RRI,U^{\text{M}})$ & $|U^{\Phi,M}|$ & $|U^{\Phi,F}|$ & $\langle \lambda^{u}_{i,j} \rangle$ & $i$ & $j$ & $RRI^{M}_{i,j}$ & severity \\
        \midrule
% UPDATED: 2018-11-07
% FILE: display_stats.py
 1 (1,32)  &   4 &   0 &  72 $\pm$  61 &        Atenolol &           Verapamil &   inf &     Major \\
 2 (1,38)  &   3 &   0 & 185 $\pm$  98 &       Phenytoin &            Levodopa &   inf &  Moderate \\
 3 (20,2)  & 280 & 244 & 243 $\pm$ 188 &     Haloperidol &           Biperiden &  1.62 &  Moderate \\
 3 (40,1)  & 553 & 696 & 141 $\pm$ 124 &             ASA &           Glyburide &  1.12 &  Moderate \\
 5 (1,43)  &   1 &   0 & 124 $\pm$   0 &   Carbamazepine &           Verapamil &   inf &     Major \\
 5 (1,43)  &   1 &   0 &   6 $\pm$   0 &    Erythromycin &       Carbamazepine &   inf &     Major \\
 5 (1,43)  &   1 &   0 &  31 $\pm$   0 &       Phenytoin &        Sulfadiazine &   inf &  Moderate \\
 8 (18,5)  & 207 & 178 & 155 $\pm$ 125 &         Digoxin &          Furosemide &  1.64 &  Moderate \\
 9 (13,9)  &  99 &  74 & 109 $\pm$  96 &         Digoxin &          Carvedilol &  1.89 &  Moderate \\
10 (8,15)  &  37 &  10 & 135 $\pm$ 109 &     Allopurinol &            Warfarin &  5.22 &  Moderate \\
11 (41,3)  & 262 & 347 &  48 $\pm$  93 &      Prednisone &                 ASA &  1.07 &  Moderate \\
12 (33,4)  & 236 & 260 & 103 $\pm$  87 &             ASA &          Gliclazide &  1.28 &      None \\
13 (17,8)  & 149 & 123 & 140 $\pm$ 114 &         Digoxin &      Spironolactone &  1.71 &     Minor \\
14 (6,24)  &  10 &   1 &  57 $\pm$  46 & Methylphenidate &       Carbamazepine & 14.12 &      None \\
15 (21,9)  &  99 &  87 & 142 $\pm$ 156 &     Haloperidol &       Carbamazepine &  1.61 &  Moderate \\
16 (7,32)  &   4 &   1 & 107 $\pm$ 123 &       Phenytoin &          Amiodarone &  5.65 &  Moderate \\
17 (12,22) &  16 &  11 &  10 $\pm$   6 &       Phenytoin &       Ciprofloxacin &  2.05 &  Moderate \\
18 (16,17) &  29 &  23 & 118 $\pm$ 114 &         Digoxin &          Amiodarone &  1.78 &     Major \\
19 (46,6)  & 158 & 219 & 143 $\pm$ 138 &      Fluoxetine &       Carbamazepine &  1.02 &  Moderate \\
20 (11,26) &   9 &   6 &  20 $\pm$  17 &      Carvedilol &           Fenoterol &  2.12 &         * \\
21 (42,7)  & 156 & 208 & 110 $\pm$ 106 &   Carbamazepine &         Simvastatin &  1.06 &  Moderate \\
22 (9,38)  &   3 &   1 &  62 $\pm$  44 &   Phenobarbital &       Aminophylline &  4.24 &  Moderate \\
23 (19,20) &  23 &  20 & 144 $\pm$ 153 &       Phenytoin &            Diazepam &  1.62 &  Moderate \\
24 (15,26) &   9 &   7 &  94 $\pm$  83 &       Phenytoin &          Furosemide &  1.82 &     Minor \\
25 (14,28) &   8 &   6 &   5 $\pm$   3 &     Doxycycline &         Amoxicillin &  1.88 &  Moderate \\
26 (34,12) &  70 &  79 & 115 $\pm$ 109 &       Phenytoin &          Omeprazole &  1.25 &  Moderate \\
27 (10,41) &   2 &   1 &   9 $\pm$   4 &       Phenytoin &        Prednisolone &  2.82 &  Moderate \\
28 (22,21) &  20 &  19 &  82 $\pm$  64 &      Gliclazide &          Carvedilol &  1.49 &      None \\
29 (36,13) &  55 &  68 &  31 $\pm$  43 &       Ibuprofen &          Carvedilol &  1.14 &  Moderate \\
30 (30,16) &  35 &  36 & 169 $\pm$ 151 &       Phenytoin &          Fluoxetine &  1.37 &  Moderate \\
31 (48,11) &  94 & 132 & 151 $\pm$ 145 &   Amitriptyline &       Carbamazepine &  1.01 &  Moderate \\
32 (32,17) &  29 &  31 &  49 $\pm$  95 &      Prednisone &            Warfarin &  1.32 &  Moderate \\
33 (31,19) &  28 &  29 &  96 $\pm$  96 &       Glyburide &          Carvedilol &  1.36 &  Moderate \\
34 (45,14) &  50 &  67 & 126 $\pm$ 127 &         Digoxin & Hydrochlorothiazide &  1.05 &  Moderate \\
35 (23,30) &   7 &   7 &   9 $\pm$   7 &      Furosemide &          Gentamicin &  1.41 &     Major \\
36 (24,32) &   4 &   4 &  82 $\pm$  54 &   Carbamazepine &       Aminophylline &  1.41 &  Moderate \\
36 (24,32) &   4 &   4 &  67 $\pm$  36 &       Diltiazem &       Carbamazepine &  1.41 &     Major \\
38 (35,23) &  11 &  13 &   3 $\pm$   2 &      Diclofenac &            Warfarin &  1.19 &     Major \\
39 (24,41) &   2 &   2 &   4 $\pm$   3 &     Doxycycline &         Clavulanate &  1.41 &      None \\
40 (24,43) &   1 &   1 & 102 $\pm$ 110 &     Propranolol &           Verapamil &  1.41 &     Major \\
40 (24,43) &   1 &   1 & 274 $\pm$ 218 &         Digoxin &    Propylthiouracil &  1.41 &  Moderate \\
40 (24,43) &   1 &   1 &   2 $\pm$   0 &       Phenytoin &      Hydrocortisone &  1.41 &  Moderate \\
43 (37,28) &   8 &  10 &  33 $\pm$  44 &     Fluconazole &            Warfarin &  1.13 &     Major \\
44 (47,24) &  10 &  14 & 100 $\pm$  85 &   Carbamazepine &            Warfarin &  1.01 &  Moderate \\
45 (37,32) &   4 &   5 &  51 $\pm$  91 &         Timolol &          Salbutamol &  1.13 &     Major \\
45 (37,32) &   4 &   5 &   7 $\pm$   1 &    Prednisolone &       Phenobarbital &  1.13 &  Moderate \\
47 (42,31) &   6 &   8 &  88 $\pm$  71 &       Phenytoin &            Warfarin &  1.06 &  Moderate \\
48 (42,38) &   3 &   4 &  62 $\pm$  97 &       Phenytoin &        Trimethoprim &  1.06 &  Moderate \\
        \bottomrule
        \end{tabular}
    %}
    \caption{
        All 48 known DDI pairs $(i,j)$ most imbalaced for \emph{males}, sorted by rank product (1\textsuperscript{st} column; individual rank in parenthesis) of $RRI^{M}_{i,j}$, the relative gender risk of DDI pair co-administration, and $|U^{\Phi,M}|$, the number of men affected by the DDI (7\textsuperscript{nd} and 2\textsuperscript{rd} columns, respectively).
        The number of women ($|U^{\Phi,F}|$) affected is shown in column 3.
        Mean ($\pm$ s.d.) co-administration length, $\langle \lambda^{u}_{i,j} \rangle$, is shown in column 4 (in days) for each DDI pair $(i,j)$ whose English drug names are shown in columns 5 and 6.
        DDI severity classification, according to \textit{Drugs.com}, shown in column 8; DDIs or drugs not found in \textit{Drugs.com} are labeled as \textit{None} or \textit{*}, respectively. 
    }
    \label{table:SI:rrr-gender-males}
\end{table}

\begin{table}[!hbt]
    \centering
    \scriptsize
        \begin{tabular}{c|rr|rr|rr}
        \toprule
        $RRI^{g}_{i,j} \geq x$ &
        $|D^{\Phi,F}|$ & $|D^{\Phi,M}|$ &
        $\Phi^{F}$ & $\Phi^{M}$ &
        $\Phi^{F}_{major}$ & $\Phi^{M}_{major}$ \\
        \midrule
% UPDATED: 2018-11-09
% FILE: display_stats_ddi.py 
1  & 68 & 46 & 133 & 48 & 31 & 10 \\
2  & 56 & 17 &  80 & 12 & 21 &  3 \\
3  & 49 & 13 &  65 &  9 & 16 &  3 \\
4  & 45 & 13 &  58 &  9 & 13 &  3 \\
5  & 38 & 11 &  49 &  8 & 13 &  3 \\
6  & 36 &  8 &  47 &  6 & 13 &  3 \\
7  & 35 &  8 &  45 &  6 & 12 &  3 \\
8  & 32 &  8 &  42 &  6 & 12 &  3 \\
9  & 31 &  8 &  41 &  6 & 11 &  3 \\
10 & 29 &  8 &  40 &  6 & 11 &  3 \\
        \bottomrule
        \end{tabular}
    \caption{    
        Number and proportions of drugs and interactions at increasing level of relative gender risk of DDI pair co-administration, $RRI^{g}_{i,j}>x$ (1\textsuperscript{st} column).
        Number of drugs by gender, is shown in columns 2 and 3.
        Number of drug pairs known to be a DDI, by gender, is shown in columns 4 and 5.
        Number of drug pairs, known to be a \emph{major} DDI, by gender, is shown in columns 6 and 7.
        See also Supplementary Table \ref{table:SI:min-rrr-gender-u}.
    }
    \label{table:SI:min-rrr-gender-i-ij}
\end{table}

\begin{table}[!hbt]
    \centering
    \scriptsize
        \begin{tabular}{c|rr|rr|rr|rr}
        \toprule
        $RRI^{g}_{i,j} \geq x$ &
        $|U^{\Phi,F}|$ & $|U^{\Phi,M}|$ &
        $|U^{\Phi,F}_{major}|$ & $|U^{\Phi,M}_{major}|$ &
        $\frac{|U^{\Phi,F}|}{|U^F|}$ & $\frac{|U^{\Phi,M}|}{|U^M|}$ &
        $\frac{|U^{\Phi,F}_{major}|}{|U^F|}$ & $\frac{|U^{\Phi,M}_{major}|}{|U^M|}$ \\
        \midrule
% UPDATED: 2018-11-09
% FILE: display_stats_ddi.py 
1  & 9,836 & 2,010 & 3,747 & 69 & 12.66\% & 03.65\% & 04.82\% & 00.13\% \\
2  & 7,089 &    91 & 2,060 &  6 & 09.12\% & 00.17\% & 02.65\% & 00.01\% \\
3  & 3,327 &    64 & 1,255 &  6 & 04.28\% & 00.12\% & 01.62\% & 00.01\% \\
4  & 1,589 &    64 &    73 &  6 & 02.05\% & 00.12\% & 00.09\% & 00.01\% \\
5  &   775 &    61 &    73 &  6 & 01.00\% & 00.11\% & 00.09\% & 00.01\% \\
6  &   744 &    20 &    73 &  6 & 00.96\% & 00.04\% & 00.09\% & 00.01\% \\
7  &   615 &    20 &    45 &  6 & 00.79\% & 00.04\% & 00.06\% & 00.01\% \\
8  &   547 &    20 &    45 &  6 & 00.70\% & 00.04\% & 00.06\% & 00.01\% \\
9  &   536 &    20 &    33 &  6 & 00.69\% & 00.04\% & 00.04\% & 00.01\% \\
10 &   441 &    20 &    33 &  6 & 00.57\% & 00.04\% & 00.04\% & 00.01\% \\
        \bottomrule
        \end{tabular}
    \caption{    
        Number and proportions of affected patients at increasing level of relative gender risk of DDI pair co-administration, $RRI^{g}_{i,j}>x$ (1\textsuperscript{st} column).
        Number of patients by gender, is shown in columns 2 and 3.
        Number of patients by gender and major DDI, is shown in columns 4 and 5.
        The relative proportion of affected patients in relation to the Pronto population is shown in columns 6 through 9.
        See also Supplementary Table \ref{table:SI:min-rrr-gender-i-ij}.
    }
    \label{table:SI:min-rrr-gender-u}
\end{table}
% Severity
\FloatBarrier
\subsection{Relative Risk per severity}
\label{ch:SI:relative-risk-ratios-severity}

\begin{table}[!hbt]
    \centering
    \scriptsize
    \begin{tabular}{l|rr|rr|r}
        \toprule
        severity $s$ & $|U^{\Phi,M}_{s}|$ &  $|U^{\Psi,F}_{s}|$ & $RRI^{F}_{s}$ \\
        \midrule
        % 
% UPDATED: 2018-04-25
% FILE: display_stats.py
\textit{Major}    &       1,433 &       3791 &    1.8739 \\
\textit{Moderate} &       3,951 &       8760 &    1.5705 \\
\textit{Minor}    &         247 &        281 &    0.8059 \\
\textit{None}     &         409 &        905 &    1.5674 \\
\textit{*}        &          39 &        140 &    2.5428 \\
        \bottomrule
    \end{tabular}
    \caption{  
        Absolute number of patients and relative risk measures per gender ($g$) and severity score ($s$, 1\textsuperscript{st} column).
        Columns 2 and 3 lists absolute number of males ($g=M$) and females ($g=F$) affected by at least one DDI for each severity score, respectively.
        Column 4 lists the relative risk of interaction given a specific severity score and gender.
        DDIs or drugs not found in \textit{Drugs.com} are labeled as \textit{None} or \textit{*}, respectively
        Notice that the same patient may have been administered DDI of more than one severity type.
    }
    \label{table:SI:rri-severity-gender}
\end{table}
% Age
\FloatBarrier
\subsection{Risk Measures per age}
\label{ch:SI:relative-risk-ratios-age}

\begin{table}[!hbt]
    \centering
    \scriptsize
    \begin{tabular}{c|rrrr|rr}
        \toprule
        $[y_1,y_2]$ & $|U^{[y_1,y_2]}|$ & $|U^{\nu \geq 2,[y_1,y_2]}|$ &  $|U^{\Psi,[y_1,y_2]}|$ &  $|U^{\Phi,[y_1,y_2]}|$ & $RC^{[y_1,y_2]}$ & $RI^{[y_1,y_2]}$ \\
        \midrule
% UPDATED: 2018-11-29
% FILE: display_stats.py
00-04 &   8,946 &    7,195 &   6,810 &      20 &  0.9465 &  0.0029 \\
05-09 &   6,390 &    4,688 &   4,362 &       7 &  0.9305 &  0.0016 \\
10-14 &   5,631 &    3,794 &   3,507 &      25 &  0.9244 &  0.0071 \\
15-19 &   8,305 &    6,094 &   5,705 &     139 &  0.9362 &  0.0244 \\
20-24 &  10,382 &    7,819 &   7,334 &     237 &  0.9380 &  0.0323 \\
25-29 &   9,725 &    7,305 &   6,835 &     301 &  0.9357 &  0.0440 \\
30-34 &   9,100 &    6,787 &   6,386 &     525 &  0.9409 &  0.0822 \\
35-39 &   8,844 &    6,696 &   6,259 &     687 &  0.9347 &  0.1098 \\
40-44 &   9,184 &    7,043 &   6,615 &   1,023 &  0.9392 &  0.1546 \\
45-49 &  10,085 &    8,039 &   7,610 &   1,426 &  0.9466 &  0.1874 \\
50-54 &  10,650 &    8,617 &   8,200 &   1,868 &  0.9516 &  0.2278 \\
55-59 &   9,236 &    7,686 &   7,386 &   1,956 &  0.9610 &  0.2648 \\
60-64 &   8,179 &    7,049 &   6,801 &   2,006 &  0.9648 &  0.2950 \\
65-69 &   6,315 &    5,572 &   5,444 &   1,794 &  0.9770 &  0.3295 \\
70-74 &   4,412 &    3,916 &   3,843 &   1,311 &  0.9814 &  0.3411 \\
75-79 &   3,398 &    3,042 &   2,968 &   1,057 &  0.9757 &  0.3561 \\
80-84 &   2,129 &    1,909 &   1,874 &     638 &  0.9817 &  0.3404 \\
85-89 &   1,174 &    1,029 &   1,007 &     349 &  0.9786 &  0.3466 \\
90+   &     637 &      531 &     515 &     158 &  0.9699 &  0.3068 \\
        \bottomrule
    \end{tabular}
    \caption{
       Absolute number of patients and risk measures per age range ($[y_1,y_2]$, 1\textsuperscript{st} column).
        Columns 2 through 5 lists, per age range, absolute numbers of:
            patients ($|U^{[y_1,y_2]}|$),
            patients with at least $2$ drug administrations ($|U^{\nu \geq 2,[y_1,y_2]}|$),
            patients with at least one co-administration ($|U^{\Psi,[y_1,y_2]}|$), and
            patients with at least one known DDI co-administration ($|U^{\Phi,[y_1,y_2]}|$).
        Per age range risk for both co-administration ($RC^{[y_1,y_2]}$) and known DDI co-administration ($RI^{[y_1,y_2]}$) are listed in columns 6 and 7, respectively.
    }
    \label{table:SI:prob-interactions-age}
\end{table}

\begin{table}[!hbt]
    \centering
    \scriptsize
    \begin{tabular}{c|rrrr|rr}
        \toprule
        $[y_1,y_2]$ &  $|U^{M,[y_1,y_2]}|$ & $|U^{\nu \geq 2,M,[y_1,y_2]}|$ &  $|U^{\Psi,M,[y_1,y_2]}|$ &  $|U^{\Phi,M,[y_1,y_2]}|$ & $RC^{M,[y_1,y_2]}$ & $RI^{M,[y_1,y_2]}$ \\
        \midrule
% UPDATED: 2018-04-25
% FILE: display_stats.py
00-04 &  4,537 &  3,664 &  3,473 &     8 &  0.9479 &  0.0023 \\
05-09 &  3,319 &  2,416 &  2,239 &     3 &  0.9267 &  0.0013 \\
10-14 &  2,932 &  1,926 &  1,776 &    14 &  0.9221 &  0.0079 \\
15-19 &  3,518 &  2,390 &  2,247 &    33 &  0.9402 &  0.0147 \\
20-24 &  4,204 &  3,020 &  2,838 &    76 &  0.9397 &  0.0268 \\
25-29 &  4,066 &  2,890 &  2,708 &    99 &  0.9370 &  0.0366 \\
30-34 &  3,692 &  2,641 &  2,500 &  1,68 &  0.9466 &  0.0672 \\
35-39 &  3,428 &  2,488 &  2,317 &  1,90 &  0.9313 &  0.0820 \\
40-44 &  3,504 &  2,559 &  2,394 &  2,79 &  0.9355 &  0.1165 \\
45-49 &  3,945 &  3,043 &  2,892 &  4,17 &  0.9504 &  0.1442 \\
50-54 &  4,142 &  3,219 &  3,048 &  5,25 &  0.9469 &  0.1722 \\
55-59 &  3,638 &  2,953 &  2,829 &  6,06 &  0.9580 &  0.2142 \\
60-64 &  3,257 &  2,731 &  2,622 &  6,26 &  0.9601 &  0.2387 \\
65-69 &  2,525 &  2,197 &  2,148 &  6,45 &  0.9777 &  0.3003 \\
70-74 &  1,729 &  1,494 &  1,461 &  4,27 &  0.9779 &  0.2923 \\
75-79 &  1,303 &  1,162 &  1,127 &  3,44 &  0.9699 &  0.3052 \\
80-84 &    718 &    649 &    637 &  1,86 &  0.9815 &  0.2920 \\
85-89 &    361 &    312 &    304 &    98 &  0.9744 &  0.3224 \\
90+   &    214 &    168 &    163 &    49 &  0.9702 &  0.3006 \\
        \bottomrule
    \end{tabular}
    \caption{
       Absolute number of \emph{male} patients and risk measures per age range ($[y_1,y_2]$, 1\textsuperscript{st} column).
        Columns 2 through 5 lists, per age range, absolute numbers of:
            male patients ($|U^{y}|$),
            male patients with at least $2$ drug administrations ($|U^{\nu \geq 2,M,[y_1,y_2]}|$),
            male patients with at least one co-administration ($|U^{\Psi,M,[y_1,y_2]}|$), and
            male patients with at least one known DDI co-administration ($|U^{\Phi,M,[y_1,y_2]}|$).
        Per age range women risk for both co-administration ($RC^{M,[y_1,y_2]}$) and known DDI co-administration ($RI^{M,[y_1,y_2]}$) are listed in columns 6 and 7, respectively.
    }
    \label{table:SI:prob-interactions-age-male}
\end{table}

\begin{table}[!hbt]
    \centering
    \scriptsize
    \begin{tabular}{c|rrrr|rr}
        \toprule
        $[y_1,y_2]$ &  $|U^{F,[y_1,y_2]}|$ & $|U^{\nu \geq 2,F,[y_1,y_2]}|$ & $|U^{\Psi,F,[y_1,y_2]}|$ &  $|U^{\Phi,F,[y_1,y_2]}|$ & $RC^{F,[y_1,y_2]}$ & $RI^{F,[y_1,y_2]}$ \\
        \midrule
% UPDATED: 2018-04-25
% FILE: display_stats.py
00-04 &  4,409 &  3,531 &  3,337 &     12 &  0.9451 &  0.0036 \\
05-09 &  3,071 &  2,272 &  2,123 &      4 &  0.9344 &  0.0019 \\
10-14 &  2,699 &  1,868 &  1,731 &     11 &  0.9267 &  0.0064 \\
15-19 &  4,787 &  3,704 &  3,458 &    106 &  0.9336 &  0.0307 \\
20-24 &  6,178 &  4,799 &  4,496 &    161 &  0.9369 &  0.0358 \\
25-29 &  5,659 &  4,415 &  4,127 &    202 &  0.9348 &  0.0489 \\
30-34 &  5,408 &  4,146 &  3,886 &    357 &  0.9373 &  0.0919 \\
35-39 &  5,416 &  4,208 &  3,942 &    497 &  0.9368 &  0.1261 \\
40-44 &  5,680 &  4,484 &  4,221 &    744 &  0.9413 &  0.1763 \\
45-49 &  6,140 &  4,996 &  4,718 &  1,009 &  0.9444 &  0.2139 \\
50-54 &  6,508 &  5,398 &  5,152 &  1,343 &  0.9544 &  0.2607 \\
55-59 &  5,598 &  4,733 &  4,557 &  1,350 &  0.9628 &  0.2962 \\
60-64 &  4,922 &  4,318 &  4,179 &  1,380 &  0.9678 &  0.3302 \\
65-69 &  3,790 &  3,375 &  3,296 &  1,149 &  0.9766 &  0.3486 \\
70-74 &  2,683 &  2,422 &  2,382 &    884 &  0.9835 &  0.3711 \\
75-79 &  2,095 &  1,880 &  1,841 &    713 &  0.9793 &  0.3873 \\
80-84 &  1,411 &  1,260 &  1,237 &    452 &  0.9817 &  0.3654 \\
85-89 &    813 &    717 &    703 &    251 &  0.9805 &  0.3570 \\
90+   &    423 &    363 &    352 &    109 &  0.9697 &  0.3097 \\
        \bottomrule
    \end{tabular}
    \caption{
       Absolute number of \emph{female} patients and risk measures per age range ($[y_1,y_2]$, 1\textsuperscript{st} column).
        Columns 2 through 5 lists, per age range, absolute numbers of:
            female patients ($|U^{y}|$),
            female patients with at least $2$ drug administrations ($|U^{\nu \geq 2,F,[y_1,y_2]}|$),
            female patients with at least one co-administration ($|U^{\Psi,F,[y_1,y_2]}|$), and
            female patients with at least one known DDI co-administration ($|U^{\Phi,F,[y_1,y_2]}|$).
        Per age range women risk for both co-administration ($RC^{F,[y_1,y_2]}$) and known DDI co-administration ($RI^{F,[y_1,y_2]}$) are listed in columns 6 and 7, respectively.
    }
    \label{table:SI:prob-interactions-age-female}
\end{table}

% Number of Drugs
\FloatBarrier
\clearpage
\subsection{Risk Ratios per number of drug}
\label{ch:SI:relative-risk-ratios-drugs}

\begin{table}[!hb]
    \centering
    \scriptsize
    \begin{tabular}{c|rrr|rr}
        \toprule
        \# of drugs $\nu$ &      $|U^{\nu}|$ &  $|U^{\Psi,\nu}|$ &  $|U^{\Phi,\nu}|$ & $RRC^{\nu}$ & $RRI^{\nu}$ \\
        \midrule
% UPDATED: 2017-01-16
% FILE: display_stats.py
1   & 27,911 &      - &     - &      - &       - \\
2   & 25,032 & 20,517 &   283 & 1.0    &  1.0    \\
3   & 19,163 & 18,468 &   677 & 1.1758 &  3.1249 \\
4   & 14,305 & 14,185 &   929 & 1.2098 &  5.7443 \\
5   & 11,026 & 11,010 & 1,208 & 1.2183 &  9.6908 \\
6   &  8,587 &  8,583 & 1,425 & 1.2195 & 14.6785 \\
7   &  6,438 &  6,438 & 1,512 & 1.2201 & 20.7735 \\
8   &  4,970 &  4,970 & 1,477 & 1.2201 & 26.2865 \\
9   &  3,877 &  3,877 & 1,417 & 1.2201 & 32.3283 \\
10  &  2,932 &  2,932 & 1,335 & 1.2201 & 40.2742 \\
11  &  2,264 &  2,264 & 1,089 & 1.2201 & 42.5462 \\
12  &  1,691 &  1,691 &   936 & 1.2201 & 48.9600 \\
13  &  1,214 &  1,214 &   754 & 1.2201 & 54.9366 \\
14  &    937 &    937 &   641 & 1.2201 & 60.5101 \\
15  &    618 &    618 &   413 & 1.2201 & 59.1113 \\
16  &    482 &    482 &   368 & 1.2201 & 67.5320 \\
17  &    366 &    366 &   285 & 1.2201 & 68.8768 \\
18  &    268 &    268 &   218 & 1.2201 & 71.9500 \\
19  &    177 &    177 &   142 & 1.2201 & 70.9617 \\
20  &    131 &    131 &   105 & 1.2201 & 70.8969 \\
>20 &    333 &    333 &   313 & 1.2201 & 83.1398 \\
        \bottomrule
    \end{tabular}
    \caption{
        Absolute number of patients, join probabilities and relative risk per number of distinct drugs dispensed, $\nu^{u}$ (1\textsuperscript{st} column).
        By definition, patients who had only one distinct drug dispensed could not have had any co-administration or interaction.
        Columns 2 through 4 lists, per distinct drugs dispensed, absolute numbers of:
            patients ($|U^{\nu}|$),
            patients with at least one co-administration ($|U^{\Psi,\nu}|$), and
            patients with at least one known DDI co-administration ($|U^{\Phi,\nu}|$).
        Per number of distinct drugs dispensed relative risks for both co-administration ($RRC^{\nu}$) and known DDI co-administration ($RRI^{\nu}$) are listed in columns 5 and 6, respectively.
        }
    \label{table:SI:prob-interactions-nr-drugs}
\end{table}

%
% DDI Network
%
\FloatBarrier
\section{DDI Networks}
\label{ch:SI:ddi-network}

For all pair of drugs known to interact we built two different networks, which are weighted versions of $\Delta$. In these networks weights are defined by either $\tau^{\Phi}_{i,j}$ or $|U^{\Phi}_{i,j}|$.
In this section we show alternatives plotting schemes, additional subgraphs, and tables containing values used for plotting or inference.
We also show a Principal Component Analysis and two clustering methods performed on the networks.

\begin{figure}[!hbt]
    \centering
    % UPDATED: 2018-11-08
    % FILE: plot_network_dist.py
    \includegraphics[width=.72\textwidth]{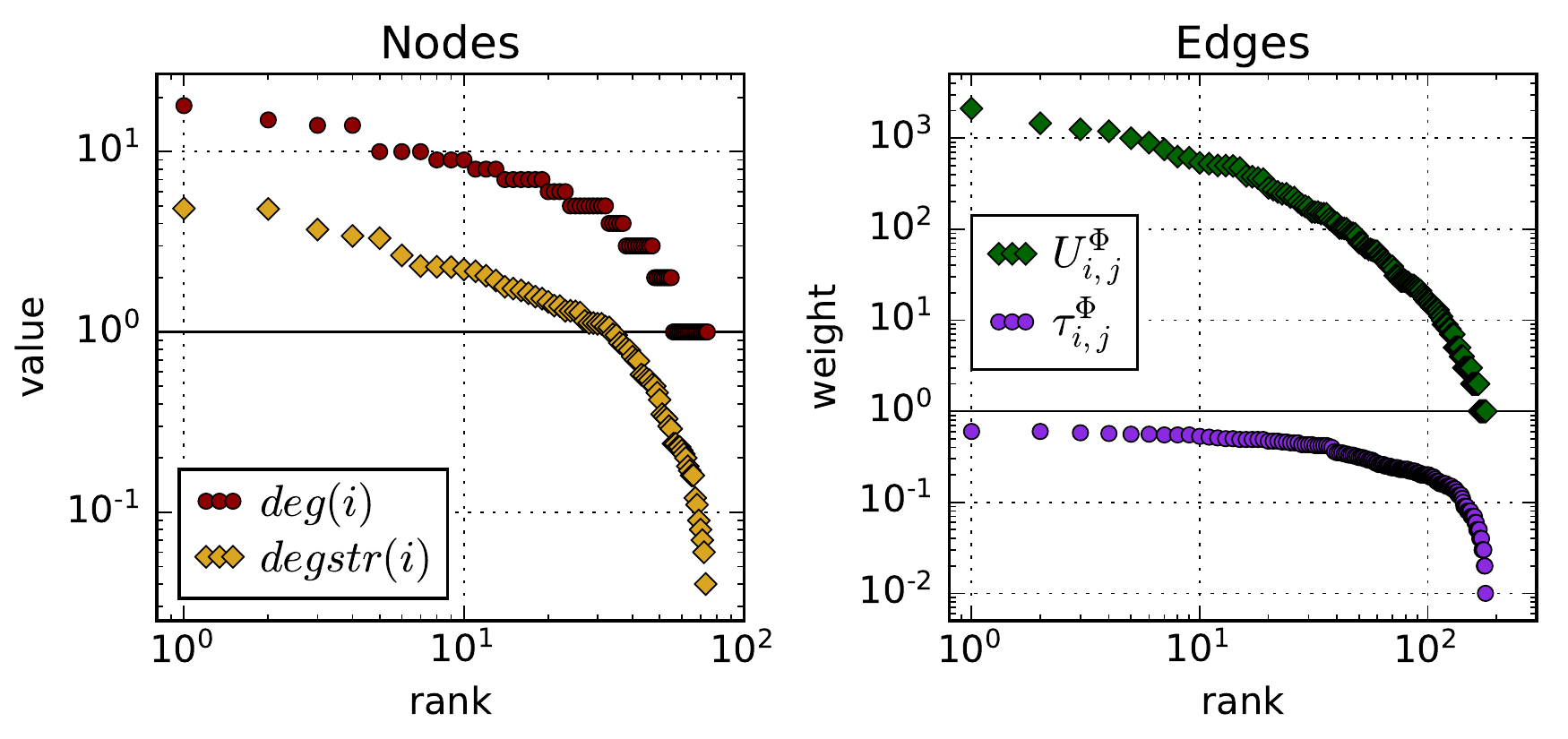}
    \caption{
        \textbf{Left.} Node degree, $deg(i)$, and node degree strength, $degstr(i) = \sum_{j} \tau^{\Phi}_{i,j}$, of weighted version of network $\Delta$ where weights are defined by $\tau^{\Phi}_{i,j}$.
        \textbf{Right.} Edge distribution of weighted version of network $\Delta$, where weights are defined by either $\tau^{\Phi}_{i,j}$ or $|U^{\Phi}_{i,j}|$.
    }
    \label{fig:SI:ddi-network-dist}
\end{figure}

\FloatBarrier

\begin{sidewaysfigure}
    \centering
    % UPDATED: 2018-11-08
    % FILE: build_ddi_network.py & /graphs/*.ai
    \includegraphics[width=.10\textwidth]{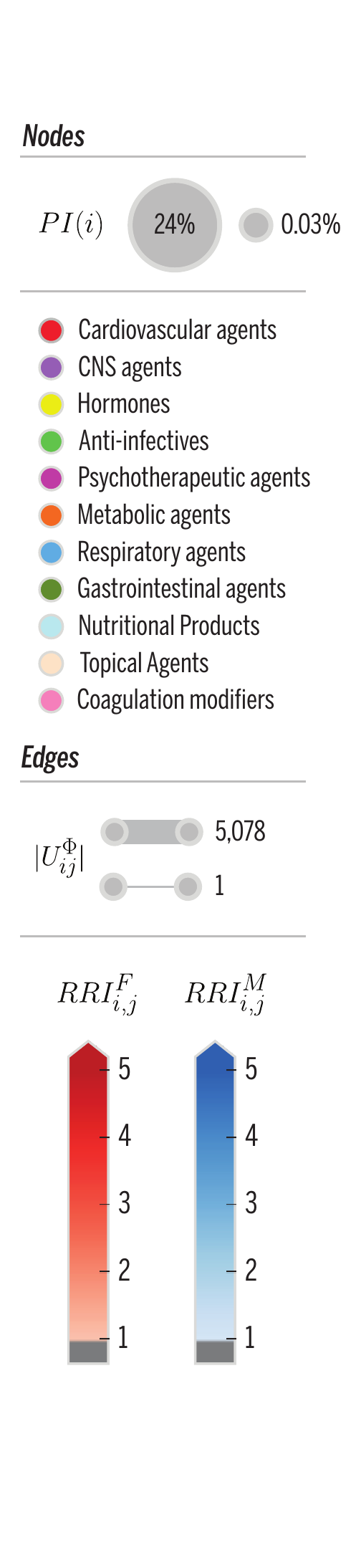}
    \includegraphics[width=.44\textwidth]{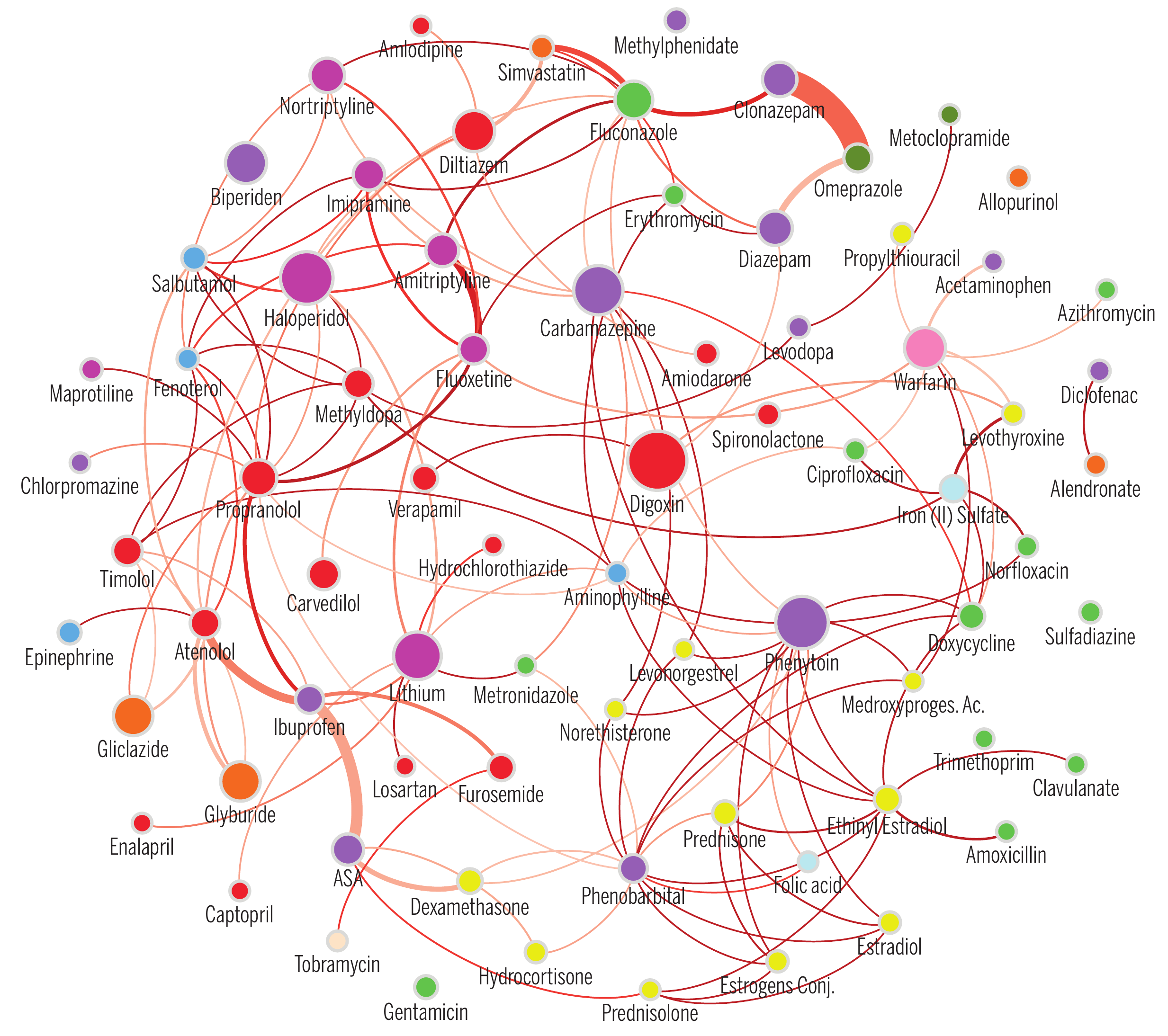}
    \includegraphics[width=.44\textwidth]{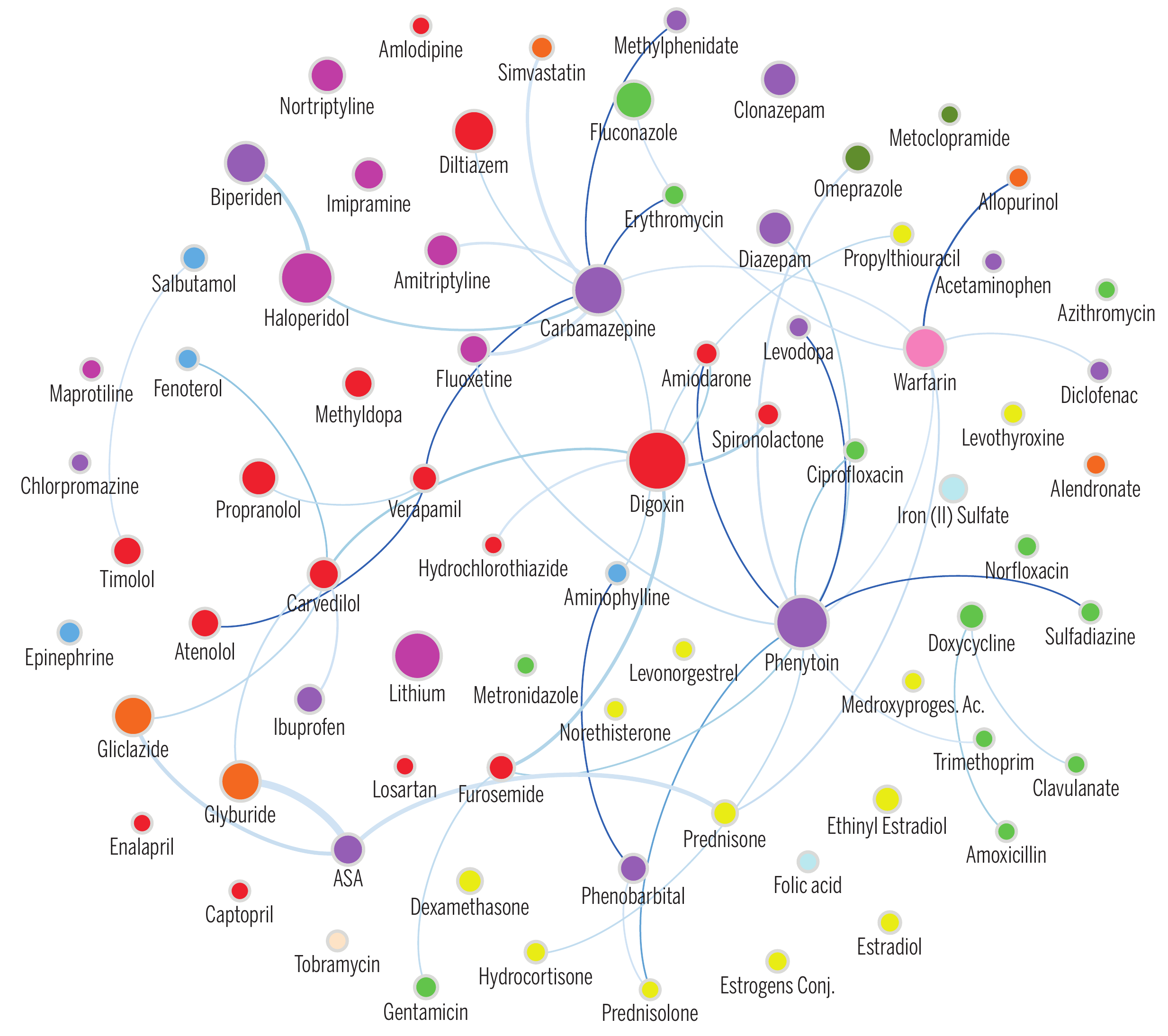} \\
    \includegraphics[width=.32\textwidth]{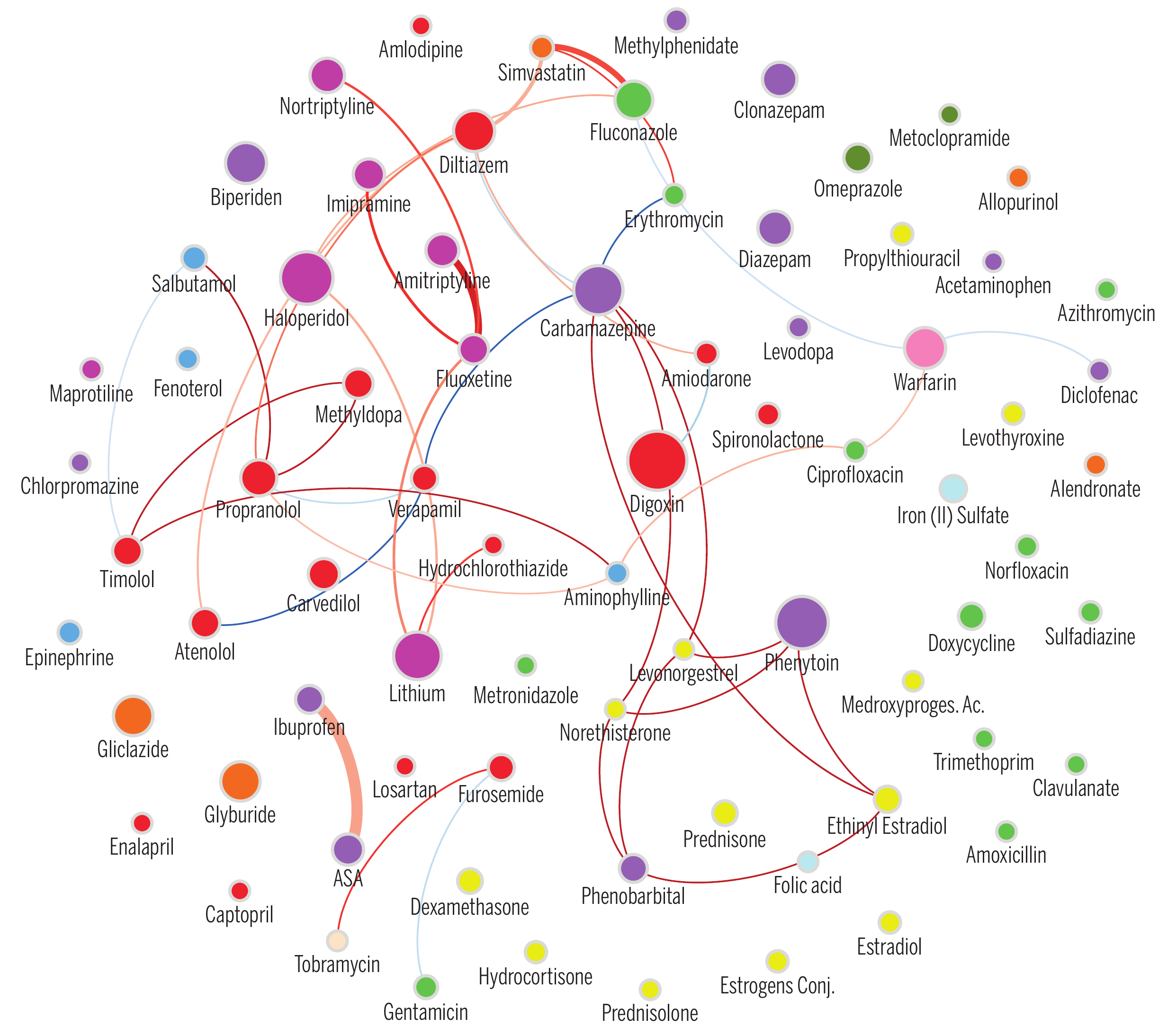}
    \includegraphics[width=.32\textwidth]{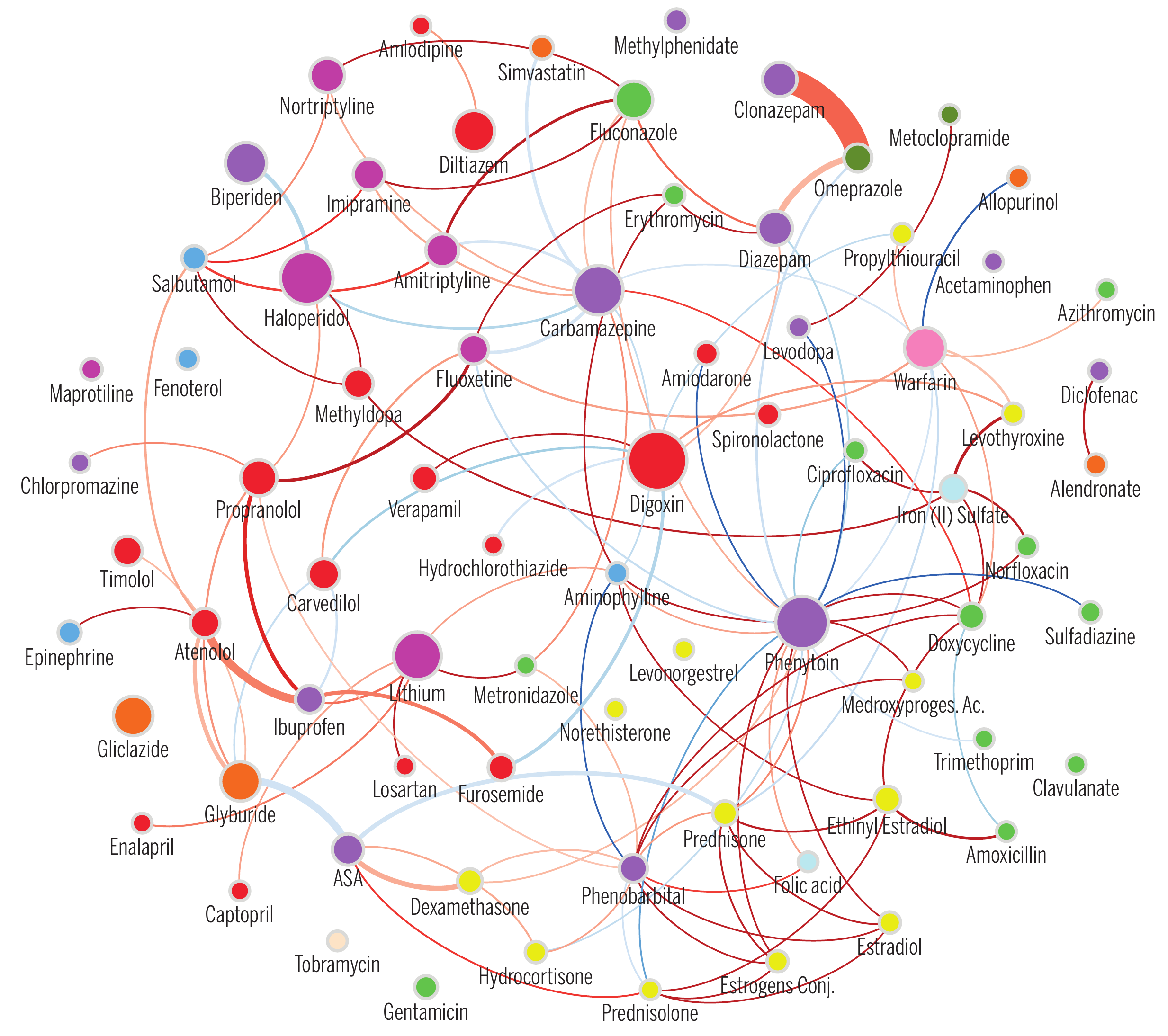}
    \includegraphics[width=.32\textwidth]{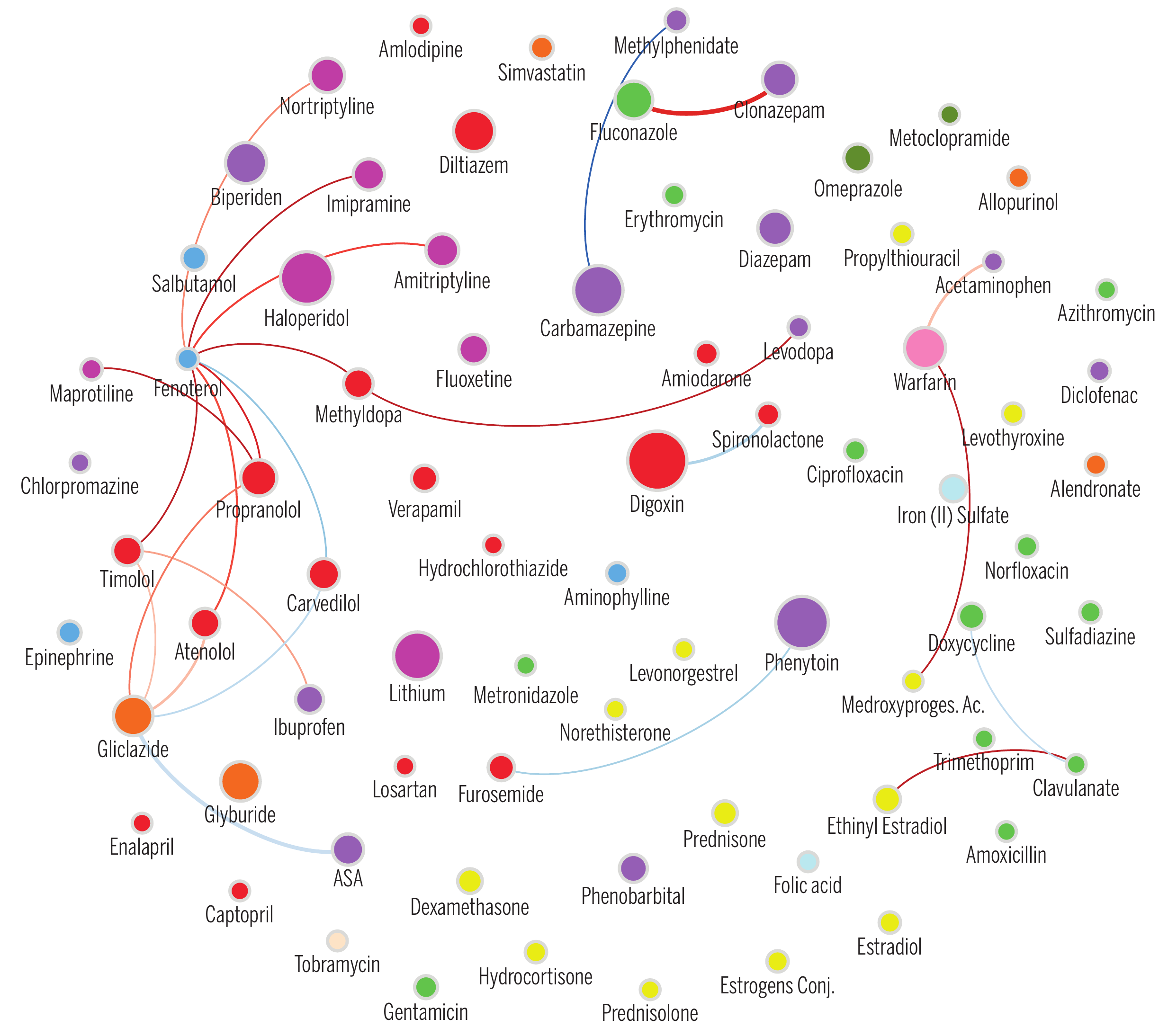}
    \caption{
        A weighted version of network $\Delta$ where weights are defined by $|U^{\Phi}_{i,j}|$.
        \textbf{Top left.} Female DDI network.
        \textbf{Top right.} Male DDI network.
        \textbf{Bottom left.} Major DDI network.
        \textbf{Bottom center.} Moderate DDI network.
        \textbf{Bottom right.} Minor DDI network.
        \textbf{Nodes} denote drugs $i$ involved in at least one co-administration known to be a DDI.
        Node color represents the highest level of primary action class, as retrieved from Drugs.com (see legend)
        Node size represents the probability of interaction, $PI(i)$, as defined in main text.
        \textbf{Edges weights} are the value of $|U^{\Phi}_{i,j}|$, the number of patients affected by the DDI.
        \textbf{Edge colors} denote $RRI^{g}_{i,j}$, where $g \in \{ \text{M},\text{F} \}$, to identify DDI edges that are higher risk for females (blue) or males (red). Color intensity for $RRI^{g}_{i,j}$ varies in $[1,5]$; that is, values are clipped at 5.
    }
    \label{fig:SI:u-ddi-network}
\end{sidewaysfigure}

\begin{sidewaysfigure}
    \centering
    % UPDATED: 2018-11-08
    % FILE: build_ddi_network.py & /graphs/*.ai
    \includegraphics[width=.10\textwidth]{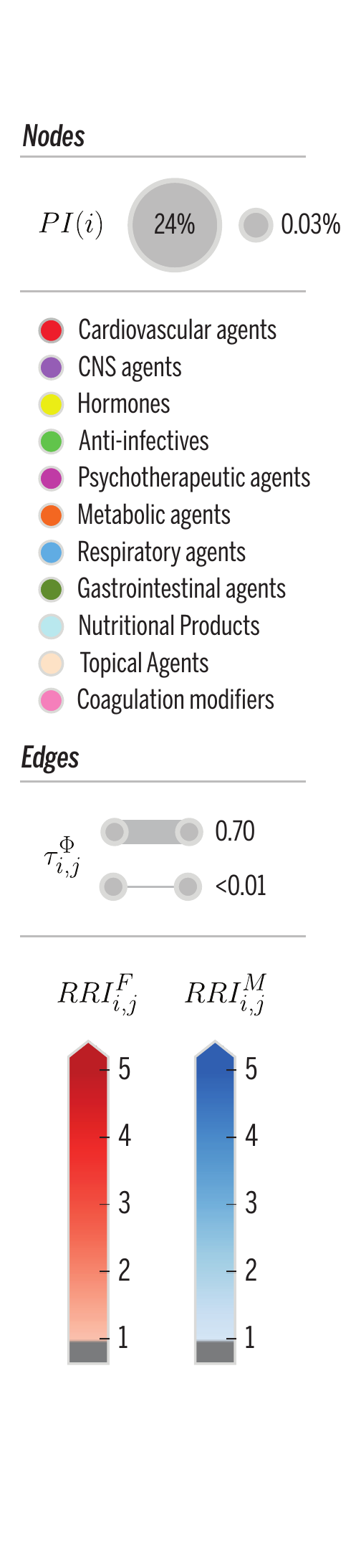}
    \includegraphics[width=.44\textwidth]{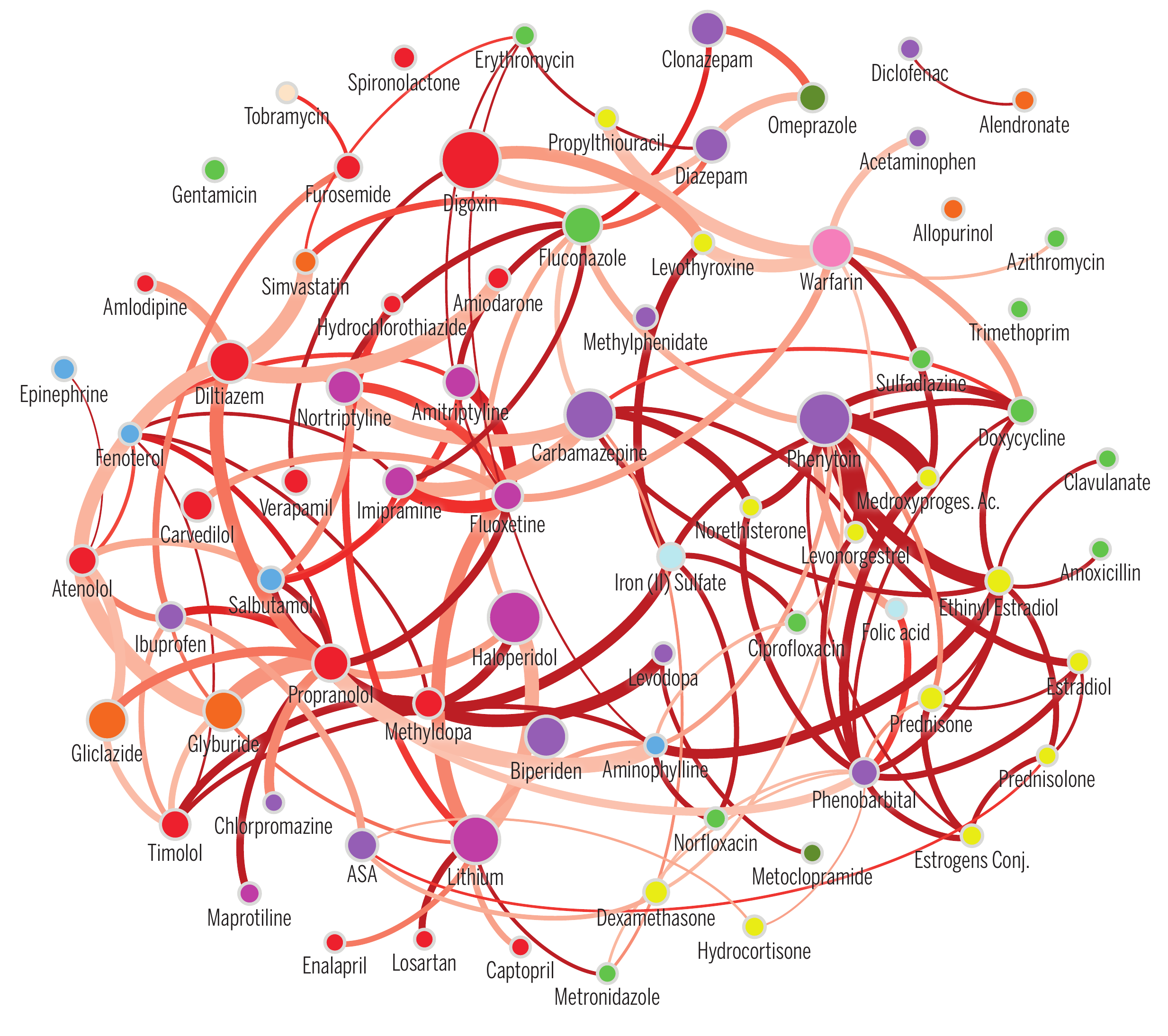}
    \includegraphics[width=.44\textwidth]{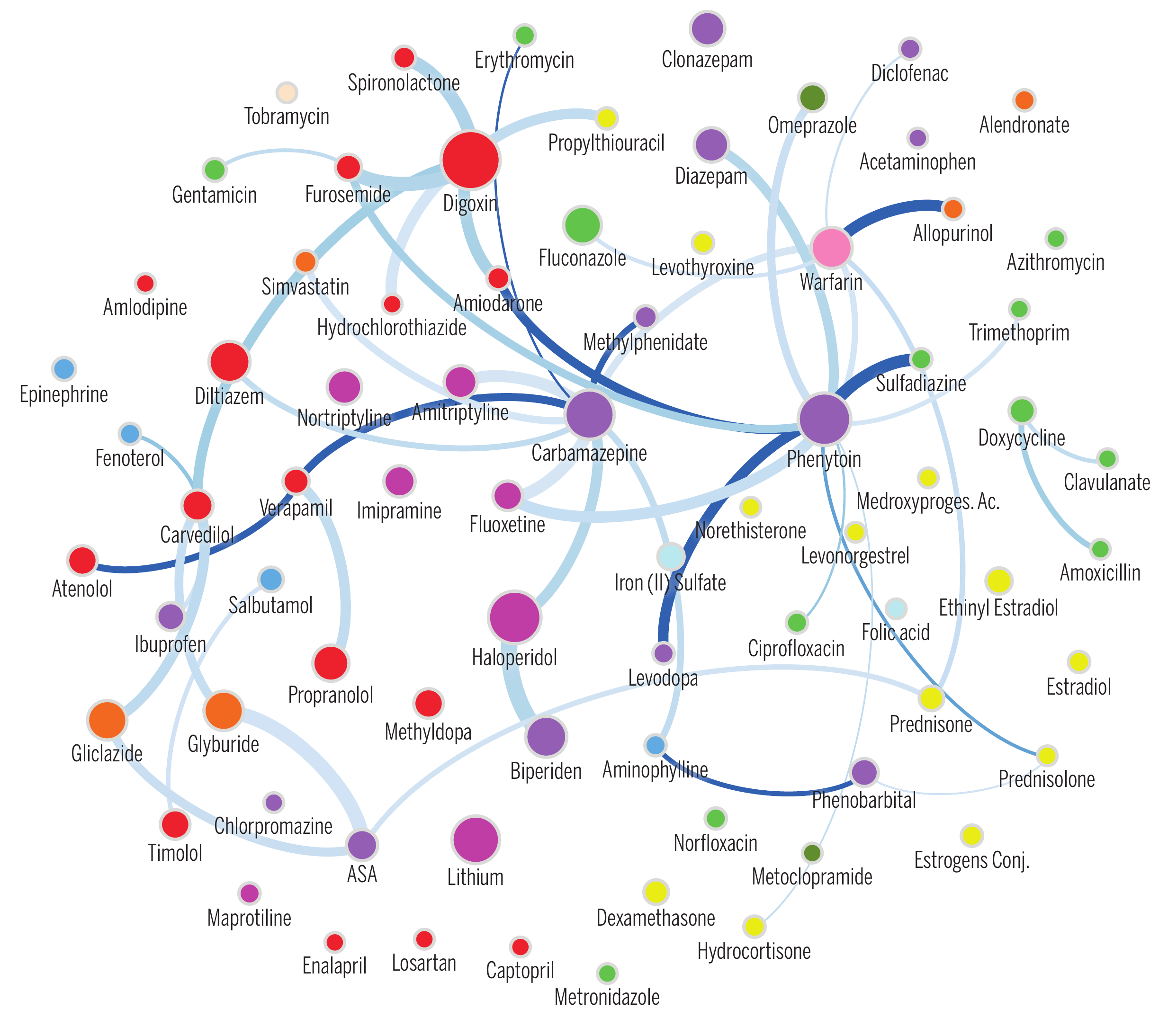} \\
    \includegraphics[width=.32\textwidth]{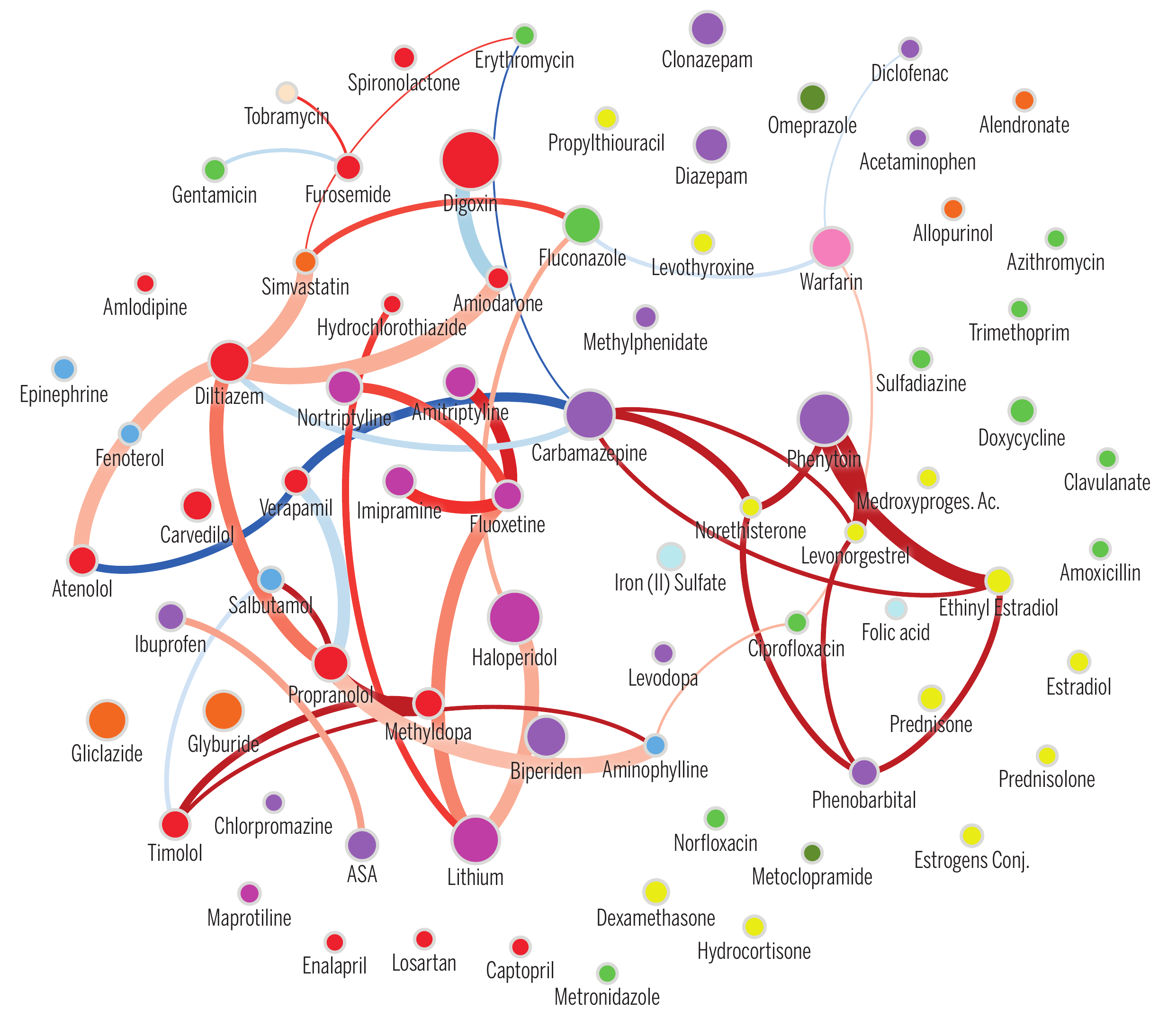}
    \includegraphics[width=.32\textwidth]{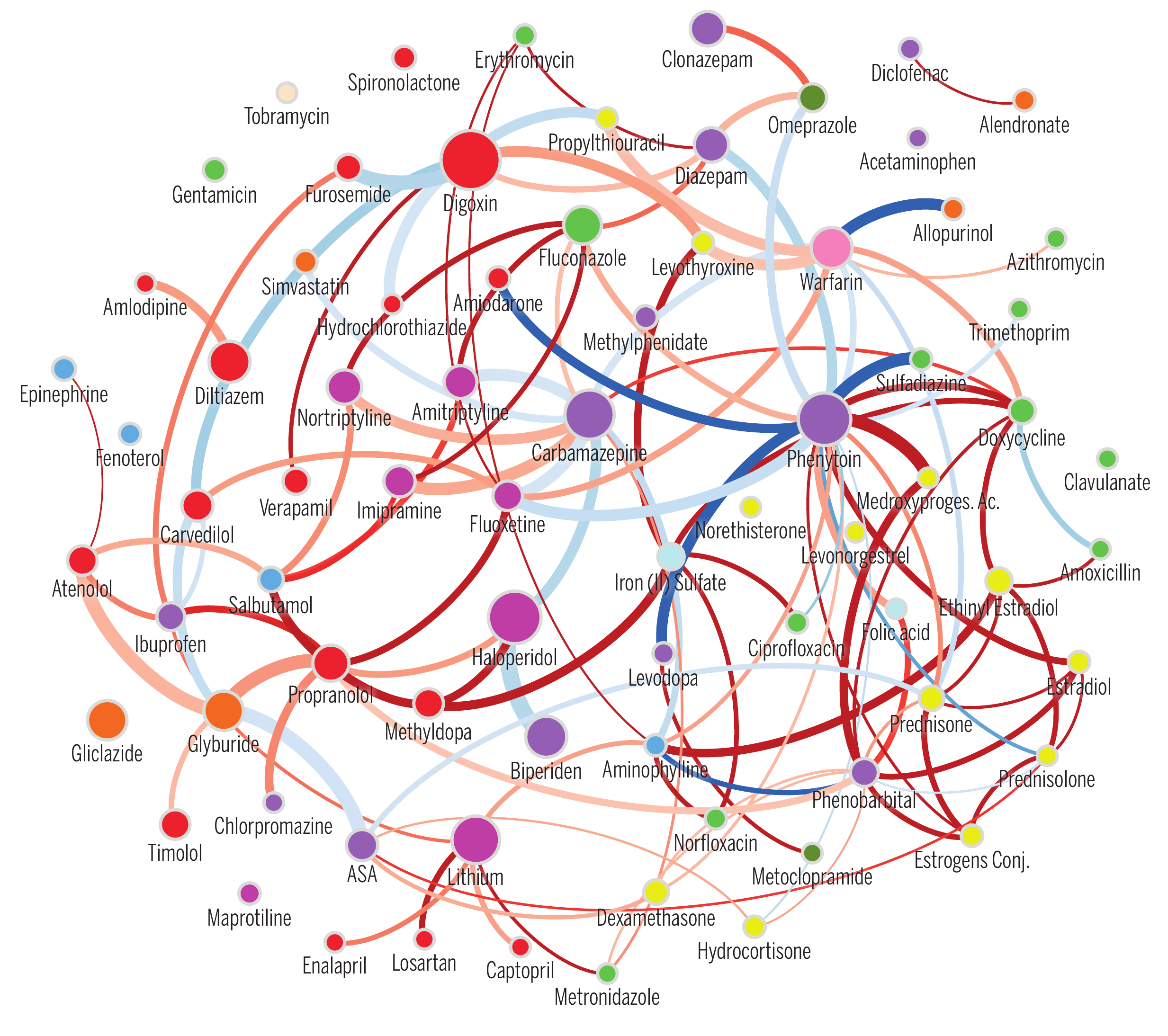}
    \includegraphics[width=.32\textwidth]{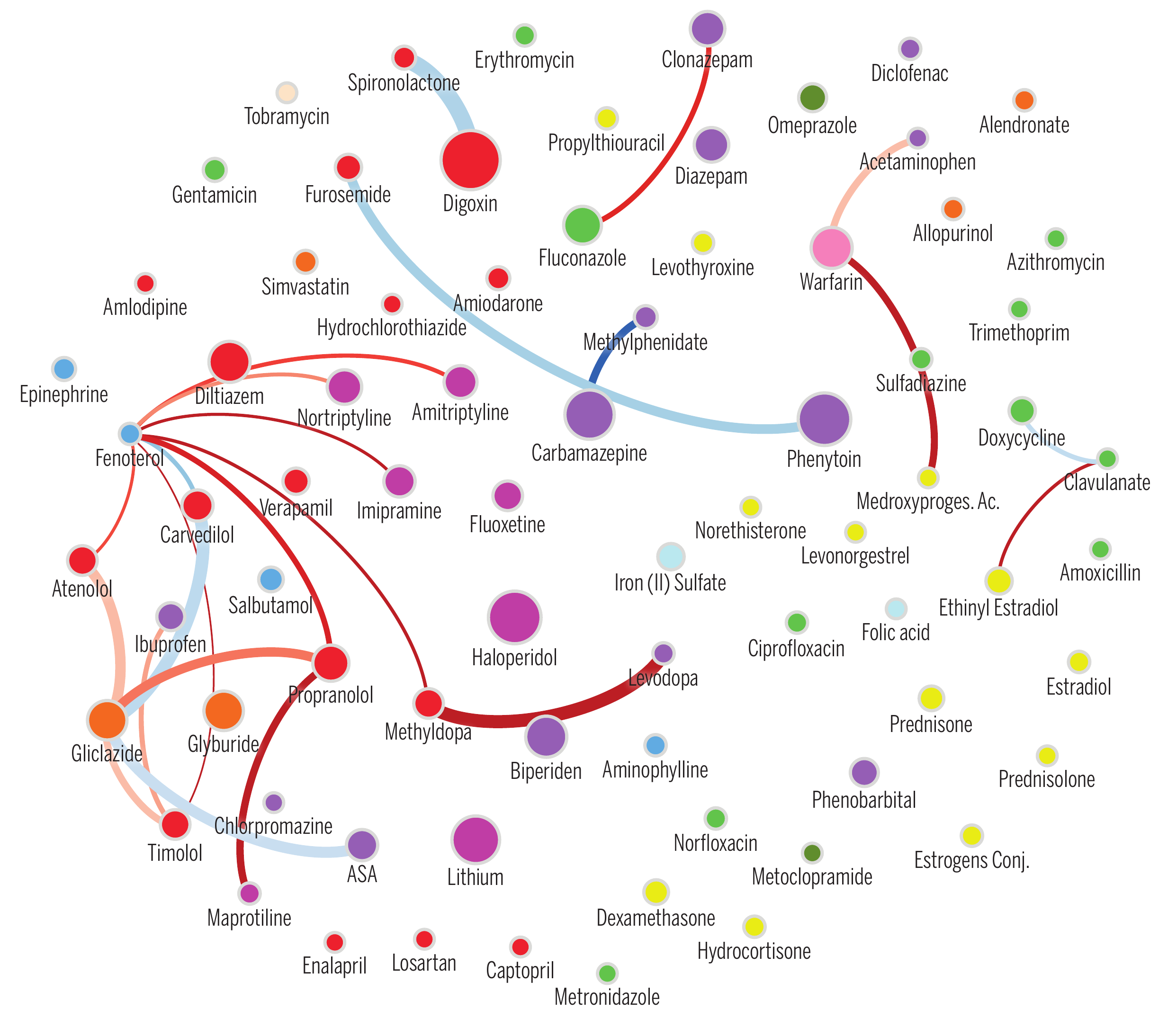}
    \caption{
        A weighted version of network $\Delta$ where weights are defined by $\tau^{\Phi}_{i,j}$.
        \textbf{Top left.} Female DDI network.
        \textbf{Top right.} Male DDI network.
        \textbf{Bottom left.} Major DDI network.
        \textbf{Bottom center.} Moderate DDI network.
        \textbf{Bottom right.} Minor DDI network.
        \textbf{Nodes} denote drugs $i$ involved in at least one co-administration known to be a DDI.
        Node color represents the highest level of primary action class, as retrieved from Drugs.com (see legend)
        Node size represents the probability of interaction, $PI(i)$, as defined in main text.
        \textbf{Edges weights} are the value of $\tau^{\Phi}_{i,j}$, obtained from Equation \ref{eq:tau_ij}, a normalized measure of the degree to which a specific DDI is co-prescribed.
        \textbf{Edge colors} denote $RRI^{g}_{i,j}$, where $g \in \{ \text{M},\text{F} \}$, to identify DDI edges that are higher risk for females (blue) or males (red). Color intensity for $RRI^{g}_{i,j}$ varies in $[1,5]$; that is, values are clipped at 5.
    }
    \label{fig:SI:tau-ddi-network}
\end{sidewaysfigure}

\FloatBarrier

\begin{figure}
    \centering
    % UPDATED: 2018-11-09
    % FILE: build_ddi_network.py / calculate_pca.py / plot_pca.py
    \includegraphics[width=.98\textwidth]{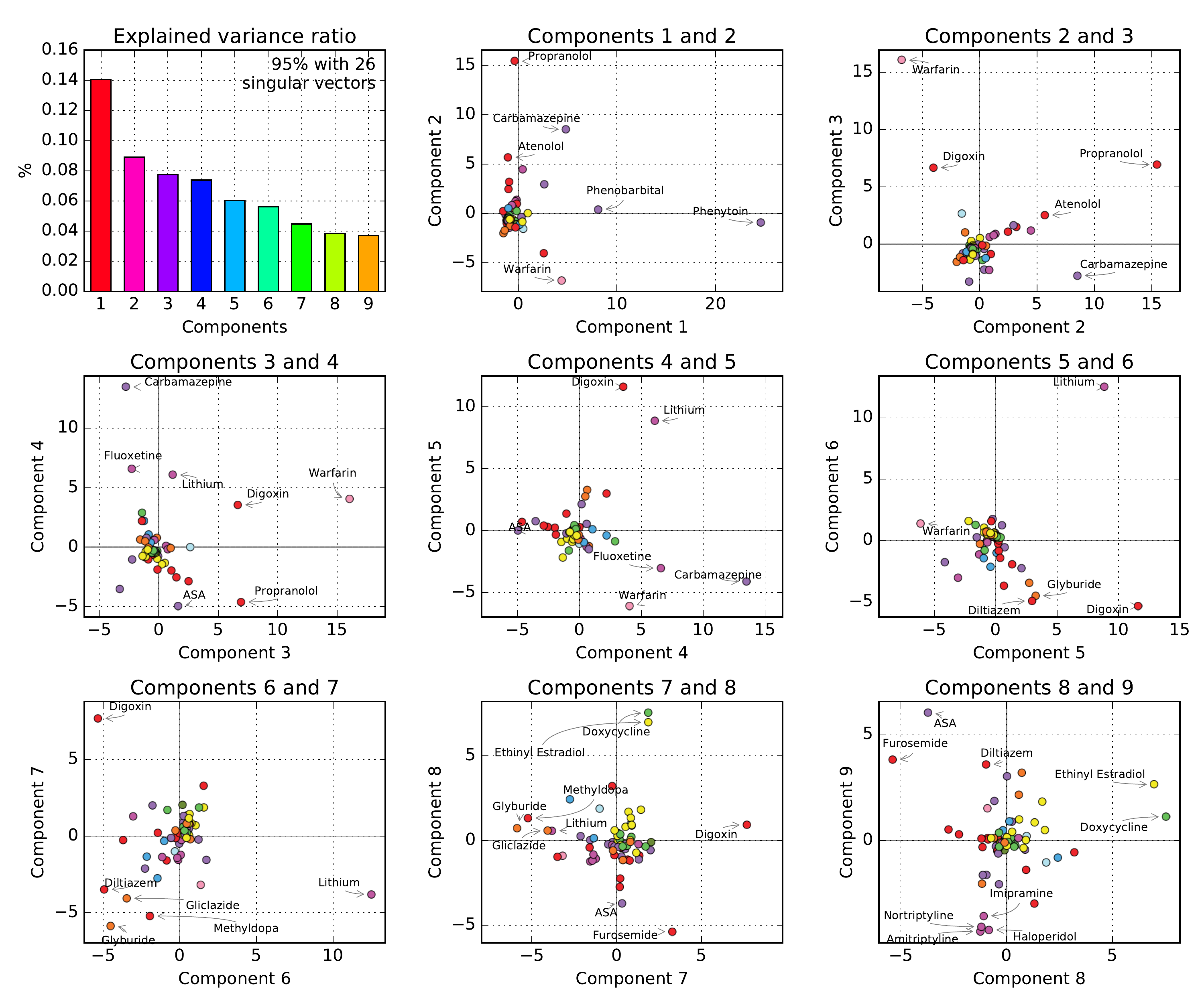}
    \caption{
        Principal Component Analysis (PCA) of network where weights are defined by $\tau^{\Phi}_{i,j}$.
        \textbf{Top Left}. Explained variance ratio for the first 9 principal components.
        \textbf{Additional plots}. Projection of network nodes (drugs) given the respective principal component.
        Nodes with loading $\geq 2$ s.d. in either component are annotated.
    }
    \label{fig:SI:ddi-network-pca-tau}
\end{figure}

\begin{figure}
    \centering
    % UPDATED: 2018-11-09
    % FILE: build_ddi_network.py / calculate_pca.py / plot_pca.py
    \includegraphics[width=.98\textwidth]{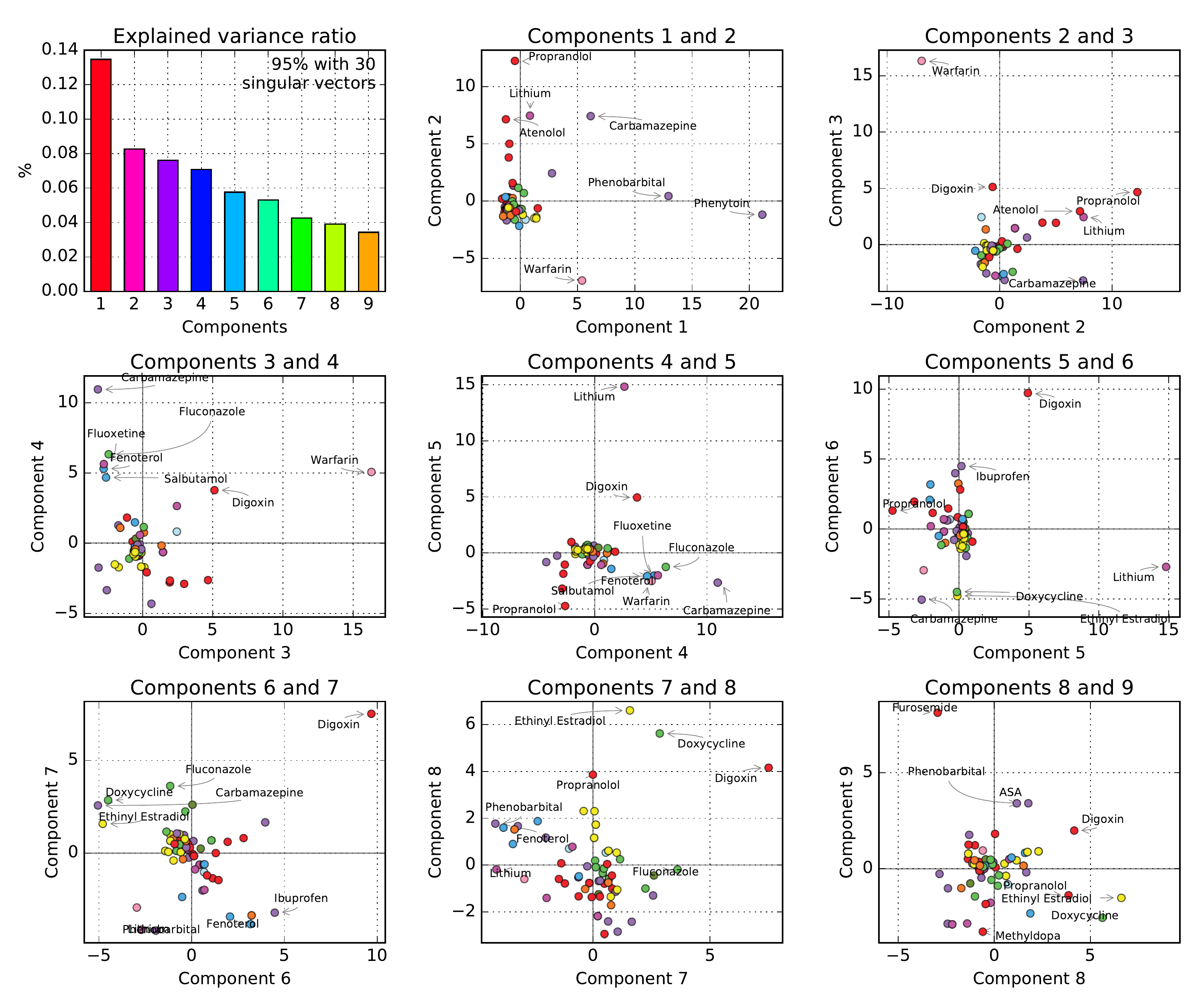}
    \caption{
        Principal Component Analysis (PCA) of network where weights are defined by $|U^{\Phi}_{i,j}|$.
        \textbf{Top Left}. Explained variance ratio for the first 9 principal components.
        \textbf{Additional plots}. Projection of network nodes (drugs) given the respective principal component.
        Nodes with loading $\geq 2$ s.d. in either component are annotated.
    }
    \label{fig:SI:ddi-network-pca-u}
\end{figure}

\FloatBarrier

\begin{table}
    \centering
    \scriptsize
        \begin{tabular}{c|c c c c|l}
        \toprule
        $i$ & $deg(i)$ & $degstr(i)$ & $betweenness(i)$ & $PI(i)$ & class \\
        \midrule
% UPDATED: 2018-11-08
% FILE: calculate_network_measures.py
         Phenytoin & 24 & 6.51 & 0.30 & 0.20 &              CNS agents \\
     Phenobarbital & 15 & 2.17 & 0.28 & 0.05 &              CNS agents \\
 Ethinyl Estradiol &  9 & 1.78 & 0.03 & 0.04 &                Hormones \\
       Doxycycline &  8 & 1.39 & 0.02 & 0.04 &         Anti-infectives \\
        Prednisone &  7 & 0.96 & 0.02 & 0.03 &                Hormones \\
      Prednisolone &  6 & 0.54 & 0.03 & 0.00 &                Hormones \\
          Diazepam &  5 & 1.12 & 0.05 & 0.09 &              CNS agents \\
      Erythromycin &  5 & 0.20 & 0.18 & 0.01 &         Anti-infectives \\
         Estradiol &  4 & 0.57 & 0.00 & 0.01 &                Hormones \\
   Estrogens Conj. &  4 & 0.58 & 0.00 & 0.01 &                Hormones \\
    Norethisterone &  3 & 0.73 & 0.00 & 0.00 &                Hormones \\
    Levonorgestrel &  3 & 0.79 & 0.00 & 0.00 &                Hormones \\
Medroxyproges. Ac. &  3 & 1.06 & 0.00 & 0.00 &                Hormones \\
        Omeprazole &  3 & 0.85 & 0.00 & 0.05 & Gastrointestinal agents \\
        Folic acid &  2 & 0.50 & 0.00 & 0.00 &    Nutritional Products \\
        Clonazepam &  2 & 0.42 & 0.00 & 0.09 &              CNS agents \\
       Amoxicillin &  2 & 0.30 & 0.00 & 0.00 &         Anti-infectives \\
       Clavulanate &  2 & 0.23 & 0.00 & 0.00 &         Anti-infectives \\
      Sulfadiazine &  1 & 0.51 & 0.00 & 0.01 &         Anti-infectives \\
      Trimethoprim &  1 & 0.16 & 0.00 & 0.00 &         Anti-infectives \\
\midrule
  Carbamazepine & 18 & 4.84 & 0.20 & 0.18 &               CNS agents \\
     Fluoxetine & 10 & 3.41 & 0.02 & 0.06 & Psychotherapeutic agents \\
    Haloperidol &  6 & 2.32 & 0.03 & 0.20 & Psychotherapeutic agents \\
        Lithium &  9 & 2.05 & 0.13 & 0.17 & Psychotherapeutic agents \\
    Fluconazole & 10 & 1.74 & 0.09 & 0.11 &          Anti-infectives \\
     Salbutamol &  7 & 1.53 & 0.00 & 0.03 &       Respiratory agents \\
  Amitriptyline &  5 & 1.47 & 0.00 & 0.08 & Psychotherapeutic agents \\
     Imipramine &  5 & 1.31 & 0.01 & 0.07 & Psychotherapeutic agents \\
  Nortriptyline &  5 & 1.30 & 0.00 & 0.09 & Psychotherapeutic agents \\
      Fenoterol &  8 & 0.81 & 0.13 & 0.01 &       Respiratory agents \\
      Biperiden &  1 & 0.70 & 0.00 & 0.13 &               CNS agents \\
Methylphenidate &  1 & 0.24 & 0.00 & 0.02 &               CNS agents \\
       Losartan &  1 & 0.21 & 0.00 & 0.00 &    Cardiovascular agents \\
      Captopril &  1 & 0.18 & 0.00 & 0.00 &    Cardiovascular agents \\
  Metronidazole &  3 & 0.17 & 0.16 & 0.00 &          Anti-infectives \\
      Enalapril &  1 & 0.16 & 0.00 & 0.00 &    Cardiovascular agents \\
\midrule
       Methyldopa & 7 & 2.30 & 0.01 & 0.06 &   Cardiovascular agents \\
Iron (II) Sulfate & 5 & 1.12 & 0.02 & 0.04 &    Nutritional Products \\
         Levodopa & 3 & 0.97 & 0.03 & 0.01 &              CNS agents \\
    Ciprofloxacin & 4 & 0.35 & 0.21 & 0.01 &         Anti-infectives \\
      Norfloxacin & 2 & 0.29 & 0.00 & 0.01 &         Anti-infectives \\
   Metoclopramide & 1 & 0.11 & 0.00 & 0.00 & Gastrointestinal agents \\
        \bottomrule
        \end{tabular}
    \caption{
        Louvain modules of weighted version of network $\Delta$ where weights are defined by $\tau^{\Phi}_{i,j}$.
        Each Louvain module is shown separated by a horizontal line.
        Drugs nodes ($i$; 1\textsuperscript{nd} column) and their respective degree, degree strength, and betweenness centrality measure, shown in columns 2, 3, and 4, respectively
        Column 5 shows the drug probability of interaction, $PI(i)$.
        Drug class is shown in column 6.
        Continues on Supplementary Table \ref{table:SI:ddi-network-louvain-2}.
        }
    \label{table:SI:ddi-network-louvain}
\end{table}

% Continues

\begin{table}
    \centering
    \scriptsize
        \begin{tabular}{c|c|cc|c|l}
        \toprule
        $i$ & $deg(i)$ & $degstr(i)$ & $betweenness(i)$ & $PI(i)$ & class \\
        \midrule
% UPDATED: 2018-11-08
% FILE: calculate_network_measures.py
            Digoxin &  9 & 3.70 & 0.03 & 0.24 & Cardiovascular agents \\
           Warfarin & 14 & 3.31 & 0.17 & 0.13 & Coagulation modifiers \\
          Diltiazem &  6 & 2.66 & 0.03 & 0.13 & Cardiovascular agents \\
         Amiodarone &  3 & 1.40 & 0.00 & 0.02 & Cardiovascular agents \\
         Furosemide &  5 & 1.31 & 0.05 & 0.04 & Cardiovascular agents \\
      Levothyroxine &  3 & 1.15 & 0.00 & 0.01 &              Hormones \\
        Simvastatin &  4 & 1.07 & 0.00 & 0.02 &      Metabolic agents \\
   Propylthiouracil &  2 & 0.87 & 0.00 & 0.01 &              Hormones \\
Hydrochlorothiazide &  2 & 0.69 & 0.00 & 0.00 & Cardiovascular agents \\
     Spironolactone &  1 & 0.55 & 0.00 & 0.02 & Cardiovascular agents \\
        Allopurinol &  1 & 0.46 & 0.00 & 0.01 &      Metabolic agents \\
         Amlodipine &  1 & 0.34 & 0.00 & 0.00 & Cardiovascular agents \\
      Acetaminophen &  1 & 0.22 & 0.00 & 0.00 &            CNS agents \\
         Gentamicin &  1 & 0.12 & 0.00 & 0.02 &       Anti-infectives \\
         Diclofenac &  2 & 0.09 & 0.03 & 0.01 &            CNS agents \\
         Tobramycin &  1 & 0.08 & 0.00 & 0.00 &        Topical Agents \\
       Azithromycin &  1 & 0.07 & 0.00 & 0.00 &       Anti-infectives \\
        Alendronate &  1 & 0.04 & 0.00 & 0.01 &      Metabolic agents \\
\midrule
 Aminophylline & 10 & 1.93 & 0.23 & 0.01 &       Respiratory agents \\
Hydrocortisone &  3 & 0.06 & 0.20 & 0.01 &                 Hormones \\
       Timolol &  7 & 1.11 & 0.16 & 0.06 &    Cardiovascular agents \\
     Ibuprofen &  7 & 1.28 & 0.06 & 0.05 &               CNS agents \\
      Atenolol &  8 & 2.22 & 0.05 & 0.06 &    Cardiovascular agents \\
   Propranolol & 14 & 4.81 & 0.06 & 0.10 &    Cardiovascular agents \\
           ASA &  7 & 1.57 & 0.01 & 0.07 &               CNS agents \\
     Verapamil &  4 & 1.11 & 0.01 & 0.04 &    Cardiovascular agents \\
     Glyburide &  5 & 2.29 & 0.00 & 0.12 &         Metabolic agents \\
    Carvedilol &  6 & 1.70 & 0.00 & 0.07 &    Cardiovascular agents \\
    Gliclazide &  5 & 1.64 & 0.00 & 0.12 &         Metabolic agents \\
Chlorpromazine &  1 & 0.33 & 0.00 & 0.00 &               CNS agents \\
 Dexamethasone &  3 & 0.24 & 0.00 & 0.03 &                 Hormones \\
   Maprotiline &  1 & 0.23 & 0.00 & 0.01 & Psychotherapeutic agents \\
\midrule
Epinephrine & 1 & 0.0 & 0.0 & 0.02 & Respiratory agents \\
        \bottomrule
        \end{tabular}
    \caption{
        Continuation.
        See Supplementary Table \ref{table:SI:ddi-network-louvain} for column description.
        }
    \label{table:SI:ddi-network-louvain-2}
\end{table}

\begin{table}
    \centering
    \scriptsize
        \begin{tabular}{c|c c c c|l}
        \toprule
        $i$ & $deg(i)$ & $degstr(i)$ & $betweenness(i)$ & $PI(i)$ & class \\
        \midrule
% UPDATED: 2018-11-08
% FILE: calculate_network_measures.py
         Phenytoin & 24 & 6.51 & 0.30 & 0.20 &           CNS agents \\
     Phenobarbital & 15 & 2.17 & 0.28 & 0.05 &           CNS agents \\
 Ethinyl Estradiol &  9 & 1.78 & 0.03 & 0.04 &             Hormones \\
       Doxycycline &  8 & 1.39 & 0.02 & 0.04 &      Anti-infectives \\
        Prednisone &  7 & 0.96 & 0.02 & 0.03 &             Hormones \\
      Prednisolone &  6 & 0.54 & 0.03 & 0.00 &             Hormones \\
         Estradiol &  4 & 0.57 & 0.00 & 0.01 &             Hormones \\
   Estrogens Conj. &  4 & 0.58 & 0.00 & 0.01 &             Hormones \\
     Dexamethasone &  3 & 0.24 & 0.00 & 0.03 &             Hormones \\
    Norethisterone &  3 & 0.73 & 0.00 & 0.00 &             Hormones \\
    Hydrocortisone &  3 & 0.06 & 0.20 & 0.01 &             Hormones \\
    Levonorgestrel &  3 & 0.79 & 0.00 & 0.00 &             Hormones \\
Medroxyproges. Ac. &  3 & 1.06 & 0.00 & 0.00 &             Hormones \\
        Folic acid &  2 & 0.50 & 0.00 & 0.00 & Nutritional Products \\
       Amoxicillin &  2 & 0.30 & 0.00 & 0.00 &      Anti-infectives \\
       Clavulanate &  2 & 0.23 & 0.00 & 0.00 &      Anti-infectives \\
      Sulfadiazine &  1 & 0.51 & 0.00 & 0.01 &      Anti-infectives \\
      Trimethoprim &  1 & 0.16 & 0.00 & 0.00 &      Anti-infectives \\
\midrule
      Propranolol & 14 & 4.81 & 0.06 & 0.10 &    Cardiovascular agents \\
       Methyldopa &  7 & 2.30 & 0.01 & 0.06 &    Cardiovascular agents \\
        Glyburide &  5 & 2.29 & 0.00 & 0.12 &         Metabolic agents \\
         Atenolol &  8 & 2.22 & 0.05 & 0.06 &    Cardiovascular agents \\
    Aminophylline & 10 & 1.93 & 0.23 & 0.01 &       Respiratory agents \\
       Carvedilol &  6 & 1.70 & 0.00 & 0.07 &    Cardiovascular agents \\
       Gliclazide &  5 & 1.64 & 0.00 & 0.12 &         Metabolic agents \\
              ASA &  7 & 1.57 & 0.01 & 0.07 &               CNS agents \\
       Salbutamol &  7 & 1.53 & 0.00 & 0.03 &       Respiratory agents \\
        Ibuprofen &  7 & 1.28 & 0.06 & 0.05 &               CNS agents \\
Iron (II) Sulfate &  5 & 1.12 & 0.02 & 0.04 &     Nutritional Products \\
          Timolol &  7 & 1.11 & 0.16 & 0.06 &    Cardiovascular agents \\
        Verapamil &  4 & 1.11 & 0.01 & 0.04 &    Cardiovascular agents \\
         Levodopa &  3 & 0.97 & 0.03 & 0.01 &               CNS agents \\
    Ciprofloxacin &  4 & 0.35 & 0.21 & 0.01 &          Anti-infectives \\
   Chlorpromazine &  1 & 0.33 & 0.00 & 0.00 &               CNS agents \\
      Norfloxacin &  2 & 0.29 & 0.00 & 0.01 &          Anti-infectives \\
      Maprotiline &  1 & 0.23 & 0.00 & 0.01 & Psychotherapeutic agents \\
   Metoclopramide &  1 & 0.11 & 0.00 & 0.00 &  Gastrointestinal agents \\
        \bottomrule
        \end{tabular}
    \caption{
        InfoMap modules of weighted version of network $\Delta$ where weights are defined by $\tau^{\Phi}_{i,j}$.
        Each InfoMap module is shown separated by a horizontal line.
        Drugs nodes ($i$; 1\textsuperscript{nd} column) and their respective degree, total degree strength (, and betweenness centrality measure, shown in columns 2, 3, and 4, respectively
        Column 5 shows the drug probability of interaction, $PI(i)$.
        Drug class is shown in column 6.
        Continues on Supplementary Table \ref{table:SI:ddi-network-infomap-2}.
        }
    \label{table:SI:ddi-network-infomap}
\end{table}

% Continues

\begin{table}
    \centering
    \scriptsize
        \begin{tabular}{c|c|cc|c|l}
        \toprule
        $i$ & $deg(i)$ & $degstr(i)$ & $betweenness(i)$ & $PI(i)$ & class \\
        \midrule
% UPDATED: 2018-11-08
% FILE: calculate_network_measures.py
  Carbamazepine & 18 & 4.84 & 0.20 & 0.18 &               CNS agents \\
     Fluoxetine & 10 & 3.41 & 0.02 & 0.06 & Psychotherapeutic agents \\
    Fluconazole & 10 & 1.74 & 0.09 & 0.11 &          Anti-infectives \\
  Amitriptyline &  5 & 1.47 & 0.00 & 0.08 & Psychotherapeutic agents \\
     Imipramine &  5 & 1.31 & 0.01 & 0.07 & Psychotherapeutic agents \\
  Nortriptyline &  5 & 1.30 & 0.00 & 0.09 & Psychotherapeutic agents \\
      Fenoterol &  8 & 0.81 & 0.13 & 0.01 &       Respiratory agents \\
Methylphenidate &  1 & 0.24 & 0.00 & 0.02 &               CNS agents \\
\midrule
            Digoxin &  9 & 3.70 & 0.03 & 0.24 & Cardiovascular agents \\
           Warfarin & 14 & 3.31 & 0.17 & 0.13 & Coagulation modifiers \\
         Furosemide &  5 & 1.31 & 0.05 & 0.04 & Cardiovascular agents \\
      Levothyroxine &  3 & 1.15 & 0.00 & 0.01 &              Hormones \\
   Propylthiouracil &  2 & 0.87 & 0.00 & 0.01 &              Hormones \\
Hydrochlorothiazide &  2 & 0.69 & 0.00 & 0.00 & Cardiovascular agents \\
     Spironolactone &  1 & 0.55 & 0.00 & 0.02 & Cardiovascular agents \\
        Allopurinol &  1 & 0.46 & 0.00 & 0.01 &      Metabolic agents \\
      Acetaminophen &  1 & 0.22 & 0.00 & 0.00 &            CNS agents \\
         Gentamicin &  1 & 0.12 & 0.00 & 0.02 &       Anti-infectives \\
         Tobramycin &  1 & 0.08 & 0.00 & 0.00 &        Topical Agents \\
       Azithromycin &  1 & 0.07 & 0.00 & 0.00 &       Anti-infectives \\
\midrule
  Haloperidol & 6 & 2.32 & 0.03 & 0.20 & Psychotherapeutic agents \\
      Lithium & 9 & 2.05 & 0.13 & 0.17 & Psychotherapeutic agents \\
    Biperiden & 1 & 0.70 & 0.00 & 0.13 &               CNS agents \\
     Losartan & 1 & 0.21 & 0.00 & 0.00 &    Cardiovascular agents \\
    Captopril & 1 & 0.18 & 0.00 & 0.00 &    Cardiovascular agents \\
Metronidazole & 3 & 0.17 & 0.16 & 0.00 &          Anti-infectives \\
    Enalapril & 1 & 0.16 & 0.00 & 0.00 &    Cardiovascular agents \\
\midrule
  Diltiazem & 6 & 2.66 & 0.03 & 0.13 &  Cardiovascular agents \\
 Amiodarone & 3 & 1.40 & 0.00 & 0.02 &  Cardiovascular agents \\
Simvastatin & 4 & 1.07 & 0.00 & 0.02 &       Metabolic agents \\
 Amlodipine & 1 & 0.34 & 0.00 & 0.00 &  Cardiovascular agents \\
\midrule
    Diazepam & 5 & 1.12 & 0.05 & 0.09 &              CNS agents \\
  Omeprazole & 3 & 0.85 & 0.00 & 0.05 & Gastrointestinal agents \\
  Clonazepam & 2 & 0.42 & 0.00 & 0.09 &              CNS agents \\
Erythromycin & 5 & 0.20 & 0.18 & 0.01 &         Anti-infectives \\
\midrule
 Diclofenac & 2 & 0.09 & 0.03 & 0.01 &       CNS agents \\
Alendronate & 1 & 0.04 & 0.00 & 0.01 & Metabolic agents \\
\midrule
Epinephrine & 1 & 0.0 & 0.0 & 0.02 &  Respiratory agents \\
        \bottomrule
        \end{tabular}
    \caption{
        Continuation.
        See Supplementary Table \ref{table:SI:ddi-network-infomap} for column description.
        }
    \label{table:SI:ddi-network-infomap-2}
\end{table}

%
% Null Models RRY^{y}
%
\FloatBarrier
\section{Null Model for \texorpdfstring{$RI^{y}$}{RIy}}
\label{ch:SI:null-models}

To test if sheer combinatorics explains the increased risk of DDI in older age, we compared the observed risk of interactions $RI^{y}$ with a random null model, $H^{rnd}_{0}$.
We separated all patients $u$ in our dataset per age range $y$.
From these subset of patients $U^{[y1,y2]}$ we also separated which drugs $d$ were prescribed in their age range as $D^{[y1-y2]}$. 
For clarity, we will refer to all measures previously reported with an added star ($\star$) in the notation to indicate that these values are calculated for the null model (e.g., $RI^{y\star}$ is the null model value of the risk of interaction per age range, $RI^{y}$).

The null model is then computed by proportionally sampling patients for each age range, $u \in U^{[y1-y2]}$.
For each drawn patient $u$ we sampled $|D^u|$ drugs available to patients in the patient's age range $D^{[y1,y2]}$, and then randomly drew $\Psi^{u}$ co-administrations from the patient's possible pairwise combinations $\binom{|D^u|}{2}$ of drugs, thus yielding random drug pairs $\psi^{u\star}_{i,j}$ that matched the observed number of co-administrations, $\Psi^{u} \equiv \Psi^{u\star}$.
To decide if a co-administration is an interaction in the null model, we compare the randomly drawn pair of drugs against DrugBank to decide if $\varphi^{u\star}_{i,j}$ is an interaction or not.

This null model allow us to measure what is the expected number of interactions given the increase of co-administrations observed with age, assuming drugs are prescribed completely at random.
In other words, it measures the risk of DDI if only age, and the drugs available to patients in these ages, were given to them at random with the same number of co-administrations.

%
% Confidence Intervals
%
To compute confidence intervals for the number of patients in the null model, we proportionally sampled the same number of patients observed in each age range, 100 times. Confidence intervals can be seen as background fills in Figures \ref{fig:rc-ri-age} and \ref{fig:rc-ri-age-gender}.
%
% Hypothesis tests
%
To measure the significance of our null models, Supplementary Table \ref{table:SI:null-models-h0-tests} shows the chi-square tests against the expected number of patients in each age bin, $|U^{[y1,y2]}|$, from our data. The null model rejects the hypothesis it was sampled from the same distribution as our data. This means the observed increase in DDI with age, seen in our data, cannot be explained alone by the increased combinatorics of drug co-administrations alone.

\begin{table}[!hbt]
    \centering
    \small
    \begin{tabular}{llrr}
    \toprule
    {} & model & chi-square & $p$-value \\
    \midrule
% UPDATED: 2018-11-15
% FILE: One of the rc/ri plotting files 
1  &  $H^{rnd}_{0}$   &  22378.5912 &  0.0 \\
    \bottomrule
    \end{tabular}
    \caption{
        Chi-square statistic when the number of patients in the null model, $|U^{y\star}|$, is compared to the observed values, $|U^{y}|$.
    }
    \label{table:SI:null-models-h0-tests}
\end{table}

%
% Maps (Geographical Results) 
%
\FloatBarrier
%\pagebreak
\section{Interactions per Neighborhood}
\label{ch:SI:neighborhood-results}

\begin{figure}[!hbt]
    \centering
    % UPDATED: 2017-01-16
    % FILE: plot_hood-disp.py
    \includegraphics[width=.45\textwidth]{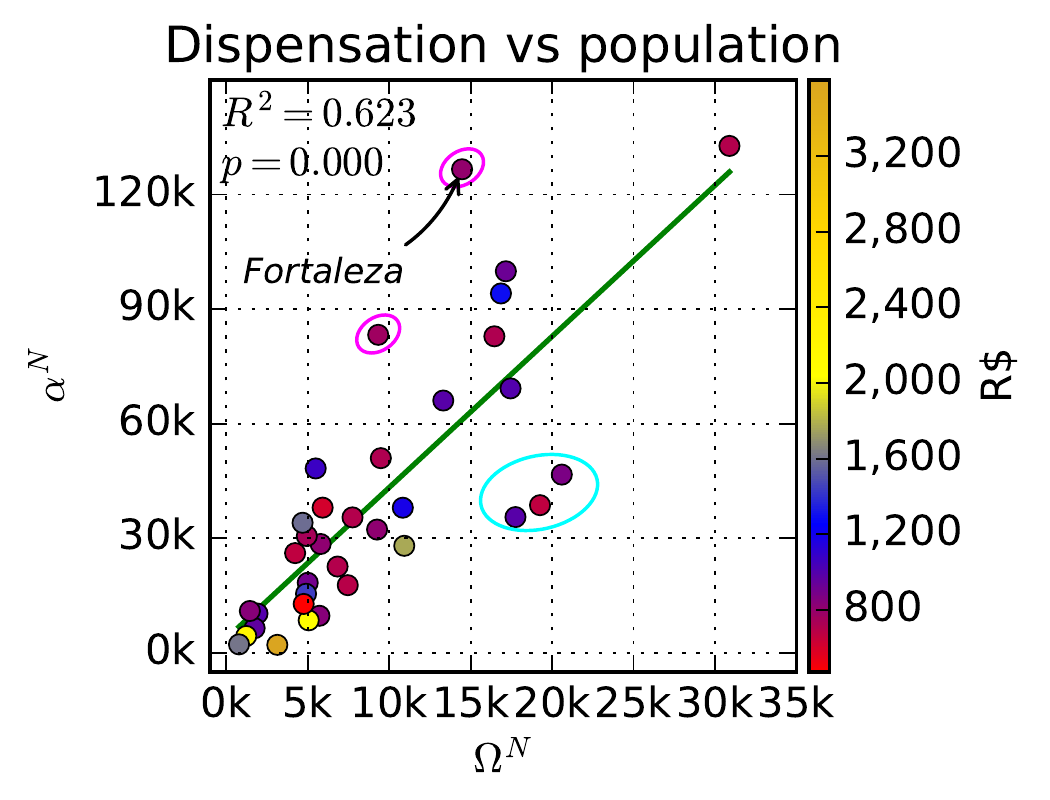}
    \includegraphics[width=.45\textwidth]{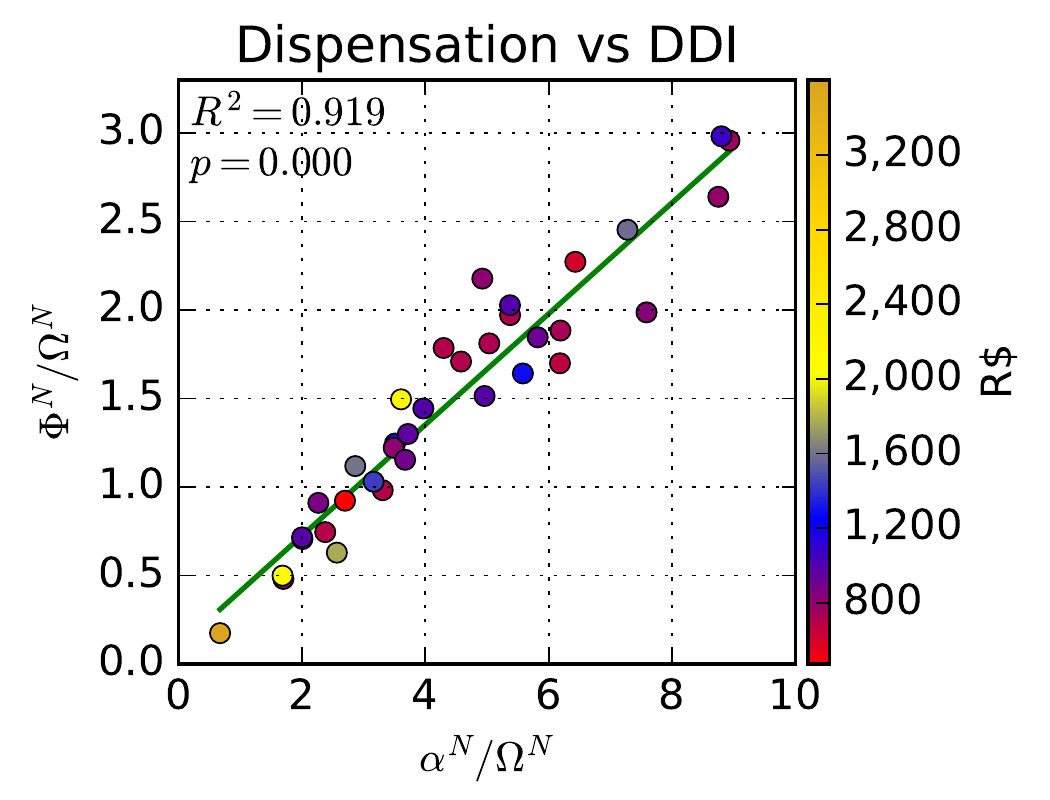}
    \caption{
        \textbf{Left.} Number of drugs intervals dispensed $\alpha^{N}$ against population $\Omega^{N}$ in each neighborhood $N$.
        \textbf{Right.} Number of drug intervals dispensed ($\alpha^{N}$) versus number of interactions ($\Phi^{N}$), per neighborhood ($N$), normalized by population ($\Omega^{N}$).
        Color denotes the average per capita income of neighborhood, in Brazilian \textit{Reais} (R\$).
        Regression line shown in green.
        Patients who reported living in neighborhood \textit{Other} were discarded from computation.
        Cartographic shapes from IBGE \cite{IBGE}.
    }
    \label{fig:SI:hood-disp}
\end{figure}

Supplementary Figure \ref{fig:SI:hood-disp} shows the number of drugs dispensed for each neighborhood $N$, colored by the average income of its residents, R\$ (\textit{Reais}).
Naturally, the larger the neighborhood population ($\Omega^N$), the more drugs are dispensed ($\alpha^{N}$), leading to a fairly clear linear relationship ($R^2=623$, $p=0.0$).
Some observed exceptions above and below the regression line are noteworthy, though.
Three neighborhoods---\textit{Itoupavazinha}, \textit{Velha Central} and \textit{Água Verde}---display dispensation levels below what is expected for their population (circled in Cyan in Supplementary Figure \ref{fig:SI:hood-disp}-left). 
On the other hand, two neighborhoods that are also not among the wealthiest---\textit{Fortaleza} and \textit{Tribess}---are well above the expected drug dispensation (circled in magenta).
Looking at these specific neighborhoods will require further work to be better understood. In any case, their identification highlights the benefits of analyzing EHR and a data science approach to support responsive public health policy.

For a visual inspection of how both official survey numbers and those analyzed in the main manuscript related geographically in the city of Blumenau, we have mapped neighborhoods to results in Supplementary Figures \ref{fig:SI:map-blumenau-ibge} and \ref{fig:SI:map-blumenau-pronto}. The first figure denotes neighborhoods mapped to official numbers from IBGE\cite{IBGE}, such as population, gender rate and income distribution. The second figure denotes dispensed drug intervals, distinct drugs, co-administrations and interactions mapped to each neighborhood.

\begin{figure}
    \centering
    % UPDATED: 2017-01-16
    % FILE: plot_map_blumenau_vars.py
    \includegraphics[width=.31\textwidth]{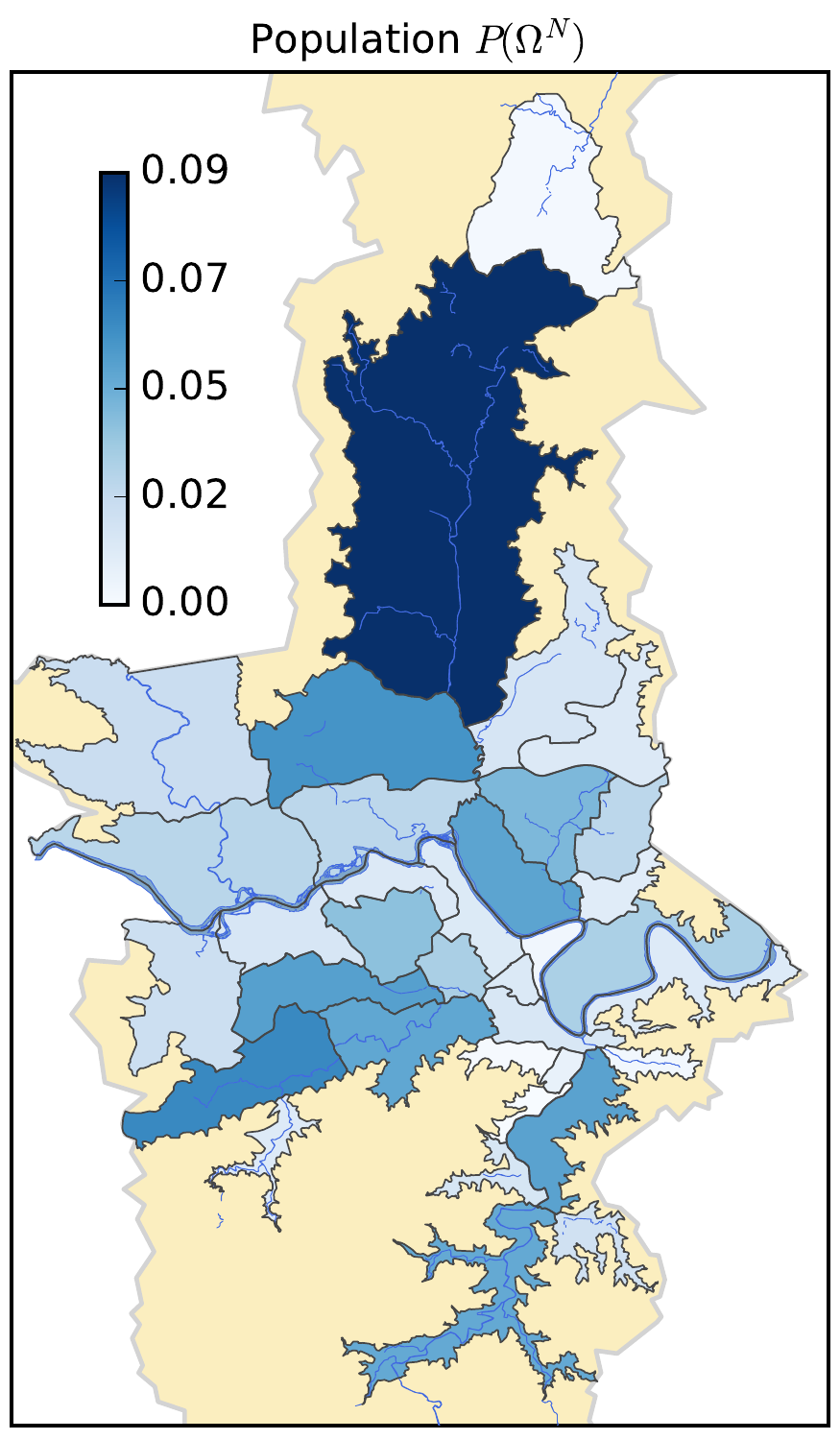}
    \includegraphics[width=.31\textwidth]{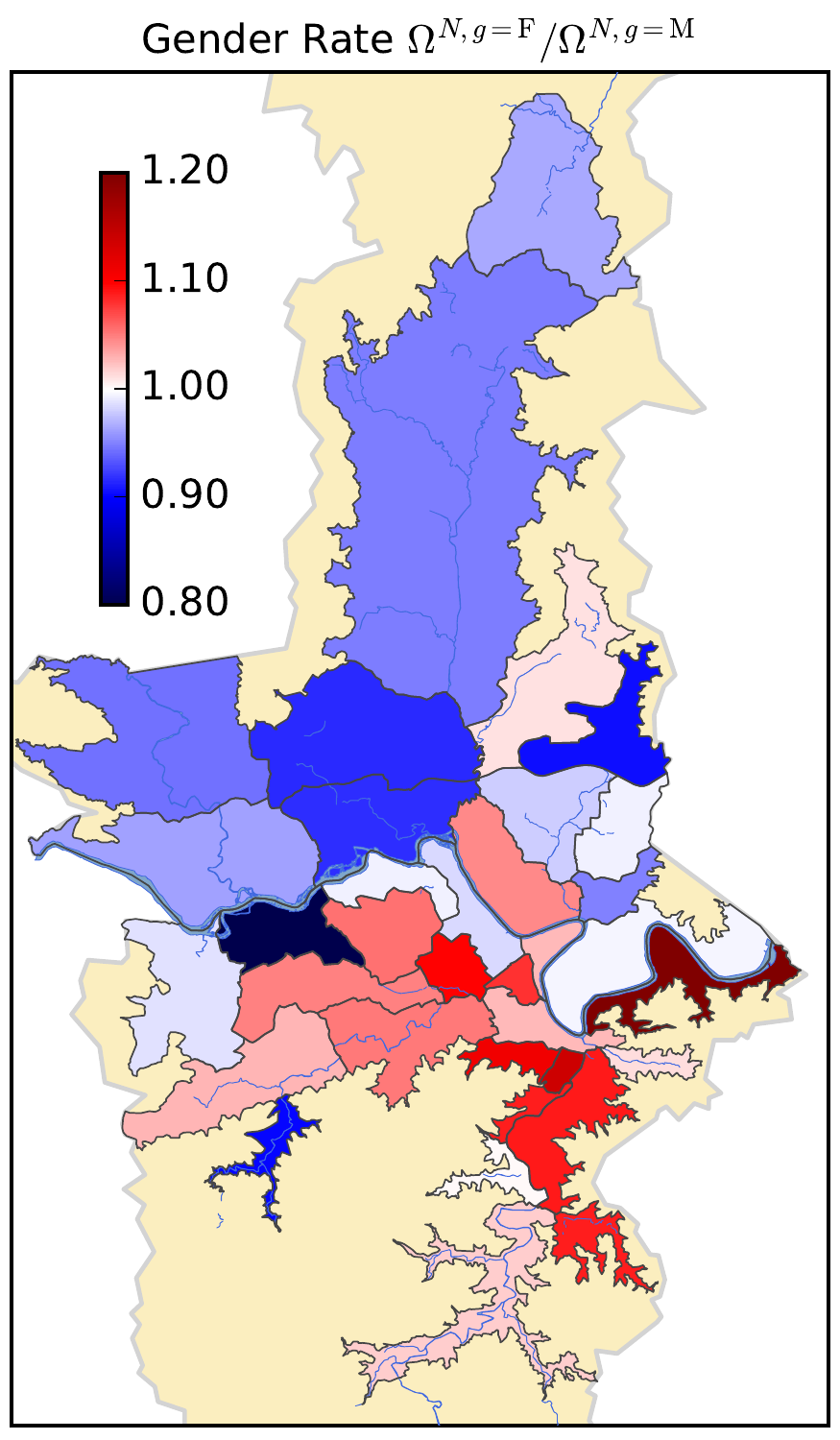}
    \includegraphics[width=.31\textwidth]{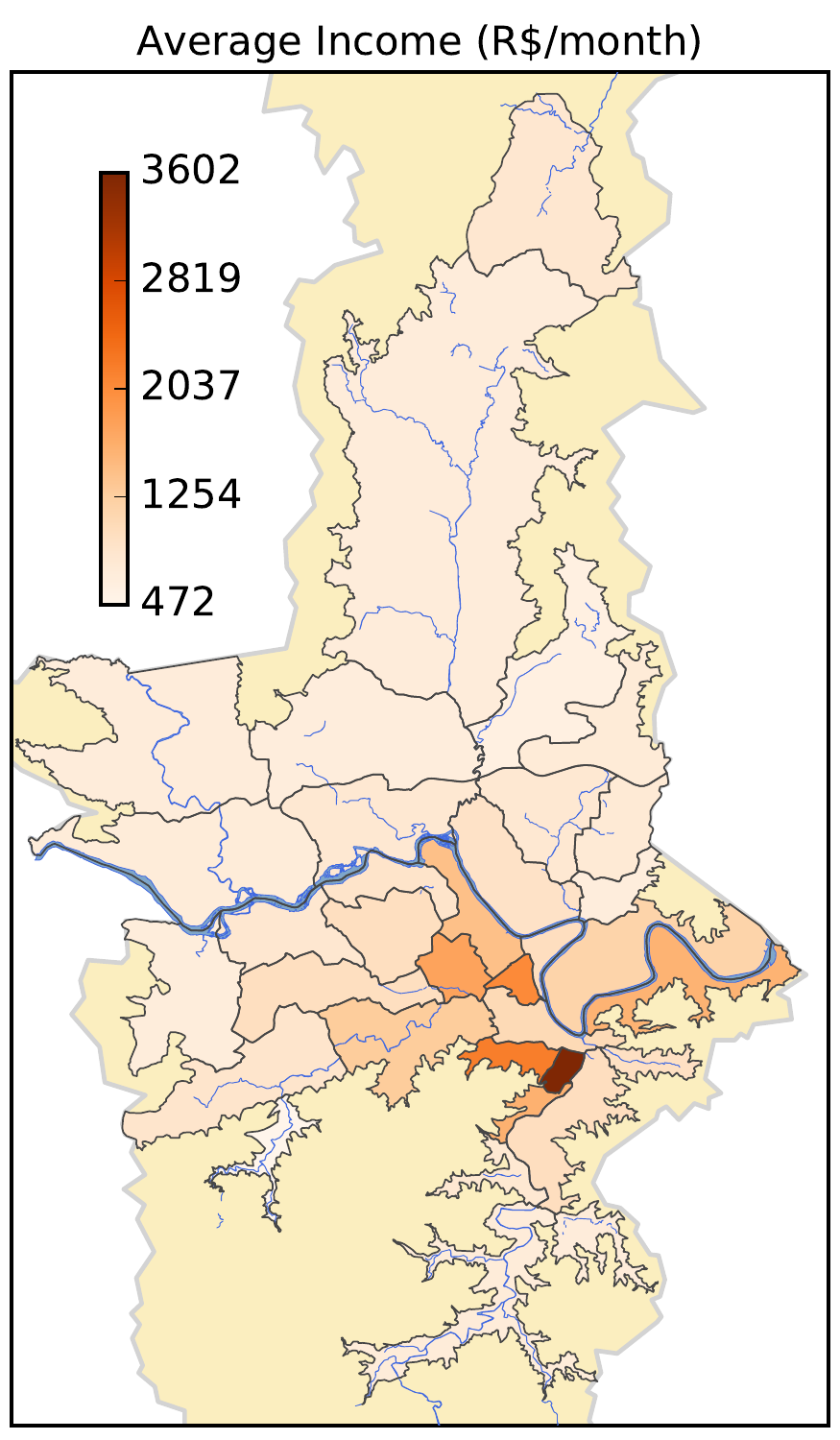}
    
    \caption{
        Data from IBGE \cite{IBGE} mapped to geographical neighbourhoods in the city of Blumenau.
        \textbf{Left.} Population probability, $P(\Omega^{N})$.
        \textbf{Center.} Gender rate, $\Omega^{N,g=F} / \Omega^{N,g=M}$.
        \textbf{Right.} Average income in Brazilian Reais (R\$/month). See Supplementary Note \ref{ch:SI:ddi-cost} for details on income.
        Cartographic shapes from IBGE \cite{IBGE}.
    }
    \label{fig:SI:map-blumenau-ibge}
\end{figure}

\begin{figure}
    \centering
    % UPDATED: 2017-01-16
    % FILE: plot_map_blumenau_vars.py
    \includegraphics[width=.31\textwidth]{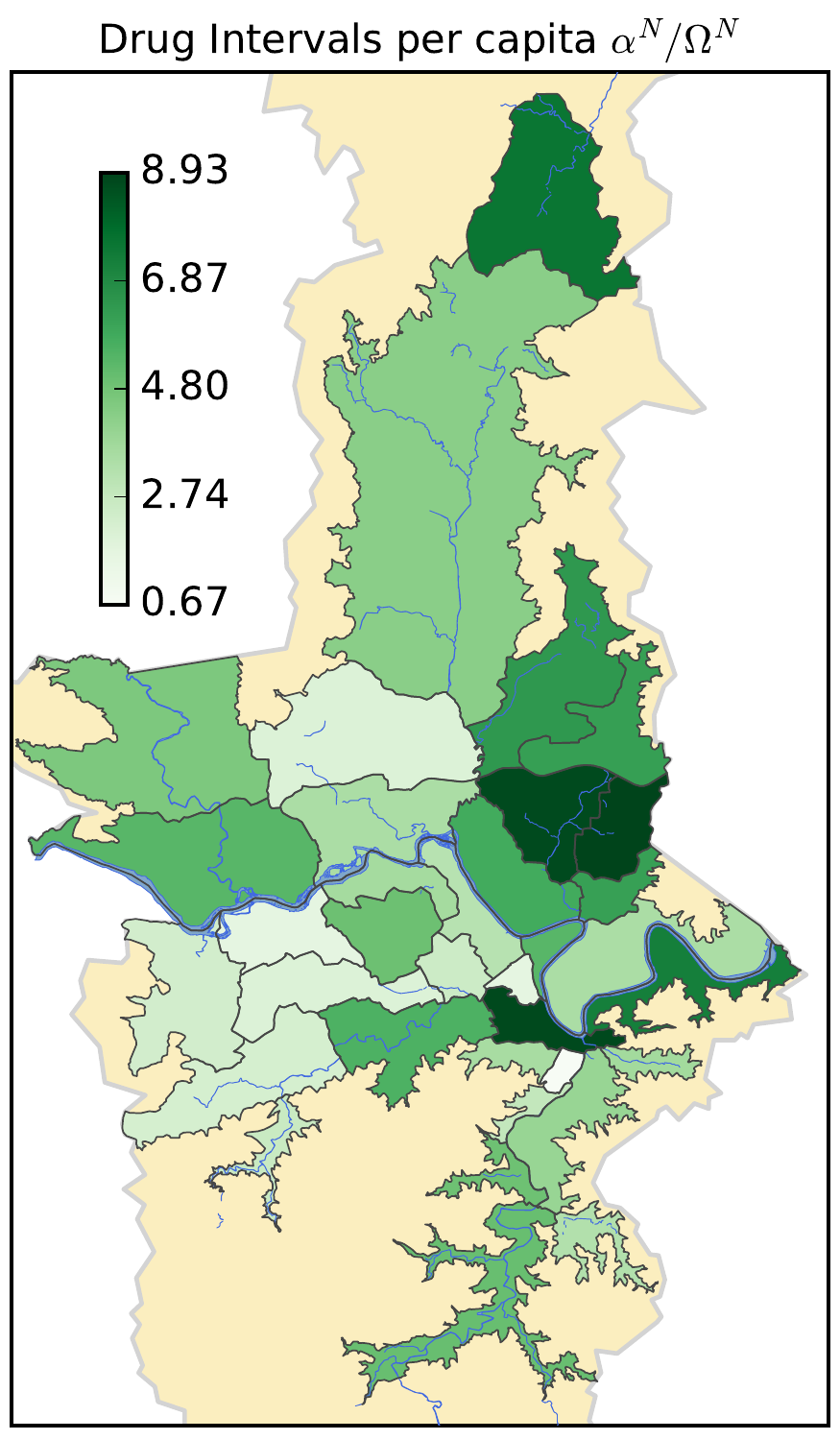}
    \includegraphics[width=.31\textwidth]{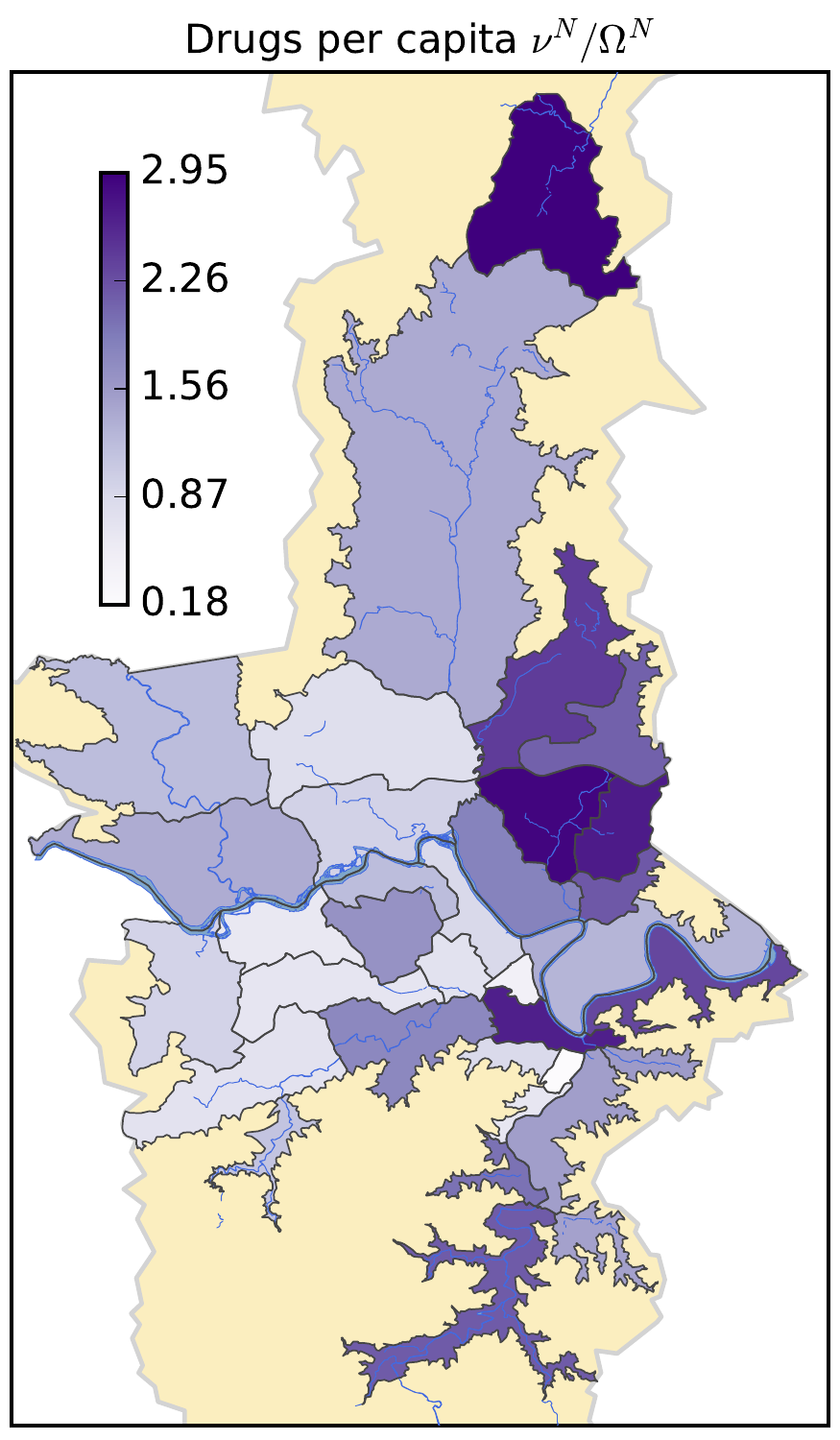}
    \\
    \includegraphics[width=.31\textwidth]{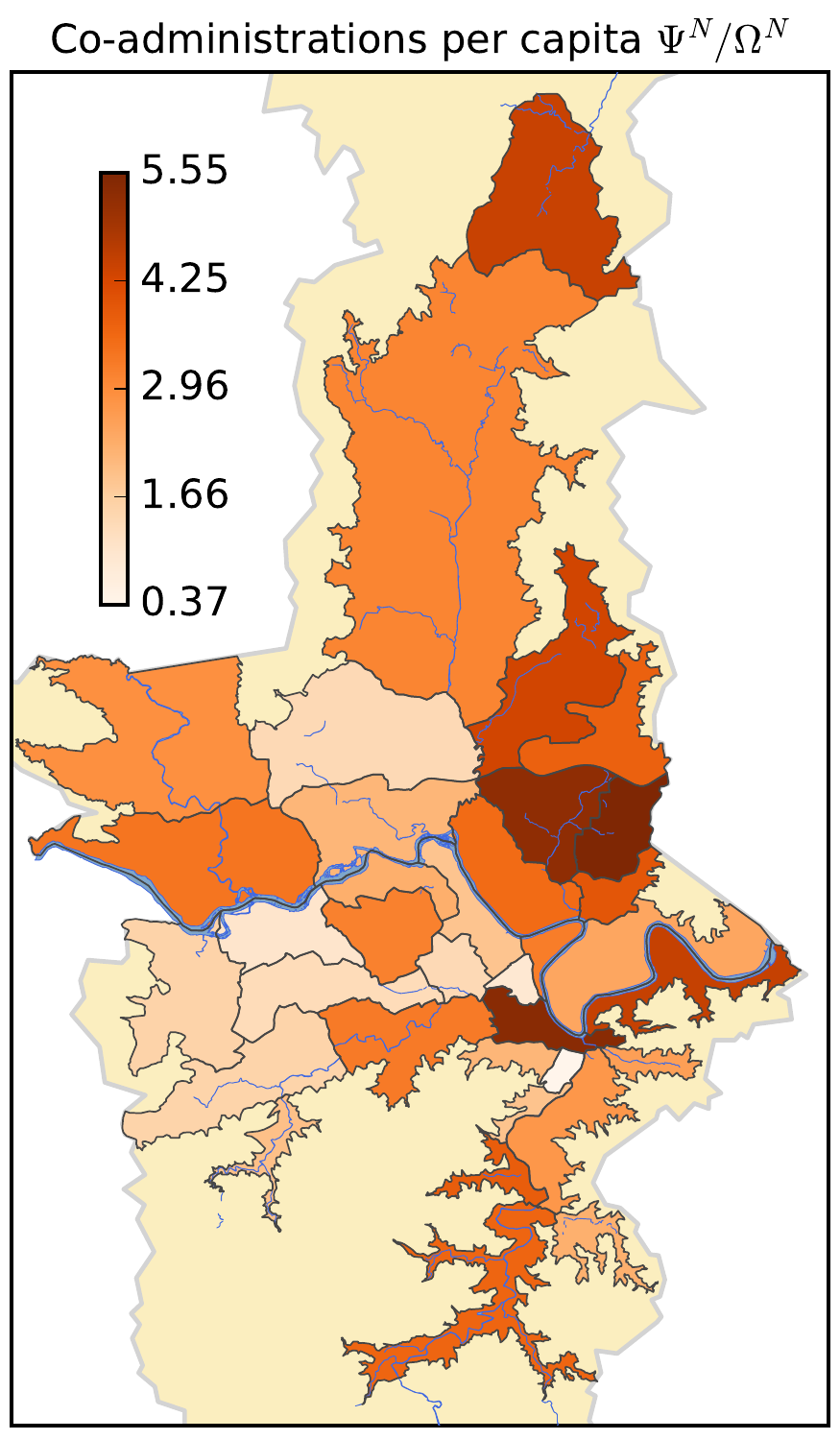}
    \includegraphics[width=.31\textwidth]{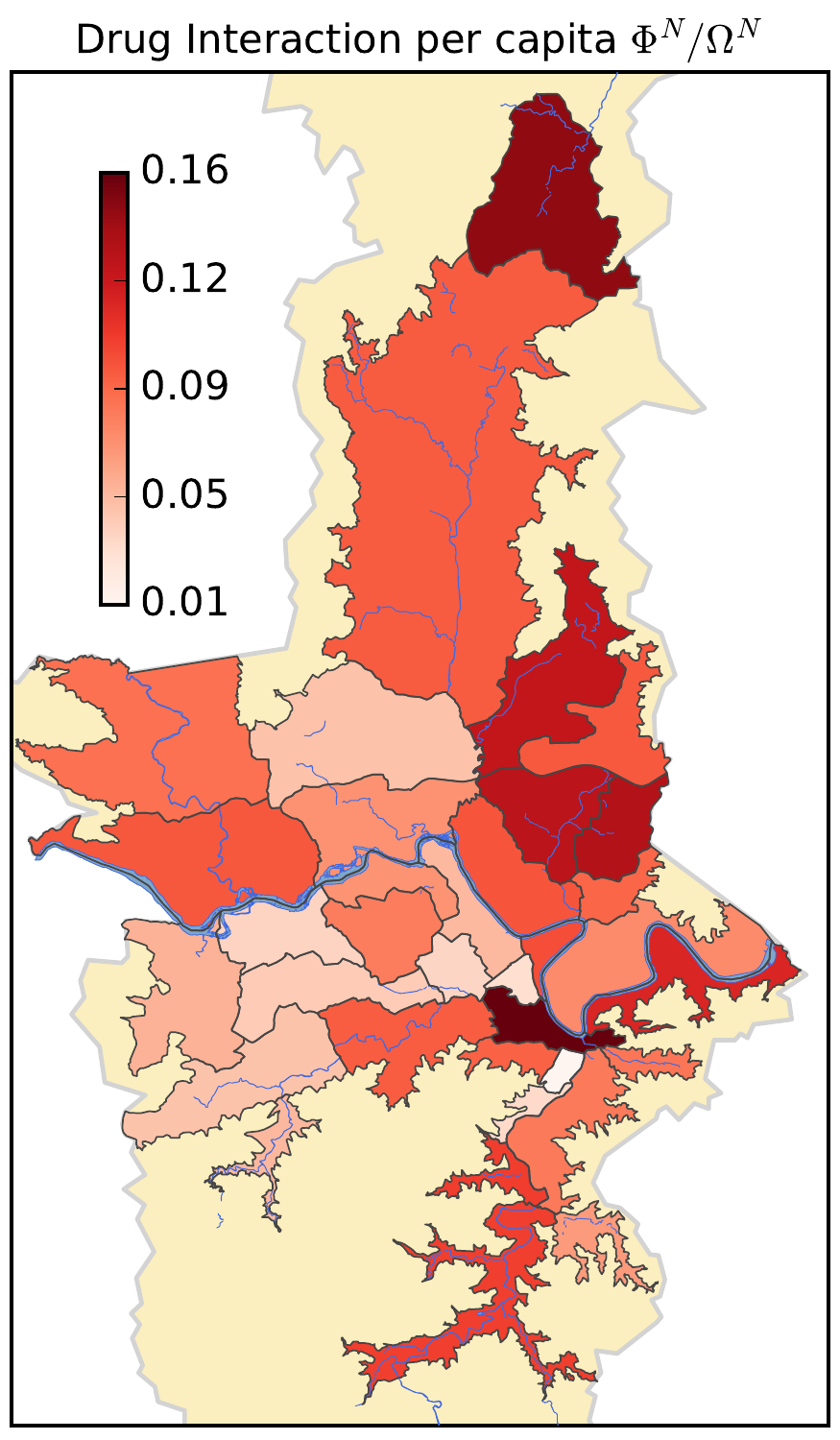}
    
    \caption{
        Results from Pronto data mapped to geographical neighbourhoods in the city of Blumenau.
        \textbf{Top left.} Number of drug interval dispensed per capita, $\alpha^{N} / \Omega^N$.
        \textbf{Top right.} Distinct drugs dispensed per capita, $\nu^{N} / \Omega^{N}$.
        \textbf{Bottom left.} Number of co-adminstrations per capita, $\Psi^{N} / \Omega^{N}$.
        \textbf{Bottom right.} Number of interactions per capita, $\Phi^{N} / \Omega^{N}$.
        Cartographic shapes from IBGE \cite{IBGE}.
    }
    \label{fig:SI:map-blumenau-pronto}
\end{figure}

%
% Cost
%
\FloatBarrier
\pagebreak
\section{Projected Cost of DDI in hospitalizations}
\label{ch:SI:ddi-cost}

Estimating the financial burden of DDI prescribed in primary and secondary care is difficult, since outcomes vary by a large margin and only few result in short-term symptoms requiring hospitalization. Measuring hospitalizations due to DDI are also strenuous, since underestimation of true risk can be masked in practitioners and pharmacists failing to recognize adverse patient outcomes caused by DDI as such.
However, drug- and cohort-focused studies have shown that the number of DDI is associated with a significantly increased risk of hospitalization \cite{Hamilton:1998, Juurlink:2003}. A review paper in 2007 \cite{Becker:2007} estimated that DDI were held responsible for 0.054\% of emergency room (ER) visits, 0.57\% of hospital admissions (4.8\% in the elderly population) and 0.12\% of re-hospitalizations.
The most common outcomes were gastrointestinal bleeding (32.8\%), hypertension/hypotension (18\%) and cardiac rhythm disturbances (18\%).

In this section a study of the financial burden of possible DDI-related hospitalizations is presented.
It considers various rates of hospitalization expected for major DDI co-administrations, and is based on a cost estimate of ADR hospitalizations in Canada \cite{Wu:2012}, and average hospitalization costs for Brazil at city, state and national levels. As average hospitalization costs were not found for the United States, results in Canadian dollars were also converted to US dollars.
Our estimation then relies on guessing what proportion of patients with major DDI co-administrations are likely to have an ADR requiring hospitalization.

\begin{table}
    \centering
    \scriptsize

    \begin{tabular}{@{}llrrrr@{}}
    \toprule
% UPDATED: 2018-04-24
% FILE: hospitalizations.xlsx
                                 &                 & Blumenau     & Santa Catarina & Brazil       & Ontario      \\
                                 &                 & city         & state          & national     & province     \\ \midrule
Population                       &                 & 338,876      & 6,819,190      & 204,450,649  & 13,680,425   \\ \midrule 
\multirow{5}{*}{Hospitalization} & Elective        & 9,761        & 146,395        & 3,391,088    & -            \\
                                 & Urgent          & 24,592 (5,808)       & 507,189 (110,748)        & 13,440,043 (2,711,527)   & -    \\ 
                                 & Work Accident   & 87           & 2,106          & 64,485       & -            \\
                                 & External Causes & 786          & 902            & 110,922      & -            \\ \cmidrule(l){2-6} 
                                 & Total           & 35,226       & 656,592        & 17,006,538   & -            \\ \midrule
\multirow{5}{*}{Avg. Cost}       & Elective        & R\$ 3,764.62 & R\$ 1,533.10   & R\$ 1,583.45 & - \\
                                 & Urgent          & R\$ 2,606.03 & R\$ 1,379.13   & R\$ 1,083.23 & C\$ 8,443.14 \\ 
                                 & Work Accident   & R\$ 1,663.27 & R\$ 2,595.45   & R\$ 1,541.38 & - \\
                                 & External Causes & R\$ 2,321.31 & R\$ 2,203.50   & R\$ 1,256.36 & - \\
%                                 & Average         & R\$ 2,588.81 & R\$ 1,927.79   & R\$ 1,366.10 & -\\ 
    \bottomrule
    \end{tabular}
    \caption{
        Population, number of hospitalizations, and average cost per hospitalization in the analyzed period shown for city, state and national levels.
        Population follows the official projections for 2015.
        Hospitalization numbers and cost shown by type.
        Urgent hospitalization values in parenthesis shown for patients over 64 years old.
        Note Blumenau has a much higher average cost per hospitalization than state and national levels.
        Brazil data from Hospitalization Information System (\textit{Sistema de Informações Hospitalares do SUS; SIH/SUS}) \cite{SIH/SUS}.
        Ontario data from \textcite{Wu:2012}, adjusted for inflation.
    }
    \label{table:SI:ddi-hospitalizations}
\end{table}

To compute costs we gathered number of public health care hospitalizations and average costs for each level (see Supplementary Table \ref{table:SI:ddi-hospitalizations}) from the national Hospitalization Information System (\textit{Sistema de Informações Hospitalares do SUS}; SIH/SUS), a data source managed by the Informatics Department under the Executive Secretary of Brazil's Ministry of Health \cite{SIH/SUS}.

As reported, the number of patients prescribed a major DDI in Blumenau (city level) was $|U^{\Phi,s=major}|=5,224$.
For state and national levels, we estimated this number from the percentage of hospitalizations it represents at city level, a reasonable assumption due the lack of data that generalizes medical practice in Blumenau for the state and country.
For example, say 261 (or 5\%) Pronto patients prescribed a major DDI had to be hospitalized. In hospitalization terms, that accounts for 1.06\% of all hospitalizations in the same period. At the state and national level, the same 1.06\% accounts for 5,376 and 142,564 patients, respectively.
Costs are then estimated by multiplying the number of patients assumed hospitalized by the average hospitalization cost in each level.

\textcite{Wu:2012} argued in 2007 that the average cost of ADR-related hospitalization for all adults over 65 in the province of Ontario (pop. 12M in 2006; 13.6M in 2014) was C\$ 7,528 (C\$ 8,443.14 or \$7,380.78 in 2014 when adjusted for inflation and exchange rate) for a total annual cost of C\$ 13.6 million (C\$ 15.2M or \$13.3M after adjusting), or estimated C\$ 35.7 million (C\$ 40M or \$35M after adjusting) in Canada.
In an attempt to compare results, we also multiplied the number of patients assumed to have been hospitalized to their average cost of ADR-relation hospitalization (see columns 6 and 7 of Supplementary Table \ref{table:SI:ddi-cost-city}).
Moreover, Supplementary Tables \ref{table:SI:ddi-cost-city}, \ref{table:SI:ddi-cost-state} and \ref{table:SI:ddi-cost-country} show the estimated costs at different percentages of hospitalizations at city, state and national levels, respectively. Costs in Brazilian Reais (columns 4 and 5) are computed based on the average cost of hospitalization in Brazil. Costs in C\$ use \textcite{Wu:2012} as reference (columns 6 and 7), and then converted to US\$ with the average exchange rate between the two currencies for the whole period of our data (columns 8 and 9). The average exchange rate in the period was C\$1.00 Canadian dollar equals to \$0.8742 US dollar, and maximum and minimum rates were $.9418$ in January 4\textsuperscript{th} 2014 and $.7821$ in March 14\textsuperscript{th} 2015, respectively.

Our cost analysis first considers two previous studies of the proportion of emergency room visits that are due to DDI and ADR. \textcite{Becker:2007} reported that 0.57\% of all hospital admissions they observed were due to DDI, which in our data would correspond to 140 patients, or $p_h = 2.68\%$ of all patients dispensed a major DDI (see Supplementary Table \ref{table:SI:ddi-cost-city}). \textcite{Wu:2012}, on the other hand, argued that 0.75\% of all hospitalizations of patients over 65 years of age were due to ADR (not only from DDI), which corresponds in our data to 436 patients, or $p_h = 8.35\%$ of all patients who were co-administered a major DDI. Both of these conjectures are likely to err on the side of under-reporting emergency room admissions due to DDI or ADR, since this a well-known problem in studies of this phenomenon \cite{Patrignani:2018, Gonzalez:2011, Ponte:2010, Alvarez:2013, Tatonetti:2012}. Indeed, the proportion of hospital admissions due to ADR has been reported in the literature to vary anywhere between 0.5\% to 12\% \cite{Ponte:2010}, with DDI reportedly being responsible for 15\% \cite{Alvarez:2013} to 30\% \cite{Iyer:2014} of all ADR. These ranges, if correct, put the proportion of hospital admissions due to DDI anywhere between $0.5 \% \times 15\% = 0.075 \%$ and $ 12\% \times 30 \% = 3.6\%$, which in our Blumenau data would mean between 18 and 885 emergency room patients, or $p_h \in [0.35, 16.95]\%$ of all patients dispensed a major DDI. These ranges could be higher, since even in the more controlled hospital environment, the proportion of patients with ADR can be as high as 41\% \cite{Cano:2009}. Therefore, in addition to the costs derived from the numbers provided by \textcite{Becker:2007} and \textcite{Wu:2012}, tables in this section also report cost estimates for various values of $p_h$, so that readers can judge what is an appropriate value to consider. Given the ranges just outlined, it is reasonable to assume, for instance,  that $p_h = 10\%$ of all patients dispensed a major DDI will have to be hospitalized (522 patients).

The lowest estimate ($p_h=2.68\%$, via \textcite{Becker:2007}) leads to a cost of DDI-related hospitalization in Blumenau of over \$1M in the 18-month period, after adjusting for Canadian cost, inflation and exchange rate to US dollar. The extrapolated costs to the state and the country are \$21M and \$565M, respectively (see Supplementary Tables \ref{table:SI:ddi-cost-state} and \ref{table:SI:ddi-cost-country}). The estimated costs obtained via \textcite{Wu:2012} ($p_h=8.35\%$) reach  \$3.2M, \$61M, and \$1.5B, for the city, state and country levels respectively. Finally, if we assume $p_h=10\%$, the estimated costs reach \$3.9M, \$79M, and \$2.1B, for the city, state and country levels respectively.

All estimations lead to very substantial costs for the various levels of government. To compare them to the costs found for Canada \cite{Wu:2012}, we computed per capita measures of the burden of DDI-related hospitalizations. For instance, the lowest estimate $p_h=2.68\%$ leads to a per capita cost for Blumenau of \$2.03, while the inflation-adjusted cost for Ontario (Canada) is \$0.97, suggesting that the financial burden of DDI is more severe than previously thought---even when considering only the lowest estimate of the proportion of hospitalizations that derive from co-administration of known major DDIs. For the state of Santa Catarina and Brazil as a whole, these numbers are \$2.09 and \$1.84, respectively. If we consider the higher estimates of $p_h=8.35\%$ or $p_h=10\%$, the per capita cost for Blumenau is \$6.33 and \$7.58, respectively.

To put these numbers in context, Brazil's minimum monthly wage was R\$724 (R\$9,412/year\footnote{Brazilians receive a 13\textsuperscript{th} salary in December. Thus, yearly gross income is calculated by a 13, and not by a 12, multiplier.}) in 2014, and workers in Blumenau received on average 2.9 wages a month \cite{IBGE}. This constitutes an average gross income of R\$2,099.60 a month (R\$27,294.80/year).
If we assume the same 140 patients were hospitalized due to ADR caused by DDI, the direct cost of such hospitalizations is equal to 3,707 lost productive worker/days (considering an 8 hour working day), with possible much higher indirect costs.

Some limitations should be noted.
When comparing to \textcite{Becker:2007}, data from \textcite{IBGE} includes patients over 64, while in their work the authors included patients over 65 years old. Our analysis then possibly contains additional patients exactly age 65, although we do not believe this affects the results presented given their large difference.
In general, other studies \cite{Becker:2007} divide hospital admissions only between two categories, emergency room (ER) visits and hospitalizations. It is not possible to conclude whether electives or external causes are included in their hospitalization numbers.
SIH/SUS data are only available for patients that were hospitalized in the public system, meaning the cost of hospitalization was billed to the public system. Therefore, if a patient was hospitalized and his/her private insurance covered the costs, the SIH/SUS would have no record of it.
Furthermore, SIH/SUS provides the number of hospitalizations broken down by type. These consist of ``electives'' (e.g., schedules cesareans), ``urgencies'', ``work accidents'', and ``other external causes'' (codes V01 to Y98 of ICD-10\footnote{\url{http://www.datasus.gov.br/cid10/V2008/WebHelp/v01_y98.htm}}; e.g., car accident, poisoning, and drowning).
To better approximate reality, we have calculated the cost of DDI-related hospitalizations only using the number of urgent hospitalizations.

\begin{table}
    \centering
    \scriptsize
    \begin{tabular}{r|rr|rrrrrr}
    \toprule
% UPDATED: 2018-04-24
% FILE: hospitalizations.xlsx
       &          &            & \multicolumn{2}{c}{Cost R\$} & \multicolumn{2}{c}{Cost CA\$} & \multicolumn{2}{c}{Cost US\$} \\
 $p_h$ & $|U^{\Phi}_{major}|$ & \% of hosp. & 18 months     & 12 months    & 18 months     & 12 months     & 18 months     & 12 months     \\ \midrule
100\%  & 5,224    & 21.24\%    & 13,613,909    & 9,075,940    & 44,106,963    & 29,404,642    & 38,557,213    & 25,704,809    \\
50\%   & 2,612    & 10.62\%    & 6,806,955     & 4,537,970    & 22,053,482    & 14,702,321    & 19,278,606    & 12,852,404    \\
30\%   & 1,567    & 6.37\%     & 4,083,652     & 2,722,434    & 13,230,400    & 8,820,267     & 11,565,688    & 7,710,458     \\
25\%   & 1,306    & 5.31\%     & 3,403,477     & 2,268,985    & 11,026,741    & 7,351,161     & 9,639,303     & 6,426,202     \\
20\%   & 1,044    & 4.25\%     & 2,720,697     & 1,813,798    & 8,814,638     & 5,876,425     & 7,705,538     & 5,137,025     \\
10\%   & 522      & 2.12\%     & 1,360,349     & 906,899      & 4,407,319     & 2,938,213     & 3,852,769     & 2,568,513     \\
5\%    & 261      & 1.06\%     & 680,174       & 453,450      & 2,203,660     & 1,469,106     & 1,926,384     & 1,284,256     \\ \midrule
2.68\% & 140      & 0.57\%     & 364,844       & 243,230      & 1,182,040     & 788,026       & 1,033,310     & 688,873       \\ \midrule
8.35\% & 436 	  & 0.75\%	   & 1,136,230 	   & 757,487 	  & 3,681,209 	  & 2,454,139 	  & 3,218,022 	  & 2,145,348     \\ \bottomrule
    \end{tabular}
    \caption{
        Projected cost of DDI for the city of Blumenau in Reais (R\$), Canadian Dollars (C\$) and US dollars (US\$) for the analysis period (18 months) and yearly (12 months).
        Each row calculates the associated cost based on different proportion of patients who had at least one major DDI and required hospitalization.
        Last row shows the projected cost when only $0.75\%$ of all hospitalizations of patients over 64 years old are considered, based on results of \textcite{Wu:2012}.
        Similarly, second-to-last row shows projected cost when only $0.57\%$ of all hospitalization are considered, based on results of \textcite{Becker:2007}.
        In the 18 month period, Blumenau had a total of 24,592 public health care emergency hospitalizations, from which 5,808 were of patients age over 64 years old.
        Average cost per hospitalization in the city is R\$ 2,606.03.
        US\$ costs were calculated based on C\$ exchange rate of $.8742$, the average rate in the study period.        
    }
    \label{table:SI:ddi-cost-city}
\end{table}

\begin{table}
    \centering
    \scriptsize
    \begin{tabular}{r|rr|rrrrrr}
    \toprule
       &          &            & \multicolumn{2}{c}{Cost in R\$} & \multicolumn{2}{c}{Cost in CA\$} & \multicolumn{2}{c}{Cost in US\$} \\
 $p_h$ & $|U^{\Phi}_{major}|$ & \% of hosp. & 18 months   & 12 months  & 18 months    & 12 months   & 18 months    & 12 months   \\ \midrule
% UPDATED: 2018-04-24
% FILE: hospitalizations.xlsx
100\%  & 107,726  & 21.24\%    & 148,567,620 & 99,045,080 & 909,545,700  & 606,363,800 & 795,102,280  & 530,068,187 \\
50\%   & 53,863   & 10.62\%    & 74,283,810  & 49,522,540 & 454,772,850  & 303,181,900 & 397,551,140  & 265,034,093 \\
30\%   & 32,307   & 6.37\%     & 44,555,391  & 29,703,594 & 272,772,524  & 181,848,349 & 238,450,972  & 158,967,314 \\
25\%   & 26,931   & 5.31\%     & 37,141,215  & 24,760,810 & 227,382,203  & 151,588,136 & 198,771,880  & 132,514,586 \\
20\%   & 21,555   & 4.25\%     & 29,727,039  & 19,818,026 & 181,991,883  & 121,327,922 & 159,092,788  & 106,061,859 \\
10\%   & 10,752   & 2.12\%     & 14,828,352  & 9,885,568  & 90,780,641   & 60,520,428  & 79,358,184   & 52,905,456  \\
5\%    & 5,376    & 1.06\%     & 7,414,176   & 4,942,784  & 45,390,321   & 30,260,214  & 39,679,092   & 26,452,728  \\ \midrule
2.68\% & 2,890    & 0.57\%     & 3,985,671   & 2,657,114  & 24,400,675   & 16,267,116  & 21,330,464   & 14,220,309  \\ \midrule
7.71\% & 8,306 	  & 0.75\%	   & 21,645,699  & 14,430,466 & 70,128,721 	 & 46,752,481  & 61,304,788   & 40,869,858  \\ \bottomrule
    \end{tabular}
    \caption{
        Projected cost of DDI for the state of Santa Catarina in Reais (R\$), Canadian Dollars (C\$) and US dollars (US\$) for the analysis period (18 months) and yearly (12 months).
        Each row calculates the associated cost based on different proportion of patients who had at least one major DDI and required hospitalization.
        Last row shows the projected cost when only $0.75\%$ of all hospitalizations of patients over 64 years old are considered, based on results of \textcite{Wu:2012}.
        Similarly, second-to-last row shows projected cost when only $0.57\%$ of all hospitalization are considered, based on results of \textcite{Becker:2007}.
        In the 18 month period, Santa Catarina had a total of 507,189 public health care emergency hospitalizations.
        Average cost per hospitalization in the state is R\$ 1,379.13.
        US\$ costs were calculated based on C\$ exchange rate of $.8742$, the average rate in the study period.
    }
    \label{table:SI:ddi-cost-state}
\end{table}

\begin{table}
    \centering
    \scriptsize
    \begin{tabular}{r|rr|rrrrrr}
    \toprule
       &           &            & \multicolumn{2}{c}{Cost in R\$} & \multicolumn{2}{c}{Cost in CA\$} & \multicolumn{2}{c}{Cost in US\$} \\
 $p_h$ & $|U^{\Phi}_{major}|$  & \% of hosp. & 18 months      & 12 months      & 18 months       & 12 months      & 18 months       & 12 months      \\ \midrule
% UPDATED: 2018-04-24
% FILE: hospitalizations.xlsx
100\%  & 2,854,665 & 21.24\%    & 3,092M         & 2,061M         & 24,102M         & 16,068M        & 21,070M         & 14,046M        \\
50\%   & 1,427,332 & 10.62\%    & 1,546M         & 1,031M         & 12,051M         & 8,034M         & 10,535M         & 7,023M         \\
30\%   & 856,130   & 6.37\%     & 927M           & 618M           & 7,228M          & 4,819M         & 6,319M          & 4,213M         \\
25\%   & 713,666   & 5.31\%     & 773M           & 515M           & 6,026M          & 4,017M         & 5,267M          & 3,512M         \\
20\%   & 571,201   & 4.25\%     & 619M           & 412M           & 4,823M          & 3,215M         & 4,216M          & 2,811M         \\
10\%   & 284,928   & 2.12\%     & 309M           & 206M           & 2,406M          & 1,604M         & 2,103M          & 1,402M         \\
5\%    & 142,464   & 1.06\%     & 154M           & 103M           & 1,203M          & 802M           & 1,051M          & 701M           \\ \midrule
2.68\% & 76,608    & 0.57\%     & 83M            & 55M            & 647M            & 431M           & 565M            & 377M           \\ \midrule
7.12\% & 203,365   & 0.75\%	    & 530M	         & 353M           &	1,717M	        & 1,145M	     & 1,501M	       & 1,001M         \\ \bottomrule
    \end{tabular}
    \caption{
        Projected cost of DDI for Brazil in Reais (R\$), Canadian Dollars (C\$) and US dollars (US\$) for the analysis period (18 months) and yearly (12 months).
        Each row calculates the associated cost based on different proportion of patients who had at least one major DDI and required hospitalization.
        Last row shows the projected cost when only $0.75\%$ of all hospitalizations of patients over 64 years old are considered, based on results of \textcite{Wu:2012}.
        Similarly, second-to-last row shows projected cost when only $0.57\%$ of all hospitalization are considered, based on results of \textcite{Becker:2007}.
        In the 18 month period, Brazil had a total of 13,440,043 public health care emergency hospitalizations.
        Average cost per hospitalization in the country is R\$ 1,083.23.
        US\$ costs were calculated based on C\$ exchange rate of $.8742$, the average rate in the study period.
    }
    \label{table:SI:ddi-cost-country}
\end{table}

%
% Statistical Modeling
%
\FloatBarrier
\pagebreak
\section{Statistical modeling}
\label{ch:SI:statistical-modeling}

Here we show the complete results from runs of simple regression (SR), polynomial regression (PR), ordinary multiple regression (OMR) and linear mixed model (LMM) that were mentioned in the original text. For further details on LMM see \cite{Woltman:2012}.

A SR is is a linear regression model with a single explanatory variable. It works by fitting a line through the data that minimizes the sum of the squared of the residual, where the residuals are differences between the observed values to the predicted values of the model.
For example, when predicting the number of interactions from the number co-administrations, the SR equation would look like

\begin{equation}
    \Phi^{u}(x) = \beta^{0} + \beta^{1} x + \epsilon^{u}
\end{equation}

where $\beta^{0}$ is the intercept (or bias), $\beta^{1}$ is the coefficient (or slope) and $\epsilon^{u}$ are the residuals.

Similarly, PR is a regression where the dependent variable being modeled is fitted with a $n$\textsuperscript{th} degree polynomial, but still a single explanatory variable. In the main manuscript we modeled $RRI^{y}$ with a cubic (3\textsuperscript{rd} degree) polynomial. In our case, the PR equation would look like

\begin{equation}
    RRI^{y}(x) = \beta^{0} + \beta^{1} (x^3) + \beta^{2} (x^2) + \beta^{3} x + \epsilon^{y}
\end{equation}

where $\beta^{0}$ is the intercept (or bias), $\beta^{1}$, $\beta^{2}$ and $\beta^{3}$ are coefficients (giving the characteristic shape of the cubic curve) and $\epsilon^{y}$ are the residuals.

An OMR is a widely used type of regression for predicting the value of one dependent variable from the value of a set of independent variables. Similar to SR, it works by fitting a hyperplane through the data that minimizes the sum of the squared of the residuals.

In our case, the OMR equation would look like
\begin{equation}
    \Phi^{u} = \beta^{0} + \beta^{1} x^{u,1} + \beta^{2} x^{u,2} + \; ... \; \beta^{j} x^{u,i} + \epsilon^{u}
\end{equation}

Where $\beta^{0}$ is the intercept (or bias), $\beta^{j}$ are the coefficients (or slopes) and each $x^{u,i}$ is a predictor (covariate, regressor) such as age or number of drugs. $\epsilon^{u}$ is the error associated with the fit.

A LMM (also known as multilevel, mixed effects, random effects or hierarchical linear model) can be seen as extensions of the OMR where instances of the data belong to certain groups---like children in classrooms or cities in states. In our case, they are patients in specific neighborhoods.

The individual levels are usually defined as level-1 (within-group), and level-2 (between-group) for a two level model. Separate level-1 models (e.g., patients) are developed for each level-2 (e.g., neighborhoods). Considering only one predictor, level-1 models take the form of simple regressions:

\begin{equation}
    \Phi^{u,n} = \beta^{0,n} + \beta^{1,n} x^{u,n} + r^{u,n}
    \label{eq:SI:HLM-1}
\end{equation}

where $\beta^{0,n}$ is the intercept for the $n$ neighborhood, $\beta^{1,n}$ is a coefficient (slope) associated with predictor $x^{u,n}$, and $r^{u,n}$ is the error.

In the level-2 models, the level-1 regression coefficients ($\beta^{0,n}$ and $\beta^{1,n}$) are used as outcome variables and are related to each of the level-2 predictors.

\begin{equation}
    \beta^{0,n} = \gamma^{0,0} + \gamma^{0,1} g^{n} + u^{0,n}
    \label{eq:SI:HLM-2}
\end{equation}
\begin{equation}
    \beta^{1,n} = \gamma^{1,0} + \gamma^{1,1} g^{n} + u^{1,n}
    \label{eq:SI:HLM-3}
\end{equation}

where $g^{n}$ is the level-2 predictor, $\gamma^{0,0}$ and $\gamma^{1,0}$ are the overall mean intercept adjusted for $g$. $\gamma^{0,1}$ ($\gamma^{1,1}$) is the regression coefficient associated with $g$ relative to level-1 intercept (slope) and $u^{0,n}$ ($u^{1,n}$) are level-2 random effects adjusted for $g$ on the intercept (slope).

A combined two-level model is created by substituting Supplementary Equations \ref{eq:SI:HLM-2} and \ref{eq:SI:HLM-3} into Supplementary Equation \ref{eq:SI:HLM-1}:

\begin{equation}
    \Phi^{u,n} = \gamma^{0,0} + \gamma^{1,0} x^{u,n} + \gamma^{0,1} g^{n} + \gamma^{1,1} g^{n} x^{u,n} + u^{1,n} x^{0,n} + u^{0,n} + \epsilon^{u,n}
\end{equation}

The combined model incorporates the level-1 and level-2 predictors ($x^{u,n}$ and $g^{n}$), a cross-level term ($g^{n}x^{u,n}$) as well as the composite error ($u^{1,n} x^{u,n} + u^{0,n} + r^{u,n}$).

In practice, LMM coefficients are estimated using maximum likelihood methods.
% SR
\subsection{Simple Regression (SR) models}
\label{ch:SI:simple-regression}

In Figure \ref{fig:coadmin-ddi-dist} of the main manuscript we show single regression models predicting the number of interactions.
Specifically, $\nu^{u}$ predicts $\Psi^{u}$ best with a quadratic regression ($R^2=.857$)
as shown in Figure \ref{fig:coadmin-ddi-dist}-left. 
When it comes to predicting number of interactions (Figure \ref{fig:coadmin-ddi-dist}, center and right), on the other hand, there is much more dispersion of the data, which leads to a relatively small linear correlation between $\Psi^{u}$ and $\Phi^{u}$ ($R^2=.487$)---though better than the linear correlation between $\nu^{u}$ and $\Phi^{u}$ ($R^2=.304$).
However, higher order regressions do not improve the prediction of the variance of $\Phi^{u}$, as demonstrated by the Pareto front in Fig. \ref{fig:coadmin-ddi-dist}-top-right (see also Supplementary Note \ref{ch:SI:classification})---thus discarding the hypothesis of a clear nonlinear relationship between co-administrations and interactions, which could explain the growth of RI with age.

Supplementary Tables below display additional regression models.

\begin{lstlisting}[caption=$\Psi^{u}$ from $\nu^u$ linear model]
===================================================
                               $\Psi^u$            
$\nu^u$                        3.891*** (0.007)        
Constant                   -8.818*** (0.037)       
---------------------------------------------------
Observations                    132,722            
R2                               0.712             
Adjusted R2                      0.712             
Residual Std. Error       8.650 (df = 132720)      
F Statistic         328,478.000*** (df = 1; 132720)
===================================================
Note:                   *p<0.1; **p<0.05; ***p<0.01
\end{lstlisting}

\begin{lstlisting}[caption=$\Psi^{u}$ from $\nu^u$ quadratic model]
===================================================
                               $\Psi^u$            
$\nu^u$                        -0.121*** (0.012)       
$(\nu^u)^2$                    0.273*** (0.001)        
Constant                    -0.023 (0.036)         
---------------------------------------------------
Observations                    132,722            
R2                               0.857             
Adjusted R2                      0.857             
Residual Std. Error       6.088 (df = 132719)      
F Statistic         399,075.300*** (df = 2; 132719)
===================================================
Note:                   *p<0.1; **p<0.05; ***p<0.01
\end{lstlisting}

\begin{lstlisting}[caption=$\Phi^{u}$ from $\nu^{u}$ linear model]
==================================================
                               $\Phi^u$           
$\nu^u$                       0.110*** (0.0005)       
Constant                  -0.267*** (0.003)       
--------------------------------------------------
Observations                   132,722            
R2                              0.304             
Adjusted R2                     0.304             
Residual Std. Error      0.580 (df = 132720)      
F Statistic         58,011.640*** (df = 1; 132720)
==================================================
Note:                  *p<0.1; **p<0.05; ***p<0.01
\end{lstlisting}

\begin{lstlisting}[caption=$\Phi^{u}$ from $\nu^{u}$ quadratic model]
==================================================
                               $\Phi^u$           
$\nu^u$                       -0.009*** (0.001)       
$(\nu^u)^2$                   0.008*** (0.0001)       
Constant                   -0.007** (0.003)       
--------------------------------------------------
Observations                   132,722            
R2                              0.372             
Adjusted R2                     0.372             
Residual Std. Error      0.551 (df = 132719)      
F Statistic         39,357.930*** (df = 2; 132719)
==================================================
Note:                  *p<0.1; **p<0.05; ***p<0.01
\end{lstlisting}

\begin{lstlisting}[caption=$\Phi^{u}$ from $\Psi^{u}$ linear model]
===================================================
                               $\Phi^{u}$            
$\Psi^{u}$                       0.030*** (0.0001)       
Constant                   -0.033*** (0.002)       
---------------------------------------------------
Observations                    132,722            
R2                               0.487             
Adjusted R2                      0.487             
Residual Std. Error       0.498 (df = 132720)      
F Statistic         126,232.900*** (df = 1; 132720)
===================================================
Note:                   *p<0.1; **p<0.05; ***p<0.01
\end{lstlisting}

%\subsubsection{$RRC^{y}$ models}
\subsubsection{\texorpdfstring{$RC^{y}$}{RCy} models}

In Figure \ref{fig:rc-ri-age} of the main manuscript two regressions were calculated to predict the growth of $RC^{y}$ and $RI^{y}$ based on age range ($y=[y1-y2]$).
Both $RC^{y}$ and $RI^{y}$ can be best approximated by a cubic polynomial regression (see Fig. \ref{fig:rc-ri-age} for $R^2$) % and SI for other regressions and details).
The regression lines show different growth processes for co-administration and interaction risks.
$RC^{y}$ first decreases in children age range $[\text{5-14}]$, followed by an almost flat level between ages $[\text{15,44}]$ before a steeper growth is observed for older age groups (see shaded area in Fig \ref{fig:rc-ri-age}-left).
In contrast, $RI^{y}$ is initially quite flat and only starts to increase after the age of 15, after which it has a much steeper growth curve than $RC^[y]$ (note the difference in scale).

In addition, Supplementary Tables below contain other regression models that were computed along with their respective ANOVA comparison, when appropriate.

A linear model is the simplest model one could fit to the increased risk of co-administration.

\begin{lstlisting}[caption=$RC^{y}$ linear model]
===============================================
                               $RC^y$            
-----------------------------------------------
$y$                      0.003*** (0.0004)     
Constant                 0.926*** (0.004)      
-----------------------------------------------
Observations                    19             
R2                             0.798           
Adjusted R2                    0.787           
Residual Std. Error       0.009 (df = 17)      
F Statistic           67.336*** (df = 1; 17)   
===============================================
Note:               *p<0.1; **p<0.05; ***p<0.01
\end{lstlisting}

A quadratic model fits slightly better but the increased model complexity is not significant.

\begin{lstlisting}[caption=$RC^{y}$ quadratic model]
===============================================
                               $RC^y$            
-----------------------------------------------
$(y)^2$                   0.0001 (0.0001)      
$y$                        0.001 (0.001)       
Constant                 0.931*** (0.005)      
-----------------------------------------------
Observations                    19             
R2                             0.820           
Adjusted R2                    0.798           
Residual Std. Error       0.009 (df = 16)      
F Statistic            36.493*** (df = 2; 16)   
-----------------------------------------------
Model 1: $RC^y$ ~ $y$
Model 2: $RC^y$ ~ $(y)^2$ + $y$
  Res.Df      RSS Df Sum of Sq      F Pr(>F)
1     17 0.0013609                            
2     16 0.0012139  1 0.00014698 1.9374  0.183
===============================================
Note:               *p<0.1; **p<0.05; ***p<0.01
\end{lstlisting}

A cubic model gives almost perfect fit while being significant for the more complex model.

\begin{lstlisting}[caption=$RC^{y}$ cubic model]
===============================================
                               $RC^y$            
-----------------------------------------------
$(y)^3$                 -0.0001*** (0.00001)    
$(y)^2$                  0.001*** (0.0003)     
$y$                      -0.008*** (0.002)     
Constant                 0.943*** (0.004)      
-----------------------------------------------
Observations                    19             
R2                             0.936           
Adjusted R2                    0.923           
Residual Std. Error       0.005 (df = 15)      
F Statistic           72.789*** (df = 3; 15) 
-----------------------------------------------
Model 1: $RC^y$ ~ $y$
Model 2: $RC^y$ ~ $(y)^2$ + $y$
Model 3: $RC^y$ ~ $(y)^3$ + $(y)^2$ + $y$
  Res.Df       RSS Df Sum of Sq       F    Pr(>F)    
1     17 0.00136086                                    
2     16 0.00121387  1 0.00014698  5.0807 0.0395787 *  
3     15 0.00043394  1 0.00077993 26.9599 0.0001094 ***
===============================================
Note:               *p<0.1; **p<0.05; ***p<0.01
\end{lstlisting}

%\subsubsection{$RI^{y}$ models}
\subsubsection{\texorpdfstring{$RI^{y}$}{RIy} models}

Similarly to how we modeled $RI^{y}$, with the risk of known DDI co-administration ($RI^{y}$) we start with the simplest linear model possible.

\begin{lstlisting}[caption=$RI^{y}$ linear model]
===============================================
                               $RI^y$            
-----------------------------------------------
$y$                      0.024*** (0.002)      
Constant                  -0.032* (0.016)      
-----------------------------------------------
Observations                    19             
R2                             0.932           
Adjusted R2                    0.928           
Residual Std. Error       0.037 (df = 17)      
F Statistic           233.631*** (df = 1; 17)  
===============================================
Note:               *p<0.1; **p<0.05; ***p<0.01
\end{lstlisting}

A quadratic model fits slightly better but the increased model complexity is not significant.

\begin{lstlisting}[caption=$RI^{y}$ quadratic model]
===============================================
                               $RI^y$            
-----------------------------------------------
$(y)^2$                  -0.0004 (0.0003)      
$y$                      0.030*** (0.006)       
Constant                 -0.050** (0.023)      
-----------------------------------------------
Observations                    19             
R2                             0.937           
Adjusted R2                    0.930           
Residual Std. Error       0.037 (df = 16)      
F Statistic           119.823*** (df = 2; 16)   
-----------------------------------------------
Model 1: $RC^y$ ~ $y$
Model 2: $RC^y$ ~ $(y)^2$ + $y$
  Res.Df      RSS Df Sum of Sq      F Pr(>F)
1     17 0.023355                           
2     16 0.021550  1 0.001805 1.3401 0.264
===============================================
Note:               *p<0.1; **p<0.05; ***p<0.01
\end{lstlisting}

Finally, a cubic model gives us almost perfect fit while being significant for the more complex model.

\begin{lstlisting}[caption=$RI^{y}$ Cubic model]
===============================================
                               $RI^y$            
-----------------------------------------------
$(y)^3$                   -0.0003*** (0.00001)    
$(y)^2$                    0.007*** (0.0004)      
$y$                        -0.019*** (0.003)     
Constant                    0.013** (0.006)      
-----------------------------------------------
Observations                    19             
R2                             0.997           
Adjusted R2                    0.997           
Residual Std. Error       0.008 (df = 15)      
F Statistic          1,927.479*** (df = 3; 15) 
-----------------------------------------------
Model 1: $RI^y$ ~ $y$
Model 2: $RI^y$ ~ $(y)^2$ + $y$
Model 3: $RI^y$ ~ $(y)^3$ + $(y)^2$ + $y$
  Res.Df       RSS Df Sum of Sq       F    Pr(>F)    
1     17 0.0233550                                  
2     16 0.0215500  1  0.001805  30.391 5.96e-05 ***
3     15 0.0008909  1  0.020659 347.842 8.66e-12 ***
===============================================
Note:               *p<0.1; **p<0.05; ***p<0.01
\end{lstlisting}

% MR
\subsection{Multiple Regression (MR) models}
\label{ch:SI:multiple-regression}

% SECTION UPDATED ON: 2017-01-18
This section displays several MR models that were generated in order to analyze the possible prediction of drug interaction based on patient demographics.
Tables below contain the model results and also their respective ANOVA comparison when appropriate.

\subsubsection{Baseline (no transformation)}

This is the baseline MR model with no transformation.

\begin{lstlisting}[caption=Baseline linear regression model]
==================================================
                               $\Phi^{u}$           
--------------------------------------------------
$\nu^{u}$                       -0.026*** (0.001)       
$\Psi^{u}$                      0.035*** (0.0002)       
Constant                   0.041*** (0.003)       
--------------------------------------------------
Observations                   132,722            
R2                              0.492             
Adjusted R2                     0.492             
Residual Std. Error      0.496 (df = 132719)      
F Statistic         64,377.810*** (df = 2; 132719)
==================================================
Note:                  *p<0.1; **p<0.05; ***p<0.01
\end{lstlisting}

\subsubsection{Baseline (transformed)}

These are other baseline MR model with transformed variables
\begin{lstlisting}[caption=Transformed baseline MR model]
==================================================
                               $\Phi^{u}$           
--------------------------------------------------
$\nu^{u}$                       -0.004*** (0.001)       
$\Psi^{u}$                      0.040*** (0.0002)       
$(\nu^{u})^2$                   -0.003*** (0.0001)      
Constant                   -0.006** (0.003)       
--------------------------------------------------
Observations                   132,722            
R2                              0.497             
Adjusted R2                     0.497             
Residual Std. Error      0.493 (df = 132718)      
F Statistic         43,696.240*** (df = 3; 132718)
--------------------------------------------------
Model 1: $\Phi^{u}$ ~ $\nu^{u}$ + $\Psi^{u}$
Model 2: $\Phi^{u}$ ~ $\nu^{u}$ + $\Psi^{u}$ + $(\nu^{u})^2$
  Res.Df   RSS Df Sum of Sq      F    Pr(>F)    
1 132719 32592                                  
2 132718 32304  1    288.37 1184.7 < 2.2e-16 ***
==================================================
                               $\Phi^{u}$           
--------------------------------------------------
$\nu^{u}$                       -0.033*** (0.001)       
$\Psi^{u}$                      0.038*** (0.0002)       
$(\Psi^{u})^2$                -0.00002*** (0.00000)     
Constant                   0.053*** (0.003)       
--------------------------------------------------
Observations                   132,722            
R2                              0.494             
Adjusted R2                     0.494             
Residual Std. Error      0.495 (df = 132718)      
F Statistic         43,145.430*** (df = 3; 132718)
--------------------------------------------------
Model 1: $\Phi^{u}$ ~ $\nu^{u}$ + $\Psi^{u}$
Model 2: $\Phi^{u}$ ~ $\nu^{u}$ + $\Psi^{u}$ + $(\Psi^{u})^2$
  Res.Df   RSS Df Sum of Sq      F    Pr(>F)    
1 132719 32592                                  
2 132718 32508  1    84.745 345.99 < 2.2e-16 ***
==================================================
                               $\Phi^{u}$           
--------------------------------------------------
$\nu^{u}$                       -0.008*** (0.001)       
$\Psi^{u}$                      0.041*** (0.0003)       
$(\nu^{u})^2$                   -0.003*** (0.0001)      
$(\Psi^{u})^2$                -0.00000*** (0.00000)     
Constant                    0.001 (0.003)         
--------------------------------------------------
Observations                   132,722            
R2                              0.497             
Adjusted R2                     0.497             
Residual Std. Error      0.493 (df = 132717)      
F Statistic         32,786.680*** (df = 4; 132717)
--------------------------------------------------
Model 1: $\Phi^{u}$ ~ $\nu^{u}$ + $\Psi^{u}$
Model 2: $\Phi^{u}$ ~ $\nu^{u}$ + $\Psi^{u}$ + $(\nu^{u})^2$ + $(\Psi^{u})^2$
  Res.Df   RSS Df Sum of Sq      F    Pr(>F)    
1 132719 32592                                  
2 132717 32297  2    295.59 607.33 < 2.2e-16 ***
==================================================
Note:                  *p<0.1; **p<0.05; ***p<0.01
\end{lstlisting}

%\pagebreak
\subsubsection{Baseline + age + gender}

This section shows the MR results when age and gender are included as dependent variables in the baseline model.

\begin{lstlisting}[caption=Baseline MR model added variables age and gender.]
==================================================
                               $\Phi^{u}$           
--------------------------------------------------
$\nu^{u}$                       -0.027*** (0.001)       
$\Psi^{u}$                      0.034*** (0.0002)       
age                       0.002*** (0.0001)       
C(gender)Male             -0.010*** (0.003)       
Constant                  -0.021*** (0.004)       
--------------------------------------------------
Observations                   132,722            
R2                              0.496             
Adjusted R2                     0.496             
Residual Std. Error      0.494 (df = 132717)      
F Statistic         32,639.900*** (df = 4; 132717)
--------------------------------------------------
Model 1: $\Phi^{u}$ ~ $\nu^{u}$ + $\Psi^{u}$
Model 2: $\Phi^{u}$ ~ $\nu^{u}$ + $\Psi^{u}$ + age + C(gender)
  Res.Df   RSS Df Sum of Sq      F    Pr(>F)    
1 132719 32592                                  
2 132717 32369  2    223.56 458.33 < 2.2e-16 ***
==================================================
Note:                  *p<0.1; **p<0.05; ***p<0.01
\end{lstlisting}

%\pagebreak
\subsubsection{Baseline (replacing \texorpdfstring{$\Psi^{u}$}{Psiu} with \texorpdfstring{$y$}{y})}

Interestingly, number of co-administrations ($\Psi^{u}$) and age ($y$) are virtually exchangeable.

\begin{lstlisting}[caption=Baseline MR model exchanging variables $\Psi^{u}$ and $y$.]
==================================================
                               $\Phi^{u}$           
--------------------------------------------------
$\Psi^{u}$                      0.029*** (0.0001)       
age                       0.002*** (0.0001)       
Constant                  -0.100*** (0.003)       
--------------------------------------------------
Observations                   132,722            
R2                              0.491             
Adjusted R2                     0.491             
Residual Std. Error      0.496 (df = 132719)      
F Statistic         63,937.920*** (df = 2; 132719)
--------------------------------------------------
Model 1: $\Phi^{u}$ ~ $\nu^{u}$ + $\Psi^{u}$
Model 2: $\Phi^{u}$ ~ $\Psi^{u}$ + age
  Res.Df   RSS Df Sum of Sq F Pr(>F)
1 132719 32592                      
2 132719 32702  0   -110.03
==================================================
Note:                  *p<0.1; **p<0.05; ***p<0.01
\end{lstlisting}

%\pagebreak
\subsubsection{Baseline + education level}

This section shows the OMR results when education level is included as one of the dependent variables in the model.

Note that this model fits a smaller dataset because the number of patients that have given their education level is smaller than the full dataset.

\begin{lstlisting}[caption=Baseline MR model added education level variable.]
================================================================
                                             $\Phi^{u}$           
----------------------------------------------------------------
$\nu^{u}$                                      -0.015*** (0.001)      
$\Psi^{u}$                                     0.033*** (0.0002)      
C(education)Cant read/write              -0.027** (0.014)       
C(education)Complete college              -0.007 (0.018)        
C(education)Complete elementary          0.037*** (0.013)       
C(education)Complete high school           0.003 (0.013)        
C(education)Doctoral                      -0.106 (0.132)        
C(education)Espec./Residency               0.009 (0.045)        
C(education)Incomplete college             0.004 (0.018)        
C(education)Incomplete elementary         0.024** (0.011)       
C(education)Incomplete high school        -0.006 (0.014)        
C(education)Masters                       -0.050 (0.119)        
Constant                                   0.018 (0.011)        
----------------------------------------------------------------
Observations                                  61,060            
R2                                             0.511            
Adjusted R2                                    0.511            
Residual Std. Error                     0.602 (df = 61047)      
F Statistic                        5,312.884*** (df = 12; 61047)
----------------------------------------------------------------
Model 1: $\Phi^{u}$ ~ $\nu^{u}$ + $\Psi^{u}$
Model 2: $\Phi^{u}$ ~ $\nu^{u}$ + $\Psi^{u}$ + C(education)
  Res.Df   RSS Df Sum of Sq      F    Pr(>F)    
1  61057 22127                                  
2  61047 22107 10    19.845 5.4801 3.472e-08 ***
================================================================
Note:                                *p<0.1; **p<0.05; ***p<0.01

\end{lstlisting}

%\pagebreak
\subsubsection{Baseline + marital status}

This section shows the OMR results when marital status is included as one of the dependent variables in the model.

\begin{lstlisting}[caption=Baseline MR model added marital status variable.]
=====================================================
                                  $\Phi^{u}$           
-----------------------------------------------------
$\nu^{u}$                          -0.027*** (0.001)       
$\Psi^{u}$                         0.035*** (0.0002)       
C(marital)Divorced            0.105*** (0.025)       
C(marital)Ignored            -0.029*** (0.008)       
C(marital)Married              -0.005 (0.008)        
C(marital)Not informed       -0.072*** (0.008)       
C(marital)Separated           0.080*** (0.011)       
C(marital)Single              -0.014* (0.008)        
C(marital)Widower              0.019* (0.011)        
Constant                      0.077*** (0.008)       
-----------------------------------------------------
Observations                      132,722            
R2                                 0.494             
Adjusted R2                        0.494             
Residual Std. Error         0.495 (df = 132712)      
F Statistic            14,420.090*** (df = 9; 132712)
-----------------------------------------------------
Model 1: $\Phi^{u}$ ~ $\nu^{u}$ + $\Psi^{u}$
Model 2: $\Phi^{u}$ ~ $\nu^{u}$ + $\Psi^{u}$ + C(marital)
  Res.Df   RSS Df Sum of Sq      F    Pr(>F)    
1 132719 32592                                  
2 132712 32464  7    128.13 74.829 < 2.2e-16 ***
=====================================================
Note:                     *p<0.1; **p<0.05; ***p<0.01
\end{lstlisting}

%\pagebreak
\subsubsection{Baseline + average neighborhood income assigned to patients}

\begin{lstlisting}[caption=Baseline MR model added average neighborhood income variable.]
==================================================
                               $\Phi^{u}$           
--------------------------------------------------
$\nu^{u}$                       -0.026*** (0.001)       
$\Psi^{u}$                      0.035*** (0.0002)       
avg_income               0.00003*** (0.00000)     
Constant                   0.016*** (0.005)       
--------------------------------------------------
Observations                   132,722            
R2                              0.493             
Adjusted R2                     0.493             
Residual Std. Error      0.495 (df = 132718)      
F Statistic         42,944.890*** (df = 3; 132718)
--------------------------------------------------
Model 1: $\Phi^{u}$ ~ $\nu^{u}$ + $\Psi^{u}$
Model 2: $\Phi^{u}$ ~ $\nu^{u}$ + $\Psi^{u}$ + avg_income
  Res.Df   RSS Df Sum of Sq      F    Pr(>F)    
1 132719 32592                                  
2 132718 32582  1    9.9727 40.622 1.853e-10 ***
==================================================
Note:                  *p<0.1; **p<0.05; ***p<0.01
\end{lstlisting}

%\pagebreak
\subsubsection{Baseline + neighborhood safety variables assigned to patients}

\begin{lstlisting}[caption=Baseline MR model added neighborhood safety variables.]
==================================================
                               $\Phi^{u}$           
--------------------------------------------------
$\nu^{u}$                       -0.026*** (0.001)       
$\Psi^{u}$                      0.035*** (0.0002)       
theft_pc                  -0.737*** (0.283)       
robbery_p1000               -0.004 (0.003)        
suicide_p1000               0.006 (0.009)         
transitcrime_p1000         0.022*** (0.002)       
traffic_p1000              0.008*** (0.002)       
rape_p1000                  -0.002 (0.004)        
Constant                   0.024*** (0.004)       
--------------------------------------------------
Observations                   132,722            
R2                              0.493             
Adjusted R2                     0.493             
Residual Std. Error      0.495 (df = 132713)      
F Statistic         16,148.060*** (df = 8; 132713)
--------------------------------------------------
Model 1: $\Phi^{u}$ ~ $\nu^{u}$ + $\Psi^{u}$
Model 2: $\Phi^{u}$ ~ $\nu^{u}$ + $\Psi^{u}$ + theft_pc +
    robbery_p1000 + suicide_p1000 + 
    transitcrime_p1000 + traffic_p1000 +
    rape_p1000
  Res.Df   RSS Df Sum of Sq      F    Pr(>F)    
1 132719 32592                                  
2 132713 32538  6    54.096 36.773 < 2.2e-16 ***
==================================================
Note:                  *p<0.1; **p<0.05; ***p<0.01
\end{lstlisting}

%\pagebreak
\subsubsection{Baseline + neighborhood}

\begin{lstlisting}[caption=Baseline MR model added neighborhood as categorical variables.]
======================================================
                                   $\Phi^{u}$           
------------------------------------------------------
$\nu^{u}$                           -0.026*** (0.001)       
$\Psi^{u}$                          0.035*** (0.0002)       
C(hood)BADENFURT                -0.021 (0.014)        
C(hood)BOA VISTA                0.009 (0.024)         
C(hood)BOM RETIRO              0.150*** (0.036)       
C(hood)CENTRO                   0.012 (0.013)         
C(hood)DA GLORIA                -0.009 (0.013)        
C(hood)DO SALTO                 -0.005 (0.016)        
C(hood)ESCOLA AGRICOLA        -0.041*** (0.012)       
C(hood)FIDELIS                  0.005 (0.013)         
C(hood)FORTALEZA              -0.030*** (0.011)       
C(hood)FORTALEZA ALTA          -0.029** (0.014)       
C(hood)GARCIA                   -0.009 (0.011)        
C(hood)ITOUPAVA CENTRAL         0.005 (0.011)         
C(hood)ITOUPAVA NORTE          -0.023** (0.011)       
C(hood)ITOUPAVA SECA           -0.037** (0.019)       
C(hood)ITOUPAVAZINHA            0.012 (0.012)         
C(hood)JARDIM BLUMENAU          -0.053 (0.047)        
C(hood)NOVA ESPERANCA         -0.055*** (0.014)       
C(hood)OTHER                  -0.067*** (0.010)       
C(hood)PASSO MANSO              0.025* (0.015)        
C(hood)PONTA AGUDA              -0.009 (0.013)        
C(hood)PROGRESSO                -0.006 (0.011)        
C(hood)RIBEIRAO FRESCO          0.010 (0.021)         
C(hood)SALTO DO NORTE           0.019 (0.015)         
C(hood)SALTO WEISSBACH          0.018 (0.018)         
C(hood)TESTO SALTO              -0.009 (0.015)        
C(hood)TRIBESS                -0.041*** (0.012)       
C(hood)VALPARAISO               -0.015 (0.014)        
C(hood)VELHA                    -0.015 (0.011)        
C(hood)VELHA CENTRAL            -0.009 (0.013)        
C(hood)VELHA GRANDE            -0.031* (0.017)        
C(hood)VICTOR KONDER            0.026 (0.024)         
C(hood)VILA FORMOSA           -0.225*** (0.053)       
C(hood)VILA ITOUPAVA            0.015 (0.017)         
C(hood)VILA NOVA              -0.041*** (0.015)       
C(hood)VORSTADT                -0.028** (0.014)       
Constant                       0.067*** (0.010)       
------------------------------------------------------
Observations                       132,722            
R2                                  0.494             
Adjusted R2                         0.494             
Residual Std. Error          0.495 (df = 132684)      
F Statistic             3,502.150*** (df = 37; 132684)
------------------------------------------------------
Model 1: $\Phi^{u}$ ~ $\nu^{u}$ + $\Psi^{u}$
Model 2: $\Phi^{u}$ ~ $\nu^{u}$ + $\Psi^{u}$ + C(hood)
  Res.Df   RSS Df Sum of Sq      F    Pr(>F)    
1 132719 32592                                  
2 132684 32486 35    106.61 12.441 < 2.2e-16 ***
======================================================
Note:                      *p<0.1; **p<0.05; ***p<0.01
\end{lstlisting}

% LMM
\subsection{Linear Mixed-Effect (LMM) models}
\label{ch:SI:liner-mixed-models}

To be sure there were not nested effects between variables gender and age, we also ran a linear mixed-model (LMM) where variable gender is nested within age. This model accounts for specific differences within patient age and across different genders. Intuitively, if there are large variations in the number of interaction that are explained by the nestedness of gender and age, these groups would help explain a large portion of the variance in the data.
The results indicate that is not the case. In fact, the variance attributed to gender (within age) group is $0.00016$ while the age group is a little higher, $0.00217$.

\begin{lstlisting}[caption=Linear Mixed Model with age nested within gender.]
Linear mixed model fit by maximum likelihood  ['lmerMod']
ForFormula: $\Phi^{u}$ ~ $\nu^{u}$ + $\Psi^{u}$ + (1 | age/gender)
   Data: data

     AIC      BIC   logLik deviance df.resid 
189314.1 189372.9 -94651.1 189302.1   132716 

Scaled residuals: 
     Min       1Q   Median       3Q      Max 
-13.1102  -0.2048  -0.0734   0.0394  19.2402 

Random effects:
 Groups     Name        Variance  Std.Dev.
 gender:age (Intercept) 0.0001645 0.01282 
 age        (Intercept) 0.0021678 0.04656 
 Residual               0.2432636 0.49322 
Number of obs: 132722, groups:  gender:age, 213; age, 109

Fixed effects:
              Estimate Std. Error t value
(Intercept)  0.0483215  0.0055929    8.64
$\nu^{u}$   -0.0262493  0.0007287  -36.02
$\Psi^{u}$   0.0343219  0.0001590  215.87

Correlation of Fixed Effects:
           (Intr) $\nu^{u}$   
$\nu^{u}$  -0.367       
$\Psi^{u}$  0.224 -0.831

\end{lstlisting}

To be sure that neighborhood did not differ in their DDI observations, we also ran a linear mixed-model (LMM) with neighborhood as a random effect. Intuitively, if there are variations in the number of interactions between neighborhoods that cannot be explained by the independent variables alone---due to, say, differences in policies or practices---we should see the random effect variable explaining a great deal of the variance in the model.
Our results indicate that is not the case. In fact, the variance attributed to the neighborhood random effect is $0.00059$ and therefore too small.
This shows that at least in predicting the number of DDI, there is neighborhood homogeneity in how they are being prescribed and thus dispensed.

\begin{lstlisting}[caption=Linear Mixed Model with neighborhood as random effect.]
Linear mixed model fit by maximum likelihood  ['lmerMod']
Formula: $\Phi^{u}$ ~ $\nu^{u}$ + $\Psi^{u}$ + (1 | hood)
   Data: data

     AIC      BIC   logLik deviance df.resid 
189980.7 190029.6 -94985.3 189970.7   132717 

Scaled residuals: 
     Min       1Q   Median       3Q      Max 
-13.1462  -0.1846  -0.0678   0.0180  19.1046 

Random effects:
 Groups   Name        Variance  Std.Dev.
 hood     (Intercept) 0.0005935 0.02436 
 Residual             0.2448642 0.49484 
Number of obs: 132722, groups:  hood, 36

Fixed effects:
              Estimate Std. Error t value
(Intercept)   0.0544948  0.0050850   10.72
$\nu^{u}$    -0.0264618  0.0007270  -36.40
$\Psi^{u}$    0.0348255  0.0001572  221.58

Correlation of Fixed Effects:
           (Intr) $\nu^{u}$   
$\nu^{u}$  -0.411       
$\Psi^{u}$  0.268 -0.841
\end{lstlisting}

%
% Classifiers
%
\FloatBarrier
\pagebreak
\section{Patient classification}
\label{ch:SI:classification}

We applied machine learning classifiers in order to predict if a specific patient had at least one DDI in the whole 18 month period.
A binary classification task.
Support Vector Machine (SVM)\cite{Boser:1992} and Logistic Regression (LR)\cite{Cox:1966} are considered both standard and reliable machine learning algorithm for binary classification problems.
We built models for each classifier considering different sets of features, including demographic (i.e., age \& gender) and drugs the patient was prescribed in the period.
For baseline comparison we also ran against three null model classifiers. One with a ``coin-toss'' probability of classification (Uniform), another with a bias with respect to class probability (Biased), and a custom made (AgeGender) which finds the best age cutoff for each gender from which it consider all patients older than the cutoff as having a DDI.
Regression and classification models were computed using \textit{R} and Python \cite{Sklearn:2011}.

We present results as measures derived from a confusion matrix, also called a contingency table\cite{Davis:2006}. The confusion matrix contains four categories: true positives ($TP$), patients correctly labeled as having a DDI; false positives ($FP$), patients incorrectly classified as having a DDI; true negative ($TN$), patients correctly labeled as not having a DDI; and finally false negatives ($FN$), patients with DDI but mislabeled as not having them. A contingency table example can be seen in Table \ref{table:SI:confusion-matrix}.

\begin{table}
    \centering
    \scriptsize
    \begin{tabular}{r|c|c}
        \toprule
        {}               & DDI  & no DDI \\
        \midrule
        predicted DDI    & $TP$ & $FP$ \\
        predicted no DDI & $FN$ & $TN$ \\
        \bottomrule
    \end{tabular}
    \caption{
        Confusion Matrix.
        }
    \label{table:SI:confusion-matrix}
\end{table}

\begin{figure*}[!htb]
    \begin{center}
    % UPDATED: 2017-01-18
    % FILE: plot_ml_auc.py
    \includegraphics[width=12cm]{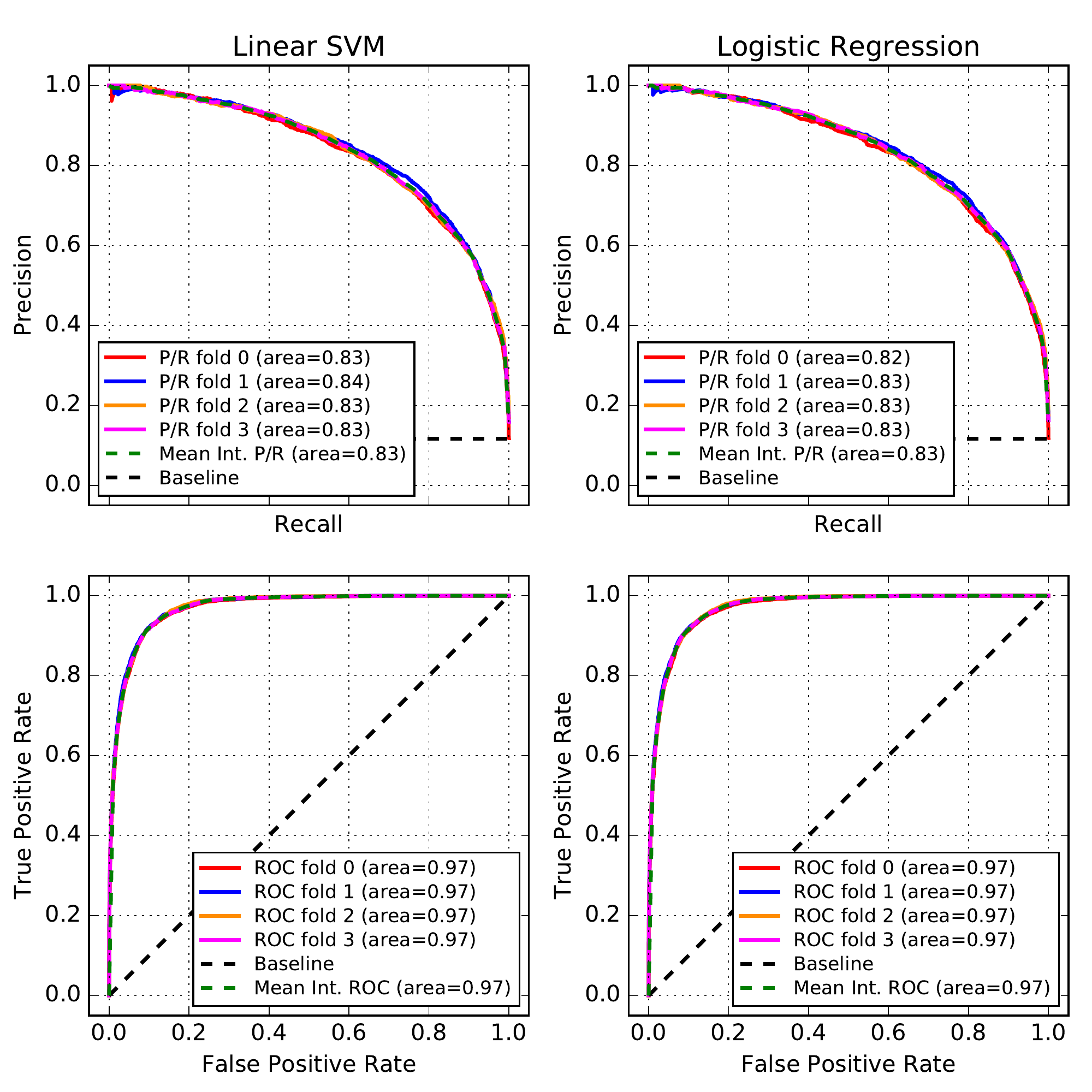}
    \caption{
        Precision and Recall (P/R) curve and Receiver operating characteristic (ROC) curve for individual cross-validation folds.
        Model containing demographic and drug features.
        Black and green dotted line shows the baseline and mean values, respectively.}
\label{fig:SI:ml-auc}
\end{center}
\end{figure*}

From the confusion matrix we compute Precision and Recall as

\begin{equation}
     \text{Precision} = \frac{TP}{TP+FP} \quad , \quad \text{Recall} = \frac{TP}{TP+FN}
     \quad ,
\end{equation}

\noindent where Precision is the fraction of patients with DDI correctly predicted, among all predicted patients with DDI; while Recall is the fraction of patients with DDI correctly predicted, among all patients with DDI.
We also compute True Positive Rate (TPR) and False Positive Rate (FPR) measures as

\begin{equation}
    TPR = \frac{TP}{TP+FN} \quad , \quad FPR = \frac{FP}{FP+TN}
    \quad ,
\end{equation}

\noindent where TPR measures the fraction of patients with DDI that are correctly classified and FPR measures the fraction of patients with no DDI incorrectly classified as having DDI.
These four measures enables the plotting of the Receiver Operator Characteristic (ROC) and the Precision and Recall (P/R) space.
In ROC space we plot FPR against TPR while in P/R space we plot Precision against Recall (Figure \ref{fig:SI:ml-auc} displays the results).
These plots are typically generated to evaluate the performance of machine learning algorithms, and to enable system users to inspect the trained algorithm's precision at a specific recall level, for example.
From both ROC and P/R curves we computes their respective interpolated area under the curve (AUC)\cite{Davis:2006}.

From Precision and Recall we also compute the $F_1\text{-score}$ (also called $F\text{-score}$ or $F\text{-measure}$) as

\begin{equation}
    F_1 = 2 \times \frac{ \text{Precision} \times \text{Recall} }{ \text{Precision} + \text{Recall} }
    \quad .
\end{equation}

We also compute Matthew's Correlation Coefficient (MCC)\cite{Matthews:1975}, which is regarded as an ideal measure of the quality of binary classification in unbalanced scenarios\cite{Baldi:2000}, as

\begin{equation}
    MCC = \frac{TP \times TN - FP \times FN}{ \sqrt{(TP+FP) + (TP+FN) + (TN+FP) + (TN+FN)} }
    \quad .
\end{equation}

Below, we display results as measures of Precision, Recall, $F_1\text{-score}$, MCC, AUC ROC curve and AUC P/R curve.

We also display the full table of feature weights for both classifiers.
Since these are both linear classifiers, one can interpret positive (negative) values as contributing to the positive (negative) class---having a DDI. The higher (smaller) the weight, the bigger (smaller) the contribution.
All results are based on 4-fold cross validation.

\begin{footnotesize}
    \noindent \textsuperscript{\dag} Gender is used as a categorical variable and expanded into the binary features ($g=M$ and $g=F$).
    
    \noindent \textsuperscript{\ddag} Education level is used as a categorical variable expanded into individuals binary features (``Cant read/write'', ``Can read/write a note'', ``Incomplete elementary'', ``Complete elementary'', ``Incomplete high school'', ``Complete high school'', ``Incomplete college'', ``Complete college'', ``Espec./Residency'', ``Masters'', and ``Doctoral'').
\end{footnotesize}
% Simple model
\subsection{Simple model}
\label{ch:SI:classification-model-simple}

\begin{description}
    \item[Patients:] 132,722
    \begin{description}
        \item[DDI (positive):] 15,527 (11.70\%)
        \item[no DDI (negative):] 117,195 (88.30\%)
    \end{description}
    \item[Features:] 127
    \begin{description}
        \item[Demographic:] gender\textsuperscript{\dag} ($g$), age ($y$), number of drugs ($\nu^u$), number of co-administrations ($\Psi^u$).
        \item[Neighborhood:] average income, number of thefts per capita, number of robberies per capita, number of suicides per capita, number of transit crimes per capita, number of traffic accidents per capita, number of rapes per capita.
        \item[Drug:] all drugs $D$.
    \end{description}
\end{description}

\begin{table}[h!]
    \centering
    \scriptsize
    \begin{tabular}{lrrrrrr}
        \toprule
        Fold &  Precision &    Recall &        $F_1$ &       MCC &   AUC ROC & AUC P/R \\
        \midrule
    1 &     0.8196 &  0.6309 &  0.7130 &  0.6877 &   0.9676 &  0.8269 \\
    2 &     0.8241 &  0.6494 &  0.7264 &  0.7011 &   0.9702 &  0.8365 \\
    3 &     0.8127 &  0.6504 &  0.7226 &  0.6957 &   0.9697 &  0.8315 \\
    4 &     0.8187 &  0.6436 &  0.7207 &  0.6949 &   0.9690 &  0.8311 \\
\midrule
 Mean &     0.8188 &  0.6436 &  0.7207 &  0.6948 &   0.9691 &  0.8315 \\
        \bottomrule
    \end{tabular}
    \caption{
        Individual fold and mean performance of Support Vector Machine (SVM) classifier on stratified 4-fold cross-validation, using demographic and drug features. Measures of performance shown are: Precision, Recall, F1 (balanced Precision and Recall), Matthew's Correlation Coefficient, the Area Under the Receiver Operating Characteristic Curve, and the Area Under the Precision and Recall Curve.
        }
    \label{table:SI:ml-simple-svm-results}
\end{table}

\begin{table}[h!]
    \centering
    \scriptsize
    \begin{tabular}{lrrrrrr}
        \toprule
        Fold &  Precision &    Recall &        $F_1$ &       MCC &   AUC ROC & AUC P/R \\
        \midrule
    1 &     0.8085 &  0.6535 &  0.7228 &  0.6953 &   0.9675 &  0.8249 \\
    2 &     0.8096 &  0.6669 &  0.7314 &  0.7037 &   0.9700 &  0.8337 \\
    3 &     0.7991 &  0.6662 &  0.7266 &  0.6977 &   0.9697 &  0.8299 \\
    4 &     0.8092 &  0.6612 &  0.7277 &  0.7002 &   0.9691 &  0.8304 \\
\midrule
 Mean &     0.8066 &  0.6619 &  0.7271 &  0.6992 &   0.9691 &  0.8297 \\
        \bottomrule
    \end{tabular}
    \caption{
        Individual fold and mean performance of Logistic Regression (LR) classifier on stratified 4-fold cross-validation, using demographic and drug features. Measures of performance shown are: Precision, Recall, F1 (balanced Precision and Recall), Matthew's Correlation Coefficient, the Area Under the Receiver Operating Characteristic Curve, and the Area Under the Precision and Recall Curve.
        }
    \label{table:SI:ml-simple-lr-results}
\end{table}

\begin{table}[h!]
    \centering
    \scriptsize
    \begin{tabular}{lrrrrrr}
        \toprule
        Classifier &  Precision &    Recall &        $F_1$ &       MCC &   AUC ROC & AUC P/R \\
        \midrule
Uniform &   0.1181 &  0.5075 &  0.1916 &  0.0035 &      0.5 &  0.5585 \\
Biased  &   0.1147 &  0.1153 &  0.1150 & -0.0026 &   0.4987 &  0.1668 \\
GenderAge   &     0.2044 &  0.8834 &  0.3320 &  0.2751 &   0.7139 &  0.5507 \\
        \bottomrule
    \end{tabular}
    \caption{
        Mean performance of Uniform (coin-toss), Biased (biased coin-toss on class distribution) and GenderAge (hard cutoff for gender and gender) classifiers on stratified 4-fold cross-validation, using demographic and drug features. Measures of performance shown are: Precision, Recall, F1 (balanced Precision and Recall), Matthew's Correlation Coefficient, the Area Under the Receiver Operating Characteristic Curve, and the Area Under the Precision and Recall Curve.
    }
    \label{table:SI:ml-simple-dummy-results}
\end{table}
% Complete model
\subsection{Complete model}
\label{ch:SI:classification-model-complete}

\begin{description}
    \item[Patients:] 132,722
    \begin{description}
        \item[DDI (positive):] 15,527 (11.70\%)
        \item[no DDI (negative):] 117,195 (88.30\%)
    \end{description}
    \item[Features:] 154
    \begin{description}
        \item[Demographic:] gender\textsuperscript{\dag} ($g$), age ($y$), number of drugs ($\nu^u$), number of co-administrations ($\Psi^u$), education levels\ddag.
        \item[Neighborhood:] average income, number of thefts per capita, number of robberies per capita, number of suicides per capita, number of transit crimes per capita, number of traffic accidents per capita, number of rapes per capita.
        \item[Drug:] all drugs $D$.
    \end{description}
\end{description}

\begin{table}[h!]
    \centering
    \scriptsize
    \begin{tabular}{lrrrrrr}
        \toprule
        Classifier &  Precision &    Recall &        $F_1$ &       MCC &   AUC ROC & AUC P/R \\
        \midrule
SVM &     0.8186 &  0.6442 &  0.7210 &  0.6951 &   0.9690 &  0.8312 \\
LR  &     0.8070 &  0.6619 &  0.7273 &  0.6994 &   0.9690 &  0.8295 \\
        \bottomrule
    \end{tabular}
    \caption{
        Mean performance of classifiers on stratified 4-fold cross-validation, using all possible features, including demographic, neighborhood and drugs dispensed. Measures of performance shown are: Precision, Recall, F1 (balanced Precision and Recall), Matthew's Correlation Coefficient, the Area Under the Receiver Operating Characteristic Curve, and the Area Under the Precision and Recall Curve.
    }
    \label{table:SI:ml-complete-results}
\end{table}
% NoDrugs model
\subsection{No Drugs model}
\label{ch:SI:classification-model-nodrugs}

This model is similar to the ``simple'' model, except no drug features are used.

\begin{description}
    \item[Patients:] 132,722
    \begin{description}
        \item[DDI (positive):] 15,527 (11.70\%)
        \item[no DDI (negative):] 117,195 (88.30\%)
    \end{description}
    \item[Features:] 5
    \begin{description}
        \item[Demographic:] gender\textsuperscript{\dag} ($g$), age ($y$), number of drugs ($\nu^u$), number of co-administrations ($\Psi^u$).
        \item[Neighborhood:] None.
        \item[Drug:] None.
    \end{description}
\end{description}

\begin{table}[h!]
    \centering
    \scriptsize
    \begin{tabular}{lrrrrrr}
        \toprule
        Classifier &  Precision &    Recall &        $F_1$ &       MCC &   AUC ROC & AUC P/R \\
        \midrule
SVM &     0.7578 &  0.3791 &  0.5053 &  0.4971 &   0.9185 &  0.6539 \\
LR  &     0.7172 &  0.4170 &  0.5273 &  0.5044 &   0.9130 &  0.6391 \\
        \bottomrule
    \end{tabular}
    \caption{
        Mean performance of classifiers on stratified 4-fold cross-validation, using only demographic features. Measures of performance shown are: Precision, Recall, F1 (balanced Precision and Recall), Matthew's Correlation Coefficient, the Area Under the Receiver Operating Characteristic Curve, and the Area Under the Precision and Recall Curve.
    }
    \label{table:SI:ml-nodrug-results}
\end{table}
% AUC
\FloatBarrier
\clearpage
\subsection{Precision \& Recall and Receiver Operating Characteristic curves}
\label{ch:SI:classification-auc}

% Features
\FloatBarrier
\clearpage
\subsection{Feature loadings}
\label{ch:SI:classification-features}

Supplementary Tables \ref{table:SI:ml-svm-features} and \ref{table:SI:ml-lr-features} shows the feature loading for both SVM and LR classifiers on model ``simple''.

\begin{table}[!htb]
\centering
\tiny
\begin{tabular}{lrlr}
\toprule
                 feature &        coef &                    feature &    coef \\
\midrule
               d=Digoxin &  1.1677 &                d=Acetaminophen & -0.1169 \\
             d=Diltiazem &  0.8718 &                   d=Tobramycin & -0.1185 \\
              d=Warfarin &  0.6938 &          d=Hydrochlorothiazide & -0.1203 \\
           d=Haloperidol &  0.6879 &               d=Norethisterone & -0.1225 \\
             d=Glyburide &  0.6681 &             d=Propylthiouracil & -0.1242 \\
         d=Pyrimethamine &  0.6549 &                d=Phenylephrine & -0.1309 \\
             d=Phenytoin &  0.6015 &         d=Estrogens Conjugated & -0.1312 \\
             d=Biperiden &  0.5807 &                 d=Trimethoprim & -0.1325 \\
         d=Carbamazepine &  0.5752 &             d=Sulfamethoxazole & -0.1325 \\
            d=Gliclazide &  0.4735 &                   d=Colchicine & -0.1333 \\
            d=Clonazepam &  0.4717 &                   d=Diclofenac & -0.1347 \\
            d=Methyldopa &  0.4617 &                   d=Ranitidine & -0.1374 \\
           d=Propranolol &  0.4487 &                     d=Neomycin & -0.1383 \\
               d=Lithium &  0.3887 &                   d=Bacitracin & -0.1383 \\
           d=Fluconazole &  0.3716 &                   d=Nimesulide & -0.1413 \\
                 $\nu_i$ &  0.3169 &                    d=Fenoterol & -0.1446 \\
  d=Acetylsalicylic Acid &  0.3119 &                     d=Nystatin & -0.1508 \\
            $\Psi_{i,j}$ &  0.3080 &                  d=Albendazole & -0.1514 \\
              d=Diazepam &  0.3038 &               d=Nitrofurantoin & -0.1514 \\
            d=Omeprazole &  0.2822 &                   d=Loratadine & -0.1611 \\
         d=Amitriptyline &  0.2810 &                   d=Metamizole & -0.1624 \\
     d=Iron (II) Sulfate &  0.2584 &               d=Spironolactone & -0.1634 \\
     d=Ethinyl Estradiol &  0.2571 &                     d=Tramadol & -0.1643 \\
             d=Ibuprofen &  0.2170 &  d=Dexchlorpheniramine maleate & -0.1664 \\
            d=Imipramine &  0.1825 &                    d=Enalapril & -0.1671 \\
            d=Fluoxetine &  0.1639 &                 d=Azithromycin & -0.1672 \\
             d=Verapamil &  0.1455 &                   d=Miconazole &  -0.169 \\
               d=Timolol &  0.1452 &     d=Scopolamine butylbromide &  -0.171 \\
              d=Atenolol &  0.1432 &                d=Metronidazole & -0.1747 \\
         d=Nortriptyline &  0.1159 &                   d=Cephalexin & -0.1767 \\
           d=Doxycycline &  0.1046 &          d=Ipratropium Bromide & -0.1779 \\
            d=Nifedipine &  0.0973 &               d=Hydrocortisone & -0.1812 \\
       d=Methylphenidate &  0.0638 &               d=Metoclopramide & -0.1832 \\
              d=Vaseline &  0.0596 &                     d=Levodopa & -0.1872 \\
                     $y$ &  0.0518 &  d=Medroxyprogesterone Acetate & -0.1877 \\
         d=Phenobarbital &  0.0274 &                    d=Doxazosin & -0.1909 \\
            d=Prednisone &  0.0232 &                   d=Amlodipine & -0.1936 \\
             d=Estradiol &  0.0181 &                     d=Losartan & -0.1937 \\
              d=Atropine &  0.0000 &                    d=Metformin & -0.1943 \\
      d=Thiocolchicoside &  0.0000 &                  d=Mebendazole & -0.1945 \\
            d=Salbutamol & -0.0071 &                 d=Fluphenazine &  -0.204 \\
         d=Dexamethasone & -0.0102 &                    d=Captopril & -0.2041 \\
 d=Penicillin G procaine & -0.0115 &                   d=Amiodarone & -0.2042 \\
           d=Simvastatin & -0.0191 &                   d=Bromazepam & -0.2063 \\
            d=Gentamicin & -0.0229 &                      d=Codeine & -0.2064 \\
           d=Epinephrine & -0.0347 &                d=Valproic acid & -0.2083 \\
            d=Furosemide & -0.0395 &      d=Penicillin G Benzathine & -0.2123 \\
            d=Carvedilol & -0.0544 &                d=Aminophylline & -0.2133 \\
          d=Erythromycin & -0.0588 &                  d=Clavulanate & -0.2141 \\
        d=Chlorpromazine & -0.0605 &                  d=Clopidogrel & -0.2162 \\
     d=Methotrimeprazine & -0.0683 &                    d=Carbidopa & -0.2269 \\
              d=Morphine & -0.0759 &                      d=Insulin &  -0.246 \\
         d=Levothyroxine & -0.0776 &       d=Isosorbide Mononitrate &  -0.269 \\
           d=Alendronate & -0.0820 &                     d=Nicotine & -0.3003 \\
           d=Amoxicillin & -0.0908 &                      d=Glucose &  -0.305 \\
         d=Ciprofloxacin & -0.0937 &                          $g=M$ & -0.3193 \\
          d=Prednisolone & -0.0944 &                          $g=F$ & -0.3213 \\
            d=Permethrin & -0.0978 &              d=Sodium chloride & -0.3474 \\
        d=Levonorgestrel & -0.0982 &         d=Isosorbide Dinitrate &  -0.351 \\
            d=Folic acid & -0.0983 &                  d=Oseltamivir & -0.3643 \\
          d=Promethazine & -0.1059 &                d=Betamethasone & -0.4765 \\
           d=Maprotiline & -0.1073 &                   d=Spiramycin &  -0.521 \\
           d=Norfloxacin & -0.1100 &                 d=Sulfadiazine & -0.5259 \\
           d=Allopurinol & -0.1148 &                              - &       - \\
\bottomrule
\end{tabular}
\caption{Feature weights for Support Vector Machine (SVM) classifier on model ``simple''.}
\label{table:SI:ml-svm-features}
\end{table}

\begin{table}[!htb]
    \centering
    \tiny
    \begin{tabular}{lrlr}
    \toprule
                 feature &    coef &                        feature &    coef \\
    \midrule
               d=Digoxin &  3.6826 &               d=Norethisterone & -0.4217 \\
             d=Diltiazem &  2.7678 &                  d=Amoxicillin & -0.4283 \\
           d=Haloperidol &  2.3874 &                 d=Promethazine & -0.4327 \\
              d=Warfarin &  2.3423 &                   d=Colchicine &  -0.434 \\
             d=Glyburide &  2.2139 &          d=Hydrochlorothiazide & -0.4526 \\
             d=Phenytoin &  2.1363 &                  d=Norfloxacin & -0.4545 \\
         d=Carbamazepine &  2.1098 &         d=Estrogens Conjugated & -0.4683 \\
             d=Biperiden &  1.9247 &                   d=Tobramycin & -0.4763 \\
            d=Clonazepam &  1.6984 &             d=Propylthiouracil & -0.4767 \\
            d=Methyldopa &  1.6363 &                 d=Trimethoprim & -0.4888 \\
           d=Propranolol &  1.5735 &             d=Sulfamethoxazole & -0.4888 \\
            d=Gliclazide &  1.5618 &                d=Acetaminophen &  -0.502 \\
                 $\nu_i$ &  1.4941 &                   d=Spiramycin &  -0.506 \\
           d=Fluconazole &  1.3668 &                   d=Ranitidine & -0.5178 \\
               d=Lithium &  1.3303 &                   d=Diclofenac & -0.5242 \\
  d=Acetylsalicylic Acid &  1.0479 &                d=Betamethasone & -0.5301 \\
              d=Diazepam &  1.0178 &                   d=Nimesulide & -0.5316 \\
            d=Omeprazole &  1.0114 &                     d=Neomycin & -0.5318 \\
         d=Amitriptyline &  0.9684 &                   d=Bacitracin & -0.5318 \\
     d=Iron (II) Sulfate &  0.8905 &                     d=Nystatin & -0.5508 \\
            $\Psi_{i,j}$ &  0.7721 &                 d=Prednisolone & -0.5531 \\
             d=Ibuprofen &  0.7282 &                    d=Fenoterol & -0.5542 \\
         d=Pyrimethamine &  0.6518 &               d=Spironolactone &  -0.564 \\
            d=Fluoxetine &  0.6245 &               d=Hydrocortisone & -0.5778 \\
            d=Imipramine &  0.6188 &                  d=Mebendazole & -0.5857 \\
              d=Atenolol &  0.5100 &                    d=Enalapril & -0.5955 \\
     d=Ethinyl Estradiol &  0.4965 &                  d=Albendazole & -0.5991 \\
             d=Verapamil &  0.3885 &               d=Nitrofurantoin & -0.6128 \\
           d=Doxycycline &  0.3681 &                   d=Miconazole & -0.6173 \\
                     $y$ &  0.3547 &          d=Ipratropium Bromide &  -0.619 \\
               d=Timolol &  0.3492 &                   d=Loratadine & -0.6196 \\
         d=Nortriptyline &  0.3217 &                   d=Metamizole & -0.6206 \\
            d=Nifedipine &  0.2797 &     d=Scopolamine butylbromide & -0.6347 \\
        d=Levonorgestrel &  0.2220 &                     d=Tramadol & -0.6364 \\
         d=Phenobarbital &  0.1465 &                d=Metronidazole & -0.6476 \\
              d=Vaseline &  0.1118 &  d=Dexchlorpheniramine maleate & -0.6534 \\
             d=Estradiol &  0.0873 &  d=Medroxyprogesterone Acetate & -0.6733 \\
            d=Prednisone &  0.0824 &                    d=Metformin & -0.6762 \\
           d=Epinephrine &  0.0357 &                 d=Azithromycin & -0.6796 \\
          d=Erythromycin &  0.0242 &                    d=Captopril & -0.6855 \\
      d=Thiocolchicoside & -0.0044 &                     d=Losartan & -0.6882 \\
              d=Atropine & -0.0128 &                   d=Amlodipine & -0.6899 \\
          d=Sulfadiazine & -0.0250 &                   d=Cephalexin & -0.6901 \\
 d=Penicillin G procaine & -0.0593 &                    d=Doxazosin & -0.6929 \\
            d=Salbutamol & -0.0845 &               d=Metoclopramide & -0.7212 \\
         d=Phenylephrine & -0.0869 &                d=Aminophylline & -0.7297 \\
           d=Simvastatin & -0.0914 &                      d=Codeine & -0.7311 \\
         d=Dexamethasone & -0.0917 &                  d=Clopidogrel & -0.7375 \\
            d=Gentamicin & -0.1000 &                   d=Amiodarone & -0.7402 \\
       d=Methylphenidate & -0.1019 &                  d=Clavulanate & -0.7449 \\
       d=Sodium chloride & -0.1856 &                d=Valproic acid & -0.7461 \\
          d=Fluphenazine & -0.2091 &                    d=Carbidopa & -0.7552 \\
            d=Furosemide & -0.2152 &                   d=Bromazepam & -0.7571 \\
     d=Methotrimeprazine & -0.2171 &                     d=Levodopa & -0.7619 \\
            d=Carvedilol & -0.2223 &      d=Penicillin G Benzathine & -0.8072 \\
        d=Chlorpromazine & -0.2356 &                      d=Insulin & -0.8443 \\
           d=Maprotiline & -0.2791 &       d=Isosorbide Mononitrate & -0.9186 \\
              d=Morphine & -0.2889 &                     d=Nicotine & -0.9342 \\
         d=Levothyroxine & -0.2929 &                      d=Glucose & -0.9742 \\
            d=Folic acid & -0.3650 &                          $g=M$ &  -1.116 \\
           d=Alendronate & -0.3683 &                          $g=F$ &  -1.132 \\
           d=Allopurinol & -0.3954 &         d=Isosorbide Dinitrate &  -1.178 \\
            d=Permethrin & -0.4101 &                  d=Oseltamivir &    -1.3 \\
         d=Ciprofloxacin & -0.4119 &                              - &       - \\
\bottomrule
\end{tabular}
\caption{Feature weights on Logistic Regression (LR) classifier on model ``simple''.}
\label{table:SI:ml-lr-features}
\end{table}

%\clearpage
\newrefcontext
\printbibliography[title={Supplementary References}]

% End.
\end{document}